%% file: Paladino_arXiv131209.tex
\documentclass[rmp,aps,twocolumn,reprint,superscriptaddress,floatfix,float]{revtex4-1}

\usepackage[usenames]{color}
\usepackage{amsbsy,bm,amsmath,amssymb}
\usepackage{graphicx}
\usepackage{bbm,bm}

\def\inbar{\,\vrule height1.5ex width.4pt depth0pt}
\def\IR{\relax{\rm I\kern-.18em R}}
\def\IC{\relax\hbox{$\inbar\kern-.3em{\rm C}$}}

\renewcommand{\Im} {\mathop{\mathrm{Im}}}
\renewcommand{\Re} {\mathop{\mathrm{Re}}}
 \newcommand{\cH}{\mathcal{H}}
 \newcommand{\cK}{\mathcal{K}}
 \newcommand{\cN}{\mathcal{N}}
 
\newcommand{\cL}{\mathcal{L}}
 
 \newcommand{\cP}{\mathcal{P}}
\newcommand{\cF}{\mathcal{F}}

 \newcommand{\bB}{\mathbf{B}}
\newcommand{\bv}{\mathbf{v}}
 \newcommand{\bM}{\mathbf{M}}
 \newcommand{\bb}{\mathbf{b}}
 \newcommand{\fl}{fluctuator }
 \newcommand{\fs}{fluctuators }
 \newcommand{\bra}[1]{\, \langle\,{#1} \,|\,}
\newcommand{\ket}[1]{\,| \, {#1} \,\rangle  \,}

\newcommand{\av}[1]{\langle #1\rangle}

\begin{document}
\title{1/f noise: implications for solid-state quantum information}

\author{E. Paladino}
\email{epaladino@dmfci.unict.it} 
\affiliation{Dipartimento di Fisica e Astronomia,
Universit\`a di Catania,  Via Santa Sofia 64, I-95123, Catania, Italy}
\affiliation{CNR - IMM - UOS Catania (Universit\'a), 
 Via Santa Sofia 64, I-95123, Catania, Italy}  

\author{Y. M. Galperin}
\email{iouri.galperine@fys.uio.no}
\affiliation{Department of Physics,
University of Oslo, PO Box 1048 Blindern, 0316 Oslo and Centre for Advanced Study, Drammensveien 78, 0271 Oslo, Norway}
\affiliation{Ioffe Physical Technical Institute, 26 Polytekhnicheskaya, St Petersburg 194021, Russian Federation}

\author{G. Falci}
\email{gfalci@dmfci.unict.it}
\affiliation{Dipartimento di Fisica e Astronomia,
Universit\`a di Catania,  Via Santa Sofia 64, I-95123, Catania, Italy}
\affiliation{CNR - IMM - UOS Catania (Universit\'a),
 Via Santa Sofia 64, I-95123,  Catania, Italy} 

\author{B. L. Altshuler}
\email{bla@phys.columbia.edu }
\affiliation{Physics Department, Columbia University, New York, New York 10027, USA}

\begin{abstract}  
\input{abstract.tex}

\end{abstract}                                                                 

\date{updated \today}

\maketitle
\tableofcontents

\input{Part_1_0901.tex}  
\input{Part_2_0901.tex} 
\input{Part_3_0901.tex}

\input{Part_4_0721.tex}

\input{conclusions.tex}

\input{acknowledgments.tex}

\input{references.bbl}

\input{abbreviations.tex}

\end{document}

%% file: abstract.tex
The efficiency of the future devices for quantum information
processing will be limited mostly by the finite decoherence rates of the 
individual qubits and quantum gates.
Recently, substantial progress was achieved in enhancing the time within
which a solid-state qubit demonstrates coherent dynamics. This progress is
based mostly on a successful isolation of the qubits from external decoherence
sources obtained by clever engineering. 
Under these conditions, the material-inherent sources of noise start to
play a crucial role. In most cases, quantum devices are affected by noise
decreasing with frequency $f$ approximately as $1/f$.  According to the present
point of view, such noise is 
due to material- and device-specific microscopic degrees of freedom interacting with 
quantum variables of the nanodevice. 
The simplest picture is  that the environment that destroys the phase
coherence of the device can be thought of as a system of two-state fluctuators,
which experience random hops between their states. If the hopping times are
distributed in a exponentially broad domain, the resulting fluctuations
have a spectrum close to $1/f$ in a large frequency range.

In this paper we review the current state of the theory of decoherence due to 
degrees of freedom producing $1/f$ noise. We discuss basic mechanisms of such 
noises in various nanodevices and then review several models describing the interaction
of the noise sources with quantum devices.  The main focus of the review is to
analyze how the $1/f$ noise destroys their coherent operation.  We start from individual qubits
concentrating mostly on the devices based on superconductor circuits, and then discuss 
some special issues related to more complicated architectures. Finally, we consider several
strategies for minimizing the noise-induced decoherence.

%% file: Part_1_0901.tex
\section{INTRODUCTION}
\label{sec:intro}

Evidence of properties that fluctuate with spectral densities varying approximately 
as $1/f$ over a large range of frequencies, $f$, has been reported in an astonishing
variety of systems.  In condensed matter physics, the difficulties in reasonably 
explaining the shape of the spectrum and in ascribing a physical origin to the noise 
in the diversity of system where it has been observed, have kept $1/f$ noise in 
the forefront of unsolved problems for a long time. 
The large theoretical and experimental effort in this direction up to the late '80s, 
with emphasis on $1/f$ conductance fluctuations in conducting materials,
has been reported in the excellent reviews ~\cite{Dutta1981,Weissman1988},  book~\cite{Kogan1996}
 and others cited therein.

With the progressive reduction of systems size, fluctuations having 
$1/f$-like spectra have been frequently observed in various mesoscopic
systems. The importance of magnetic flux noise in Superconducting Quantum Interference 
Devices (SQUIDs) was recognized already in the 1980s ~\cite{Koch1983,Weissman1988}
thus opening 
the debate about its physical origin - noise from the substrate/mount or
noise from trapped flux in the SQUID - and temperature 
dependence~\cite{Wellstood1987,Savo1987}.
Single-electron and other tunneling devices have provided a compelling evidence that
fluctuating background charges, either within the junctions or in the insulating substrate,
are responsible for low-frequency polarization fluctuations, see, e.~g., 
 \cite{Zorin1996,Krupenin1998,Krupenin2001,Wolf1997}.    

Nanodevices are the subject of intense research at present also because of their long-term
potential for quantum information. Similarly to atomic systems, the quantum nature of 
nanocircuits, despite being hundreds of nanometers wide and containing 
at huge number of electrons, is observable.
Because of these characteristics solid state quantum bits (qubits)  can
be relatively easily addressed to perform desired quantum operations. The drawback of 
tunability is sensitivity to fluctuations of control parameters. Fluctuations are
partly extrinsic, like those due to the local electromagnetic environment. This source of noise
has been greatly reduced via clever engineering.  
In almost all quantum computing nanodevices fluctuations with $1/f$ spectral density 
of different variables and of different physical origin have been observed.
There is clear evidence that $1/f$ noise is detrimental to the required maintenance of 
quantum coherent dynamics and represents the main source of decoherence.
This fact has stimulated a large effort of both the experimental and theoretical communities
aimed on one side at a characterization of the noise, on the other  at
understanding and eventually reducing noise effects.  On a complementary perspective, nanodevices
are sensitive probes of the noise characteristics and therefore may provide important insights
into its microscopic origin.

In this review we describe the current state of theoretical  work on $1/f$ noise in nanodevices
with emphasis on implications for solid state quantum information.  We will  focus 
on superconducting systems and refer to other implementations, in particular, those based on 
semiconductors, whenever physical analogies and/or formal similarities are envisaged.

Previous reviews on $1/f$ noise mainly focused on resistivity fluctuations of conducting materials 
\cite{Dutta1981,Weissman1988,Kogan1996}.
Important questions have been addressed, like universality of the mechanisms leading to
conductivity fluctuations and of the kinetic patterns leading to the $1/f$ spectral form. 
Detailed investigations in different materials (metals and semiconductors) failed 
to confirm the appealing impression of universality and lead instead to the conclusion of the
existence of a variety of origin of $1/f$ conductance noise in diverse materials.

Recent experiments with superconducting circuits evidenced 
$1/f$ low-frequency fluctuations of physically different observables, thus 
providing new important insights into noise microscopic sources.
This is due to the fact that three fundamental types of superconducting qubits exist: 
flux, charge and phase, for a recent review see~{\cite{Clarke2008,Ladd2010,You2011,Steffen2011}. The main difference 
between them is the physical observable
where information is encoded: superconducting current, excess charge in a superconducting island
or the superconducting phase difference across a Josephson junction. Different observables couple
more strongly to environmental variables of different nature and therefore
are sensitive probes of different noise sources.
As a result, magnetic flux noise, polarization or ``charge" noise and critical current noise with
$1/f$ spectrum in some frequency range are presently routinely measured in the three
implementations. 

One scope of this review is to present the current state of understanding of the
microscopic sources of $1/f$ noise in superconducting nanocircuits. Despite relevant mechanisms 
have been largely identified, in most solid state nanodevices this problem cannot be considered
as totally settled (Section \ref{sec:origin}). In some cases, available experiments do not allow drawing solid
conclusions and further investigation is needed.  A number of basic features of the phenomenon are however
agreed upon. According to previous reviews, $1/f$ noise results from a superposition of 
fluctuators (whose nature has to be specified case by case) having switching times distributed in a very broad domain
\cite{Dutta1981,Weissman1988,Kogan1996}. 
Statistical properties of $1/f$ noise have been discussed in depth in the book by \textcite{Kogan1996}.
These properties are at the origin  of 
$1/f$ noise-induced loss of coherence of solid-state qubits. For the sake
of clarity, here we recall the basic definitions and specify their use in the context of the present review.

For a Gaussian random process all nonzero $n$-th order moments can be expressed in terms of the second-order moments, 
i.~e., pair correlations. In general, statistical processes producing $1/f$ noise are non-Gaussian. This fact has 
several important implications. On one side, statistical correlations higher than the power spectrum should be 
considered in order to characterize the process, see (\textcite{Kogan1996}, paragraph 8.2.2).
On the other side, deviations from Gaussian behavior are also expected to show up in the coherent quantum 
dynamics of solid state qubits.
In relevant regimes for quantum computation, where effects of noise are weak, it can be described by linear
coupling to one or more operators of the quantum system. Under this condition, for Gaussian noise,
random noise-induced phases acquired by a qubit 
obey the Gaussian distribution.
We will refer to a process as \textit{Gaussian} whenever this situation occurs.  
As we will see explicitly in Section \ref{sec:decoherence}, even in cases where the noise can be considered 
as a sum of many statistically-independent contributions, the distribution of the phases can be essentially 
non-Gaussian.
A number of investigations aimed at predicting decoherence due to non-Gaussian $1/f$ noise:
\cite{Paladino2002,Grishin2005,Galperin2006,Bergli2006,Galperin2007,Bergli2009,
Burkard2009,Yurkevich2010} (more references can be found in Section \ref{sec:decoherence}).  

The other crucial property of $1/f$ noise is that it cannot be considered a Markovian
random process. A statistical process is \textit{Markovian} if one can make predictions for the future of the process based 
solely on its present state, just as well as one could do knowing the process's full history.  
We will see that even if the noise  can be considered  as a sum of Markovian contributions, the overall 
phase fluctuations of a qubit can be essentially non-Markovian. 
This is the case when non-Gaussian effects
may be important,  see, e. g., \textcite{Laikhtman1985} where this issue was analyzed for the case of spectral diffusion in 
glasses. 
The consequence of $1/f$ noise being non-Markovian is that the effects of $1/f$ noise on the system
evolution depend on the specific ``quantum operation" and/or measurement protocol.
In this review we will illustrate various approaches developed in recent years to deal 
with the non-Markovian nature of $1/f$ noise starting both from microscopic quantum models
and from semi-classical theories.  
We will discuss the applicability range of the Gaussian approximation as well as 
deviations from the Gaussian behavior in connection with the problem of qubit dephasing.

A statistical process is \textit{stationary} if all joint probability distributions
are invariant for translations in time. 
%remain invariable under identical shifts of all time points.
To the best of our knowledge, at present time there is no clear evidence of non-stationarity of the processes 
leading to $1/f$ noise (see the discussion in Section \ref{sec:decoherence}).

Low-frequency noise is particularly harmful since it is difficult to filter 
it out by finite-band filters. In recent years different techniques have been proposed, and
sometimes experimentally tested, in order to limit the effect of low frequency fluctuations. 
One successful strategy to increase phase-coherence times is to operate qubits at working
points where low-frequency noise effects vanish to the lowest-order; such
operating conditions are called ``optimal point" or ``magic point"~\cite{Vion2002}.
Further substantial improvement resulted from the use of dynamical
techniques inspired to Nuclear Magnetic Resonance (NMR)~\cite{Schlichter}.
In a quantum information perspective,
any approach aimed at limiting decoherence should be naturally integrated with 
other functionalities,  as quantum gates. In addition, achieved fidelities should
be sufficiently high to allow for the successful application of Quantum Error
Correction codes.  The question about the best strategy to limit $1/f$ noise 
effects via passive or active stabilization  is still open.  
We will review the current status of the ongoing research along this direction in Section \ref{DD}.

Building scalable multi-qubit systems is presently the main challenge
towards the implementation of a solid-state quantum information processor~\cite{Nielsen1996}.
The effect of $1/f$ noise in solid state complex architectures is a subject of 
current investigation.
Considerable improvement in minimizing sensitivity to charge noise has been reached
via clever engineering. 
A new research area named \textit{circuit quantum electrodynamics} (cQED) 
recently developed from synergy of superconducting circuits technology  and 
phenomena of the  atomic and quantum optics realm.
In this framework important steps further have been done.
Among the newest we mention the achievement of three-qubit entanglement with superconducting 
nano-circuits~\cite{Dicarlo2010,Neeley2010}  which, in combination with longer qubit coherence, 
illustrate a potentially viable approach to factoring numbers \cite{Lucero2012}, implementing 
quantum algorithms \cite{Mariantoni2011,Fedorov2012} and simple 
quantum error correction codes \cite{Reed2012,Chow2012,Rigetti2012}.

\subsection{General features and open issues}

Low frequency noise is commonly attributed to so-called \textit{fluctuators}.
In a ``minimal model", reproducing main features of $1/f$ noise,
fluctuators are dynamic defects, which randomly
switch between two metastable states ($1$ and $2$), see, e.~g., \cite{Dutta1981,Weissman1988,Kogan1996}.
Such a switching produces Random  Telegraph (RT) noise.  
The process is characterized by the switching rates $\gamma_{1 \to 2}$ and $\gamma_{2 \to 1}$ 
for the transitions $1 \to  2$ and $2 \to 1$.
Only the fluctuators with energy splitting $E$ less than a few  $ k_B T$ ($T$ is temperature)  contribute to the 
dephasing of a qubit,  since the fluctuators with large level splitting are frozen in their ground states.
As long as $E \lesssim k_B T$ the rates $\gamma_{1 \to 2}$ and $\gamma_{2 \to 1}$ 
are close in magnitude, and to describe general features of decoherence one can assume that 
$\gamma_{1 \to 2} = \gamma_{2 \to 1} \equiv \gamma$, i.~e., the fluctuations can be
described as a RT process,  see~ \cite{Kogan1996,Buckingham1989,Kirton1989} for reviews. A set of random
telegraph fluctuators with exponentially broad distribution of relaxation rates, $\gamma$, produces
noise with a $1/ f$ power spectrum at $\gamma_{\min} \ll \omega= 2 \pi f \ll \gamma_{\max}$. Here,
 $ \gamma_{\min}$ is the switching rate
of the ``slowest" fluctuator affecting the process, whereas $ \gamma_{\max}$  is the maximal switching rate 
for fluctuators with energy difference $E \sim k_B T$.
Random telegraph noise has been observed in numerous nano-devices based on semiconductors, normal 
metals and superconductors~\cite{Parman1991,Ralls1984,Rogers1984,Rogers1985,Duty2004,
Peters1999,Eroms2006}.  Multi-state fluctuators with number of states greater than two were also 
observed~\cite{Bloom1993,Bloom1994}.

Various microscopic sources can produce classical random telegraph noise.
Here we briefly summarize some of them in connection with charge, flux and 
critical current noise and mention some recent key references; a detailed
discussion will be presented in Section \ref{sec:origin}.

The obvious source of RT charge-noise is a charge which jumps between two different locations
in space. Various hypotheses about the actual location of these charges and the nature of
the two states are still under investigation. The first attempt at constructing such a model in
relation to qubit decoherence appeared in~\cite{Paladino2002}, where electron tunneling between
a localized state in the insulator and a metallic gate was studied. 
The quantitative importance (in explaining observed spectra) of  effects of hybridization between 
localized electronic  states (at the trap) and electrodes extended states  was pointed out by
~\textcite{Grishin2005,Abel2008}.
Models considering the actual  superconducting
state of the electrodes have been studied in ~\cite{Faoro2005,Faoro2006}, leading to predictions 
in agreement with the experimental observations of charge noise  based on measurements of relaxation
rates in charge qubits reported by~\textcite{Astafiev2004}. 

Studies of flux noise have a long history. 
Already in ~\cite{Koch1983}
it was demonstrated that  flux rather than  critical current noise  limits the sensitivity of dc SQUIDs. 
The interest in this problem was recently renewed when it was realized that flux noise can 
limit the coherence in flux and phase superconducting qubits~\cite{Yoshihara2006,Harris2008}. 
Two recent models for fluctuators producing low-frequency noise were suggested. In ~\cite{Koch2007}
flux noise is attributed to electrons hopping between traps,  with spins having fixed,
 random orientations. In \cite{Sousa2007} it is 
proposed that electrons flip their spins due to interaction with tunneling 
two-level systems (TLSs) and phonons. 
A novel mechanism - based on independent spin diffusion along the surface of a superconductor - was
suggested by~\textcite{Faoro2008b}.  It seems to agree with  experiments on measurements 
of the $1/ f$ flux noise reported by~\textcite{Bialczak2007,Sendelbach2008}.
Instead, recent measurements in \textcite{Anton2013} appear to be incompatible with 
random reversal of independent surface spins, possibly suggesting a non-negligible spin-spin
interaction \cite{Sendelbach2009}.}

The microscopic mechanism and the source of the fluctuations of the critical current in a
Josephson junction are long-standing open problems. These fluctuations were initially attributed
to  charges tunneling or hopping between different localized states inside the barrier, forming
glass-like TLSs. However, a more detailed comparison with experiments revealed an important
problem: the noise spectrum experimentally observed by
\textcite{Wellstood2004,VanHarlingen2004} 
was proportional to $T^2$, which is incompatible with the assumption of constant TLS density 
of states, or equivalently with any power-law dependence of relaxation rates for $E \lesssim k_B T$.
The experiments by~\textcite{Eroms2006} on fluctuations in the normal state  in small Al junctions - 
similar to those used in several types of qubits~\cite{Chiorescu2003,Martinis2002,Vion2002} -
brought a new puzzle. It turned out that the temperature dependence of the noise power spectrum in the normal 
state is linear, and the noise power is much less than that reported for large superconducting contacts.
A plausible explanation of such behavior was proposed by~\textcite{Faoro2007}, who
suggested that the critical current noise is due to electron trapping in shallow subgap
states that might be formed at the superconductor-insulator boundary.
Recently, measurements on Al/AlO$_x$/Al junctions reported in \textcite{Nugroho2013}
have shown an equivalence between the critical-current and normal-state resistance fractional noise power spectra, 
both scaling $\propto T$, suggesting the possibility of an upper limit to the additional  noise contribution from electrons tunneling
between weak Kondo states at subgap energies.

The dramatic effect of $1/f$ charge noise was pointed out already in the breakthrough experiment 
performed at NEC and reported by~\textcite{Nakamura1999}, where time-domain 
coherent oscillations of a superconducting charge qubit were observed for the first time.
This experiment has renewed the interest for charge noise and 
has stimulated a number of investigations aimed at explaining decoherence due to noise sources
characterized by a $1/f$ power spectrum in some frequency range. 
The theoretical understanding of decoherence in solid-state single qubit gates is presently
quite well established. 
Quite often in the literature the statistics of the fluctuations
of the qubit parameters displaying $1/f$ spectrum is assumed to be Gaussian. This assumption is 
not a priori justified.  
In order to discuss the applicability range of the Gaussian approximation as well as 
deviations from the Gaussian behavior in connection with the problem
of qubit dephasing, we will consider in detail some of the above mentioned models
where the qubit response to typical manipulation protocols can be solved exactly. 

Peculiar features originated from fluctuations with $1/f$ spectrum show up in the qubit dynamics
superposed to effects due to high frequency fluctuations.  
The intrinsic high-frequency cut-off of  $1/f$ noise is in fact hardly detectable, measurements
typically extending to $100$~Hz. Recently, charge noise up to $10$~MHz has been 
detected in a single-electron transistor (SET)
 by \textcite{Kafanov2008} and flux noise in the $0.2$ - $20$ MHz range has been measured by \textcite{Bylander2011}
by proper pulse sequences.  
Incoherent energy exchanges between system and environment, leading to relaxation and
decoherence, occur at typical operating frequencies (about $10$ GHz).
Indirect measurements of noise spectrum in this frequency range (quantum
noise) often 
suggest a ``white" or Ohmic   behavior~\cite{Astafiev2004,Ithier2005}.
In addition,  narrow resonances at selected frequencies (sometimes 
resonant with the nanodevice-relevant energy scales) have been
observed~\cite{Simmonds2004,Cooper2004,Eroms2006}.  
In certain devices they originate from the circuitry~\cite{VanderWal2000} 
and may eventually be reduced by improving filtering.
More often, resonances are signatures of the
presence of spurious fluctuators which also show up in the
time resolved evolution, unambiguously proving the discrete nature of 
these noise sources~\cite{Duty2004}. 
Fluctuators may severely limit the reliability
of nanodevices~\cite{Falci2005,Galperin2006}.
Explanation of this rich physics is beyond  phenomenological theories 
describing the environment as a  set of harmonic oscillators.
On the other side, an accurate characterization 
of the noise sources might be a priori inefficient, since a microscopic description
would require a huge number of parameters.  
A road-map to treat broadband noise which allows 
to obtain reasonable approximations by systematically 
including only the relevant information on the environment, 
out of the huge parametrization needed to specify it has been proposed in ~\cite{Falci2005}.
The obtained predictions for the  decay of the coherent signal are 
in agreement with observations in various superconducting implementations and in  
different protocols, like the decay of Ramsey fringes in 
charge-phase qubits~\cite{Vion2002}.

One successful strategy to increase phase-coherence times in the presence of
$1/f$ noise is to operate close to qubits' ``optimal points" 
where low-frequency noise effects vanish to the lowest-order. 
This strategy was first implemented in a device named ``quantronium"~\cite{Vion2002}
and it is now applied to all types of superconducting qubits (except for phase qubits). 
Further substantial improvement resulted from the use of charge- or flux-echo 
techniques~\cite{Nakamura2002,Bertet2005}.
In NMR the spin-echo removes the inhomogeneous broadening that is associated with, for example, 
variations of static local magnetic field over the sample, changing the NMR frequency. 
In the case of qubits, the variation is in their energy-level 
splitting frequency from measurement to measurement.  
For some qubits defocusing is strongly suppressed by combining optimal point and echo techniques, 
thus providing further evidence of the fundamental role of $1/f$ noise.
Dynamical decoupling, 
which uses sequences of spin flips to average out effectively 
the coupling to the environment, is another promising 
strategy~\cite{Falci2004,Faoro2004,Bylander2011}.
Optimized sequences limiting the blowup in  
resources involved in $1/f$ noise suppression 
\cite{ka:207-uhrig-prl-UDD,ka:211-biercuk-jpb-ddfilter,Lee2008,Du2009} 
as well as ``optimal control theory" design of quantum 
gates \cite{Montangero2007,Rebentrost2009} have been recently explored.
The experimental realization of optimal dynamical decoupling in 
solid-state systems, and the implementation of quantum 
gates with integrated decoupling in a scalable
and/or hybrid architecture is an open problem.

\subsection{Why $1/f$-noise is important for qubits?} \label{why}

Let us briefly discuss in which way noise influences the operation of an elementary part of a quantum 
computer -- a quantum bit (qubit). This topic will be addressed in detail in Section \ref{sec:decoherence}.
The qubit can be described as a two-level system with effective Hamiltonian 
\begin{equation}\label {qb1}
\hat{H}_q=\frac{\hbar}{2}\left(\epsilon \, \sigma_z + \Delta \,  \sigma_x \right) \, .
\end{equation}
Here $\hbar \epsilon$ represents diagonal splitting of the individual levels, $\hbar \Delta$ represents their 
tunneling coupling, 
$\sigma_{x,z}$ are the Pauli matrices.  The physical meaning of the quantities 
$\epsilon$ and $\Delta$ depends on the specific implementation of the qubit.
In its diagonalized form  the qubit Hamiltonian (\ref{qb1}) reads
\begin{equation}\label {qb1_diagonal}
\hat{H}_q=   \frac{\hbar\Omega}{2} \, \sigma_{z^\prime}
\equiv \frac{\hbar\Omega}{2}\left(\cos \theta \,  \sigma_z + \sin \theta \,  \sigma_x \right) \, ,
\end{equation}
where  $\Omega= \sqrt{\epsilon^2+\Delta^2}$ and the quantization axis $ \sigma_{z^\prime}$ forms an angle $\theta$ with $\sigma_z$. 
The  Hamiltonian~\eqref{qb1} corresponds to a pseudospin 1/2 in a ``magnetic field" $\mathbf{B}$, which can be 
time-dependent:
\begin{equation} \label{qb2}
\hat{H}_q=(\hbar/2)\mathbf{B}\cdot \bm{\sigma}\, , \ B_z\equiv  \epsilon(t), \ B_x \equiv \Delta (t)\, .
\end{equation}
Any state vector, $|\Psi \rangle$, of the qubit determines the Bloch vector $\mathbf{M}$ through the 
density matrix
\begin{equation} \label{dm1}
\rho =|\Psi \rangle \langle \Psi |=(1+\mathbf{M}\cdot \bm{\sigma})/2\, .
\end{equation}
The Schr{\"o}dinger equation turns out to be equivalent to the precession equation for the
Bloch vector:
\begin{equation} \label{bv1}
\dot{\mathbf{M}} = \mathbf{B} \times \mathbf{M} \, .
\end{equation}
 The problem of decoherence arises when the ``magnetic field" is a sum of a controlled
part $\mathbf{B}_0$ and a fluctuating part $\mathbf{b}(t)$ which represents the noise, i.~e., the field is a stochastic
process, $\bB (t) = \bB_0 +\bb(t)$, determined by its statistical properties. The controlled part $\bB_0$
is not purely static -- to manipulate the qubit one has to apply certain high-frequency
pulses of $\bB_0$ in addition to the static fields applied between manipulation steps.  
In this language, the role of the environment is that it creates a stochastic field $\bb (t)$, i. e., 
stochastic components of $\epsilon(t)$ and $\Delta(t)$.  These contributions destroy coherent 
evolution of the qubit making its \textit{coherence time} finite.

\begin{figure}[b]
\centerline{\includegraphics[width=4.5cm]{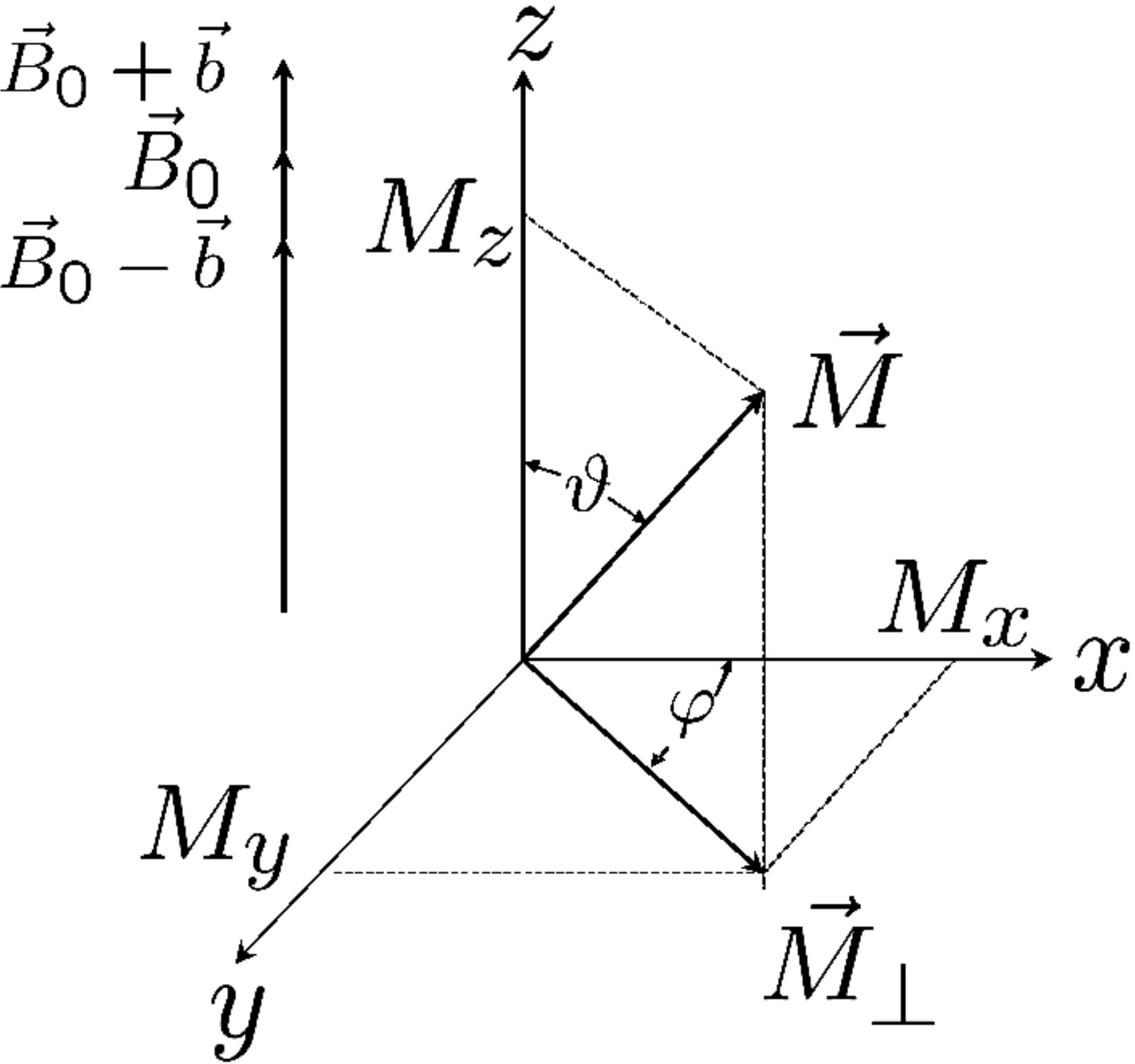}}
\caption{Bloch vector representing the state of a qubit in the rotating 
(with angular frequency $B_0$) frame of reference.  
In the laboratory frame of reference it precesses around the $z$-axis
 with (time-dependent) angular velocity $B$. Fluctuations of 
 this velocity is $b(t)$ \label{figAGB1}}
\end{figure}

Let us consider first the simple case  of ``longitudinal noise", 
where $\bb \parallel \bB_0$, and let the $z$-axis lie along the common
direction of $\bB_0$ and $\bb$, see Fig.~\ref{figAGB1}. 
In the physics of magnetic resonance, this situation is called \textit{pure dephasing} because $z$-component of the Bloch 
vector, $M_z$, is conserved during the process. As long as the time
evolution of $\mathbf{M}$ is governed by Eq.~(\ref{bv1}), the length $|\mathbf{M}|=\sqrt{M_x^2+M_z^2} \equiv M$ is 
also conserved, while the length
$|\langle \mathbf{M}\rangle |$ of the vector $\mathbf{M}$ averaged over the stochastic process $\bb (t)$ decays. 
Description of this decay is the main objective of the decoherence theory. In the case of pure dephasing, this will 
be the decay of the components $M_x$ and $M_y$. It is convenient to introduce a complex combination 
$m_+ = (M_x + i M_y)/M$. Equation (\ref{bv1}) can be written in terms of $m_+$ as $\dot{m}_+=iB m_+$ with solution
\begin{equation} \label{mplus}
m_+(t)=e^{i\phi(t)}m_+(0), \quad \phi(t) \equiv \int_0^t B(t')\, dt' \, .
\end{equation}
Here we have assumed that the variable $B(t)$ is a classical one, i.~e., the order of times in the 
products $B(t_1) B(t_2) \cdots B(t_n) $ is not important.  The solution has to be averaged over the
 stochastic process $b(t)$. We define the phase $\phi (t)$ accumulated by  $m_+$ during the time $t$  
 as a sum of a regular, $\phi_0$,  and stochastic, $\varphi (t)$, parts:
$$\phi (t)=\phi_0 (t)  + \varphi (t), \quad \phi_0(t) = B_0t, \ \varphi(t) =\int_0^t b(t')\, dt' $$
and obtain: $\langle m_+ (t)\rangle = e^{i\phi_0} \, \av {e^{i\varphi(t) }} m_+(0)$.
 The stochastic phase $\varphi$ is the integral of the random process $b(t)$. The Bloch vector precesses 
 around the $z$-axis with the angular velocity that has random modulation $b(t)$.  In the Gaussian 
 approximation the only relevant statistical characteristics of $b(t)$ is the correlation function 
 $\av{b(t_1)b(t_2)}=S_b(|t_1-t_2|)$ (we assume that $b(t)$ is a stationary random process).
  The function $S_b(t)$ decays at $|t| \to \infty$ and the scale of this decay is the correlation time.  
  If the integration time, $t$,  strongly exceeds the correlation time the random phase $\varphi$ is 
  a sum of many uncorrelated contributions. According to the central limit theorem such a sum 
  has Gaussian distribution,
\begin{equation}\label{Gauss1}
p(\varphi) =\frac{1}{\sqrt{2\pi \av{\varphi^2}}}\exp\left(-\frac{\varphi^2}{2\av{\varphi^2}}\right)\, ,
\end{equation}
 independently of the details of the process.  Therefore, the Gaussian distribution should be valid as
  soon as $t$ exceeds the correlation time of the noise.  In Sec.~\ref{sec2E}  we will further discuss 
  this conclusion.  As follows from Eq.~(\ref{Gauss1}),
\begin{eqnarray}
\av{e^{i\varphi}}&=&\int p(\varphi) e^{i\varphi}\, d\varphi=e^{-\av{\varphi^2}/2}\, , \label{avfi} \\
\av{\varphi^2}&=& 
\int_0^t dt_1 \int_0^t dt_2\, S_b(|t_1-t_2|)\, . \label{vf2}
\end{eqnarray}
Representing $S_b(\tau)$ by its Fourier transform,
\begin{equation} \label{ns1}
S_b(\omega)=\frac{1}{\pi}\int_0^\infty dt \,  S_b(t)\cos \omega t \, ,
\end{equation}
that is just the \textit{noise spectrum}, and using Eq.~(\ref{vf2}) we obtain 
\begin{equation}\label{vf3}
\av{\varphi^2(t)}=2\int_{0}^\infty d\omega\, \left( \frac{\sin \omega t/2 }{\omega /2}\right)^2 S_b(\omega) \, .
\end{equation}
Therefore, the signal decay given by Eq.~(\ref{avfi}) is determined (in the Gaussian approximation) 
only by the noise spectrum, $S_b(\omega)$.
For large $t$, the identity $\lim_{a \to \infty} (\sin^2 ax/\pi ax^2) =\delta (x)$ implies that 
$\av{\varphi^2(t)}=2\pi t S_b(0)$ and thus
\begin{equation} \label{t21}
\av{e^{i\varphi(t)}}=e^{-t/{T_2^*}}\, , \quad {T_2^*}^{-1}=\pi S_b(0)\, . 
\end{equation}
Thus, the Gaussian approximation leads to exponential decay of the signal at large times, 
the decrement being given by the noise power at zero frequency. The above expression shows that 
the pure dephasing is determined by the noise spectrum at low frequencies.  In particular, 
at $S_b(\omega) \propto 1/\omega$ the integral in Eq.~(\ref{vf3}) diverges, cf. with further discussion of spin echo.

The time dependence of $\av{m_+}  \propto \av{e^{i\varphi}}$ characterizes decay of the so-called free induction 
signal~\cite{Schlichter}. The free induction decay (FID) is the observable NMR signal generated by non-equilibrium 
nuclear spin magnetization precessing about the magnetic field. This non-equilibrium magnetization can be induced, 
generally by applying a pulse of resonant radio-frequency close to the Larmor frequency of the nuclear spins.  
In order to extract it in qubit experiments, one usually has to average over many repetitions of the same qubit 
operation. Even in setups that allow single-shot measurements \cite{Astafiev2004} each repetition gives one of the 
two qubit states as the outcome. Only by averaging over many repeated runs can one see the decay of the average as 
described by the free induction signal. The problem with this is that the environment has time to change its
state between the repetitions, and thus we average not only over the stochastic dynamics of the environment 
during the time evolution of the qubit, but over the initial states of the environment as well. As a result, the 
free induction signal decays even if the environment is too slow to rearrange during the operation time. This is 
an analogue of the inhomogeneous broadening of spectral lines in magnetic resonance experiments. This analogy also 
suggests ways to eliminate the suppression of the signal by the dispersion of the initial conditions. One can use 
the well-known echo technique [see, e.~g., \cite{Mims1972}] when the system is subjected to a short manipulation pulse
(the so-called $\pi$-pulse) with duration $\tau_1$ at the time $\tau_{12}$. The duration $\tau_1$ of the pulse is 
chosen to be such that it switches the two states of the qubit. This is equivalent to reversing the direction of 
the Bloch vector and thus effectively reversing the time evolution after the pulse as compared with the initial
 one. As a result, the effect of any static field is canceled and decay of the echo signal is determined only 
 by the dynamics of the environment during time evolution. 
 The decay of two-pulse echo can be expressed as $\av{m_+^{(e)}(2\tau_{12})}$
  \cite{Mims1972}, where
\begin{equation}\label{echo1}
\av{m_+^{(e)}(t)}\equiv \av{e^{i\psi(t)}}, \ \psi(t) =\left(\int_0^{\tau_{12}} \! \! \! - \int_{\tau_{12}}^t\right)b(t')\, dt'.
\end{equation}
Finite correlation time of $b(t)$ again leads to the Gaussian distribution of $\psi (t)$ at large enough $t$ with
\begin{equation}\label{echo2}
 \av{\psi^2(2\tau_{12})}=8 \int_0^\infty d\omega \left(\frac{\sin^2 (\omega \tau_{12}/2)}{\omega/2} \right)^2 S_b(\omega) \, .
\end{equation}
This variance can be much smaller than $\av{\varphi^2}$ given by Eq.~(\ref{vf3}), expecially if $S_b(\omega)$ is
singular at $\omega \to 0$.  Though the integral (\ref{echo2}) is not divergent in the case of $1/f$-noise, the time dependence 
of the echo signal is sensitive to the low-frequency behavior of $S_b$.  Therefore, the low-frequency noise strongly 
affects coherent properties of qubits.

Along the simplified model discussed above, the component $\mathbf{M}_z$ of the magnetic moment 
which is parallel to the ``magnetic field" $\mathbf{B}$ does not decay because longitudinal 
fluctuations of the magnetic field do not influence its dynamics.  
In a realistic situation it also decays in time, the decay time being referred 
to as relaxation time and conventionally denoted  $T_1$. 
This occurs when the stochastic field $\bb (t)$ has a component perpendicular to the controlled part $\bB_0$. 
Relaxation processes also induce another decay channel for the $M_x$ and $M_y$ components
on a time scale denoted $T_2$. 
Let us suppose that stochastic field is $\bb(t) = b(t) \hat z$, while $\bB_0= \Delta \hat x + \epsilon \hat z$.
Due to the non-isotropic interaction term, the effect of noise on the qubit phase-coherent dynamics
depends on the  angle $\theta$. 
When $\theta=0$ the interaction is longitudinal and we obtain the already discussed pure dephasing condition.
When $\theta \neq 0$ the stochastic field also induces transitions between the qubit 
eigenstates. As a result, if the noise has spectral component at the qubit energy 
splitting, this interaction induces inelastic transitions between 
the qubit eigenstates, i.~e., incoherent emission and absorption processes.
This is easily illustrated for a weak amplitude noise treated in the Markovian approximation.
Approaches developed in different areas of physics, as the Bloch-Redfield theory \cite{Bloch1957,Redfield1957}, 
the Born-Markov Master Equation \cite{CohenTannoudji1992}, and the systematic weak-damping approximation in a 
path-integral approach \cite{Weiss2008}, lead to exponential decay with time scales
\begin{eqnarray}
1/T_1 &=&   \pi \, \sin^2 \theta  \, S_b(\Omega)\, ,
 \label{T1} \\
1/T_2&=& 1/2 T_1+ 1/T_2^*
\label{T2}
\end{eqnarray} 
where the adiabatic or pure dephasing term of Eq. (\ref{t21}), in the general case reads $1/T_2^*= \pi \cos^2 \theta S_b(0)$.
Even if the above formulas do not hold for $1/f$ noise, which would lead to a singular dephasing time,
they indicate that the diverging adiabatic term containing $S_b(0)$ may be eliminated (in lowest order) if
the qubit operates at $\theta= \pi/2$, or equivalently when the noise is ``transverse",  $\bb \perp \bB_0$. 
In the context of quantum computing with superconducting systems, this condition -  of reduced sensitivity to 
$1/f$ noise - is usually referred to as ``optimal point"~\cite{Vion2002}. 
In this case $T_2= 2T_1$, which is the upper limit to the dephasing time.
We will come again to this issue in Sec.~\ref{subsection4B}.   

%% file: Part_2_0901.tex
\section{PHYSICAL ORIGIN OF $1/f$ NOISE IN NANODEVICES}
\label{sec:origin}

\subsection{Basic models for the $1/f$ noise} \label{subsec:models}

Studies of the noise with the spectral density $\propto \omega^{-1}$ have a long history,
 see, e.~g., \cite{Kogan1996} for a review.  This behavior at low frequencies is typical 
 for many physical systems, such as bulk semiconductors, normal metals and superconductors, 
 strongly disordered conductors as well as devices based on these materials.  One observes, 
 in practically all cases, an increase of the spectral density of the noise with decreasing 
 frequency $f \equiv \omega/2\pi$ approximately proportional to  $1/f$ down to the
  lowest experimentally achievable frequencies.  Therefore, the noise of this type 
  is referred to as $1/f$ noise (the term `flicker noise' proposed by
 Schottky is now more rarely used).

$1/f$ noise poses many puzzles. Is the noise spectrum decreasing infinitely as $f \to 0$, or 
does it saturate at small frequencies? What are the sources of the $1/f$ noise? Why many systems 
have very similar noise spectra at low frequencies? Is this phenomenon universal and does a unified 
theory exist? Those and many other questions stimulated interest to the $1/f$ noise from a fundamental 
point of view.  This interest is also supported by the crucial importance of such a noise for all 
the applications based on dc and low-frequency response. 

Usually, the observed $1/f$ noise of the electrical current is a quadratic function of the applied voltage 
in uniform Ohmic conductors.  This indicates that the noise is caused by fluctuation in the sample 
resistance, which are independent of the mean current.  Therefore, the current just `reveals' the fluctuations.  
Though the low-frequency noise spectrum seems to be rather universal the noise intensity differs very much 
in different systems depending not only on the material, but also on the preparation technology, heat 
treatment, etc. These facts lead to the conclusion that in many cases the $1/f$ noise is \textit{extrinsic}, i.~e., 
caused by some dynamic defects. 

In the simplest case, when the kinetics of fluctuations is characterized by a single relaxation rate,
 $\gamma$, the correlation function of a fluctuating quantity $x(t)$ is proportional to $e^{-\gamma |t|}$. 
 Then the spectral density is a Lorentzian function of frequency, 
\begin{equation} \label{Lorentzian}
S_x (\omega)  \propto \mathcal{L}_\gamma (\omega) \equiv \frac{1}{\pi} \frac{\gamma}{\omega^2 +\gamma^2}.
\end{equation}
In more general cases, the kinetics of $x(t)$ is a superposition of several, or even many, relaxation 
processes with different rates. In general, a continuous distribution of the relaxation rates, 
$\mathcal{P}_x(\gamma)$ may exist. Then the noise spectral density can be expressed as 
\begin{equation} \label{total-spectrum}
S_x (\omega)  \propto\int_0^\infty d\gamma \, \mathcal{P}_x(\gamma)  \mathcal{L}_\gamma (\omega) .
\end{equation}
Equation (\ref{total-spectrum}) describes a non-exponential kinetics, $\mathcal{P}_x (\gamma)\, d\gamma$ being
 the contribution of the processes with relaxation rates between $\gamma$ and $\gamma + d\gamma$ to the variance
\begin{equation} \label{variance}
\overline{(\delta x)^2} =2\int_0^\infty d\omega  S_x(\omega) \propto \int_0^\infty d\gamma \, \mathcal{P}_x (\gamma)\, .
\end{equation}
If $\mathcal{P}_x(\gamma) \propto \gamma^{-1}$ in some window $\gamma_0 \gg \gamma_{\min}$, but very small outside 
this interval, then, according to Eq.~(\ref{total-spectrum}), one gets $S_x(\omega) \propto \omega^{-1}$ 
in the frequency domain $\gamma_0 \gg \omega \gg \gamma_{\min}$ \cite{Surdin1939}. Following a model of this type, one 
has to specify the processes responsible for the noise, which depend on the specific properties 
of the system under consideration.  Several types of 
kinetic processes were indicated as able to produce the 
$1/f$ noise.  Among them are activated processes with different relaxation rates exponentially dependent on the 
inverse  temperature: $\gamma =\gamma_0 e^{-E/k_B T}$, where $\gamma_0$ is some attempt frequency and the 
distribution of  activation energies, $\cF(E)$,  is smooth  in a sufficiently broad domain. Then the 
distribution of the relaxation rates 
has the required form since 
$\cP_x(\gamma)=\cF_x(E) |\partial \gamma/\partial E|^{-1}=\cF_x(E)(k_B T/\gamma)$~\cite{VanDerZiel1950}.

If the kinetics of the fluctuations  is controlled by tunneling processes then the relaxation rates depend 
approximately exponentially on the width and the height of the tunneling barrier. If the distribution of these 
parameters is almost constant in a wide interval, than again the distribution of relaxation rates will be 
proportional to $1/\gamma$.  In particular, \textcite{McWhorter1957} suggested that fluctuation of number of 
carriers in a surface layer of a semiconductor arise from exchange of electrons between this layer and the 
traps lying in the oxide layer covering the surface, or on the outer surface of the oxide.  Since the electron 
transfer takes place via tunneling, the characteristic relaxation rate exponentially depends on the distance 
$x$ between the surface and trap: $\gamma =\gamma_0e^{-x/\lambda}$.  Since the distances $x$ vary with a scatter 
$\gg \lambda$, the distribution of relaxation times is exponentially wide.  This model has 
been extensively used to interpret $1/f$ noise in field-effect transistors. However,  the
model failed to explain observed very small rates $\gamma$ in the devices with relatively thin
insulator layer.

\begin{figure}[t]
\centerline{
\includegraphics[width=4cm]{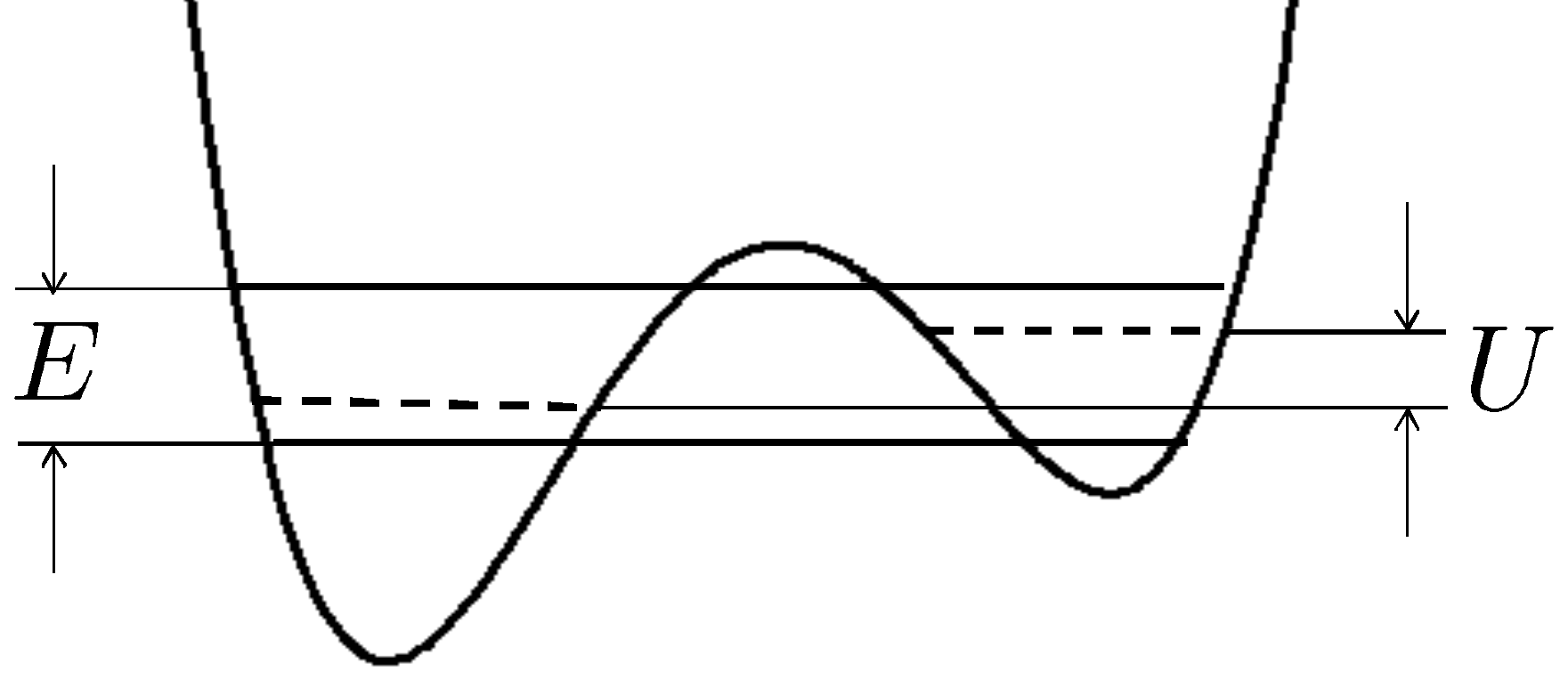}
}
\caption{Schematic diagram of the potential energy of  a two-level tunneling system versus configuration 
coordinate. $U$ is the distance between the single-well energy levels, characterizing asymmetry of the 
potential. $E$ is the difference between the lowest energy levels in a two-well potential with the account 
of tunneling below the barrier.\label{fig:dw}}
\end{figure}

There exist special low-energy excitations in  amorphous materials, in particular, in all dielectric and 
metallic glasses.  They result in anomalous temperature dependencies of the heat capacity and the thermal
conductivity at low temperatures, as well as specific features of the sound absorption, see, e. g., 
\cite{Hunklinger1981,Black1981,Galperin1989} for a review.  According to this model,  atoms or groups of atoms 
exists that can occupy two positions. Therefore, their energy as a function of some configuration coordinate 
can be represented as a double-well potential, as shown in Fig.~\ref{fig:dw}.

The model of two-level systems (TLSs) formulated by \textcite{Anderson1972} and \textcite{Phillips1972} contains 
two parameters - the asymmetry, $U$, of the potential (which is approximately equal to the difference 
between the minima) and the tunnel matrix element, $\lambda$, characterizing the strength of the barrier
which can be estimated as 
\begin{equation} \label{Lambda01}
\Lambda = \hbar \omega_0 e^{-\lambda}
\end{equation}
where $\omega_0$ is the frequency of the intra-well vibrations. Thus the inter-level spacing $E$ of a TLS 
(which is its excitation energy) is given by
\begin{equation} \label{Espacing}
E = \sqrt{U^2 + \Lambda^2}\, .
\end{equation}
The  transitions of atoms or group of atoms between these levels, and the change in the levels' relative occupancy with 
varying temperature or under acoustic vibrations are responsible for the low-temperature properties of 
structural glasses. The same type of  low-energy excitations were also  found in amorphous metals and ionic conductors, 
see \cite{Black1981} for a review.

Due to disorder, the TLSs have different values of $U$ and $\lambda$.  Physical considerations lead one to  assume 
that the distribution of $U$ and $\lambda$, $\cP (U,\lambda)$, is almost constant in the region 
$\lambda \gg 1$, $U \ll \hbar \omega_0$ important for the effects observed in the experiments.  
Therefore it is assumed that in this region $\cP (U,\lambda)=P_0$. Here $P_0$ is a constant, which can be 
found by comparison with experiments  \cite{Halperin1976,Black1978a}.

The rate of transitions between the two levels of a TLS is determined by interactions with phonons 
(in insulating solids), or with electrons (in metals). Assuming that fluctuations of the diagonal 
splitting $U$ are most important, we can describe the interaction between the TLS and the environment as 
\begin{equation} \label{TLS-env}
\cH_{\text{TLS-env}}=g'\hat{c}\tau_z \, ,
\end{equation}
where $\hat{c}$ is an operator in the Hilbert space of the environment depending on the specific interaction mechanism.
It is convenient to diagonalize the TLS Hamiltonian, 
\begin{equation}\label{pP01a}
\cH_{\text{TLS}}=\frac{1}{2}\left(U\tau_z+\Lambda \tau_x \right)\, , 
\end{equation}
where $\tau_i$ are the Pauli matrices,  by rotating the TLS Hilbert space. Then 
\begin{eqnarray}
\cH_{\text{TLS}}&=&(E/2)\tau_z \, , \label{TLS-envd} \\
\cH_{\text{TLS-env}}&=&g'\hat{c}\left(\frac{U}{E} \tau_z + \frac{\Lambda}{E} \tau_x \right)\, . \label{TLS-envd1}
\end{eqnarray}
The interlevel transitions are described by the second item in the interaction Hamiltonian (\ref{TLS-envd1}).
Therefore, the  relaxation rate for the deviation of the occupancy numbers of the levels from the equilibrium 
ones is proportional 
to $(\Lambda/E)^2$  \cite{Jackle1972,Black1978}:
\begin{equation}\label{relrates}
\gamma=\gamma_0 (E) \left(\frac{\Lambda}{E}\right)^2\, , \quad \gamma_0 (E) \propto E^a \coth  \left(\frac{E}{2 k_B T}\right)\, .
\end{equation}
The quantity $\gamma_0 (E)$ has the meaning of a \textit{maximal} relaxation rate for the TLSs with given 
inter-level spacing, $E$. The exponent $a$ depends on the details of the interaction mechanism, its typical 
values are 3 (for the interaction with phonons) and 1 (for the interaction with electrons). Using 
Eq.~(\ref{relrates}) one can obtain the distribution of the relaxation rates as
\begin{equation} \label{df001}
\cP(\gamma,E)= \frac{E}{2U \gamma} \cP(U, \lambda) = \frac{P_0}{2\gamma \sqrt{1-\gamma/\gamma_0}}\approx \frac{P_0}{2\gamma} 
\end{equation}
(here we have taken into account that small relaxation rates require small tunnel coupling of the wells, 
$\Lambda \ll U \simeq E$).  Therefore, owing to the exponential dependence of $\gamma$ on the tunneling 
parameter $\lambda$ ($\gamma \propto e^{-2\lambda}$) and approximately uniform distribution in $\lambda$,
 the distribution with respect to $\gamma$ is inversely proportional to  $\gamma$ in an exponentially broad 
 interval, as is characteristic of systems showing $1/f$ noise.

Spontaneous transitions between the levels of the TLSs can lead to fluctuations of macroscopic properties, 
such as resistance of disordered metals \cite{Ludviksson1984,Kogan1984a}, density of electron states in 
semiconductors and metal-oxide-semiconductor structures \cite{Kogan1984b}, etc.  These fluctuations have
 $1/f$ spectrum.

%Two-level systems can be formed in hopping insulators by different electronic configurations having 
%close energies \cite{Burin2006}. Giant relaxation times necessary for $1/ f$ noise are provided by 
% New text by Yuri
The mechanism of $1/f$ noise in hopping insulators has been investigated by several theoretical groups. It was 
first suggested \cite{Shklovskii1980,Kogan1981} that the $1/f$ noise in the nearest-neighbor-hopping transport 
is associated with electronic traps, in a way similar to McWorter’s idea of $1/f$ noise in metal-oxide-semiconductor 
field-effect transistors \cite{McWhorter1957}. Each trap consists of an isolated donor within a spherical pore of the 
large radius $r$. Such rare configurations form fluctuators, which have two possible states (empty or occupied) 
switching back and forth with the very slow rate defined by the tunneling rate of electron out or into the pore. 

According to \cite{Burin2006}, two-level fluctuators can be formed also
by different many-electron configurations having  close energies. In this case, giant  relaxation times necessary for $1/ f$ noise are provided by 
a slow rate of simultaneous tunneling of many localized electrons and by large activation barriers 
for their consecutive rearrangements.  The model qualitatively agrees with the low-temperature observations 
of $1/ f$ noise in $p$-type silicon and GaAs.
Several other models of the $1/f$ noise are briefly reviewed in the book by \textcite{Kogan1996}.  

\subsection{$1/f$ noise and random telegraph noise}
\label{subsection2B}

In many systems comprising such semiconductor devices as $p-n$ junctions, metal-oxide-semiconductor 
field-effect transistors,  point contacts and small tunnel junctions between the metals, small 
semiconductor resistors or small metallic samples, the resistance switches at random between two 
(or several) discrete values.  The time intervals between switchings are random, but the two values 
of the fluctuating quantity are time-independent.  This kind of noise is now usually 
called \textit{random telegraph noise} (RTN), see \cite{Kogan1996} for a review.

Statistical properties of RTNs in different physical systems are rather 
common. Firstly, the times spent by the device in each of the states 
are much longer than the microscopic relaxation times.  Therefore, 
the memory of the previous state of the system is erased, and the
 random process can be considered as a discrete Markovian process.  
 Such a process (for a two-state system) is characterized by the 
 equilibrium probabilities $p_1$ and $p_2=1-p_1$ of finding the 
 system  respectively in the first or in the second state, 
 as well as transition probabilities per unit time, 
 $\gamma_{1\to 2}$ and  $\gamma_{2\to 1}$. The spectral density of a 
 random quantity, which switches between the two 
 states $x_1$ and $x_2$ can be easily derived as, cf. with \cite{Kogan1996}, 
\vspace*{-0.1cm}
\begin{equation}\label{sd1}
S_x(\omega)= \frac{ (x_1-x_2)^2}{4 \cosh^2(E/2k_BT)}  \cL_{\gamma} (\omega)
\end{equation}
where $\gamma  \equiv \gamma_{1\to 2} +\gamma_{2\to 1}$.
Therefore each RT process contributes a Lorentzian line to the noise spectrum. 

RTN was observed in numerous small-size devices.  At low temperatures, usually only one or few telegraph 
processes were observed.  However, at higher temperatures (or at higher voltages applied to the device) 
the number of contributing telegraph processes increased. Typically, at high enough temperatures discrete 
resistance switching is not observed. Instead a continuous $1/f$ noise is measured.  This behavior was 
interpreted by \textcite{Rogers1984,Rogers1985} as a superposition of many uncorrelated two-state 
telegraph \textit{fluctuators} with various relaxation rates $\gamma$. The increase of temperature leads 
to an increase of the number of contributing fluctuators and discrete switchings become indistinguishable. 

A different interpretation was suggested by \textcite{ Ralls1991}, see also references therein.  This interpretation is based on 
interaction between the fluctuators leading to a deviation from the simple Lorentzian spectrum.  
Moreover, the system of interacting defects may  pass to another metastable state where different defects 
play role of  active fluctuators.  This interpretation is based on the observation of RTN in metallic 
nanobridges where the record of resistance at room temperature is still composed by one or two telegraph 
processes,  but their amplitudes and characteristic switching rates change randomly in time.  According 
to this interpretation, the systems of interacting dynamic defects is similar to a glass, particularly, to
a spin glass with a great number of metastable states between which it is incessantly wandering.

Despite great progress of the $1/f$ noise physics, for the major part of systems showing $1/f$ or/and random 
telegraph noise the actual sources of the low-frequency  fluctuations remain unknown: this is the main 
unsolved problem.  Below we will discuss some simple models in connection with devices for quantum computation.

\subsection{Superconducting qubits and relevant noise mechanisms\footnote{More detailed description can be 
found in the reviews by \textcite{Makhlin2001,Clarke2008,Xiang2013}.}
}\label{subsectionIIC}

Circuits presently being explored combine in variable ratios the Josephson effect and single Cooper-pair 
charging effects. When the Coulomb energy is dominant, the ``charge circuits" can decohere from charge noise 
generated by the random motion of offset charges ~\cite{Zimmerlie1992,Zorin1996,Nakamura2002}. 
Conversely, when the Josephson energy is dominant, these ``flux circuits" are sensitive to external flux and 
its noise \cite{Friedman2000,Mooij1999,Wellstood1987}.
For the intermediate-energy regime, a circuit designed to be insensitive to both the charge and flux
bias has recently achieved long coherence times ($\lesssim \! 500$~ns), demonstrating the potential of superconducting
circuits~\cite{Vion2002,Cottet2002}. The third type is the phase qubit, which consists of a single Josephson 
junction current biased in the zero voltage state~\cite{Martinis2002,Yu2002}. In this case, the two quantum 
states are
the ground and first excited states of the tilted potential well, between which Rabi oscillations have been 
observed. 

In the case of charge qubits, the coherence times
have been limited by low-frequency fluctuations of background
charges in the substrate which couple capacitively to
the island, thus dephasing the quantum state~\cite{Nakamura2002}.
Flux and phase qubits are essentially immune to fluctuations of charge
in the substrate, and, by careful design and shielding, can
also be made insensitive to flux noise generated by either the
motion of vortices in the superconducting films or by external
magnetic noise. The flux-charge hybrid, operating at a proper working point,
is intrinsically immune to both charge and flux fluctuations. However, all of these qubits
remain sensitive to fluctuations in the Josephson coupling
energy and hence in the critical current of the tunnel junctions
at low frequency $f$. These fluctuations lead to variations
in the level splitting frequency over the course of the measurement
and hence to dephasing.

		\subsubsection{Charge noise in Josephson qubits } \label{cnjq}
The importance of the charge noise was recognized after careful spin-echo-type experiments applied to an 
artificial TLS utilizing a charge degree of freedom of a small superconducting electrode -- so-called
 single-Cooper-pair box (CPB)~\cite{Nakamura1997,Bouchiat1998}.

To explain the main principle behind this device, let us consider a small superconductor grain 
located close a (gate) metallic electrode. The ground state energy of such a grain depends in 
an essential way on the number of electrons on it. Two contributions to such a dependence are 
given by the electrostatic Coulomb energy $E_C(n)$ determined by  the extra charge accumulated 
on the superconducting
grain and by the so-called parity term $\Delta_n$ 
\cite{Tuominen1992,Tuominen1993,Hergenrother1994,Lafarge1993,Eiles1993,Joyez1994,Matveev1993,Hekking1993,
Matveev1994,Glazman1994}. 
The latter originates from the fact that
only an even number of electrons can form a BCS ground state of a superconductor 
(which is a condensate of paired electrons) and therefore in the case of an odd number of electrons $n$ one
unpaired electron should occupy one of the quasiparticle states \cite{Averin1992}.\footnote{Reference 7 in \cite{Averin1992}
indicates that the original idea belongs to K. A. Matveev.} 

\begin{figure}[b]
\centerline{
\includegraphics[width=5cm]{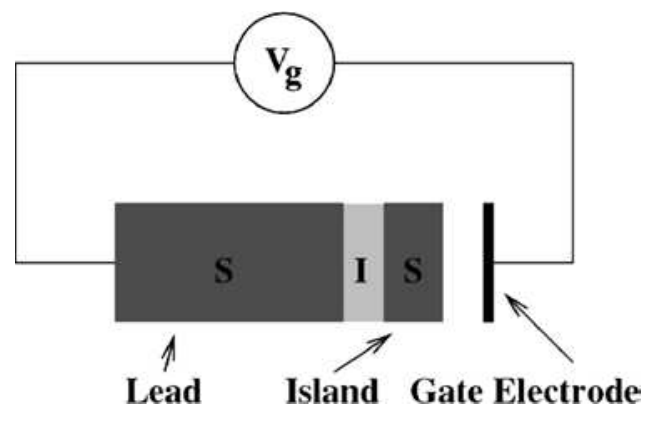}
}
\caption{A schematic diagram of a single-Cooper-pair box. An island of superconducting
material is connected to a larger superconducting lead via a weak link. This allows coherent
tunneling of Cooper pairs between them. For a nanoscale system, such quantum fluctuations of
the charge on the island are generally suppressed due to the strong charging energy associated with
a small grain capacitance. However, by appropriate biasing of the gate electrode it is possible to
make the two states, $|2n\rangle$ and $|2(n+1)\rangle$, differing by one Cooper pair, have the same
 energy (degeneracy of the ground state). This allows the creation of a hybrid state 
 $|\Psi\rangle =\alpha_1 |2n\rangle + \alpha_2 |2(n+1)\rangle$.  \label{figscpb1b}}
\end{figure}

The energy cost of occupying
a quasiparticle state, which is equal to the superconducting gap 
$\Delta_0$, brings a new scale to bear
on the number of electrons that a small superconducting grain can hold. Taking the above into
account, one presents the ground state energy $U_0(n)$ in the form, see
 \cite{Matveev1993,Hekking1993,Matveev1994,Glazman1994},
\begin{equation} \label{scpb1} \hspace*{-0.01in}
	U_0(n)=E_C\left(n-\frac{q}{e}\right)^2 \! \! =\Delta_n, \ \Delta_n=\left\{ \begin{array}{ll}
0 & \text{even} \ n, \\
\Delta_0& \text{odd} \ n.  \end{array} \right.
\end{equation}
\begin{figure}[t!]
\centerline{
\includegraphics[height=4.0cm]{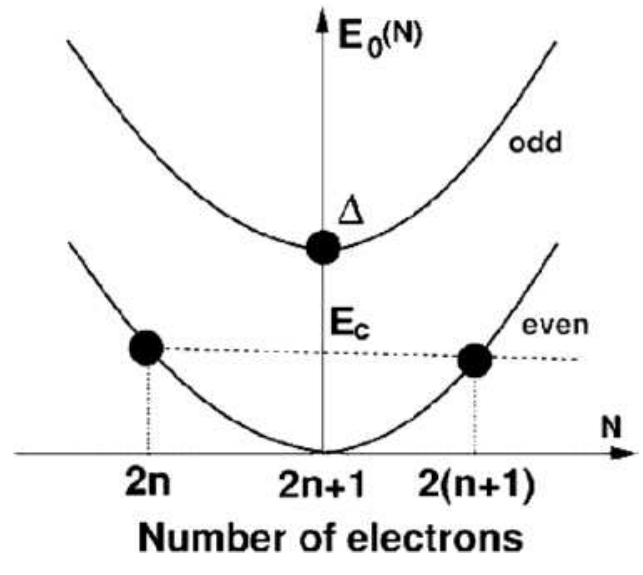}\hfill  \includegraphics[height=4.5cm]{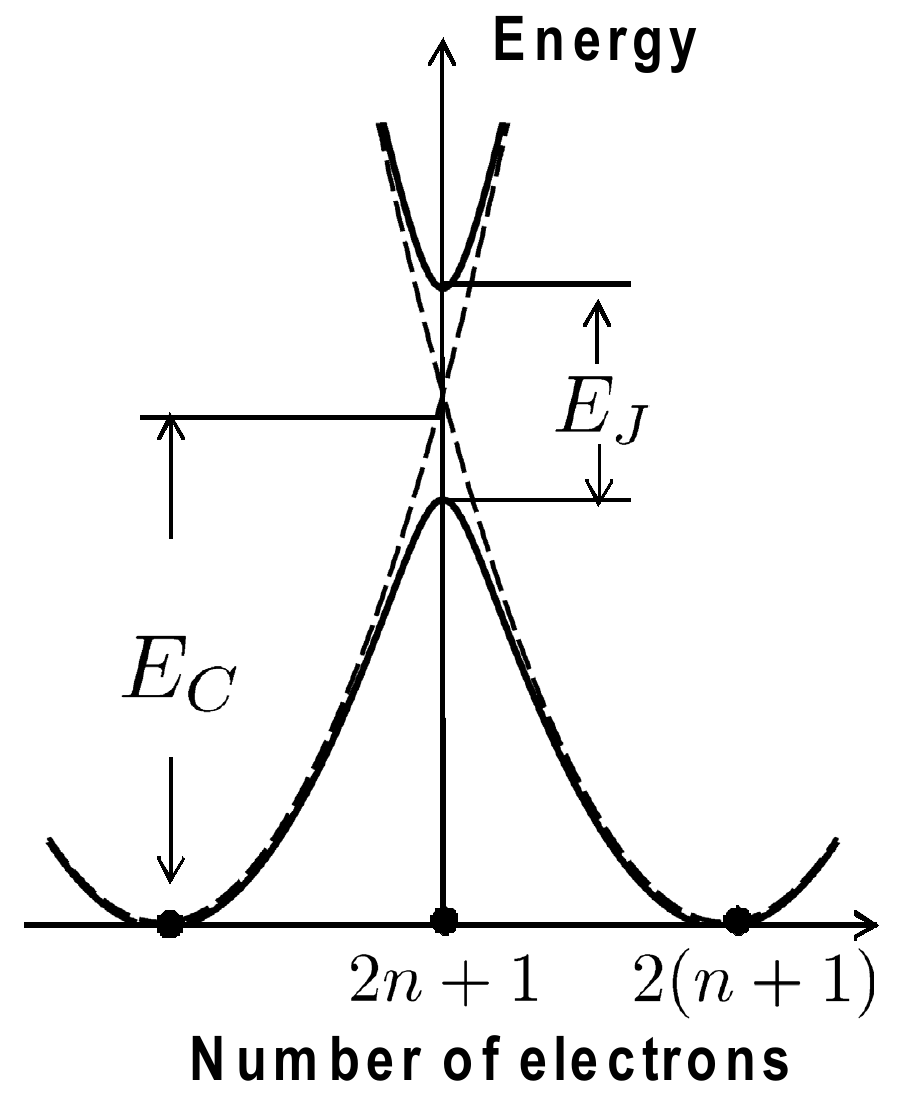} 
}
\caption{ Left panel: The energy diagram for the ground state of a superconducting grain with respect to
charge for the case $\Delta_0 > E_C$. For a certain bias voltage when $q = (2n + 1)e$, ground
states differing by only one single Cooper pair become degenerate. Right panel: Energy of a Cooper-pair 
box with account of the Josephson tunneling. \label{figscpb1a}}
\end{figure} 
Here  $q$ is the charge induced on the grain by the gate electrode. One can see from Eq.~(\ref{scpb1}) 
that if $\Delta_0 > E_C$ only an even number of electrons can be accumulated in
the ground state of the superconducting grain. Moreover, for special values of the gate voltage
corresponding to  $q = (2n + 1)e$ a degeneracy of the ground state occurs with respect to changing
the total number of electrons by one single Cooper pair. An energy diagram illustrating this
case is presented in Fig.~\ref{figscpb1a}.
The occurrence of such a degeneracy brings about an important
opportunity to create a quantum hybrid state at low temperatures which will be a coherent
mixture of two ground states, differing by a single Cooper pair:
\begin{equation}\label{hsscp}
|\Psi\rangle =\alpha_1 |2n\rangle +\alpha_2 |2(n+1)\rangle\, .
\end{equation}
The idea of the device is presented in Fig.~\ref{figscpb1b}
where the superconducting dot is shown to be in tunneling contact with
a bulk superconductor. A gate electrode is responsible for lifting the Coulomb blockade of
Cooper-pair tunneling (by creating the ground state degeneracy discussed above). This allows
the delocalization of a single Cooper pair between two superconductors. Such a hybridization
results in a certain charge transfer between the bulk superconductor and the grain.  
At the charge degeneracy point, $q=(2n+1)e$, the Josephson tunneling produces an avoided crossing
 between the degenerate levels corresponding to  the symmetrical and anti-symmetrical superpositions,
  $|2n\rangle \pm  |2(n+1)\rangle$. As a result, the terms are split by an energy $E_J \ll E_C$. Far 
  from this point the eigenstates are very close to being charge states. 

\begin{figure}[t]
\centering
\includegraphics[width=.6\columnwidth]{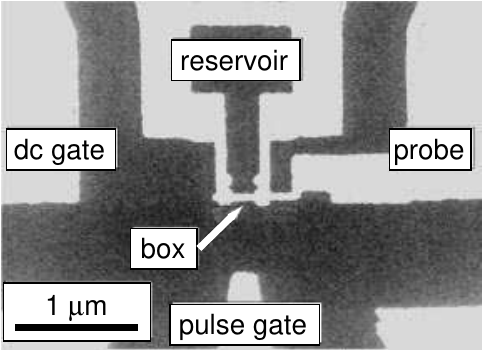}
\caption{Single-Cooper-pair box with a probe junction -- Micrograph of the sample.  The electrodes 
were fabricated by electron-beam lithography and
shadow evaporation of Al on a SiN$_x$ insulating layer (400-nm thick) above a gold
ground plane (100-nm thick) on the oxidized Si substrate. The `box' electrode is a
$700 \times 50 \times 15$ nm Al strip containing $\sim 10^8$  conduction electrons. 
Adapted by permission from Macmillan Publishers Ltd.: \cite{Nakamura1999}, copyright (1999).
\label{fig:Nak1999}}
\end{figure}

To summarize, in a single-Cooper-pair box all electrons form Cooper pairs and condense in a 
single macroscopic ground state separated from the quasiparticle states by the superconducting 
gap, $\Delta_0$. The only low-energy excitations are transitions between the charge number states, 
$|2n\rangle$, which are the states with excess number of Cooper pairs in the box due to Cooper-pairs 
tunneling if $\Delta_0$ is larger than the single-electron charging energy of the box, $E_C$. The 
fluctuations of $n$ are strongly suppressed if $E_C$ exceeds both the Josephson energy, $E_J$, and the 
thermal energy, $k_B T$. Then we come back to the Hamiltonian (\ref{qb1}) with $\hbar \epsilon=E_C(n-q/e)^2$ (where $q$ 
is the induced charge)  and $\hbar \Delta=E_J$, i.~e., the Josephson energy of the split Josephson junction 
between the box and the superconducting reservoir, see  Fig.~\ref{figscpb1b}. Therefore, 
$\epsilon$ can be tuned through the gate voltage determining the induced charge. The Josephson junction is 
usually replaced by a dc SQUID with low inductance. $E_J$ (and consequently, $\Delta$)  is then adjusted 
by applying  the appropriate magnetic flux.
A realistic device is shown  Fig.~\ref{fig:Nak1999} adapted from \cite{Nakamura1999}.

\begin{figure}[t]%[b]
\centering
\includegraphics[width=6cm]{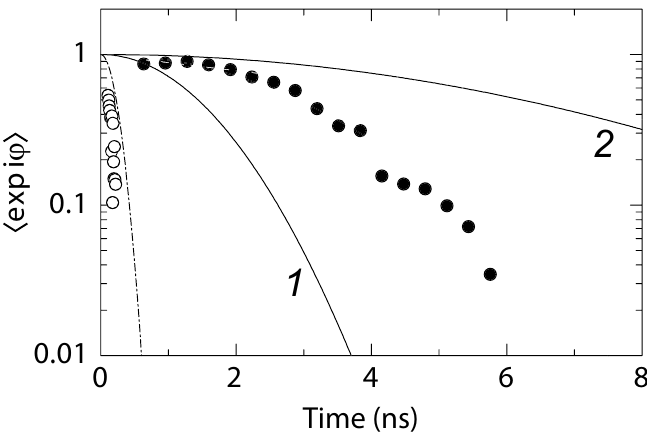}
\caption{Decay of the normalized amplitude of the echo signal
(filled circles) and the free induction decay signal (open circles) compared
with estimated decoherence factors $\langle e^{i\varphi}\rangle$  due to charge noise with the 
spectrum $\alpha/\omega$.  
Here $\sqrt{\alpha}\times 10^3e^{-1}$ is $1.3$ for line $1$ and $ 2 - 0.3$ for $2$. 
Adapted from  \cite{Nakamura2002}.
\label{fig:Nak2002}}
\end{figure}

\textcite{Nakamura2002} have compared the decay of the normalized echo signal, Fig.~\ref{fig:Nak2002}, with 
the expression by \textcite{Cottet2001}
\begin{equation} \label{eg1}
\langle e^{i\varphi}\rangle= \exp\left[- \frac{1}{2\hbar^2}\int_{\omega_{\min}}^\infty \! \! 
d\omega S_{\epsilon}(\omega)\left(\frac{\sin^2(\omega\tau/4)}{\omega/4}\right)^2\right]
\end{equation}
where $S_{\epsilon}(\omega)$ is the spectrum of the noise in the interlevel spacing, $\hbar \epsilon$, of the qubit. The 
latter is expressed through the charge noise, $S_e(\omega)$ as $S_{\epsilon}(\omega) = (4E_C/e)^2S_e(\omega)$.   
The charge noise spectrum was determined by a standard noise measurement on
the same device used as a single-electron transistor. It can be expressed as $S_e (\omega)=\alpha/\omega$ 
with $\alpha =(1.3\times 10^{-3}e)^2$. The estimate following from Eq.~(\ref{eg1}), with the mentioned value of 
$\alpha$ and  $\omega_{\min}=2\pi/t_{\max}$ being the low-frequency cutoff due to the finite data-acquisition time 
$t_{\max}$ (20 ms), is shown in Fig.~\ref{fig:Nak2002} (solid line 1). Solid line 2 in the same figure corresponds 
to $\alpha=(3.0 \times 10^{-4} e)^2$.
Note that Eq.~(\ref{eg1}) is based on the assumption that the fluctuations are Gaussian; it predicts that 
at small delay time, $\tau$, the echo signal decays as $\ln \langle e^{i\varphi}\rangle \propto -\tau^2$. 
In Section \ref{sec:decoherence} this assumption and the ensuing prediction will be further discussed.

\begin{figure}[b]%[t]
\centering
\includegraphics[width=.6\columnwidth]{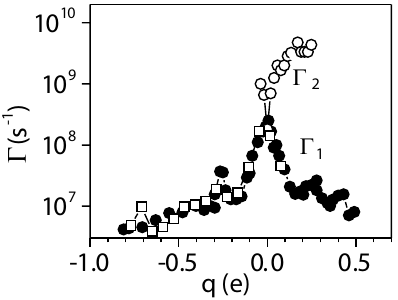}
\caption{Energy relaxation rate, $\Gamma_1$ (closed circles and open squares),  and phase decoherence 
rate $\Gamma_2=T_2^{-1}$ (open
circles) versus gate induced charge $q$.    Adapted from \cite{Astafiev2004}
\label{fig:Ast2004}}
\end{figure}

\textcite{Astafiev2004} studied decoherence of the Josephson charge qubit by measuring energy relaxation 
and dephasing with help of single-shot readout. Both quantities were measured at different charges induced at the
 single-Cooper-pair box by the gate electrode. The decoherence was determined from decay of the coherent 
 oscillations related to the noise spectrum as
\begin{equation} \label{eg2} \hspace*{-2mm}
   \ln \langle e^{i\varphi}\rangle = \!  - 
 \frac{\epsilon^2}{2[(\hbar\epsilon)^2+E_J^2]} \!
   \int_{\omega_{\min}}^\infty \! \! \! \! d\omega \,S_\epsilon(\omega )
\!\left[ \frac{\sin(\omega t/2)}{\omega/2}\right]^2 \! \! .
\end{equation}
Based on the dependence of the decoherence rate on the induced charge shown in Fig.~\ref{fig:Ast2004}
and on estimates of the noise, authors conclude that the source of decoherence 
is charge noise having $1/f$ spectrum.
Another conclusion is that the energy relaxation rate, $\Gamma_1$, is also determined by 
low-frequency noise of the same origin.  This 
conclusion is drawn from the observed $\Gamma_1 \propto E_J^2/[(\hbar\epsilon)^2+E_J^2]$ dependence.  
The importance of $1/f$ charge noise for decoherence in the so-called quantronium quantum bit 
circuit \cite{Cottet2002} was emphasized by \textcite{Ithier2005}.

\begin{figure}[t]%[h]
\centering
\includegraphics[width=.6\columnwidth]{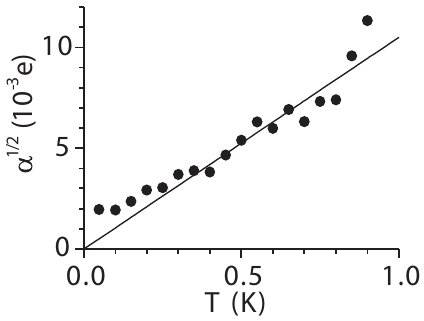}
\caption{Temperature dependence of the amplitudes $\alpha^{1/2}$, which can be approximated as 
$\alpha^{1/2}=(1.0 \times 10^{-2} \ e/\text{K}) T$. Adapted from  \cite{Astafiev2006}.
\label{fig:Ast2006}}
\end{figure}

To verify the hypothesis about the common origin of the low-frequency $1/f$ noise and the quantum  
$f$ noise recently measured in the Josephson charge qubits,
\textcite{Astafiev2006} studied the temperature dependence of the $1/f$ noise amplitude and decay of coherent 
oscillations. The $1/f$ noise was measured in the single-electron tunneling (SET) regime. In the temperature 
domain 50~mK - 1~K it demonstrated $\propto T^2$ dependence, see
 Fig.~\ref{fig:Ast2006} adapted from \cite{Astafiev2006}. 
The measurements of the noise were accompanied by measurements of the decay rate of the coherent 
oscillations away from the degeneracy point ($\hbar \epsilon \gg E_J$).  The decay of the oscillations 
was fitted according to Eq.~(\ref{eg2}) yielding at small times the dependence 
$\langle e^{- i \varphi} \rangle \propto e^{-t^2/2T_2^{*2}}$.  The results turned out to be consistent with the 
strength of the $1/f$ noise observed in transport measurements.  To explain the quadratic temperature 
dependence of the $1/f$ noise the authors assumed that this dependence originates from two-level 
fluctuators with the density of states linearly dependent on the inter-level spacing.  We will 
discuss this assumption later while considering models for the noise-induced decoherence.
Recently, measurements of charge noise in a SET shown a linear increase with temperature (between 50 mK and 1.5 or 4 K)
above a voltage-dependent threshold, with a low temperature saturation below 0.2 K
\cite{Gustafsson2012}. Authors conclude that this result is consistent with 
thermal interaction between SET electrons and TLSs residing in the immediate vicinity of the device.
Such possible defects include residue of Al grains formed around the perimeter of the SET island and leads
during the two-angle evaporation \cite{Kafanov2008}, as well as interface states between metal of the SET and its surrounding
oxides \cite{Choi2009}.

Though the obvious source of RT charge noise is a charged particle
which jumps between two different locations in space, less clear is where these charges are actually located 
and what are the two states.
The first attempt of constructing such a model in relation to qubit decoherence appeared in 
\cite{Paladino2002}, where electrons tunneling between a localized state in the insulator and a metallic 
gate was studied. This model has been further studied in \cite{Grishin2005,Abel2008,Yurkevich2010}. 
Later, experimental results \cite{Astafiev2004} indicated a linear dependence of the relaxation rate 
on the energy splitting of the two qubit states. One also has to take into account that in the experimental 
setup there is no normal metal in the vicinity of the qubit: all gates and leads should be in the 
superconducting state at the temperatures of experiment. These two facts suggest that the model~\cite{Paladino2002} 
was not directly applicable to explain
the decoherence in charge qubits~\cite{Nakamura2002} and favored a model with superconducting 
electrodes \cite{Faoro2005}.
\begin{figure}
\begin{center}
\includegraphics[width=8.0cm]{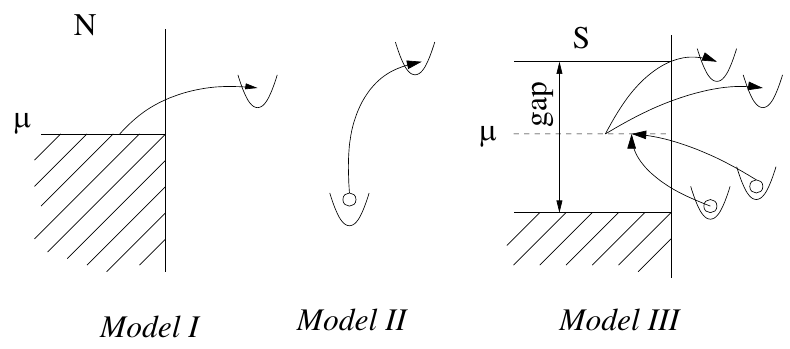}
\end{center}
\caption{ Three possible models for the fluctuating charges: Model I, electrons jumping between a 
localized state and a normal metal, as
discussed in \cite{Paladino2002,Grishin2005,Abel2008}. Model II, electrons
jumping between localized states. Model III, electrons jumping between localized states and a 
superconductor, as discussed in \cite{Faoro2005}.
\label{fig:models}}
\end{figure}
In this model, the two electrons of a Cooper pair are split and tunnel separately to  some 
localized states in the insulator (see
Fig.~\ref{fig:models} for an illustration of this (Model III) and
other models). A constant density of these localized states
gives a linearly increasing density of occupied pairs, in
agreement with experiments \cite{Astafiev2004}.  This model was
criticized \cite{Faoro2006} because it required an unreasonably high
concentration of localized states, and a more elaborate model 
involving Kondo-like traps
was proposed. However, it was shown~\cite{Grishin2005,Abel2008} that allowance
for quantum effects of hybridization between the electronic states
localized at the traps and extended states in the electrodes relaxes
the above requirement. At present it seems that no solid conclusions
can be drawn based on the available experiments.

As we have seen, the ``standard" Cooper-pair boxes are rather sensitive to low-frequency noise 
from electrons moving among defects. This problem can be partly relaxed in more advanced 
charge qubits, such as transmon~\cite{Koch2007} and quantronium~\cite{Vion2002}. The transmon
 is a small Cooper-pair box where the Josephson junction is shunted by a large external 
 capacitor to increase $E_C$ and by increasing the gate capacitor to the same size. The 
 role of this shunt is played by a transmission line, and therefore the qubit is called 
 the \textit{transmon}. 
The main idea is to increase the ratio $E_J/E_C$ making the energy bands shown in the 
Fig.~\ref{figscpb1a} (right panel) almost flat.  For this reason, the transmon is weakly sensitive 
to low-frequency charge noise at all operating points. This eliminates the need for individual
 electrostatic gate and tuning to a charge degeneracy point. A complementary proposal for using 
 a capacitor to modify the $E_J /E_C$ ratio in superconducting flux qubits is put forward by \textcite{You2007}

\begin{figure}[t]
\centerline{
\includegraphics[width=4.5cm,angle=-90]{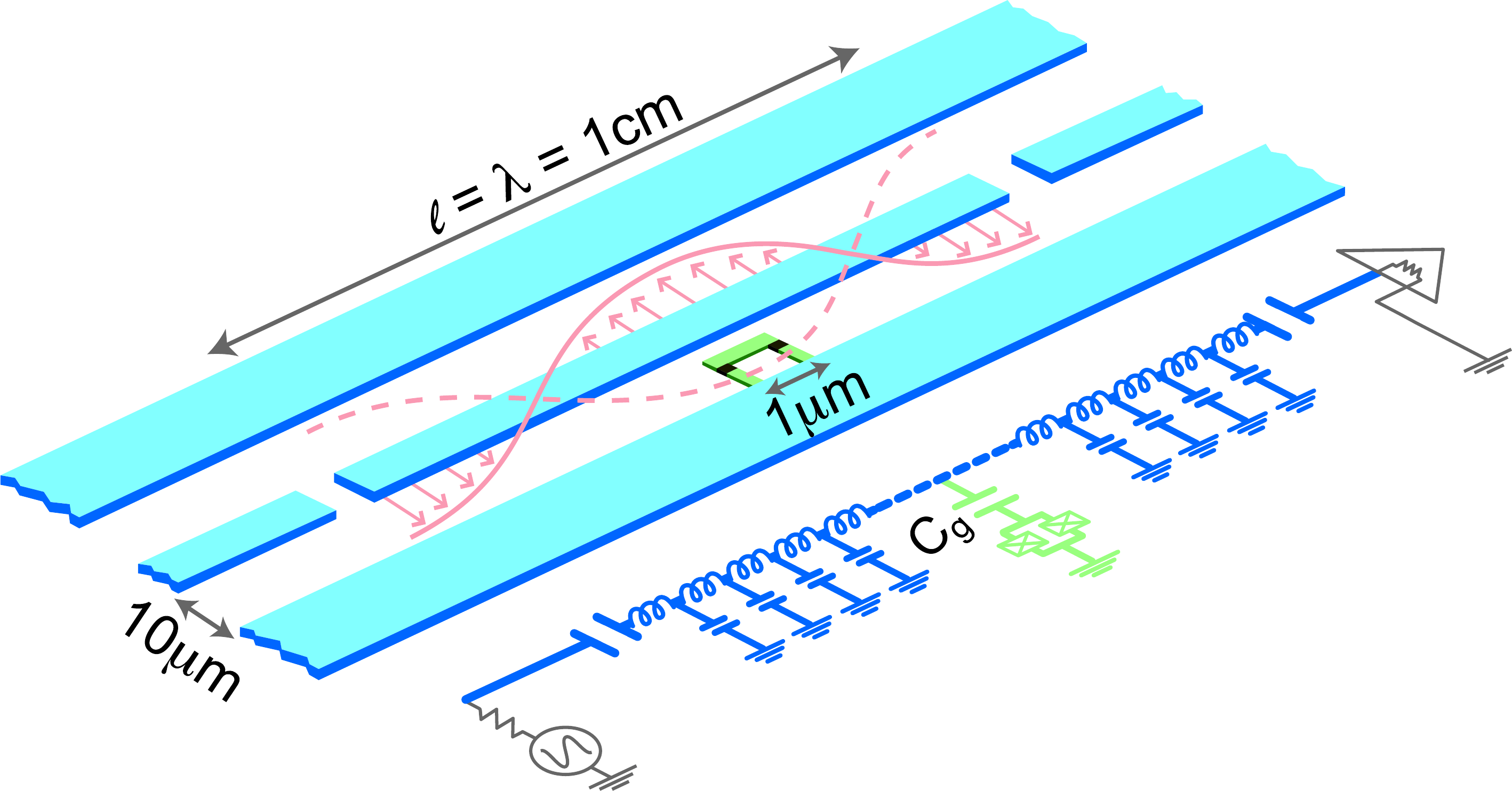}
}
\caption{(Color online) Schematic layout and equivalent lumped circuit representation of proposed implementation of 
cavity QED
using superconducting circuits. The 1D transmission line resonator consists of a full-wave section of 
superconducting coplanar waveguide,
which may be lithographically fabricated using conventional optical lithography. A Cooper-pair box qubit 
is placed between the
superconducting lines and is capacitively coupled to the center trace at a maximum of the voltage standing 
wave, yielding a strong electric
dipole interaction between the qubit and a single photon in the cavity. The box consists of two small 
($\approx 100 \times 100$ nm$^2$) Josephson junctions, configured in a $\approx 1$ $\mu$m loop to permit 
tuning of the effective Josephson energy by an external flux $\Phi_{\text{ext}}$. Input and output
 signals are coupled to the resonator, via the capacitive gaps in the center line, from 50 $\Omega$ 
 transmission lines, which allow measurements of the amplitude and phase of the cavity transmission, 
 and the introduction of dc and rf pulses to manipulate the qubit states.
Reprinted from \cite{Blais2004}. \label{figBlais2004}}
\end{figure}

At the same time, the large gate capacitor provides strong coupling to external microwaves
 even at the level of a single photon, greatly increasing coupling for circuit quantum electrodynamics (cQED).
  The schematics of the suggested device is shown in Fig.~\ref{figBlais2004} reprinted from \cite{Blais2004} 
  where a detailed discussion of cQED devices is presented.
\textcite{Koch2007} have considered possible sources of dephasing and energy relaxation in  transmon devices.  
In particular, the contributions of the charge, flux, and critical current to the dephasing rate, 
$T_2^{-1}$, were estimated.  These (quite favorable) estimates are based on Eqs.~(\ref{vf3}) and (\ref{t21}),
 which follow from  the assumption of the Gaussian statistics of the noise. In some cases, the decoherence 
 is predicted to be limited by the energy relaxation, $T_2 \sim 2T_1$, which occurs due to dielectric 
 losses \cite{Martinis2005} and quasiparticle tunneling~\cite{Lutchin2005,Lutchin2006}. 
 \textcite{Catelani2011} developed a general theory for the qubit decay rate induced by quasiparticles. 
 They  studied its dependence on the magnetic flux used to tune the qubit properties in devices such 
 as the phase and flux qubits, the split transmon, and the fluxonium. 
Recently \textcite{Hassler2011} proposed to use a transmon qubit to perform parity-protected rotations and 
read-out of a topological qubit. The advantage over an earlier proposal using a flux qubit is that the 
coupling can be switched on and off with exponential accuracy, promising a reduced sensitivity to charge noise.

Interestingly, transmon qubits allow to perform flux noise spectroscopy at frequencies near 1 GHz using the phenomenon
of measurement-induced qubit excitation in circuit QED~\cite{Slichter2012}. The extracted values agree with a $1/f^\alpha$
power-law fit below 1 Hz extracted from Ramsey spectroscopy and  around  1-20 MHz deduced from Rabi oscillation 
decay. The above technique  can be used to measure different types of qubit dephasing noise (charge, flux, or critical 
current noise, depending on the type of qubit used) at frequencies  ranging from a few hundred MHz to several GHz, depending 
on the system parameters chosen.

\textcite{Barends2013} have demonstrated a planar, tunable superconducting qubit with energy relaxation times up to
44~$\mu$s. This was achieved by using a geometry based on a planar transmon \cite{Koch2007,Houck2007} and
designed to both minimize radiative loss and reduce coupling to materials-related defects. The authors report  a fine structure 
in the qubit energy lifetime as a function of frequency,  indicating the presence of a sparse population of incoherent, 
weakly coupled two-level defects.  The suggested   qubit (called `Xmon' because of its special geometry)  combines facile 
fabrication, straightforward connectivity, fast control, and long coherence.

Finally, we mention an alternative single Cooper-pair circuit based on a superconducting loop
coupled to an LC resonator used for dispersive measurement analogously to 
cQED qubits and insensitive to offset charges. 
The circuit, named fluxonium, consists of a small junction  shunted  with the Josephson kinetic inductance of a series 
array of large-capacitance tunnel junctions, thereby ensuring that all superconducting islands are connected to the circuit
by at least one large junction \cite{Manucharyan2009}. The  array of Josephson junctions with appropriately chosen
parameters can perform two functions simultaneously: short-circuit the offset charge variations
of a small junction and protect the strong nonlinearity of its Josephson inductance from quantum
fluctuations.

\subsubsection{Flux and phase qubits} 
\begin{figure}[t]
\centerline{
\includegraphics[width=6cm]{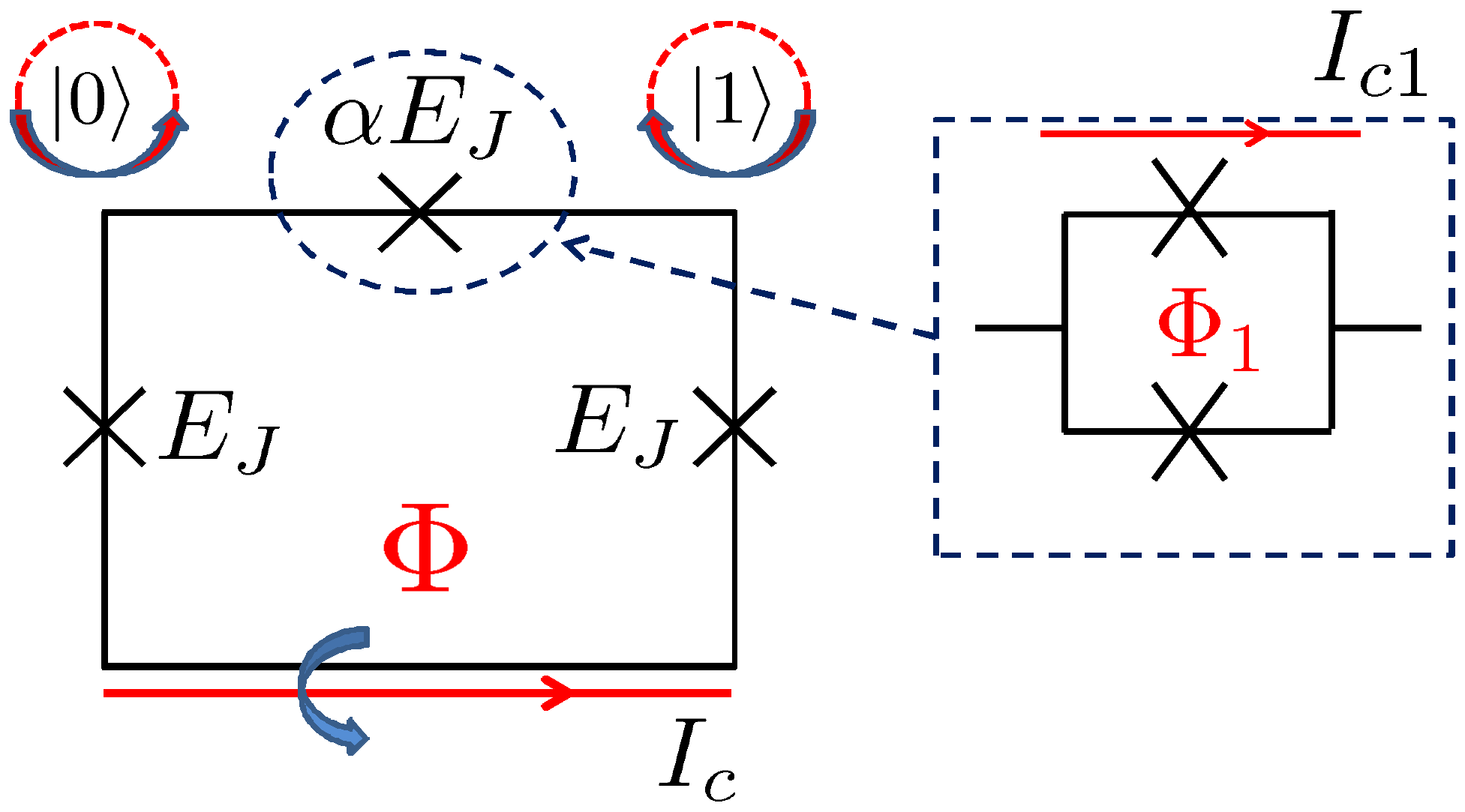}
}
\caption{(Color online) A superconducting loop with three Josephson junctions (indicated with crosses) encloses a 
flux $\Phi$ that is supplied by an
external magnet. Two junctions have a Josephson coupling energy $E_J$, and the third junction has 
$\alpha E_J$, where $\alpha = 0.75$. This system has two (meta)stable states, $|0\rangle$ and $|1\rangle$, with
opposite circulating persistent current. The level splitting is determined by the offset
of the flux from $\Phi_0/2$. The barrier between the states depends on the value of $\alpha$. The qubit is operated by
resonant microwave modulation of the enclosed magnetic flux by a superconducting control line $I_c$
(indicated in red). 
From \cite{Mooij1999}. Reprinted with permission from AAAS.
\label{figMoij99}}
\end{figure}

\begin{figure}[t]
\centerline{
\includegraphics[width=6cm]{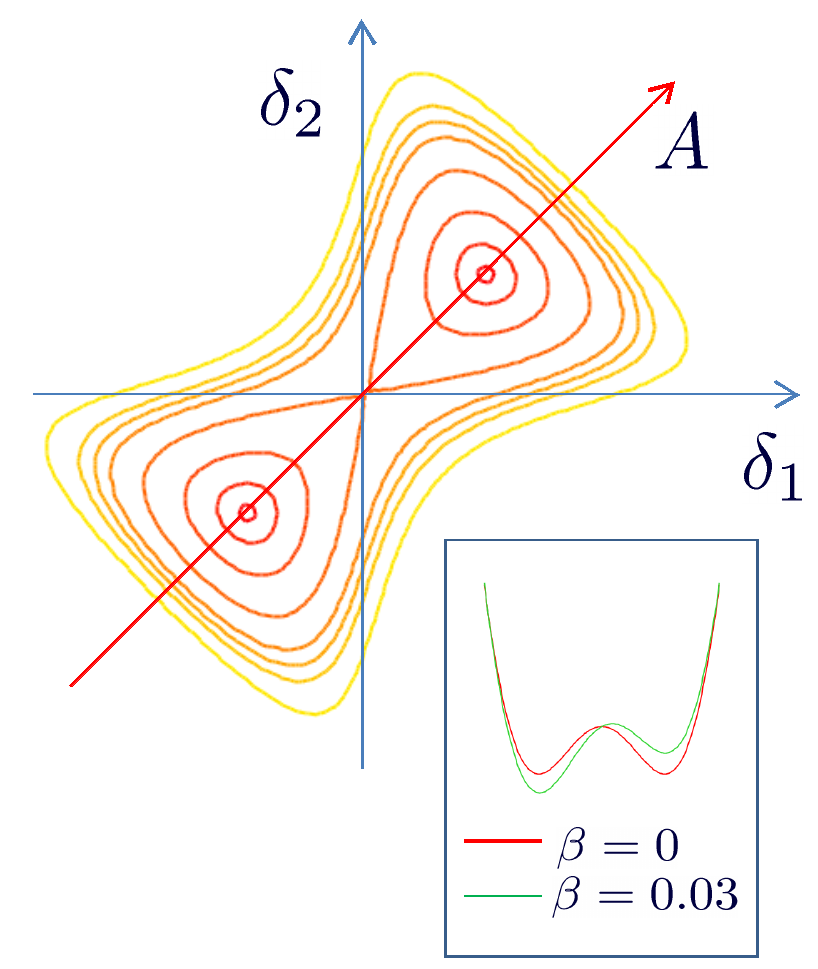}}
\caption{(Color onine) Plot $u(\bm{\delta})$ for $\alpha=0.75$. The inset show the potential profile along the line 
$A$ for different $\beta \equiv (2\Phi -\Phi_0)/\Phi_0$, symmetric potential for $\beta=0$. \label{figdpf}}
\end{figure}

A flux qubit, see, e.~g., Fig.~\ref{figMoij99} adapted from~\cite{Mooij1999},  consists of a micrometer-sized 
loop with three or four Josephson junctions.. 
The energy of each Josephson junction can be expressed as $E_J^{(i)}(1-\cos \delta_i)$ where $E_J^{(i)}$ 
is the Josephson energy of $i$-th junction while $\delta_i$ is the phase drop on the junction. The phase 
drops on the junctions are related as $\sum_i \delta_i +2\pi (\Phi/\Phi_0) = 2\pi n$ where $n$ is an integer 
number. Here $\Phi$ is the magnetic flux embedded in the loop while $\Phi_0=\pi \hbar c/e$ is the magnetic 
flux quantum.  To get the total energy one should add the charging energy of each junction, $Q_i^2/2C_i$, 
and magnetic energy, $(\Phi-\Phi_b)^2/2L$ where $\Phi_b$ is the bias flux created by external sources. 
Neglecting charging energies and magnetic energy of the loop end assuming that $E_J^{(1)}=E_J^{(2)}\equiv E_J$, 
$E_J^{(3)}=\alpha E_J$ one can express the total energy of the qubit as
\begin{equation}  \label{en1}
u(\bm{\delta}) =2-\cos \delta_1-\cos \delta_2 
+\alpha \left[1+\cos \left(\varphi-\delta_1-\delta_2\right)\right]
\end{equation}
where $u(\bm{\delta}) \equiv U(\delta_1,\delta_2)/E_J$, $\varphi \equiv \pi(2 \Phi-\Phi_0)/\Phi_0$. 
The energy is a periodic function of $\delta_1$ and $\delta_2$ with a period of $2\pi$. At $\varphi \ll 1$ and
 $\alpha  >1/2 $ 
each elementary cell contains two close minima, the barrier between them being small. In the eigen basis 
of the states representing these minima the qubit is then described by the effective Hamiltonan (\ref{qb1}) 
where the asymmetry $\hbar \epsilon$ can be tuned by the embedded magnetic flux, $\Phi$. Shown in Fig.~\ref{figdpf} is the 
plot $u(\bm{\delta})$ for $\alpha=0.75$. The inset shows the potential profile along the line $A$ for different 
$\beta \equiv (2\Phi -\Phi_0)/\Phi_0$.
To tune the tunneling parameter, $\Delta$, one can replace the third junction by two Josephson junctions 
connected in parallel, as it is shown in Fig.~\ref{figMoij99} (inset). The Josephson energy of this circuit 
can be tuned by the magnetic flux through the second loop, which in turn can be tuned by an external 
current-carrying line. 
Quantum superpositions of these states are obtained by pulsed microwave modulation of the enclosed magnetic 
flux by currents in control lines. Such a superposition has been demonstrated by \textcite{Friedman2000,VanderWal2000}.

Though fabricated Josephson circuits exhibit a high level of static and dynamic charge 
noise due to charged impurities, while the magnetic background is much more clean and stable. 
The flux qubits can be driven individually by magnetic microwave pulses; measurements can 
be made with superconducting magnetometers (SQUIDs). They are decoupled from charges and 
electrical signals, and the known sources
of decoherence allow for a decoherence time of more than 1~ms. Entanglement is achieved by 
coupling the flux, which is generated by
the persistent current, to a second qubit. The qubits are small (of order 1 mm), they can be  
individually addressed and  integrated
into large circuits.  However, they are slower than the charge qubits.
As it follows from Eq.~(\ref{en1}), fluctuations of two parameters -- magnetic flux, $\Phi$, and Josephson 
energies $E_J^{(i)}$ (or Josephson critical currents $J_C^{(i)}$) -- are important for operation of flux qubits. 

\begin{figure}[t]
\centerline{
\includegraphics[width=6cm]{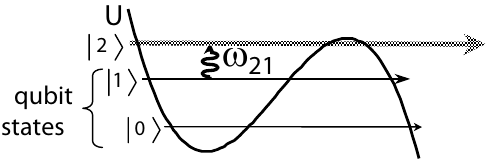}
}
\caption{Cubic potential U showing qubit states and measurement
scheme. Adapted from \cite{Martinis2002}.\label{figMartinis2002}}
\end{figure}

Phase qubits, see, e.~g., \cite{Martinis2002},  are designed around a 10 $\mu$m scale Josephson junction
in which the charging energy is very small, thus providing immunity to charge noise. Although still 
sensitive to flux, the circuit retains the quality of being tunable, and calculations indicate that
decoherence from flux noise is small.  The main part of a phase qubit is a current-biased Josephson 
junction, which can be characterized by the potential energy
\begin{equation} \label{pot1}
U(\delta)=-E_J [\cos \delta +(I/I_J)\delta]\, .
\end{equation}
Because the junction bias current $I$ is typically driven close to the critical current $I_J$ the tilted 
washboard potential (\ref{pot1}) can be well approximated by a cubic potential with the barrier height 
$$\Delta U (I)= (4\sqrt{2}/3) E_J\left[1-I/I_J\right]^{3/2}\, .$$ 
Therefore, the barrier can be tuned by the bias current; at $I \to I_J$ it vanishes. The bound quantum
 states $|n\rangle$
with energy $E_n$, see Fig.~\ref{figMartinis2002}, can be observed spectroscopically by resonantly inducing 
transitions with 
microwaves at frequencies $\omega_{mn}=(E_m-E_n)/\hbar$.
The qubit state can be manipulated with dc and microwave pulses at frequency $\omega_{10}$ of bias current. 
The measurement of the qubit state utilizes the escape  from the cubic potential via tunneling. 
To measure the occupation probability $p_1$ of state $|1\rangle$,  microwave pulses at frequency $\omega_{21}$  
driving a $1 \to 2$  transition were used. The large tunneling rate then causes state $|2\rangle$ to rapidly tunnel. 
Since the potential profile depends on $E_J$, i.~e., on the Josephson critical current, as well as on the 
bias current through the junction, their fluctuations are important. 
Unlike the other qubits, the phase qubit does not have a degeneracy point.
Below we will briefly discuss  main noise sources in flux and phase qubits.

\paragraph{Flux noise  --}		
The origin of magnetic flux noise in SQUIDs with a power
spectrum of the $1/f$ type has been a puzzle for over 20 years. The noise magnitude, a
few $\mu \Phi_0$ Hz$^{1/2}$ at 1 Hz,  scales slowly with the SQUID area and
does not depend significantly on the nature of the thin film superconductor or the substrate
on which it is deposited. The substrate is typically silicon or sapphire, which are insulators
at low temperature~\cite{Wellstood1987}. Flux noise of similar magnitude is observed in 
flux~\cite{Yoshihara2006,Kakuyanagi2007} and phase \cite{Bialczak2007} qubits.  
The near-insensitivity to device area of the noise magnitude (normalized by the device area)
~\cite{Wellstood1987,Bialczak2007,Lanting2009} suggests that  the origin of  the noise is local.  

\textcite{Koch2007} proposed a model in which electrons hop stochastically between traps with different
 preferential spin orientations.  They found that the major noise contribution arises from electrons 
 above and below the superconducting loop of the SQUID or qubit, and that an areal density of about 
 $5 \times 10^{13}$ $\text{cm}^{-2}$ unpaired spins is required to account for the observed noise magnitude.  
 \textcite{Sousa2007} proposed that the noise arises from spin flips of paramagnetic
dangling bonds at the Si-SiO$_2$ interface.  Assuming an array of localized electrons, 
\textcite{Faoro2008} suggested that the noise results from electron spin diffusion. 
The model was extended in \cite{Faoro2012a} where it was shown that in a typical random configuration  
some fraction of spins form strongly coupled pairs behaving as two-level systems. Their switching 
dynamics is driven by the high-frequency noise from the surrounding spins, resulting in low-frequency
 $1/f$ noise in the magnetic susceptibility and other physical quantities. 

\textcite{Sendelbach2008}  showed that thin-film SQUIDs are paramagnetic, with a Curie ($\propto T^{-1}$) susceptibility.
Assuming the paramagnetic moments arise from localized electrons, they deduced an areal
density of $5 \times 10^{13}$~cm$^{-2}$. Subsequently, \textcite{Bluhm2009}
used a scanning SQUID microscope
to measure the low-$T$ paramagnetic response of (nonsuperconducting) Au rings deposited
on Si substrates, and reported an areal density of $4 \times 10^{13}$ cm$^{-2}$ for localized electrons. 
Paramagnetism was not observed on the bare Si substrate.

\begin{figure}[t]
\centerline{
\includegraphics[width=8cm]{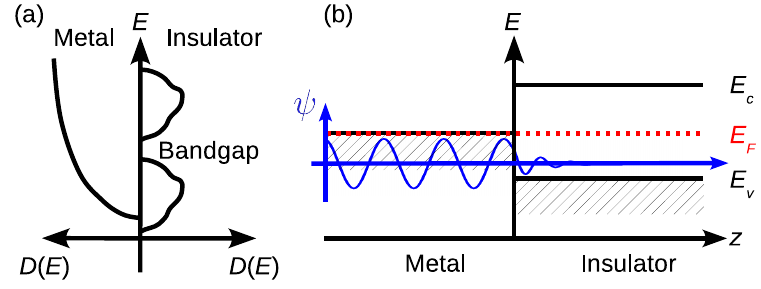}
}
\caption{(Color onine) (a) Schematic density of states. (b) MIGS at a perfect interface with energy
in the band gap are extended in the metal and evanescent in the insulator. Adapted from
 \cite{Choi2009}\label{figChoi2009}}
\end{figure}

\textcite{Choi2009} have proposed that the local magnetic moments originate in metal-induced
gap states (MIGS)~\cite{Louie1976} localized by potential disorder at the metal-insulator interface. At
an ideal interface, MIGS are states in the band gap that are evanescent in the insulator and
extended in the metal, see  Fig.~\ref{figChoi2009} adapted from \cite{Choi2009}. 
At a metal-insulator interface there are inevitably random fluctuations in the electronic potential. 
The MIGS are particularly sensitive to these potential fluctuations, and a significant fraction of  
them with single occupancy becomes strongly localized near the interface, producing the observed 
paramagnetic spins.   The local moments interact via mechanisms such as direct superexchange and the 
RKKY interaction between themselves, and Kondo exchange with the quasiparticles in the superconductor. 
 This system, in principle, can exhibit a spin-glass transition. However, experiments~\cite{Harris2008} 
 suggest that at $T > 55$ mK the spins are
in thermal equilibrium and exhibit a $1/T$ (Curie law) static susceptibility.  To explain the observed 
$1/f^{\alpha}$ ($0.6 <\alpha <1$) noise spectrum of the magnetic flux   the authors suggest that in 
this region one can use the fluctuation-dissipation theorem leading to the conclusion that fluctuations 
of the electronic momenta have also $1/f^\alpha$ spectrum.  Unfortunately, without knowing the form of 
the interaction between the spins, one cannot derive this behavior theoretically -- this is still an 
open question.

Recently \textcite{Sank2012}  measured the dependence of qubit phase coherence and flux noise on inductor 
loop geometry. They concluded that while
wider inductor traces change neither the flux noise power spectrum nor the qubit dephasing time,
increased inductance leads to a simultaneous increase in both. Another important result is 
the absence of scaling with the trace aspect ratio. \textcite{Anton2013} performed flux noise measurements as a
function of temperature in ten dc SQUIDs with systematically varied geometries.
Measurements have shown that $\alpha$ increases as the temperature is lowered.
Moreover, for a given SQUID the spectrum pivoted about a nearly fixed frequency as the temperature was changed.
The  mean-square flux noise, inferred by integrating the power spectra, was found to grow rapidly with temperature and,
at given $T$ to be approximately independent of the outer dimension of a given SQUID.
Authors argued that those results are incompatible with a model based on the random reversal of independent, 
surface spins and considered the possibility that the spins form clusters \cite{Sendelbach2009} - see Section 
\ref{subsub_complex}. An interpretation in terms of a spin-diffusion constant increasing with temperature
is instead proposed in \cite{Lanting2013}.

\begin{figure}[t]
\centerline{
\includegraphics[width=4cm,angle=-90]{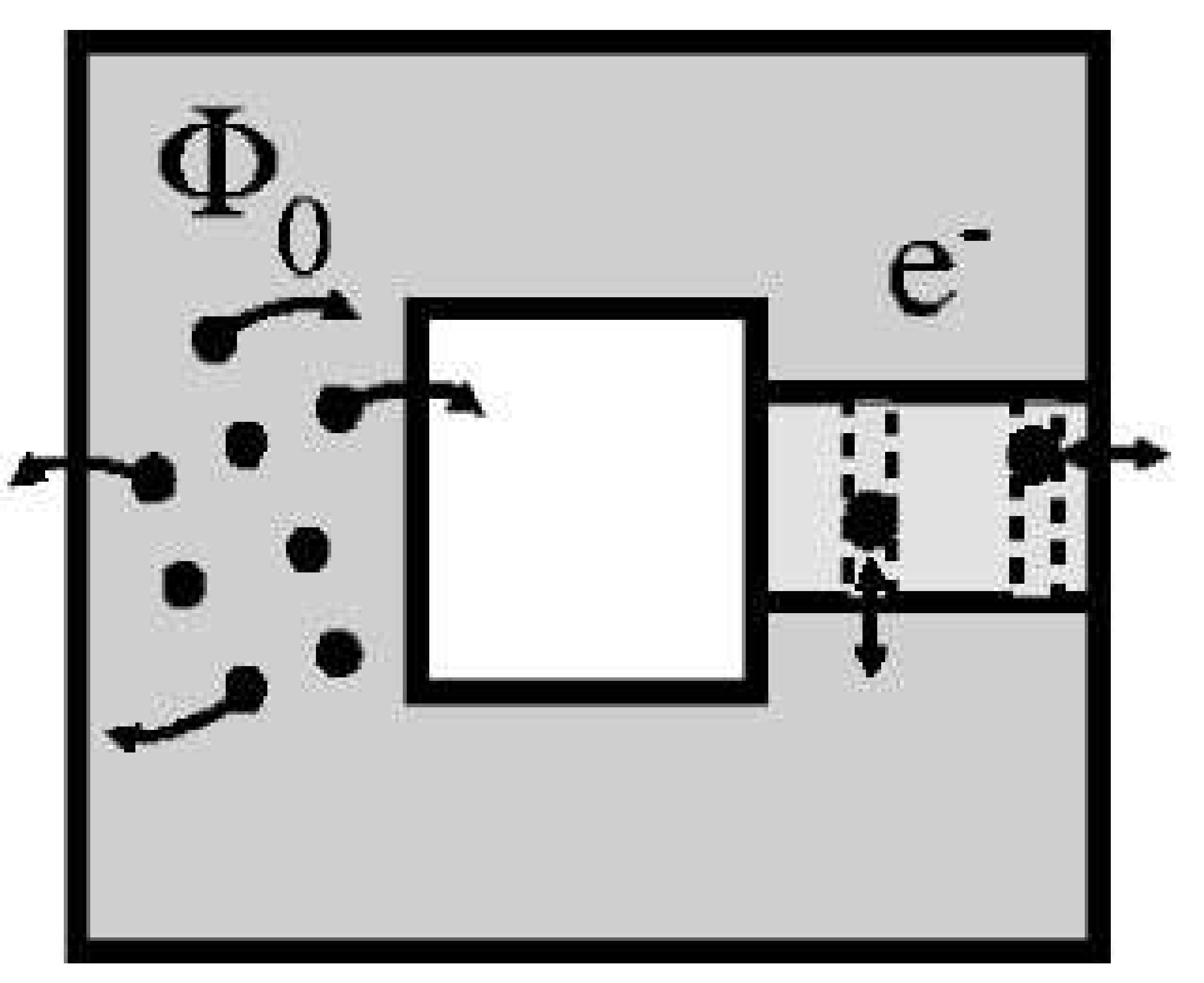}
}
\caption{Flux modulation from vortices
hopping into and out of a loop, and critical-current modulation
from electrons temporarily trapped at defect sites in the junction
barrier.  Adapted from \cite{VanHarlingen2004}. \label{figVanHarlingen2004}}
\end{figure}

\paragraph{Critical current noise --}
Noise of the Josephson critical current in various superconducting qubits incorporating Josephson 
junctions has been investigated in detail by \textcite{VanHarlingen2004}. They consider critical-current 
fluctuations caused by charge trapping at defect sites in the tunneling
barrier and compare their contribution to the dephasing time with that of the flux noise due to hopping
 of the vortices through the SQUID loop, see Fig.~\ref{figVanHarlingen2004} adapted from~\cite{VanHarlingen2004}.  
This mechanism can usually be made negligible in devices
fabricated with linewidths less than approximately $\sqrt{\Phi_0/B}$ for which vortex trapping in the 
line is suppressed~\cite{Dantsker1996}; here
$B$ is the field in which the device is cooled.
The trapped charges block tunneling through a region of the junction due
to the Coulomb repulsion, effectively modulating the junction
area. In general, a single-charge fluctuator produces a
two-level, telegraph noise in the critical current of a junction
characterized by the lifetimes $\tau_u$ in the untrapped state (high critical-current),
and $\tau_t$ in the trapped state (low critical-current).
This produces a Lorentzian peak in the power spectral
density with a characteristic rate $\gamma=\tau_u^{-1}+ \tau_t^{-1}$. 
Experiments on dynamics of such fluctuators and their life times \cite{Wakai1986,Rogers1984,Rogers1985}
provide strong evidence 
that the dominant charges enter the barrier from one electrode and exit to the
other (voltage dependence), and that the fluctuators exhibit a crossover from thermal
activation to tunneling behavior at about 15 K. In the
tunneling regime, the fluctuating entity has been shown to
involve an atomic mass, suggesting that ionic reconfiguration
plays an important role in the tunneling process, see also discussion in  \cite{Galperin1989}.
Though interactions between traps resulting in multiple-level hierarchical kinetics
have been observed \cite{Wakai1987}, usually the traps can be considered
to be local and noninteracting. In this limit, the coexisting
traps produce a distribution of Lorentzian features that superimpose
to give a $1/ f$-like spectrum.  Careful analysis of the influence of the noise in the Josephson 
critical current on different qubit designs 
and various data acquisition schemes,
has led \textcite{VanHarlingen2004} to the conclusion that although there is strong evidence that the
 noise derives from a superposition of random telegraph signals produced by charge trapping and untrapping 
 processes, the
origin of $1/ f$ noise in the critical current of Josephson junctions is still not fully understood.  
In particular, the origin of the $\propto T^2$ dependence of the noise power observed  by 
\textcite{Wellstood1987,Wellstood1987a,Wellstood2004} remains puzzling. To account for this behavior 
\textcite{Shnirman2005} suggested that the density of states for two-level fluctuators is proportional
to their inter-level spacing, $E$.  This assumption was supported by a microscopic model by \textcite{Faoro2005}.
However, detailed measurements by \textcite{Eroms2006} showed that in shadow-evaporated Al/AlO$_x$/Al tunnel 
junctions utilized in many superconductor-based qubits, the noise power is  $\propto T$ between 150 and 1 K rather than 
$\propto T^2$. The observed spectral density saturates below 0.8 K due to individual strong two-level
fluctuators.  Interestingly, the noise spectral density at 4.2 K is two orders of magnitude lower 
than expected from the literature survey of \cite{VanHarlingen2004}.
Recently, measurements in Al/AlO$_x$/Al Josephson junctions reported in \textcite{Nugroho2013} have shown an equivalence 
between the fractional noise power spectra of the critical current, $S_{I_c}/I^2_c$, and of the normal-state resistance, 
$S_{R_n}/R^2_n$, with a linear temperature dependence down to the lowest temperatures measured,
consistent with \textcite{Eroms2006}.
Both fractional power spectra displayed an inverse scaling with the junction area down to $A \lesssim 0.04 \, \mu$m$^2$ at $T=2$~K.
The estimated TLS density is consistent with observations from qubit energy spectroscopy \cite{Martinis2005} and glassy 
systems \cite{Phillips1987}.
Similar noise characteristics have been observed in junctions with AlO$_x$  and Nb electrodes \cite{Pottorf2009}.
These properties suggest that the noise sources are insensitive to the barrier interfaces and that 
the main contribution comes from TLSs buried within the amorphous AlO$_x$ barrier.

The role of the critical current noise is conventionally allowed for by introducing a coupling term into
 the Hamiltonian \cite{Simmonds2004,Ku2005,Shnirman2005,Faoro2005,Faoro2006}.
A simple microscopic model relevant to Al/AlO$_x$/Al was developed by \textcite{Constantin2007}.  This model 
leads to the scaling of the $1/f$ noise with the junction thickness as $\propto L^5$. The results are in a 
reasonable agreement with corresponding experimental values of \cite{Eroms2006,Zimmerlie1992,Zorin1996}. However,
 to the best of our knowledge, the predicted scaling has not yet been verified.

\paragraph{Decoherence in Josephson qubits from dielectric losses --}
 As it was mentioned in Sec.~\ref{subsec:models}, amorphous materials contain low-energy excitations behaving 
 as two-level tunneling systems.  These states are responsible for low-temperature thermal and kinetic 
 properties of structural glasses, in particular, for specific heat and dielectric losses.  Due to interaction 
 with their environment, the TLSs switch between their states producing low-frequency noise. These noises act 
 on a qubit reducing its coherence time. 

 \textcite{Martinis2005} pointed out that the noises produced by TLSs in amorphous parts of qubit devices are
  of primary importance. 
The reason is that crossover wiring in complex superconducting devices requires an insulating spacer that 
is typically
made from amorphous SiO$_2$ deposited by chemical vapor deposition (CVD). They performed a variety of 
microwave  qubit measurements and showed that the results  are well modeled by loss from resonant 
absorption of two-level defects.  Dielectric loss (loss tangent) in a system formed by a superconductor 
lead and a 300 nm thick CVD SiO$_2$ layer was measured at $f \sim 6$ GHz and $T = 25 \ \text{mK}  \ll \hbar \omega/k_B$. 
Generally, two mechanisms contribute to the dissipation induced by the two-level defects. The first -- resonant -- 
 is due to direct microwave-induced transitions between the TLS's level with subsequent emission of phonons.  
The second one is due to the relaxation-induced lag in phase between the nonequilibrium level populations and 
the driving ac electric field. One can expect that at $\hbar \omega \gg k_B T$ the first mechanism should dominate, 
see, e. g., \cite{Hunklinger1981}.  A hallmark of the resonant absorption is its strong dependence on the
 amplitude of the applied ac electric field.  This dependence is due to decrease of the difference between 
 the occupancies of the upper and lower level with the field amplitude increase.  The theoretical prediction 
 for the loss tangent  $\beta$ is, see, e. g.,~\cite{Schickfus1977}
\begin{equation} \label{res1}
\beta=\frac{\pi \bar{P}(ed)^2}{3\varkappa} \frac{\tanh (\hbar \omega/2k_BT)}{\sqrt{1+ \omega_R^2 T_1T_2}}\, .
\end{equation}
Here $\bar{P}$ is the TLS density of states each having a fluctuating dipole moment $ed$ and relaxation 
times $T_1$ and $T_2$, $\varkappa$ is the dielectric constant, while $\omega_R =e\mathcal{E} d/\hbar$ is 
the TLS  Rabi frequency corresponding to the ac field amplitude, $\mathcal{E}$. This expression is derived 
under the assumption that distributions of inter-level spacings and logarithms of the relaxation times of 
the TLSs are smooth.  The theory fits the experimental  data well with parameters compatible with previous 
measurements of bulk SiO$_2$~\cite{Schickfus1977}. The above results are consistent with previous measurements
 of an AlO$_x$ capacitor by \textcite{Chiorescu2004}.

A key difference between tunnel junctions and bulk materials  is that tunnel junctions have small volume, 
and the assumption
of a continuous distribution of defects is incorrect. Instead, dielectric loss must be described by a 
sparse bath of discrete defects. 
Indeed, individual defects were measured spectroscopically with the phase 
qubit~\cite{Simmonds2004,Cooper2004,Lisenfeld2010,Lisenfeld2010b}. They are observed as avoided crossings 
in the plots of the occupation probability versus qubit bias.  A qualitative trend is that small-area 
qubits show fewer splittings than do large-area qubits, although larger splittings are observed in the
 smaller junctions.  It follows from quantitative analysis of the number of resonances that couple to the 
 qubit that at large area the decoherence  rate is compatible with the loss tangent of a bulk material. 

Based on the obtained results, \textcite{Martinis2005} have formulated the following trends for making 
devices with long coherence times
(i) usage small-area junctions where number of two-level defects is small; (ii) usage of simple designs 
 with no lossy
dielectrics directly connected to the qubit junction; (iii) trying to find insulating materials with low 
dielectric losses.

A way to eliminate the effects of low-frequency charge (and flux) noise is to operate the system close 
to the degeneracy
point [$\epsilon=0$ in the Hamiltonian (\ref{qb1})]. Indeed, since the distance between the energy levels is
 $\hbar \Omega= \hbar \sqrt{\epsilon^2+\Delta^2}$, the  fluctuation part $\hbar \delta \epsilon(t)$ of the potential energy 
 $\hbar \epsilon(t)$  creates the additional shift 
\begin{equation} \label{deU}
\delta \Omega(t)=\frac{\epsilon}{\Omega}\, \delta \epsilon(t) +\frac{1}{2}\frac{\Delta^2}{\Omega^3}\,  [ \delta \epsilon(t)]^2 \, .
\end{equation}
Therefore, at $\epsilon=0$ the charge noise vanishes in the linear approximation and only second-order 
contributions are important.  For a charge qubit, the degeneracy point corresponds to the induced charge 
$q=e$ while for a flux qubit this point corresponds to an integer number of the half-flux 
quanta in the loop. This idea of tuning the device to the \textit{double} degeneracy point where the qubit is 
insensitive both to the charge and flux noise was implemented in the \textit{quantronium} device 
\cite{Vion2002,Cottet2002,Ithier2005}. 
\begin{figure}[b]%[h]
\centerline{
\includegraphics[width=\columnwidth]{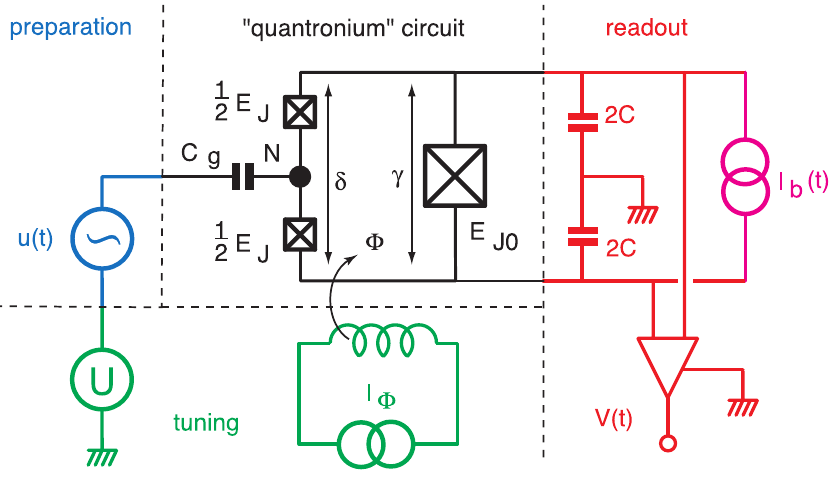}
}
\caption{(Color online) Idealized circuit diagram of the quantronium, a quantum-coherent circuit with its tuning,
preparation, and readout blocks. The circuit consists of a CPB island (black node)
delimited by two small Josephson junctions (crossed boxes) in a superconducting loop. The loop
also includes a third, much larger Josephson junction shunted by a capacitance $C$. The Josephson
energies of the box and the large junction are $E_J$ and $E_{J0}$. The Cooper pair number $N$ and the phases
$\delta$ and $\gamma$ are the degrees of freedom of the circuit. A dc voltage $U$ applied to the gate capacitance
$C_g$ and a dc current $I_\Phi$ applied to a coil producing a flux $\Phi$ in the circuit loop tune the quantum
energy levels. Microwave pulses $u(t )$ applied to the gate prepare arbitrary quantum states of the
circuit. The states are read out by applying a current pulse $I_b(t )$ to the large junction and by
monitoring the voltage $V(t )$ across it.
From  \cite{Vion2002}. Reprinted with permission from AAAS. 
\label{figVion2002}}
\end{figure}

The main principle behind the device is shown in Fig.~\ref{figVion2002} reprinted from \cite{Vion2002}. 
The CPB involves \textit{two} Josephson junctions with a capacitance $C_g$ connected to the island 
separating them. The junctions are connected to a third, larger junction, with a larger Josephson energy, 
to form a 
superconducting loop threaded by a magnetic flux $\Phi$. To achieve 
insensitivity to the charge noise the qubit is operated at 
$N_g\equiv C_g U/(2e)=0.5$,  where the energy levels  have zero 
slope and the energy-level splitting is $E_J$. Insensitivity to the flux noise is achieved by applying 
an integer number of half-flux quanta to the loop.   Therefore, a double degeneracy point can be achieved by tuning dc
 gate voltage $U$ and the current $I_{\Phi}$. To measure the qubit state one has to shift the qubit from 
 this point.  This was achieved by the current pulse $I_b(t)$ applied to the loop, which produces a clockwise 
 or counterclockwise current in the loop, depending on the state of the qubit. The direction of the current 
 is determined by the third (readout) junction since the circulating  current either adds or subtracts from 
 the applied current pulse. As a result, the read-out junction switches out of the zero-voltage state at 
 different values of the bias current.  Thus, the state of the qubit was determined by measuring the 
 switching currents. In the quantronium, much longer relaxation and decoherence times can be achieved 
 compared with conventional CPB.

\subsection{Semiconductor-based qubits}

Here we will briefly discuss main trends in designing the qubits based on semiconductors and semiconductor 
devices.  One can find more detailed discussion in the reviews \cite{Hanson2007,Chirolli2008,Liu2010,Zak2010}.

\subsubsection{Spin qubits}

It is natural to choose the electron spin as the two-state system that encodes the qubit.  In modern 
semiconductor structures the spin of the electron can have a much longer coherence time than the charge 
degrees of freedom. However, it is not easy to isolate, control, and manipulate the spin degree of freedom 
of an electron to a degree required for quantum computation. A successful and promising device for the 
physical implementation of electron spin-based qubits is the semiconductor quantum dot~\cite{Loss1998}.

The quantum dots (QDs) considered for implementation of quantum algorithms are confined regions of semiconductor 
materials coupled with reservoirs by tunable tunnel barriers.  The height of the barriers, and consequently 
the rates for tunneling through the barriers on and off the dot, can be controlled via the application of
 gate electrodes.  The dots are actually quantum boxes having discrete energy levels, their positions with 
 respect to the chemical potential of the reservoir can be also tuned by electrostatic potentials. Therefore, 
 QDs can be considered as tunable artificial atoms. 
Coulomb interaction between the electrons (or holes) occupying the dot's levels determine the energy cost for 
adding an extra electron. Because of this cost, the electron transport through the dot can be strongly 
suppressed at low temperatures (the so-called Coulomb blockade).  Since the energy cost can be tuned by 
gate electrodes, the QD devices are promising for many applications.

\begin{figure}[t]
\centerline{
\includegraphics[width=4cm]{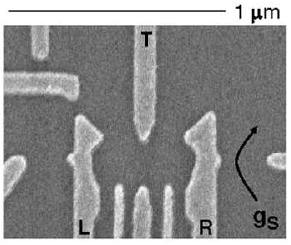}
}
\caption{SEM image of a double quantum dot device . A quantum point contact  with conductance, $g_S$, senses
charge on the double dot. Reprinted from \cite{Petta2008}. \label{figPetta2008}}
\end{figure}

Among many types of quantum dots, 
devices based on lateral III-V semiconductor QDs are of special interest.  Such devices  are usually  
fabricated from heterostructures of GaAs and AlGaAs grown by  molecular beam epitaxy.  In such heterostructures 
the electron motion can be confined to a thin layer 
 along the interface forming a two-dimensional electron gas (2DEG) with high mobility 
 ($\sim 10^5-10^7$ cm$^2$/V$\cdot$ s) and low density ($\sim 10^{11}$ cm$^{-2}$).  The low density results in a 
 relatively long Fermi wavelength ($\sim$40  nm) and a large screening length.  Therefore, the 2DEG can be 
 locally depleted by an electrostatic field applied to a metal gate electrode allowing to design quantum devices
  similar to that  shown in Fig.~\ref{figPetta2008}.
When the lateral size of the dot is comparable with the Fermi wavelength the distance between the 
discretized energy levels becomes larger than the temperature (at temperatures of tens millikelvin), 
and quantum phenomena become important.

\paragraph*{Spin relaxation and dephasing in quantum dots  --}
Two kinds of environment turn out to mainly affect the dynamics of an electron spin in a quantum dot, 
the phonons in the lattice, and the spins of atomic nuclei in the quantum dot.  

Starting from~\cite{Khaetskii2000,Khaetskii2001}, the phonon-induced relaxation was extensively studied. 
It turns out that the lattice phonons do not couple directly to the spin degree of freedom. However, 
even without the application of external electric fields, the breaking of inversion symmetry in GaAs gives
rise to spin-orbit (SO) interaction, which couples the spin and the orbital degrees of freedom. These 
orbital degrees of freedom, being coupled to the phonons, provide an indirect coupling between the electron 
spin and the phonons, which constitute a large dissipative bosonic reservoir and provide a source of 
decoherence and relaxation. Short time correlations in the phonon bath induce a Markovian dynamics of 
the electron spin, with well-defined relaxation and decoherence times $T_1$ and $T_2$.  

As we discussed in Sec.~\ref{why}, in the Bloch picture, pure dephasing arises from longitudinal 
fluctuations of the magnetic field, while a perturbative treatment of the SO interaction gives rise, 
within first order, to a fluctuating magnetic field perpendicular to
the applied magnetic field. As a consequence the decoherence time $T_2$ is limited only by its upper 
bound $T_1$, $T_2=2T_1$.

Hyperfine interaction was first taken into consideration as a source of decoherence for an electron 
spin confined in a quantum dot by \textcite{Burkard1999}. This interaction is very important since in a 
$\sim \! 40$~nm GaAs quantum dot the wavefunction of an electron overlaps with approximately 10$^5$ nuclei.  
The electron spin and the nuclear spins in the dot couple via the Fermi contact hyperfine interaction, 
which creates entanglement between them and strongly influences the electron spin dynamics. It turns out 
that long-time
correlations in the nuclear spin system induce a non-Markovian dynamics of the electron spin, with 
non-exponential decay in time of the expectation values of the electron spin components.

Relative importance of the above mechanisms of decoherence depends on the external magnetic field:  
the phonon-induced relaxation rate of the electron spin is enhanced by an applied magnetic field, 
whereas the influence of the hyperfine interaction is reduced by
a large Zeeman splitting.  

In this review,
we will not focus on decoherence in semiconductor spin qubits, which has been extensively reviewed, 
e.~g., by \textcite{Chirolli2008}. 
More recent papers related to random telegraph or $1/f$ noise address charge traps near the interface
 of a Si heterostructure \cite{Culcer2009}, as well as nearby two-level charge fluctuators in a double 
 dot spin qubit \cite{Ramon2009}.

\subsubsection{Charge qubits}

\begin{figure}[t]
\centerline{
\includegraphics[width=4.3cm]{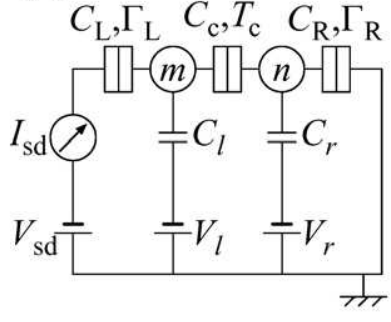}
}
\caption{Electric circuit model of a DQD containing $m$ and $n$ electrons in the left and right
dot, respectively. The two QDs are coupled to each other via a tunnel junction, and each dot is
connected to an electron reservoir via a tunnel junction. The electrostatic potential of the left (right)
quantum dot is controlled by the gate voltage $V_l$ ($V_r$ ) through the capacitance $C_l$ ($C_r$). 
Adapted from \cite{Fujisawa2006}.\label{figEc} }
\end{figure}

In semiconductor systems, a charge qubit can be formed by isolating an electron in a tunnel-coupled 
double quantum dot (DQD)~\cite{Hayashi2003,Fujisawa2006}.  Here we will discuss a DQD consisting of two 
lateral QDs, which are coupled to each other through a tunnel barrier. Each QD is also connected to an 
electron reservoir via a tunnel junction. The operation of such device can be analyzed from its electric
 circuit model shown in Fig.~\ref{figEc}.
\begin{figure}[b]
\centerline{
\includegraphics[width=5cm]{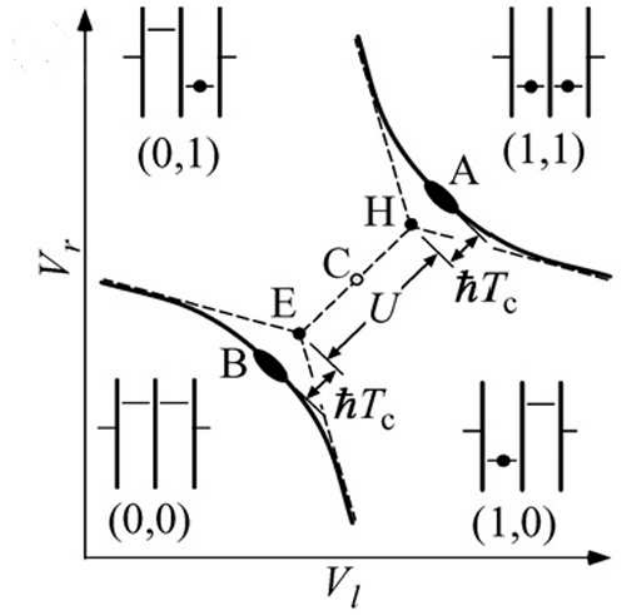} }
\caption{Fragment from the stability diagram. Adapted from \cite{Fujisawa2006}. \label{figFujisawa1}}
\end{figure}
The transport properties of a semiconductor DQD have been studied extensively, see reviews 
\cite{Kouwenhoven1997,Grabert1991,vanderWiel2003}. 
Each tunnel barrier has a small coupling capacitance, $C_i$, as well as a finite tunneling coupling,
 $T_c$, and single-electron transport through the DQD can be measured.  The tunneling rates for the 
 left and right tunnel barriers are denoted as $\Gamma_L$ and $\Gamma_R$, respectively. In addition, 
 the DQD is connected to gate voltages $V_l$ and $V_r$ via capacitors $C_l$ and $C_r$, respectively, so 
 that the local electrostatic potential of each dot can be controlled independently. 
The energy difference between the electrochemical potentials $\mu_L$ and $\mu_R$ of the left and right 
reservoirs corresponds to the applied source~-~drain voltage $V_{sd}$.  The stable charge configuration 
$(m,n)$, with $m$ electrons on the left QD and $n$ electrons on the right one, minimizes 
the total energy in all capacitors minus  the work that has been done by the voltage sources. 
This energy can be estimated from the equivalent circuit Fig.~\ref{figEc}, see, e.~g., analysis in the 
book by \textcite{Heinzel2007}, Sec. 9.3. A fragment of the stability diagram in the $V_l-V_r$ plane is
 shown in Fig.~\ref{figFujisawa1}.
%\begin{figure}[b]
%\centerline{
%\includegraphics[width=5cm]{paladino_fig19} }
%\caption{Fragment from the stability diagram. Adapted from \cite{Fujisawa2006}. \label{figFujisawa1}}
%\end{figure}

When the tunneling coupling, $T_c$, is negligibly small, the boundaries of the stable charge states
appear as a honeycomb pattern, a part of which is shown by dashed lines.  The triple points, $E$
and $H$, of three charge states are separated by a length corresponding to the inter-dot Coulomb
energy $U$. Electrons pass through three tunnel barriers sequentially in the vicinity of triple
points.  On the other hand, the tunneling process at $H$ can be viewed as hole
transport, as the unoccupied state (hole) moves from the right to the left (not shown in the
diagram).

When the tunneling coupling is significantly large, the charging diagram deviates from the honeycomb 
pattern as shown by solid lines in Fig.~\ref{figFujisawa1}. Due to quantum repulsion of the levels, 
the minimum distance between $A$ and $B$ is increased by the coupling energy $\hbar \Delta= \hbar T_c$ from its
 original value $U$.

The idealized dynamics of the qubit can be described assuming that each dot has a single energy level, 
$\epsilon_i$. Then the effective Hamiltonian can be expressed in the form (\ref{qb1}) with 
$\hbar \epsilon=\epsilon_L -\epsilon_R$ and $\Delta=T_c$. The parameters of the Hamiltonian can be tuned by 
the gate voltages. 

Coherent control
of a GaAs charge qubit has been demonstrated~\cite{Hayashi2003,Fujisawa2004}, along with correlated two-qubit
 interactions~\cite{Shinkai2009}.  In these experiments each dot contained a few tens of electrons, 
 potentially complicating the qubit level structure.  In addition, the dots were strongly coupled with 
 the leads typically limiting coherence times to $\sim\!  1$~ns (due to quantum co-tunneling).

\begin{figure}[t]
\centerline{
\includegraphics[width=5cm]{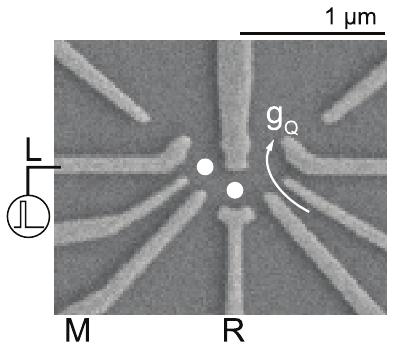}
}
\caption{Scanning electron microphotograph of a device similar to the one measured. Adapted from 
\cite{Petersson2010}. \label{figPetersson2010}}
\end{figure}

A coherent control of tunable GaAs charge qubit containing a single electron was demonstrated by 
\textcite{Petersson2010}, see Fig.~\ref{figPetersson2010}. 
The gate electrodes are arranged in a triple QD geometry and deplete 2DEG in the GaAs/AlGaAs heterostructure 
\cite{Petta2010}. The DQD was formed using the left and middle dots of the structure while the right side of 
the device was configured as a non-invasive quantum point contact (QPC) charge detector. The device operated 
near the $(1,0)-(0,1)$ charge transition, where $(n_L,n_R)$ denote the absolute number of electrons in the left 
and right dots. The level detuning,  $\hbar \epsilon$, was adjusted by the voltage $V_R$ on the gate $R$ while the tunnel 
splitting $\hbar \Delta$ was adjsuted by the voltage $V_M$ on the gate $M$. 
The coherence time was extracted as a function of the detuning (from the charge degeneracy point) voltage, 
the maximal value being $\sim 7$ ns.  The result is ascribed to $1/f$ noise, whose influence is analyzed along
the Gaussian assumption.

In spite of the successful manipulation of a single-charge qubit, the qubit is actually influenced by
 uncontrolled decoherence, which is present even in the Coulomb blockade regime. Several possible decoherence 
 mechanisms were
discussed.

First, background charge ($1/f$)  noise in the sample and electrical noise in the gate
voltages cause fluctuation of the qubit parameters $\epsilon$ and $\Delta$, which gives rise 
to decoherence of the system ~\cite{Hayashi2003,Itakura2003,Paladino2002}. 
The amplitude of low-frequency fluctuation in $\hbar \epsilon$ is estimated to be about 1.6 $\mu$eV, 
which is obtained from low-frequency noise in the single-electron current, or 3 $\mu$eV,
which is estimated from the minimum line width of an elastic current peak at the weak  coupling
limit~\cite{Fujisawa1998}. Low-frequency fluctuations in $\hbar \Delta$ are relatively small and 
estimated to be about 0.1  $\mu$eV  for  $\hbar \Delta=10$ $\mu$eV, assuming  local potential
fluctuations in the device~\cite{Jung2004}. Actually,  the $\epsilon$ fluctuation explains the 
decoherence rate observed at the off-resonant condition ($\epsilon \gtrsim \Delta$).  
$1/f$ is usually attributed to a set of bistable fluctuators, each of which produced 
Lorentzian spectrum. However, the microscopic origin of the charge fluctuators is not fully understood, 
and their magnitude differs from sample to sample, even when samples are 
fabricated in the same batch. It was shown \cite{Jung2004} that fluctuations in
 $\epsilon$ can be reduced by decreasing the temperature as suggested by a simple phenomenological 
 model where the activation energy of the traps is uniformly distributed in the energy range 
 of interest.  Cooling samples very slowly with positive gate voltage is sometimes effective in 
 reducing charge fluctuation at low temperature \cite{Pioro2005}. Authors suggest that the noise
originates from a leakage current of electrons that tunnel through the Schottky barrier
under the gate into the conduction band and become trapped near the active region of the
device.  According to  \textcite{Buizert2008},
an insulated electrostatic gate can strongly suppress ubiquitous 
background charge noise in Schottky-gated GaAs/AlGaAs devices.  
This effect is  explained by reduced leakage of electrons from the Schottky gates into the semiconductor 
through the Schottky barrier. 
A similar result has been recently reported in  Schottky gate-defined QPCs 
an DQDs in Si/SiGe heterostructures with a global top gate voltage \cite{Takeda2013}.  
By negatively biasing the top gate, $1/f^2$ switching noise is suppressed in a homogeneous 
 $1/f$ charge noise background.
It is suggested that this technique may be useful to eliminate dephasing of qubits due to charge noise via 
the exchange interaction \cite{Culcer2009}.

In contrast, the decoherence at the resonant condition ($\epsilon=0$) is dominated by other
mechanisms. Although the first-order tunneling processes are forbidden in the Coulomb
blockade regime, higher-order tunneling, namely co-tunneling, processes can take place and
decohere the system \cite{Eto2001}. Actually, the co-tunnelling rate estimated from the tunneling rates
is close to the observed decoherence rate and may thus be a dominant mechanism in the
present experiment \cite{Eto2001}. However, since one can reduce the co-tunneling effect by making
the tunneling barrier less transparent, it is possible  to eventually eliminate it.

The electron-phonon interaction is an intrinsic decoherence mechanism in semiconductor
QDs. Spontaneous emission of an acoustic phonon persists even at zero temperature and causes
an inelastic transition between the two states \cite{Fujisawa1998}. The phonon emission rate at the resonant 
condition $\epsilon=0$ cannot be directly estimated from the experimental data on the FID-type protocols, but it 
may be comparable to the observed decoherence rate.   Strong electron-phonon coupling is related to the
fact that the corresponding phonon wavelength is comparable to the size of the QD \cite{Brandes1999,Fujisawa1998}. 
In this sense, electron-phonon coupling may be reduced by 
optimizing the size of QD structures. In addition, polar semiconductors, such as GaAs, exhibit a piezoelectric
type of electron-phonon coupling, which is significant for low-energy excitations ($ < \!0.1$ meV
for GaAs). Non-polar semiconductors, such as Si or carbon-based molecules, may be
preferable for reducing the phonon contribution to the decoherence.

%% file: Part_3_0901.tex
\section{DECOHERENCE DUE TO $1/f$ NOISE}
\label{sec:decoherence}

During the last decade we  witnessed of 
an extraordinary progress in quantum devices engineering 
reaching a high level of isolation from the local electromagnetic environment. 
 Under these conditions, the material-inherent sources 
of noise play a crucial role. While the microscopic noise sources may have different physical
origin, as elucidated in  Section \ref{sec:origin}, their noise spectral densities show similar
$1/f$-type behavior at low frequencies. 
Material-inherent fluctuations with $1/f$ spectrum represent the main limiting factor to quantum 
coherent behavior of the present generation of nanodevices.
This fact has stimulated a great effort in understanding and predicting decoherence due to
$1/f$ noise and to the closely related RT noise. 

There are two characteristic features which make any prediction of decoherence due to 
$1/f$ noise quite complicated.
Firstly, stochastic processes with $1/f$ spectrum are long-time correlated.
The spectral density of the noise increases with decreasing frequency down to
the lowest experimentally accessible frequencies. The measurement frequency band is 
limited either by frequency filters or simply by the finite duration of each realization 
of the random process, set by the measurement time, $t_m$.
In particular, during  $t_m$ some of the excitations responsible for the noise 
may not reach the equilibrium. For this reason $1/f$ noise is considered a non-equilibrium
phenomenon for which fluctuation-dissipation relations may not hold~\cite{Galperin2003a}\footnote{The typical 
example are $1/f$ voltage fluctuations in uniform conductors resulting from 
resistance fluctuations which are detected by a current and are proportional to the current squared.
To observe them one has to bring the device out of equilibrium where the 
fluctuation-dissipation theorem is not necessarily valid.
This may not be the case in  some magnetic systems. For instance, experiments in spin glasses  evidenced 
magnetic noise with $1/f$ spectrum satisfying, within experimental accuracies, 
fluctuation-dissipation relations. Despite being non-ergodic, in these systems magnetic noise
is with satisfactory accuracy a thermal and equilibrium one.  The mechanism of magnetic
fluctuations is, however, not yet clear. 
The problem of kinetics of spin glasses, determined both by many-body 
competing interactions and disorder, is extremely complicated and beyond the scope of this review. 
In connection with $1/f$ noise it is addressed in \cite{Kogan1996}.}.

The existence of relaxation times longer than any finite measurement time $t_m$ 
also poses the question of stationarity of $1/f$-type noise, a problem which 
has attracted much attention~\cite{Kogan1996}. 
For a stationary process repeated measurements yield the same power spectrum $S_x(\omega)$, 
within experimental accuracy. In some systems variations of the spectrum or its ``wandering" 
have been reported, see \cite{Kogan1996,Weissman1993}.
However this effect could be  attributed to the finite measuring
time, or to non-equilibrium initial conditions for different measurements due to the
fact that some degrees of freedom do not completely relax between successive measurements.
The long relaxation times may correspond to rare transitions of the system, 
overcoming high energy barriers, from one ``valley" in phase space to another one in 
which the spectrum of relaxation times $\tau < t_m$ is different.
Several experiments tried to reveal a possible nonstationarity of $1/f$ noise
(see \textcite{Kogan1996} and references therein), but no clear
manifestation of nonstationarity has been found.

Secondly, in general $1/f$ noise cannot be assumed to be a Gaussian random process.
Even if the probability density function of many $1/f$ processes resembles a Gaussian form 
(which is necessary, but not sufficient to guarantee Gaussianity), clear evidences demonstrating 
deviations from a Gaussian random behavior have been reported,
see for a review  \cite{Weissman1988}, Sec. IIIA, and \cite{Kogan1996}, Sec. 8.2.2.
The explanation of the variety of observed behaviors stems from the fact
that the mechanisms of $1/f$ noise may be different in various physical systems, 
implying that their statistical properties may strongly differ. 

As a consequence, the two standard approximations allowing simple predictions for
the evolution of open quantum systems, namely the Markovian approximation 
and the modelization of the environment as a bath of 
harmonic oscillators, cannot be straightforwardly applied when the power spectrum of the noise is
of $1/f$ type.\footnote{We remind that the {\em Markovian approximation} for the reduced dynamics of a 
quantum system  is applicable  provided that the noise correlation time, $\tau_c$, and the system-bath 
coupling strength, $v$, satisfy the condition $v \tau_c \ll 1$~\cite{CohenTannoudji1992}.
Physically, the quantum system perturbs very weakly the environment, thereby memory of its previous states
is quickly lost. If $1/f$ noise is produced by superposition of RT processes, each of which being a discrete {\em Markov process}, 
some of them should have very long correlation times.
``Sufficiently slow"  fluctuators would violate the above inequality (see the discussion in 
Sec. \ref{subsub:SPpuredeph}). 
Therefore, the Markovian approximation may be not applicable to the reduced system evolution.
For $1/f$ noise, a signature of the failure of the Markovian approximation is the divergence of the adiabatic
decoherence rate $1/T_2^*$, Eq. (\ref{T2}). For more general situations see, e.~g., 
\cite{Laikhtman1985}.}
In the context of quantum computation, the implication of long-time correlations of the
stochastic processes is that 
the effects of $1/f$ noise on the system evolution depend on the specific quantum operation performed
and/or on the measurement protocol.
Some protocols show signatures of the non-Gaussian nature of the process, whereas for others
a Gaussian approximation captures the main effects at least on a short time
scale~\cite{Falci2005,Makhlin2004,Rebenstein2004}. 

In this Section we will illustrate various approaches developed in recent years to 
address the problem of decoherence due to noise sources having $1/f$ spectrum, 
considering both  microscopic quantum models and  semi-classical theories.  
We will start addressing the decohence problem in single-qubit gates driven by dc-pluses or
by ac-fields, then we will consider more complex architectures needed to implement the set of universal
gates. To cast these problems in a general framework,
we introduce here the  Hamiltonian of a nanodevice plus environment on a phenomenological basis. 
In some cases this general structure is derived from  a microscopic description
of the device, including the most relevant environmental degrees of freedom.
% as illustrated in the following. 
A convenient general form for the Hamiltonian is
\begin{equation}
\hat{H}_{\text{tot}} = \hat{H}_{0} +  \hat H_c(t) + \hat H_n(t) + \hat{H}_R + \hat{H}_I \, ,
\label{H_tot}
\end{equation} 
where $\hat{H}_{0} +  \hat H_c(t)$ describes the driven closed system, classical noise affecting the system is
included in $\hat H_n(t)$ and $\hat{H}_R + \hat{H}_I$ represents quantum environment and its interaction with the system.
The structure of this general Hamiltonian can be justified as follows.
The  macroscopic Hamiltonian of the device $\hat{H}_{0}[{\bf q}]$ is an operator acting onto
a $N$-dimensional Hilbert space, $\mathbbm{H}$. It depends on a set of parameters ${\bf q}$, which  fix 
the bias (operating) point and account for tunability of the device. 
The eigenstates of $\hat{H}_0[\bf q]$, $\{ \ket{\phi_i(\bf q)}:\;
 i=1,\dots, N \}$,  form the ``local basis" of 
the ``laboratory frame" where $\hat H_0 = \sum_{i=1}^N E_i(\bf q)\, \ket{\phi_i(\bf q)}\bra{\phi_i(\bf q)}$.
External control is described by a time-dependent term. In 
a one-port design, the driving field $A(t)$ couples to a single time-independent system operator  $\hat{Q}$, 
\begin{eqnarray}
\label{eq:control-hamiltonian-singleport}
\hat{H}_c(t)  = A(t) \, \hat{Q}\, ,
\end{eqnarray}
which is Hermitian and traceless. 
In general, control is operated via the same ports  used for biasing the system and accounted for
by time-dependent parameters ${\bf q}$. 
%, i.~e.,  by allowing  the bias in $\hat H_{0}[{\bf q}]$ to depend on time. This suggests 
It is convenient to split ${\bf q}$ 
in a slow part,
${\bf q}(t)$, which includes  static bias, and the fast control parameter, ${\bf q}_c(t)$,
as 
${\bf q} \to {\bf q}(t)+ {\bf q}_c(t)$. Accordingly, we write
\begin{equation}
\label{eq:sys-hamiltonian-lab}
\hat{H}_{0}[{\bf q}(t)+ {\bf q}_c(t)] := \hat{H}_{0}[{\bf q}(t)] + \hat{H}_{c}(t) 
\end{equation}
where  $\hat{H}_{c}(t)$ describes (fast) control; in relevant situations it can be linearized 
in ${\bf q}_c(t)$, yielding  the structure of Eq.~(\ref{eq:control-hamiltonian-singleport}).\footnote{Notice
 that while $\hat{Q}$ is an observable and does not depend on the local basis, its matrix 
representation does. The physical consequence is that the effectiveness 
of the fast control ${\bf q}_c(t)$ in triggering transitions depends also 
on the slow ${\bf q}(t)$, a feature of artificial atoms which is the counterpart of tunability.}
At this stage the interaction with the complicated environment of microscopic degrees 
of freedom in the solid state can be introduced in a phenomenological way. 
We first consider classical noise usually acting through 
the same ports used for control and that can be modeled by adding a  
stochastic component $\delta{\bf q}(t)$ to the drive. Again, we split
slow and fast noise, $\delta{\bf q}(t) \to \delta{\bf q}(t)+ {\bf q}_f(t)$, and include the slow part
in $\hat H_{0}[{\bf q}] \to \hat{H}_{0}[{\bf q}(t)+ \delta{\bf q}(t)] $. The same steps leading 
to Eq.~(\ref{eq:sys-hamiltonian-lab}) yield the noisy Hamiltonian
\begin{eqnarray}
\label{eq:sys-noise-hamiltonian-lab}
\hat H_{\text{tot}} = \hat{H}_{0}[{\bf q}(t)+ \delta{\bf q}(t)] + \hat H_c(t) + H_{f}(t)
\end{eqnarray}
where $H_{f}(t)$ describes a short-time correlated stochastic process. % Markovian classical noise. 
``Quantization'' of this term, $H_{f}(t) \to  \hat{E} \,\hat{Q} + \hat H_R= \hat H_I+ \hat H_R$,  
yields the phenomenological system-environment Hamiltonian in the form of Eq.~(\ref{H_tot}), where
$\hat{H}_{0}[{\bf q}(t)]+ H_n(t) \equiv \hat{H}_{0}[{\bf q}(t)+ \delta{\bf q}(t)]$, 
$\hat{E}$ operates on the environment and  $\hat H_R$ is its Hamiltonian (plus possibly suitable 
counter-terms).

When the nanodevice operates as a qubit, $\hat{H}_{0}[{\bf q}]$ can be projected onto the eigenstates and it can be cast in 
the form Eq.~(\ref{qb1}), where  both the level splitting $\Omega$ and the polar angle $\theta$ shown in Fig.~\ref{figAGB1} 
depend on the set of parameters ${\bf q}$.
For simplicity, we suppose that a single parameter is used to fix the bias point, $q$. We can write
Eq.~(\ref{eq:sys-noise-hamiltonian-lab}) as 
\begin{equation}
\hat H_{\text{tot}} = \frac{\hbar}{2}  \, \vec \Omega[q+ \delta q(t)] \cdot \vec \sigma + \hat H_c(t) + \hat H_I + \hat  H_R \, .
\label{H_general}
\end{equation}
Expanding $\vec \Omega[q+ \delta q(t)]$ about the fixed bias $q$, we obtain
\begin{equation}
\hat H_{\text{tot}} = \frac{\hbar}{2} \, \vec \Omega(q) \cdot \vec \sigma 
+   \delta q(t) \frac{\partial \hbar \vec \Omega}{\partial q} \, \cdot \vec \sigma
+ \hat H_c(t) + \hat H_I + \hat  H_R\, ,
\label{H_tot_linearized} 
\end{equation}
and if bias $q$ controls only one qubit component, for instance $\Omega_z(q)$,
then Eq.~(\ref{H_tot_linearized}) reduces to the commonly used form  
\begin{equation}
\hat H_{\text{tot}} = \frac{\hbar\Omega_x}{2} \sigma_x +
\frac{\hbar}{2} \left[ \Omega_z(q)+E(t)+\hat{E} \right] \sigma_z 
 +  \hat H_c(t) + \hat H_R \, .
\label{H_qb_linearized}
\end{equation}
Here the qubit working point is parametrized by the angle $\theta_q$ which is tunable via the bias $q$,
$\tan \theta_q = \Omega_x/\Omega_z(q)$, the classical noise term is 
$E(t) =   2\,  \delta q(t) (\partial  \Omega_z/\partial q) $, 
and  we have consistently set $\hat Q = \sigma_z$ in $\hat H_I$.

From the physical point of view, the phenomenological Hamiltonian (\ref{H_qb_linearized}) treats on 
different footings fast environmental modes exchanging energy  with the system and and slow modes 
responsible for dephasing. The fast modes, being responsible for spontaneous decay,
must be treated quantum mechanically and included in $ \hbar \hat E \sigma_z/2 + \hat H_R$.
Slow modes can be  accounted for classically and included in 
$H_n(t) \equiv \hbar E(t) \,  \sigma_z/2$.
This term yields the longitudinal fluctuating part, $\mathbf{b}(t)$, of a ``magnetic field" formed by 
an external part $\mathbf{B}_0$ plus an ``internal" classical stochastic component, as introduced in Eq.~(\ref{qb2}). 
Results of measurements involve both quantum and classical ensemble averaging.
From the technical point of view, effects of quantum noise described by $\hat H_I + \hat  H_R$, can studied by 
weak coupling Master Equations \cite{Bloch1957,Redfield1957,CohenTannoudji1992,Weiss2008},
which lead to exponential decay
both of the diagonal  (populations) and of 
the off-diagonal (coherences) elements  of qubit density matrix elements in the $\hat{H}_q$  eigenbasis.
The corresponding time scales, 
$T_1$ and $T_2$, are given by Eqs.~\eqref{T1} and \eqref{T2}, respectively.

The  weak coupling Master Equation approach fails in dealing with slow noise as
%This approach fails for the classical stochastic term 
$H_n(t)$, which describes low-frequency (e.~g.,  $1/f$) noise. An important feature of superconducting nanodevices is that 
the Hamiltonian  $\hat H_0[q]$ can be tuned in a way such that  symmetries  (usually parity) are enforced.
At such symmetry points  $\partial E_i/\partial q = 0$ and selection rules hold for the matrix elements 
$Q_{ij}$ in the local basis. In these symmetry points the device is well protected 
against  low frequency noise, and the system is said to operate
at ``optimal points"~\cite{Vion2002,Chiorescu2004}.
In general, noise affecting solid-state devices,
% suffer from broadband and non-monotonic noise formally 
as described by $H_n(t)$ and $\hat H_I + \hat  H_R$ in Eq.~(\ref{eq:sys-noise-hamiltonian-lab}), has a
broadband colored spectrum.
Therefore,  approaches  suitable to deal with noise acting on very different time scales are required. 
This topic will be discussed in Sec.~\ref{paragraph:multistage}. 

To start with, we analyze the effect of $1/f$-noise on the qubit's evolution under pure dephasing conditions. 
As we discussed in Sec.~\ref{why}, in some situations the Gaussian approximation 
does not apply. In these cases knowledge of only the noise power 
spectrum $S(\omega)$  is not sufficient since noise sources with identical power spectra can have 
different decohering effects on the qubit. Therefore, it is necessary to specify the model for 
the noise source in  more detail. In Sec. \ref{sub:SF} we illustrate dephasing  
by the {\em spin-fluctuator} model, which  can be solved exactly under general conditions.
When we depart from pure dephasing, i.~e., when the noise-system interaction
is not longitudinal, no exact analytic solution for the time evolution of the system is available even for the
spin-fluctuator model. Different exact methods have been proposed which lead to approximate
solutions in relevant regimes.

In Sec.~\ref{subsection4B} we present approximate approaches proposed to predict dephasig due
to $1/f$ noise described as a classical stochastic process.  Approaches based on the adiabatic 
approximation \cite{Falci2005,Ithier2005}, allow simple explanations of
peculiar non-exponential decay reported in different experiments with various 
setups~\cite{Cottet2001,Vion2002,VanHarlingen2004,Martinis2003,Ithier2005,Bylander2011,Chiarello2012,Sank2012,Yan2012}. 
These approaches also predict the existence of operating conditions 
where leading order effects of $1/f$ fluctuations are eliminated also for complex architectures,
analogously to the single qubit ``optimal point".
The effect of $1/f$ noise in solid state complex architectures is a subject of 
current investigation \cite{Bellomo2010,Brox2012,Hu2007,Storcz2005,D'Arrigo2008,Paladino2009,Paladino2010,Paladino2011,DArrigo2012}. 
Considerable improvement  in minimizing sensitivity to charge noise has been reached via clever engineering, 
in particular in the cQED architecture, see Sec.~\ref{cnjq}.
Very recently, highly sensitive superconducting circuits have been used as ``microscopes" for probing
characteristic properties of environmental fluctuators. Recent progress in the ability of direct
control of these  microscopic quantum TLS has opened a new research scenario where 
they may be used as naturally formed qubits. These
issues will be addressed in Sec. \ref{qcimp}. 

In the  final part of this Section we will present current strategies to reduce effects
of $1/f$ noise  based on techniques developed in NMR~\cite{Schlichter}.
The open question about the best strategy to limit $1/f$ noise 
effects via open or closed loop control will be discussed and   
we will review the current status of the ongoing research along this direction.

\subsection{Spin-fluctuator model}
\label{sub:SF}

In the following we will use a simple classical  model according to which the quantum system -- qubit --  is 
coupled to a set of two-state \textit{fluctuators}.  The latter randomly switch between their states due to 
interaction with a thermal bath, which can be only weakly directly coupled to the qubit. Since we are interested 
only in the low-frequency noise generated by these switches, they will be considered as classical. 
(The situations where quantum effects are of importance will be discussed separately). 
The advantage of this approach, which is often referred to as the  \textit{spin-fluctuator} (SF) model, is that 
the system qubit+fluctuators can be described by relatively simple set of stochastic differential equations,
 which in many cases can be exactly solved. In particular, many results can just be borrowed from much earlier 
 papers on magnetic resonance \cite{Klauder1962,Hu1997,Maynard1980}, on spectral diffusion in glasses \cite{Black1977},
  as well as works on single-molecule spectroscopy \cite{Moerner1994,Moerner1999,Geva1996,Barkai2001}.

The interaction of electrons with two-state \fs was previously used for evaluation of the effects of noise on various 
systems \cite{Ludviksson1984,Kogan1984a,Kozub1984,Galperin1989,Galperin1991,Galperin1994,Hessling1995}.  
It was recently applied to the analysis of decoherence in qubits 
\cite{Paladino2002,Paladino2003b,Falci2003,Galperin2003a,Falci2004,Falci2005,Galperin2006,Martin2006,Bergli2006,Bergli2009}.   
Various quantum and non-Markovian aspects of the model were addressed by 
\textcite{DiVincenzo2005,Lutchyn2008,Coish2008,Grishin2005,Abel2008,Burkard2009,Culcer2009,Yurkevich2010,Paladino2002,Galperin2003,Sousa2005}.

\subsubsection{Exact results at pure dephasing}
\label{subsub:SPpuredeph}

In Sec.~\ref{why} we discussed a simple model of so-called pure dephasing, when the diagonal splitting
$\epsilon$ of a qubit represented by the Hamiltonian (\ref{qb1}) fluctuates in time and $\Delta=0$. 
The resulting expression (\ref{vf3}) for
 the FID was obtained assuming that the fluctuations obey the Gaussian statistics.  To approach specific 
  features of the $1/f$ noise let us take into account that such a noise can be considered as superposition 
  of random telegraph processes. To begin with let us consider decoherence due to a single random telegraph 
  process.

\paragraph{Dephasing due to a single RT fluctuator --} \label{sec2E}
A random telegraph process is defined as follows \cite{Kirton1989,Buckingham1989}. Consider a classical stochastic 
variable $\chi (t)$, which at any time takes the values $\chi (t) = \pm 1$.  It is thus suitable for describing 
a two-state fluctuator that can find itself in one of two (meta)stable states, 1 and 2, and once a while it 
makes a switch between them.  The switchings are assumed to be uncorrelated random events with rates 
$\gamma_{1\to 2}$ and $\gamma_{2\to 1}$, which can be different.  For simplicity, we will limit ourselves to 
symmetric RT process: $\gamma_{1\to 2}=\gamma_{2\to 1}=\gamma/2$.  The extension to the general case can be easily 
made, see, e. g., \cite{Itakura2003}. The number $k$ of switches that the fluctuator experiences within a time 
$t$ follows a Poisson distribution
\begin{equation} \label{Poi1}
P_k(t)=\frac{(\gamma t)^k}{2^k k!}e^{-\gamma t/2}\, .
\end{equation}
The number of switches, $k$, determines the number of times the function $\chi (t)$ changes its sign 
contributing $(-1)^k$ to the correlation function, $C(t) \equiv \av{\chi(t) \chi (0)}$. Therefore, 
\begin{equation} \label{cf01}
C(t)=e^{-\gamma t/2}\sum_{k=0}^\infty (-1)^k\frac{(\gamma t)^k}{2^k k!}=e^{-\gamma t}, \quad t \ge 0\, .
\end{equation}
The RT process results in a fluctuating field $\bb (t) = \bb \chi (t)$ applied to the qubit. The magnitude, 
$b=|\bb|$, together with the switching rate, $\gamma$, characterizes the fluctuator.  Using Eq.~(\ref{ns1}) we 
find power spectrum of the noise generated by $i$-th fluctuator:
\begin{equation} \label{ns2}
S_i(\omega)=b_i^2 \cL_{\gamma_i} (\omega)\, ,
\end{equation}
with $\cL_{\gamma_i} (\omega)$ given by Eq.~(\ref{Lorentzian}).
This expression corresponds to the high temperature limit, $E \ll k_BT$, of Eq.~\eqref{sd1} with
$x_1-x_2=2 b_i$.
Thus, together with the considerations reported in Sec.~\ref{subsection2B}, we conclude that telegraph 
fluctuators provide a reasonable model for the $1/f$ noise.

\vspace{-0.1in}
\paragraph*{{Single shot measurements and FID} --}
Using Eq.~(\ref{bv1}) let us now discuss how a single RT fluctuator effects a qubit.
We assume that the \fl does not feel any feedback from the qubit and thus the RT function $\chi (t)$ 
equals to $+1$  or $-1$ with the probability 1/2 regardless to the direction of the Bloch vector 
$\mathbf{M}$. One can show  that under this assumption the probability to find the angle $\varphi$ 
(see Fig.~\ref{figAGB1}) at time $t$, $p (\varphi,t)$,  satisfies the second-order differential 
equation~\cite{Bergli2009}: 
\begin{equation} \label{prob01}
 \ddot{p} +\gamma \dot{p} = b^2 \partial^2_\varphi p\, ,
\end{equation}
which is known as the telegraph equation. We can always choose the $x$-direction such that $\varphi =0$ at 
$t=0$, so $p(\varphi, 0)=\delta (\varphi)$.  The second initial condition, 
$$\dot{p}(\varphi, 0)=\pm 2b \, \partial_\varphi p(\varphi,0), $$ can be derived from the integral equation for 
$p(\varphi, t)$, see \cite{Bergli2009}. After averaging over the fluctuator's initial state, $\dot{p}(\varphi, 0) = 0$. 

The FID is given as $\av{m_+}=\av{e^{i\varphi}} = \int d\varphi \, p(\varphi,t) e^{i\varphi}$.  The differential 
equation and initial conditions for this quantity can be obtained multiplying Eq.~(\ref{prob01}) by $e^{i \varphi}$ 
and integrating over $\varphi$. In this way we obtain:
\begin{equation} \label{FID01}
\av{\ddot{m}_+} + \gamma \av{\dot{m_+}} = -b^2  \av{m_+} 
\end{equation}
with initial conditions $  \av{ m_+ (0)}=1$, $\av{\dot{m}_+(0)}=0$.  The solution of Eq.~(\ref{FID01}) with these 
initial conditions is
\begin{eqnarray} \label{FID02}
\av{m_+}&=&(2\mu)^{-1}e^{-\gamma t/2}\left[(\mu +1)e^{\gamma \mu t/2}+ (\mu -1)e^{-\gamma \mu t/2}\right]\, , \nonumber \\
\mu &\equiv& \sqrt{1-(2b/\gamma)^2}\, .
\end{eqnarray}
In the context of decoherence due to discrete noise sources affecting superconducting qubits
this expression was firstly reported in \cite{Paladino2002} where it has been derived as the high-temperature 
limiting form of a real-time path-integral result. Equation~(\ref{FID02}) also  follows by
direct averaging the qubit coherence over the stochastic bistable process $\bb (t) = \bb \chi (t)$,
i.~e.,  by evaluation of the time average $\mathcal Z(t)= \langle \langle \exp{[-i \int_0^{t} dt' b(t')]} \rangle
\rangle$. 
The last two approaches do not necessarily assume a thermal equilibrium initial condition for the fluctuator at the
initial time $t=0$. Thus, results also depend on the initial population difference between the two
states $\chi= \pm 1$, $\delta p_0$. When this quantity is fixed to one of the classical values $+1$ or $-1$, 
we obtain the effect of noise only {\em during} the qubit quantum evolution, i.~e., in an ``ideal" single shot measurement.
The result takes the form given by Eq.~(\ref{FID02}) with the prefactors of the two exponentials, 
$\mu \pm 1$, replaced by $\mu \pm 1 \mp i \delta p_0 2b/\gamma$.
The FID results by further averaging over the initial conditions, i.~e., by replacing $ \delta p_0$ with
$\langle \delta p_0 \rangle= \delta p_{\text{eq}}$, where the thermal equilibrium value $\delta p_{\text{eq}}$ is
consistently set to zero in the regime $E \ll k_B T$.
Both the single shot and the free induction signals demonstrate qualitatively different behaviors
for large and small values of the ratio $2b/\gamma$. This is easily seen from  Eq.~(\ref{FID02}).
At $b \gg \gamma$ one can consider the qubit as a quantum system
experiencing beatings between the states with different splittings, $B_0 \pm b$, the width of these
states being $\gamma/2$. In the opposite limiting case, $b \ll \gamma$, the energy-level splitting is 
self-averaged to
a certain value, the width being $b^2/\gamma$. This situation was extensively discussed in connection
with the magnetic resonance and is known as the \textit{motional narrowing of spectral lines} \cite{Klauder1962}. 
\begin{figure}[t]
\centerline{
\includegraphics[width=5cm]{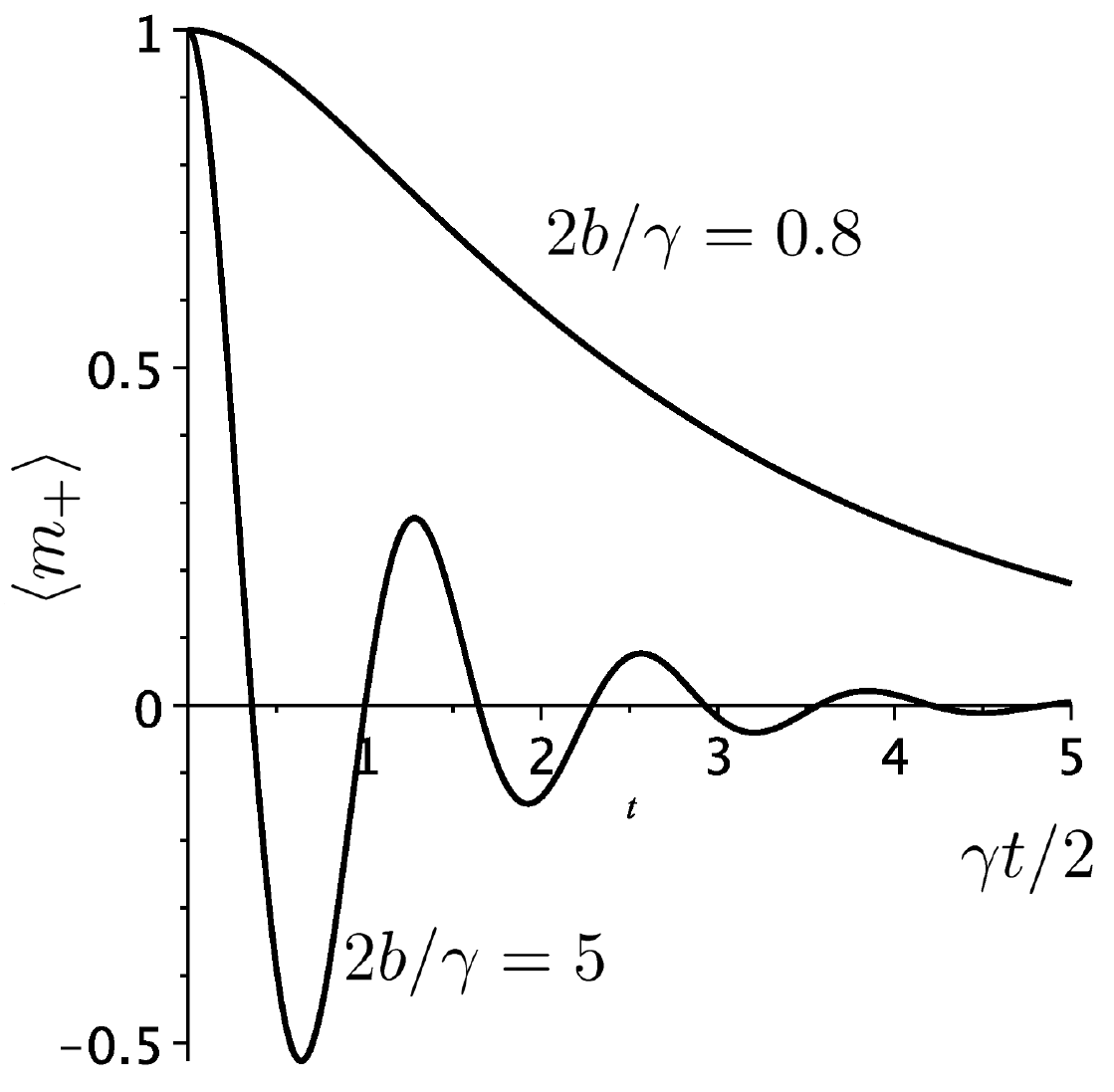}
}
\caption{Time dependence if the FID amplitude for $2b/\gamma = 0.8$ and 5.\label{figFID}}
\end{figure}
Different behaviors of the FID amplitudes depending on the ratio $2b/\gamma$ are  illustrated in 
Fig.~\ref{figFID}. 

\begin{figure}[t]
\centerline{
\includegraphics[width=0.9\columnwidth]{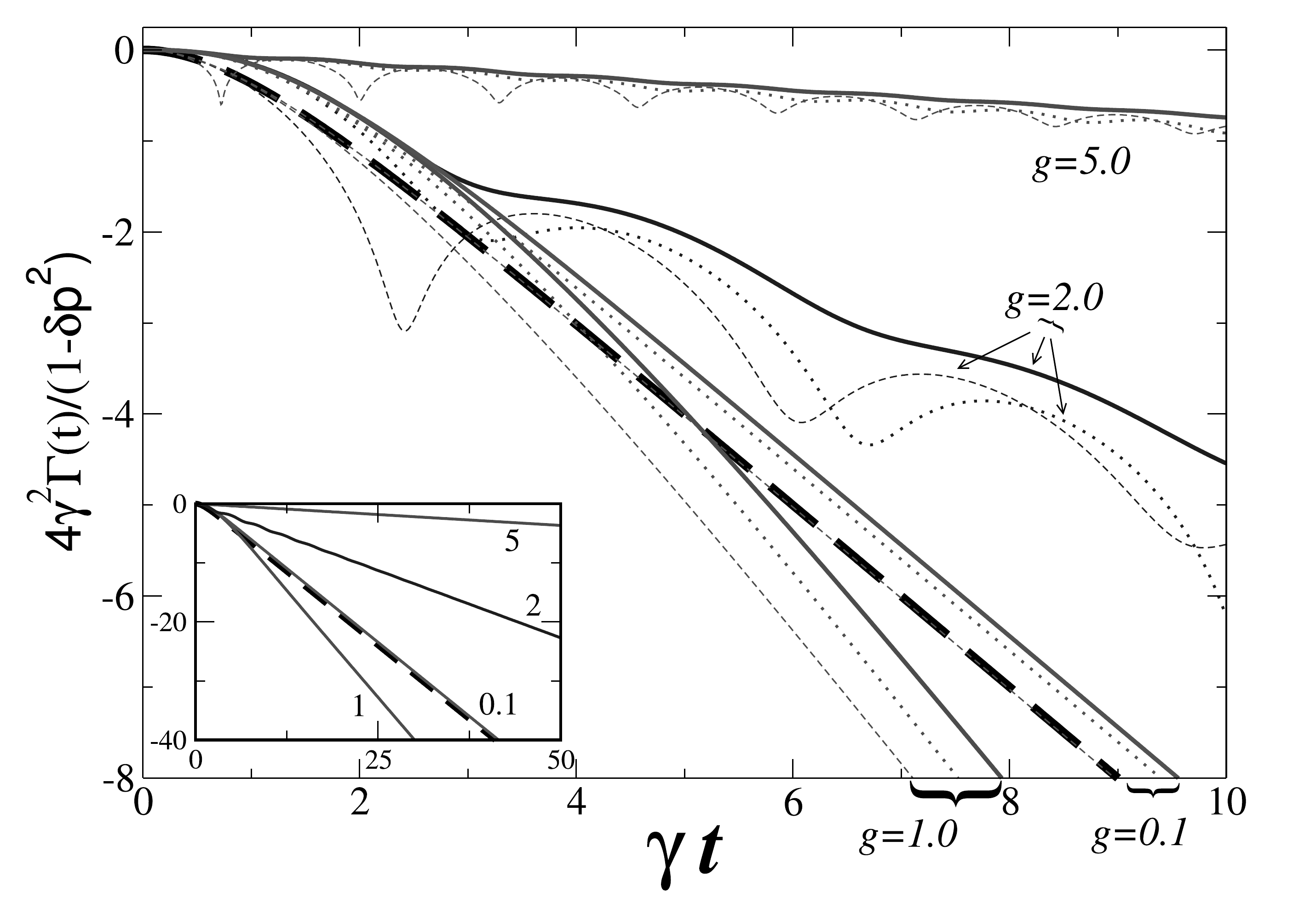}
}
\caption{Reduced $\Gamma(t)$ due to \fs prepared in a stable state ($\delta p_0=1$ solid lines and 
$\delta p_0=-1$ dotted lines) and in a thermal mixture ($\delta p_0=0$ dashed lines) for the indicated
values of $g=2 b/\gamma$. Inset: longer time behavior for stable state preparation. The curves are
normalized in a such a way that the oscillator approximation for all of them coincides (thick dashed line).
Adapted from \cite{Paladino2003b}, with kind permission of Springer Science+Business Media.}
\label{fig2_AdvSolStPh2003}
\end{figure}

Thus, one can discriminate between {\em weakly coupled} fluctuators
$2b/\gamma \ll 1$ and {\em strongly coupled} fluctuators (in the other regimes)~\cite{Paladino2002},
which influence the qubit in different ways. 
In Fig.~\ref{fig2_AdvSolStPh2003}, adapted from \cite{Paladino2003b}, we 
show the decay factor 
$\Gamma(t) = - \ln[|\av{m_+}(t)|]$ for different values of $2b/\gamma$ normalized in such a way that 
the Gaussian decay factor $\Gamma^{G}(t)\equiv \av{\varphi^2}/2$, see Eqs.~\eqref{avfi} and \eqref{vf3},
is the same for all curves. We report both the single-shot and the FID signals.
Substantial deviations are clearly observed, except in the presence of a weakly coupled fluctuator. In
particular, a \fl with $2b/\gamma > 1$ induces a slower dephasing compared to an oscillator environment
with the same $S_b(\omega)$, a sort of saturation effect. Recurrences at times comparable with $1/2b$ are
visible in $\Gamma(t)$. In addition, strongly coupled \fs show memory effects. This is clearly seen
considering different initial states for the fluctuator, corresponding to the single-shot and FID measurement
schemes. 
This is already seen in the short-times behavior, $\gamma t \ll 1$, relevant for quantum operations which
need to be performed before the signal decays to a very low value. 
In the limit $\gamma t \ll 1$, 
$\Gamma^{G}(t) \approx b^2 (1- \delta p_{\text{eq}}^2) t^2/2$, whereas 
\begin{equation}
\Gamma(t) \approx b^2\left[\frac{1- \delta p_{0}^2}{2}\, t^2+ \frac{1+2 \delta p_{0} \delta p_{\text{eq}} - 
3 \delta p_{0}^2}{6}\,  \gamma  t^3\right] .
\label{gamma_short_times}
\end{equation}
In a single-shot process, $\delta p_{0}=\pm 1$, and $\Gamma(t) \propto t^3$, showing that a fluctuator is stiffer
than a bath of oscillators. On the other side, for repeated measurements $\delta p_{0}= \delta p_{\text{eq}}$ for very
short times we  recover the Gaussian result. This is due to the fact that inhomogeneous broadening due
to the uncontrolled preparation of the \fl at each repetition adds to the effect of decoherence during the
time evolution and results in a faster decay of $\Gamma(t)$. 

\begin{figure}[t]
\centerline{
\includegraphics[width=0.6\columnwidth]{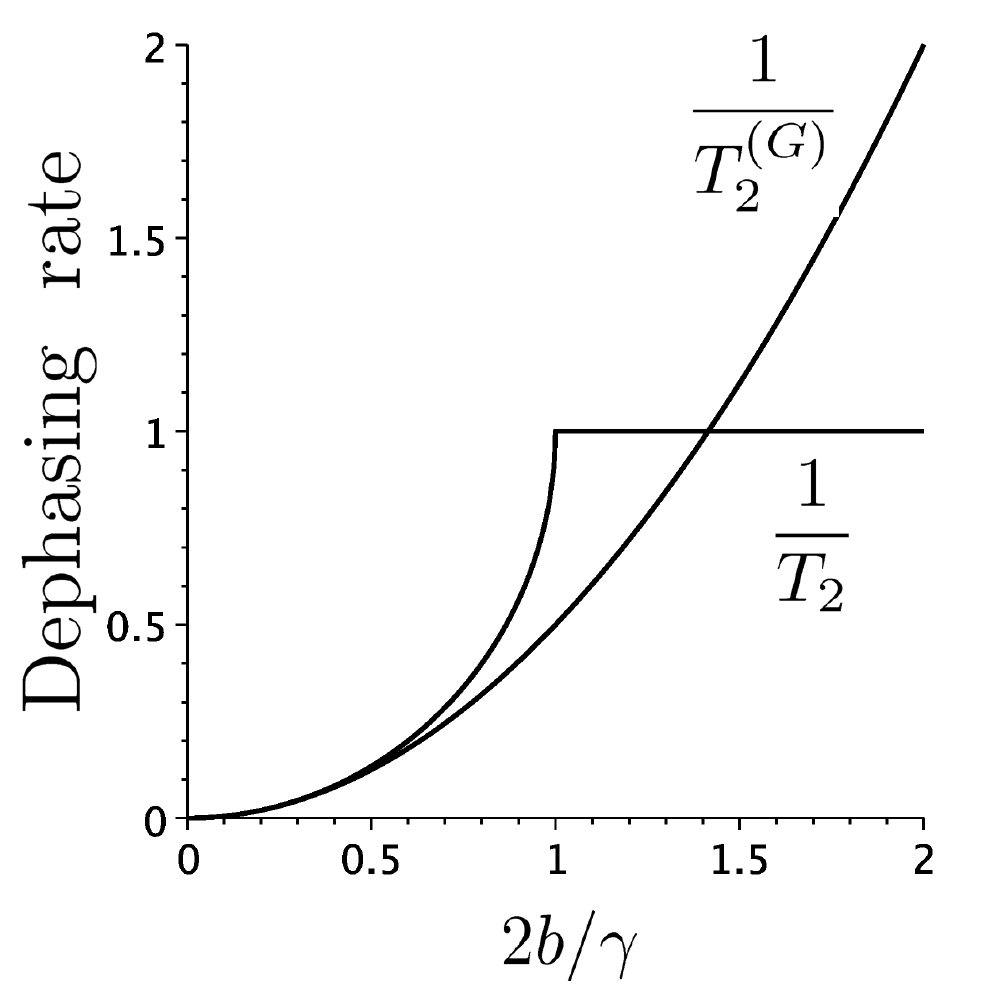}}
\caption{Comparison of the dephasing rate $T_2^{-1}$ for a single random telegraph process and the corresponding 
Gaussian approximation. Adapted from \cite{Bergli2009}.\label{figGnG}}
\end{figure}

\begin{figure}[t]
\centerline{
\includegraphics[width=0.6\columnwidth]{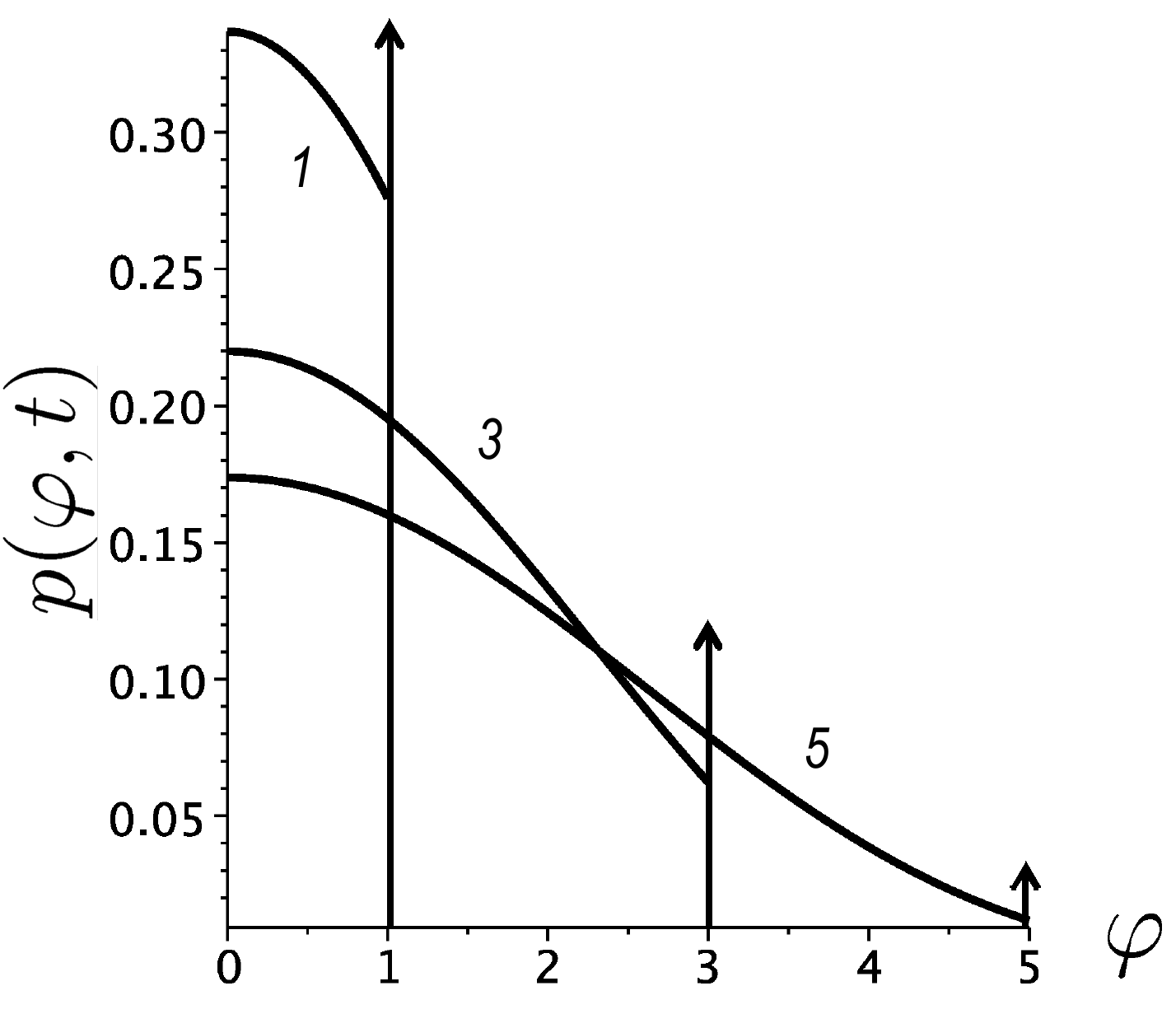}
}
\caption{The distribution (\ref{df2}) for $2b/\gamma =1$ and $\gamma t/2 = 1$, $3$, and $5$ (numbers  at the curves). 
Only the part with positive $\varphi$ is shown; the function is symmetric.  The arrows represent the 
delta-functions (not to scale). Adapted from \cite{Bergli2009}. \label{figBergli2006}}
\end{figure}

The difference between Eq.~(\ref{FID02}) and the result (\ref{t21}) based on the Gaussian 
assumption is further elucidated considering the long-time behavior.  
Substituting Eq.~(\ref{ns2}) for the noise spectrum into Eq.~\eqref{t21} one obtains
\begin{equation} \label{t2G}
1/T_2^{*(G)}=b^2/\gamma \, .
\end{equation} 
This result should be valid for times much longer than the correlation time of the noise, which is 
$\gamma^{-1}$.  Expanding Eq.~(\ref{FID02}) at long times we find that FID also decays exponentially (or the 
beatings decay exponentially at $b > \gamma$). However, the rate of decay is parametrically different from
 Eq. (\ref{t2G}):
\begin{equation} \label{t2nG}
\frac{1}{T_2^*}=\frac{\gamma}{2}\left(1- \Re \sqrt{1-\frac{4b^2}{\gamma^2}}\right)\, .
\end{equation}
At $b \ll \gamma$ Eq.~(\ref{t2nG}) coincides with the Gaussian result. Shown in Fig.~\ref{figGnG} are the 
dephasing rates $1/T_2^{(G)}$ and $1/T_2$ given by equations (\ref{t2G}) and  (\ref{t2nG}), respectively. 
Again we see that the Gaussian approximation is valid only in the limit $b \ll \gamma$. 
The main effect of a strongly coupled fluctuator is a static energy shift, the contribution to
the qubit decoherence rate saturates at $\sim \gamma$ and at $b \gtrsim \gamma/2$ 
the Gaussian assumption overestimates the decay rate.
Apparently, this conclusion is in contradiction with the discussion in Sec. \ref{why} following from the central limiting theorem.  According to this theorem $p(\varphi , t)$ 
\textit{always} tends to a Gaussian distribution with time-dependent variance provided that the time exceeds 
the correlation time of the noise.
To resolve this apparent contradiction, let us analyze the shape of the distribution function, 
$p(\varphi , t)$, following from the telegraph equation (\ref{prob01}) with initial conditions
$p(\varphi,0)=\delta(\varphi)$, $\dot{p}(\varphi,0)=0$.  This solution is \cite{Bergli2006}:
\begin{eqnarray} \label{df2}
p(\varphi, t)&=& (1/2)e^{-\gamma t/2} \left[ \delta (\varphi + bt)+  \delta (\varphi - bt)\right] \nonumber \\
&&+[\gamma/b\nu(t)] e^{-\gamma t/2}\left[ \Theta (\varphi + bt) -  \Theta (\varphi - bt)\right] \nonumber \\
&& \quad \times \left\{ I_1[\nu (t) \gamma t/2]+ \nu (t) I_0[\nu (t) \gamma t /2] \right\} \, .
\end{eqnarray}
Here $I_\beta (x)$ is the modified Bessel function, $\nu (t)=\sqrt{1-(\varphi /bt)^2}$, while 
$\Theta (x) =1$ at $x>0$ and 0 at $x <0$ is the Heaviside step function.  This distribution for 
various $t$ shown in Fig.~\ref{figBergli2006} consists of two delta-functions and a central peak.
The delta-functions represent the finite probability for a \fl to remain in the same state during time 
$t$. As time increases, the weight of the delta-functions decreases and the central peak broadens. 
 At long times, this peak acquires a Gaussian shape. Indeed, at $\gamma t \gg 1$ one can use asymptotic 
 form of the Bessel function, $I_\beta (z)  \approx(2\pi z)^{-1/2} e^z$, as $z \to \infty$. In addition, 
 at $t \gg \varphi/b$  we can also expand $\sqrt{1-(\varphi/bt)^2}$ and convince ourselves that the central 
 peak is described by the Gaussian distribution (\ref{Gauss1}) with $\av{\varphi^2}=2b^2t/\gamma$.  
 If the qubit-\fl coupling is weak, $b \ll \gamma$, this Gaussian part of $p(\varphi,t)$ dominates the average 
 $\av{e^{i\varphi}}$ and the Gaussian approximation is valid.  Contrary, when the coupling is strong, 
 $b > \gamma/2$, the average is dominated by delta-functions at the ends of the distribution and the decoherence 
 demonstrates a pronounced non-Gaussian behavior, even at long times ($t >2/\gamma$).

Unfortunately, we are not aware of a way to measure the distribution $p(\varphi,t)$ in experiments with a 
single qubit.
The reason is in the difference between the qubit that can can be viewed as a pseudospin 1/2 and a classical 
Bloch vector $\mathbf{M}$. According to Eq.~(\ref{dm1}) the components $M_x$, $M_y$, $M_z$ of $\mathbf{M}$ are 
connected with the mean component of the final state of the pseudospin. Therefore, to measure the
value of the phase $\varphi$  (argument of $m_+$) that corresponds to a given realization of the noise 
one should repeat the experimental shot with the \textit{same realization of the noise} many times.
 This is impossible because each time the realization of noise is different. Therefore, the only observable 
 in decoherence experiments is the average $\av{e^{i\varphi}}$. 

\begin{figure}[t]
\centerline{
\includegraphics[width=0.55\columnwidth]{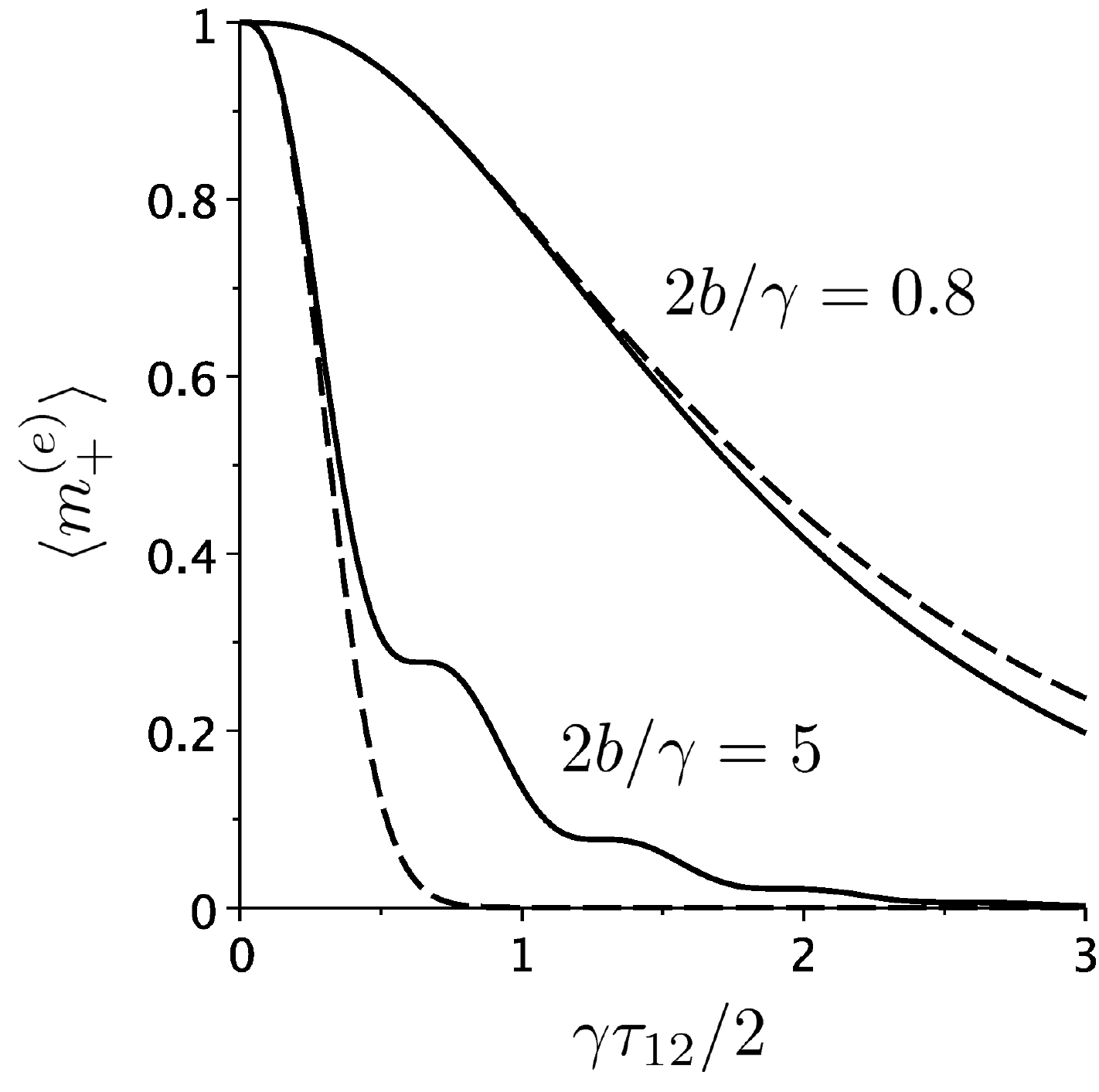}
}
\caption{Solid lines -- echo signal for different values of the ratio $2b/\gamma$, Eq.~(\ref{Ec001}). 
Dashed lines -- calculations along the Gaussian approximation, Eq.~(\ref{Gauss003}). 
 Adapted from \cite{Bergli2009}. \label{figEcho2}}
\end{figure}

\vspace*{-0.1in}
\paragraph{Echo --}
The analysis of the echo signal is rather similar: one has to replace $\av{m_+(t)}$ taken from Eq.~(\ref{FID02}) 
by $\av{m_+^{(e)} (2\tau_{12})}$ where, cf. with  \cite{Laikhtman1985},
\begin{eqnarray} \label{Ec001}
\av{m_+^{(e)}(t)}&=&\frac{e^{-\gamma t/2}}{2\nu^2}\left[ (\nu +1)e^{\gamma \nu t/2} 
\right. \nonumber \\ && \left.
- (\nu -1)e^{-\gamma \nu t/2}-8b^2/\gamma^2 \right] \, .
\end{eqnarray}
This result can also be obtained by direct calculation of the function 
$\mathcal Z(2t|\eta)= \langle \langle \exp{[-i \int_0^{2t} dt' \eta(t') b(t')]}\rangle \rangle$
where $\eta(t')= 1$ for $0<t'<t$ and $\eta(t')= -1$ for $t<t'<2t$ \cite{Falci2003}.
To demonstrate non-Gaussian behavior of the echo signal  let us evaluate from Eq.~(\ref{echo2}) the 
variance $\av{\psi^2(2\tau_{12})}$. We obtain, cf. with \cite{Bergli2009},
\begin{equation}\label{Gauss003}
\av{\psi^2}=(2b^2/\gamma^2)
\left( 2\gamma \tau_{12}-3 +4e^{-\gamma \tau_{12}}-e^{-2\gamma \tau_{12}}\right)
\end{equation}
with $\av{m_+^{(e)}}=e^{-\av{\psi^2}/2}$ as before. 
The comparison between the two results is shown in Fig.~\ref{figEcho2} for a weakly ($2b/\gamma=0.8$) 
and a strongly ($2b/\gamma=5$) coupled fluctuator.
As discussed above, the Gaussian approximation is accurate at $2b \lesssim \gamma$, while at
 $2b > \gamma$ the results are qualitatively different. In particular, the plateaus in the time dependence
  of the echo signal shown in Fig.~\ref{figEcho2} are beyond the Gaussian approximation. We believe that 
  such a plateau was experimentally observed by \textcite{Nakamura2002} (see Fig.~3 from that paper partly 
  reproduced in Fig.~\ref{fig:Nak2002}).  In the limit $b \gg \gamma, \sqrt{\gamma /\tau }$ equation (\ref{Ec001})
   acquires a simple form:
\begin{equation} \label{Ec003}
\av{m_+^{(e)}(\tau)}=e^{-\gamma \tau}\left[ 1+(\gamma/2b)\sin 2b \tau\right]\, .
\end{equation}
According to this expression, the plateau-like features ($|\partial \av{m_+^{(e)}}/\partial \tau | \ll 1$) 
occur at $b\tau \approx k\pi$ (where $k$ is an integer) and their heights 
$\av{m_+^{(e)}} \approx e^{-\pi k \gamma/b}$ exponentially decay with the number $k$. Measuring experimentally
 the height and the position of the first plateau, one can determine both the \fl coupling strength $b$ and 
 its switching rate $\gamma$. For example, the echo signal measured by \textcite{Nakamura2002} shows a plateau-like 
 feature at $\tau_{12}= 3.5$ ns at the height $\av{m_+^{(e)}}=0.3$, which yields $b \approx 143$ MHz and
  $\gamma/2 \approx 27$ MHz. If the \fl is a charge trap near a gate producing a dipole electric field, 
  its coupling strength is $b=e^2(\mathbf{a \cdot r})/r^3$. Using the gate-CPB distance $r=0.5$ $\mu$m, 
  we obtain a reasonable estimate for the tunneling distance between the charge trap and the gate, 
  $a \sim 20$ {\AA}. A more extensive discussion can be found in \cite{Galperin2006}.
A similar analysis for arbitrary qubit working point has been reported in \cite{Zhou2010}.
The effect of RTN for arbitrary working point will be discussed in Sec.~\ref{subsub:SFgeneral}.

\paragraph{Telegraph noise and Landau-Zener transitions --}

Driven quantum systems are exceedingly more complicated to study than stationary systems, and only few 
such problems have been solved exactly. An important exception is the Landau-Zener (LZ) transitions 
\cite{Landau1932,Zener1932,Stueckelberg1932}. In the conventional LZ problem, a TLS is driven by changing an 
external parameter in such a way that the level separation $\hbar \epsilon$ is a linear function of time, 
$\epsilon(t)\equiv a^2t$. Close to the crossing point of the two levels, an interlevel tunneling matrix element
 $\hbar \Delta$ lifts the degeneracy in an avoided level crossing. When the system is initially in the ground 
 state, the probability to find it in the excited state after the transition is $e^{-\pi \Delta^2/2a^2}$. 
 Hence, a fast rate drives the system to the excited state, while the system ends in the ground state when driven slowly.

In connection with decoherence of qubits, there has recently been increased interest in Landau-Zener 
transitions in systems coupled to an environment. This problem is both of theoretical interest and of 
practical importance for qubit experiments \cite{Sillanpaa2006}. 
The noisy Landau-Zener problem has been discussed by several
authors \cite{Kayanuma1984,Kayanuma1985,Shimshoni1991,Shimshoni1993,Wubs2006,Saito2007,Nishino2001,Pokrovsky2007,Pokrovsky2003}. 
Here we will discuss the role of a classical telegraph fluctuator following \textcite{Vestgarden2008}.

Let us model a driven qubit by the Hamiltonian (\ref{qb2}) with $\epsilon(t)=a^2t$ and $\Delta (t)\equiv \Delta + v\chi(t)$
 where $\chi(t)$ is a RT process. Representing the density matrix as in Eq.~(\ref{dm1}) one can
  study the dynamics of the Bloch vector averaged over the realizations of the telegraph noise.  To this end, one has 
  to study the solution of the Bloch equation (\ref{bv1}) in the presence of the ``magnetic fields'' 
  $\bB_\pm = \bB_0 \pm \mathbf{v} (t)$. 
  Introducing the partial probabilities $p_\pm(\bM,t)$  to be in state $\bM$ at time $t$ under rotations around 
  $\bB_+$ and $\bB_-$, respectively and defining 
$$\bM_\pm \equiv \int d^3 M\, [p_+(\bM,t)\pm  p_-(\bM,t)] $$
one arrives at the set of equations, cf. with \cite{Vestgarden2008}
\begin{eqnarray} \label{LZ01}
\dot{\bM}_+&=&\bM_+\times \bB_0 + \bv\times \bM_-\, , \nonumber \\
\dot{\bM}_-&=&\bM_-\times \bB_0 + \bv\times \bM_+ -\gamma \bM_- \, .
\end{eqnarray}
Here $\bM_+$ is the average Bloch vector whose $z$-component is related to the occupancy of the qubit's levels. 
To demonstrate the role of  telegraph noise, let us consider a simplified problem putting $\Delta =0$. In 
this way we will study the LZ transitions induced solely by the telegraph process.  In this case, 
the Eqs.~(\ref{LZ01})  reduce to an integro-differential equation for $M_z \equiv M_{+z}$:
\begin{equation}\label{LZ02} \!
\dot{M}_z(t)=- v^2\! \int_{0}^\infty  \! \!  \! \! dt_1 \cos \left[\frac{a^2t_1(2t - t_1)}{2}\right] e^{-\gamma |t_1|} M_z(t-t_1).
\end{equation}
This equation can be easily analyzed for the case of \textit{fast} noise, $\gamma \gg a$. A series expansion in 
$t_1$ leads to the following solution of Eq.~\eqref{LZ02} with  initial condition $M_z(0)=1$:
\begin{equation}\label{LZ03}
M_z (\infty)=e^{-\pi v^2/a^2}\, .
\end{equation}
This is the usual expression for the LZ transition probability with replacement $\Delta \to \sqrt{2} v$. 
This result holds for any noise correlated at short times ($\ll a^{-1}$), as shown by \textcite{Pokrovsky2003}.
 Note that the result (\ref{LZ03}) can be obtained using the Gaussian approximation. 

\begin{figure}[t!]
\centerline{
\includegraphics[width=0.7\columnwidth]{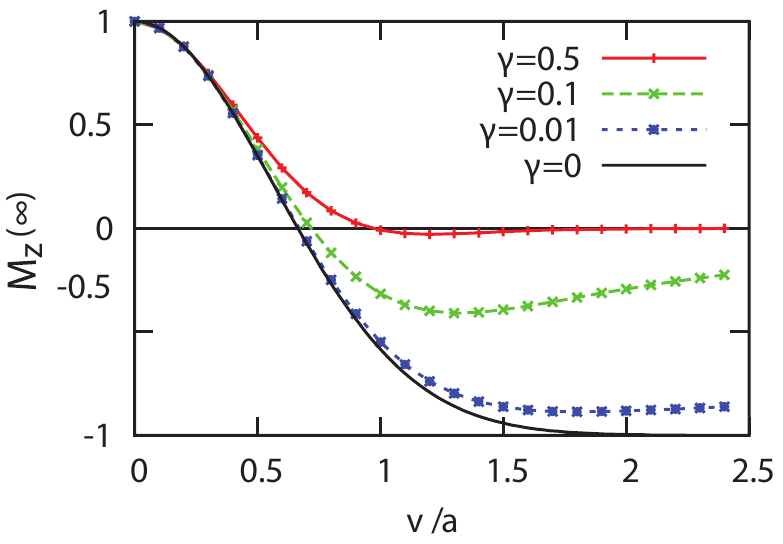}
}
\caption{(Color online) The $M_z(\infty)$ versus the ratio $v/a$ for different switching rate $\gamma$ (shown in units of $a$). 
 Adapted from \cite{Vestgarden2008}. \label{figLZ1}}
\end{figure}

The case of a ``slow" \fl with $\gamma \lesssim a$ leads to very different results which, for an arbitrary 
ratio $v/a$, require a numerical solution of the integro-differential equation (\ref{LZ02}). The result of 
such analysis is shown in Fig.~\ref{figLZ1} adapted from \cite{Vestgarden2008}.
The static case $\gamma =0$  is equivalent to the standard Landau-Zener transition with the noise 
strength $\sqrt{2}v$  replacing the tunnel coupling $\Delta$ between the diabatic levels. In the 
adiabatic limit, $v \gg a$, we see that $M_z(\infty) \to - 1$, which corresponds to the transition to 
the opposite diabatic state.  In this case the dynamics is fully coherent, but since we
average over the fluctuator's initial states, the Bloch vector asymptotically lies on the $z$ axis of the Bloch sphere.
At finite $\gamma$, the noise also stimulates Landau-Zener transitions, however
 the transition probability decreases with increasing switching rate. Note that the curves cross the line
  $M_z (\infty)=0$ at some $v/a$,  which depends on $\gamma$. For this noise strength, the final state is 
  at the  center of the Bloch sphere, corresponding to full decoherence. However,  increasing 
  the noise strength beyond this point results in a final state with negative $M_z(\infty)$. Thus, we 
  have the surprising result that under some conditions, increasing the noise strength will also increase 
  the system purity after the transition.  The results for slow and strong noise cannot be obtained in the
   Gaussian approximation.  
These considerations point out that telegraph noises can facilitate LZ transitions.  
This processes may  prevent the implementation of protocols for adiabatic quantum computing.

\paragraph{Ensemble of fluctuators: effects of $1/f$ noise --}
As a result of the above considerations, we conclude that the role of a \fl in decoherence of a qubit 
depends on the ratio between the interaction strength, $b$, and the correlation time, $\gamma^{-1}$,
 of the random telegraph process. 
 {\em Weakly coupled} fluctuators, i.~e., ``weak" and relatively ``fast" \fs for which $2b/\gamma \ll 1$, 
 can be treated as a Gaussian noise acting on the qubit.  Contrary, the influence of {\em strongly coupled} 
 fluctuators, i.~e., ``strong" and ``slow" 
 \fs  for which $2b/\gamma \gtrsim 1$, is characterized by quantum beatings in the qubit evolution.  

As we have seen, a set of \fs characterized by the distribution function $\cP(\gamma) \propto 1/\gamma$ of
 the relaxation rates provides a realistic model for $1/f$-noise.  Therefore, it is natural to study in 
 which way the qubit is decohered by a sum of the contributions of many fluctuators, $b(t)=\sum_i b_i \chi_i (t)$.  
 A key question here is ``Which \fs -- weak or strong -- are responsible for the qubit decoherence?" In the 
 first case the noise can be treated as Gaussian while in the second case more accurate description is needed.

Here we will analyze this issue using a simple model assuming  that the dynamics of different \fs are 
not correlated, i.~e., 
$\av{\chi_i(t) \chi_j (t')}=\delta_{ij}e^{-\gamma_i |t-t'|}$.  Under this assumption, the average 
 $\av{m_+}$ is the product of the partial averages,
\begin{equation}\label{m+001}
\av{m_+ (t)}=\prod_i \av{m_{+i} (t)}=\exp \left(\sum_i \ln \av{m_{+i} (t)}\right) .
\end{equation}
Since the logarithm of a product is a self-averaging quantity, it is natural to approximate the sum 
of logarithms, $\sum_i \ln \av{m_{+i} (t)}$ by its average value, $-\cK_m(t)$, where
\begin{equation} \label{m+002}
\cK_m (t) \equiv - \overline{\sum_i \ln \av{m_{+i} (t)}}\, .
\end{equation}
Here, the bar denotes the average over both the coupling constants, $b$, of the \fs and their 
transition rates, $\gamma$. This expression can be further simplified  when the total number 
$\cN_T$ of thermally excited fluctuators is very large. Then we can replace $ \overline{\sum_i \ln \av{m_{+i} (t)}}$ 
by $\cN_T \overline{ \ln \av{m_{+} (t)}}$. Furthermore, we can use the so-called Holtsmark 
procedure~\cite{Chandrasekhar1943}, i.~e., to replace $ \overline{ \ln \av{m_{+} (t)}}$ by $\overline{\av{m_+}}-1$ 
assuming that in the relevant  time domain each $\av{m_{+i}}$ is close to 1.
Thus, $\cK_m(t)$ is approximately equal to \cite{Klauder1962,Galperin2003,Galperin2003a,Laikhtman1985}
\begin{eqnarray}\label{m+003}
\cK_m(t)&\approx& \cN_T \left[1- \overline{ \av{m_+ (t)}}\right] 
\nonumber \\ &=&
\int db\, d \gamma \, \cP(b,\gamma) \left[ 1-\av{m_+(b,\gamma|t)}\right] \, .
\end{eqnarray}
Here $\av{m_{+i}}$ is specified as $\av{m_+(b,\gamma|t)}$, which depends on the parameters $b$ and $\gamma$ 
according to Eq.~(\ref{FID02}). The free induction signal is then 
$\exp[-\cK_m(t)]$. 
Analysis of the echo signal is rather similar: one has to replace $\av{m_+(t)}$ taken from Eq.~(\ref{FID02})
with Eq.~(\ref{Ec001}). 

To evaluate the time dependence of the free induction or echo signal, one has to specify the distribution 
function of the coupling constants $b$, i.~e., partial contributions of  different \fs to the random 
``magnetic field" $b(t)$.  
To start with, we consider the situation where the couplings $b_i$ are
distributed with a  small dispersion around an average value $\overline b $.
Under these conditions the total power spectrum reads
\begin{equation} \label{ns_narrow_b}
S(\omega) = \overline{b^2} \int_{\gamma_m}^{\gamma_M} d\gamma \, \frac{P_0}{2 \gamma} \,  \cL_{\gamma} (\omega) 
\approx    \frac{{\mathcal A}}{\omega}\, ,
\end{equation}
where the amplitude  $\mathcal A$ can be expressed in terms of the number of \fs per noise
decade, $n_d= {\mathcal N}_T \ln{(10)}/\ln{(\gamma_M/\gamma_m)}$, as follows 
${\mathcal A} = \overline{b^2} P_0/4 = \overline{b^2}  n_d/(2 \ln(10))$.
The spectrum exhibits a crossover to $\omega^{-2}$ behavior at $\omega \sim \gamma_M$. 
\begin{figure}[h!]
\centerline{
\includegraphics[width=0.8\columnwidth]{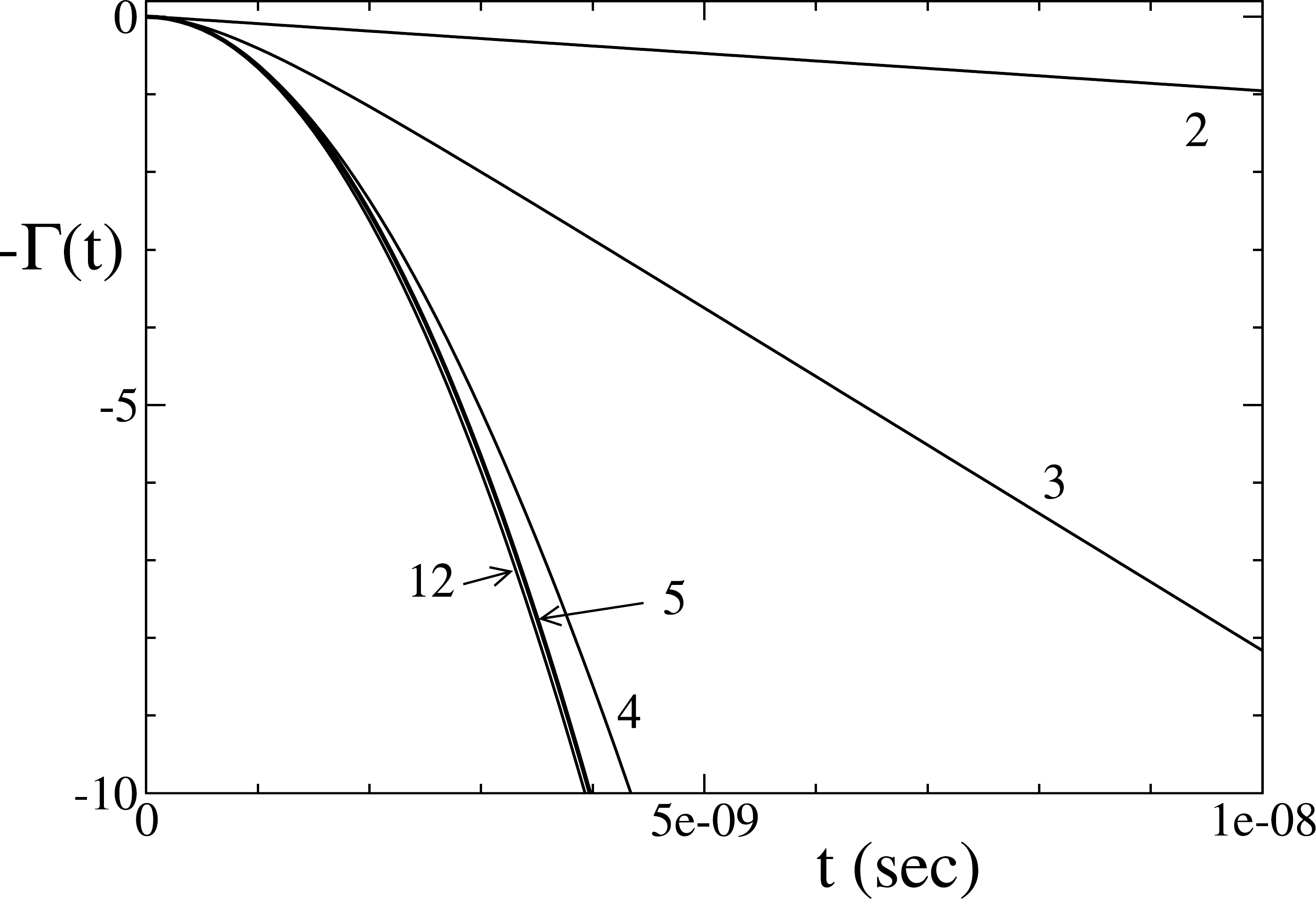}
}
\caption{Saturation effect of slow \fs for a 
$1/f$ spectrum and coupling distributed with $\langle \Delta b \rangle/\langle 
b \rangle =0.2$. Labels indicate the number of decades included.
Adapted from \cite{Paladino2002}.}
\label{fig:saturation}
\end{figure}
In Fig.~\ref{fig:saturation} we show the results for a sample with a number of \fs per decade
$n_d = 1000$ and with couplings distributed around $\overline{b}/(2 \pi) =  4.6 \times 10^7$~Hz. 
Initial conditions $dp_{0j}= \pm 1$ are distributed according to $\langle dp_{0j} \rangle= dp_{\text{eq}}$, 
the equilibrium value. The different role played by weakly and strongly coupled \fs 
is illustrated considering sets with $\gamma_M/(2 \pi) = 10^{12}$~Hz and different $\gamma_m$. 
In this case the dephasing is given by \fs with 
$\gamma_j > 2 \pi \times 10^7\ {\rm Hz} \approx  2 \overline{b} /10$. 
The main contribution comes from three decades at frequencies
around $\approx  2 \overline{b}$. The overall effect of the strongly coupled
\fs ($\gamma_j<   2 \overline{b}/10$) is minimal, despite their large number.
We remark that, while saturation of dephasing
due to a single \fl is physically intuitive, it is
not a priori clear whether this holds also for the $1/f$
case, where a large number ($\sim 1/\gamma$) of slow 
(strongly coupled)  fluctuators
is involved.
The decay factor $\cK_m(t)$ can be easily compared with the Gaussian approximation
(\ref{vf3})~\cite{Paladino2002}.
This approximation fails to describe \fs with $2b / \gamma \gg 1$. For instance,
$\langle \varphi^2 \rangle$ at a fixed $t$ scales with the number of decades
and does not show saturation. On the other side the Gaussian approximation
becomes correct if the environment has a very
large number of extremely weakly coupled fluctuators.
This is shown in Fig. \ref{fig:gaussian}
where the power spectrum is identical for all the curves but
it is obtained by sets of \fs with different $n_d$ and $b_i$.
The Gaussian behavior is recovered for $t \gg 1/\gamma_m$ if $n_d$ 
is large (all the \fs are weakly coupled). 
If in addition we take $\langle dp_{0j} \rangle=  dp_{\text{eq}}$, $\cK_m(t)$ approaches $\langle \phi^2 \rangle$ 
also at short times. Hence decoherence depends separately on $n_d$ and $\overline b$, whereas in the Gaussian 
approximation only the combination $n_d \overline{b^2}$, which enters $S(\omega)$, matters. 
In other words, the characterization of the effect of slow sources
of $1/f$ noise requires knowledge of moments of the bias
fluctuations higher than $S(\omega)$.
\begin{figure}[t]
\centerline{
\includegraphics[width=0.8\columnwidth]{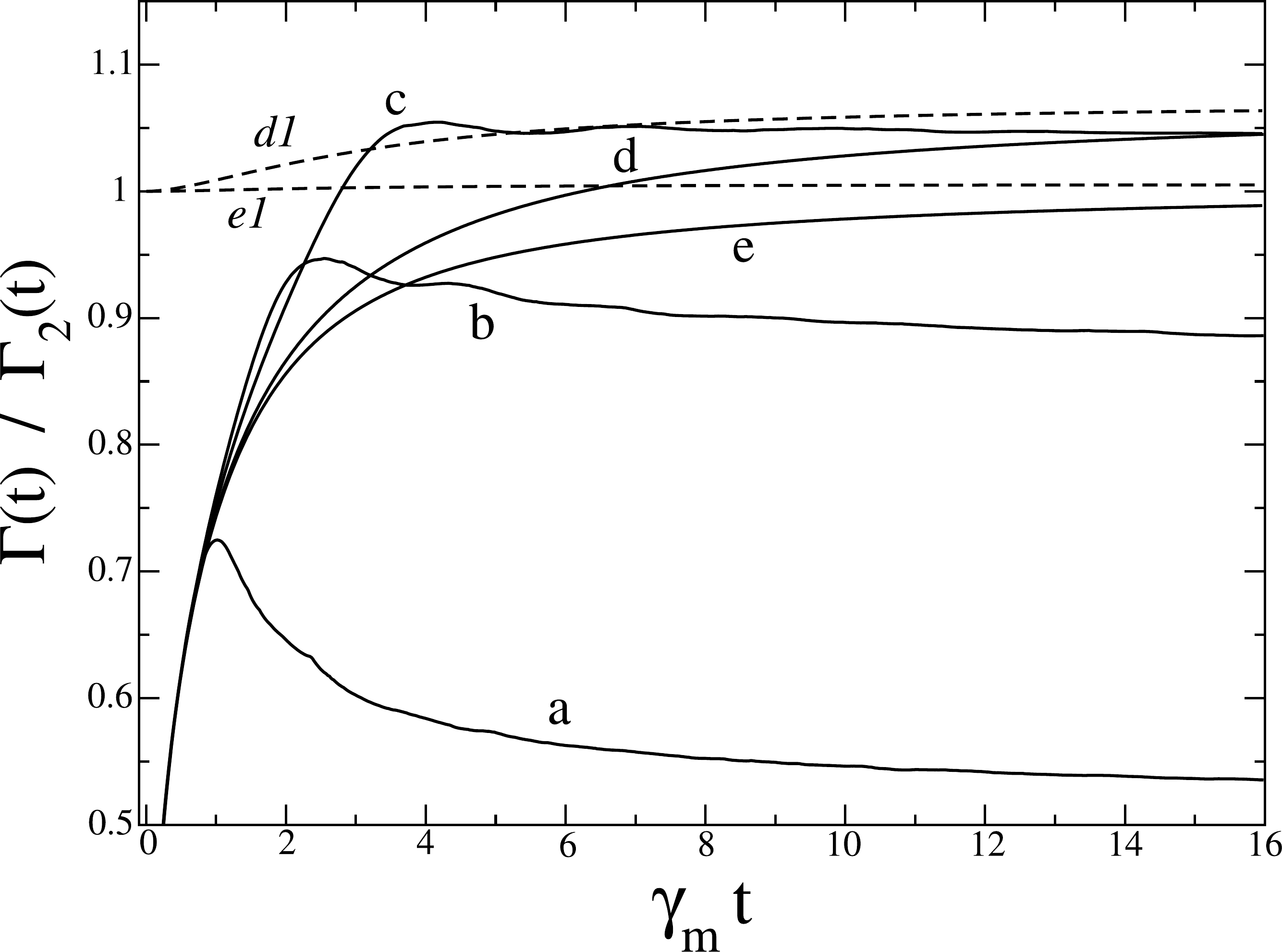}
}
\caption{Ratio $\cK_m(t)/\langle \phi^2 \rangle \equiv \Gamma(t)/\Gamma_2(t)$ for $1/f$ spectrum between 
$\gamma_m/(2 \pi)= 2 \times 10^7$~Hz
and $\gamma_M /(2 \pi)= 2 \times 10^9$~Hz with different numbers of \fs per decade:
(a) $n_d= 10^3$,(b) $n_d= 4 \times 10^3$, (c) $n_d= 8 \times 10^3$, (d) and (d1) $n_d= 4 \times 10^4$,
(e) and (e1) $n_d= 4 \times 10^5$. Solid lines correspond to $\delta p_{0j}= \pm 1$,
dashed lines correspond to equilibrium initial conditions (FID). Adapted from \cite{Paladino2002}. }
\label{fig:gaussian}
\end{figure}
A more accurate averaging procedure considering $\langle \delta p_{0j} \rangle = \pm 1$
for \fs  with $\gamma_i t_m <1$ and $\langle \delta p_{0j} \rangle = \delta p_{\text{eq}}$ for $\gamma_i t_m >1$,
shows that, for long measurement times $ \overline b  t_m  \gg 1$, 
\fs with $\gamma < 1/t_m$ are saturated and therefore not effective, whereas other fluctuators, being averaged, 
give  for short enough times
\begin{equation}
\label{Gamma_Cottet}
\Gamma(t) \approx \int_{1/t_m}^\infty d \omega \,S(\omega) \left ( \frac{ \sin \omega t/2}{\omega/2}  \right )^2 \, .
\end{equation}
This is illustrated in Fig. \ref{fig4_AdvSolStPh2003}.
\begin{figure}[t!]
\centerline{
\includegraphics[width=0.8\columnwidth]{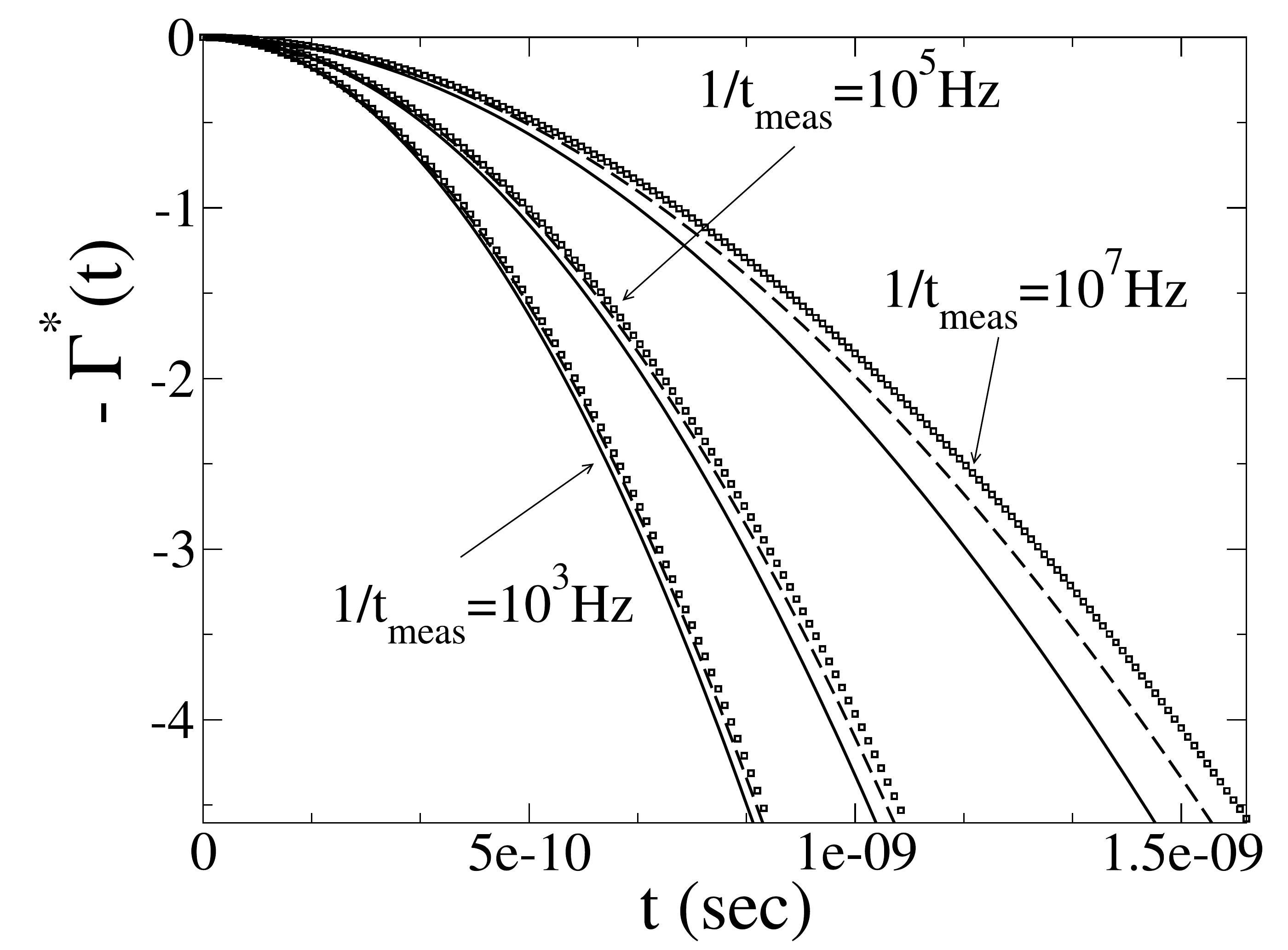}
}
\caption{Different averages over $\delta p_{0j}$ for $1/f$ spectrum reproduce the effect of repeated
measurements. They are obtained by neglecting (dotted lines) or account for (solid lines)
the strongly correlated dynamics of $1/f$ noise. The noise level of \cite{Nakamura2002} is used, by setting
$|\overline b|/(2 \pi)=9.2 \times 10^6$~Hz, $n_d=10^5$, $\gamma_m/(2 \pi)=1$~Hz, $\gamma_M/(2 \pi)=10^9$~Hz. Dashed lines are the
oscillator approximation with a lower cutoff at $\omega={\rm min}\{ |\bar b|,1/t_m\}$.
Adapted from \cite{Paladino2003b}, with kind permission of Springer Science+Business Media.}
\label{fig4_AdvSolStPh2003}
\end{figure}
 The above considerations explain the
often used approximate form (\ref{Gamma_Cottet}) 
obtained from Eq. (\ref{vf3}) with low-frequency cut-off at $1/t_m$
which was proposed in \cite{Cottet2001} and that we already mentioned in Section \ref{subsectionIIC}, Eq.~(\ref{eg2}).

The echo signal obtained from Eq. (\ref{m+003}) with (\ref{Ec001})
assuming that $\mathcal{P} (b,\gamma) \propto \gamma^{-1} \delta(b-\bar{b}) \Theta(\gamma_m-\gamma)$ leads to the following time
dependence when $\bar b \ll \gamma_M$~\cite{Galperin2007}
\begin{equation}
\label{echo_SF1}
{\mathcal K}^{(e)}(t) = \left\{ 
\begin{array}{ll}
 \mathcal A  \gamma_M \, t^3/6 &  \quad t \ll \gamma_M^{-1} \\
 \ln{(2)} \mathcal A  \,  t^2  &  \quad \gamma_M^{-1} \ll t \ll \bar b^{-1} \\
 \alpha (\mathcal A/ \bar{b^2}) \, \bar b t & \quad  \bar b^{-1} \ll t 
\end{array} \right. 
\end{equation}
where $\alpha \approx 6$. 
At small times, the echo signal behaves as in the  Gaussian approximation,
Eq.~(\ref{echo2}), where the $t^3$ dependence follows from the crossover of the spectrum from
$\omega^{-1}$ to $\omega^{-2}$. On the other side, at large times $ \bar b^{-1} \ll t$  the
SF model dramatically differs from the Gaussian result which predicts $\mathcal K^{(e)}_G(t) =  \mathcal A \ln(2) t^2$.
The origin of the non-Gaussian behavior comes from the already observed fact that
decoherence is dominated by the fluctuators with $\gamma \approx 2 \bar b$. Indeed, 
very ``slow" fluctuators produce slowly varying fields, which are effectively refocused in the course of the echo experiment. 
As to the ``too fast" fluctuators, their influence is reduced due to the effect of motional narrowing.
Since only the fluctuators with $\bar b \ll \gamma $ produce Gaussian noise, 
 the noise in this case is essentially non-Gaussian. 
 Only at times $ t \ll \bar b^{-1}$, which are too short for these most important fluctuators to switch, the decoherence 
 is dominated by the faster fluctuators contributions, and the Gaussian approximation turns out to be valid.
 Instead when  $\bar b \gg \gamma_M$ all fluctuators are strongly coupled and the long-time echo decay is 
 essentially non-Gaussian~\cite{Galperin2007}
\begin{equation}
\label{echo_SF2}
{\mathcal K}^{(e)}(t) = \left\{ 
\begin{array}{ll}
 \mathcal A   \, \gamma_M \, t^3/6   &  \quad  t \ll \bar b^{-1}, \\
4 (\mathcal A/ \bar{b^2}) \, \gamma_M  \, t & \quad  t \gg \bar b^{-1} .
\end{array} \right. 
\end{equation}
Thus, the long-time behavior of the echo signal depends on the details of the noise model, the dependence on
the high-frequency behavior of the spectrum being the first manifestation. Unfortunately,  the measured
signal is usually very weak at long times, making it difficult to figure out the characteristics of the noise
sources in the specific setup. Indeed, echo signal data measured on a flux qubit in the experiment by
\textcite{Yoshihara2006}  can be equally well fit assuming either  a Gaussian statistics of the noise
or a non-Gaussian model. The two models can in principle be distinguished  analyzing the different dependence
on the average coupling $\bar b$.
Details of this analysis are given in \cite{Galperin2007} where 
a fit with the SF model for different working points indicates that equations (\ref{echo_SF1}) for the case
$\bar b \ll \gamma_M$ give an overall better fit. A similar experiment on a flux qubit has been
reported in \cite{Kakuyanagi2007}. Data for the echo decay rate in that case have been interpreted 
assuming Gaussian fluctuations of magnetic flux, and consistency with this model has been observed
by changing the qubit  working point. 
In general, echo procedures allow extraction of relevant information about the
noise spectrum, like the noise amplitude~\cite{Yoshihara2006,Kakuyanagi2007,Galperin2007,Zhou2010}, the noise sensitivity 
defined by Eq.~(\ref{sensitivities}) \cite{Yoshihara2006} or the average change of the flux in the qubit loop
due to a single \fl flip \cite{Galperin2007}.
Similar possibilities can be provided by the real time qubit tomography~\cite{Sank2012}.

We now consider the role of the interaction strengths distribution,  $b_i$.
Using the same model as for TLS in glasses (see Sec.~\ref{subsec:models}) we  consider each \fl 
as a TLS with partial Hamiltonian
\begin{equation} \label{pH01}
\hat \cH_{\text{F}}^{(i)}=\frac{1}{2} \left( U_i \tau_z + \Lambda_i \tau_x \right)\, ,
\end{equation}
where $\tau_i$ is the set of Pauli matrices describing $i$-th TLS. The energy-level splitting for 
this \fl is $E_i=\sqrt{U_i^2+\Lambda_i^2}$. Fluctuators switch between their states due  to the interaction 
with the environment, which is modeled as thermal bath. It can represent a phonon bath as well as, e.~g., 
electron-hole pairs in the conducting part of the system. Fluctuations of the environment affect the \fl 
through the parameters $U_i$ or $\Lambda_i$. Assuming that fluctuations of the diagonal splitting $U$ are most 
important, we describe \fs as TLSs in glasses, see Sec.~\ref{subsec:models}.

For the following, it is convenient to characterize \fs by the parameters $E_i$ and $\theta_i=\arctan (\Lambda_i/U_i)$. 
The mutual distribution of these parameters can be written as \cite{Laikhtman1985}
\begin{equation} \label{df002}
P(E,\theta)=P_0/\sin \theta\, , \quad 0 \le \theta \le \pi/2\, ,
\end{equation} 
which is equivalent to the distribution (\ref{df001}) of relaxation rates. To normalize the distribution one 
has to cut it off at small relaxation rates at a minimal value $\gamma_{\min}$ or cut off the distribution 
(\ref{df002}) at $\theta_{\min}=\gamma_{\min}/\gamma_0 \ll 1$. The distributions given by Eqs.~(\ref{df001}) 
and (\ref{df002}) lead to the $\propto 1/\omega $ noise spectra at $\gamma_{\min} \ll \omega \ll \gamma_0$.

The variation of the qubit's energy-level spacing can be cast in the Hamiltonian, which [after a rotation 
similar to that leading to  Eq.~(\ref{TLS-envd1})] acquires the form
\begin{equation} \label{efH01}
\hat \cH_{\text{qF}}= \hbar \sum_i b_i\sigma_z\tau_z^{(i)}\, , \quad b_i=g(r_i) A(\mathbf{n}_i)\cos \theta_i\, .
\end{equation}
Here   $\mathbf{n}_i$ is the direction of the elastic or electric 
dipole moment of $i$-th fluctuator, and $r_i$ is the distance between the qubit and the  $i$-th fluctuator. 
Note that in Eq.~(\ref{efH01}) we neglected the term $\propto \sigma_z\tau_x$.  This can be justified as long 
as the fluctuator is considered to be a classical system. The functions $A(\mathbf{n}_i)$ and $g(r_i)$ are 
not universal.

The coupling constants, $b_i$,  defined by Eq.~(\ref{efH01}) contain $\cos  \theta_i$. Therefore they are 
statistically correlated with $\theta_i$.  It is convenient to introduce an uncorrelated random coupling
parameter, $u_i$, as
\begin{equation}\label{uncoru}
u_i\equiv g(r_i)A(\mathbf{n}_i), \quad b_i=u_i\cos \theta_i \, .
\end{equation}
Let us also  assume for simplicity that the direction $\mathbf{n}_i$ of a fluctuator is correlated neither
 with its distance from the qubit, $r_i$, nor to  the tunneling parameter represented by the 
 variable $\theta_i$. This assumption allows us to replace $A(\mathbf{n}_i)$ by its average over the angles,
  $\bar{A} \equiv \av{A(|\mathbf{n}|)}_\mathbf{n}$.

Now we are ready to analyze the decoherence using Eq.~(\ref{m+003}).  The question we will address in this 
Section is whether a special group of \fs responsible for decoherence does exist.

An interesting feature of the problem is that the result strongly depends on the decay of the coupling
 parameter, $g(r)$, with  the distance $r$, which usually can be described by a power law: $g (r)=\bar{g}/r^s$.  
 To illustrate the point, let us compare two cases: (i) the \fs are distributed in three-dimensional space ($d=3$) 
 and (ii) the \fs are located in the vicinity of a two-dimensional manifold, e.~g., in the vicinity of the
  interface between an insulator and a metal ($d=2$). Using the distribution (\ref{df002}) of the relaxation 
  rates one can express the distribution $\cP(u,\theta)$ as
\begin{equation} \label{putheta}
\hspace*{-0.05in}
\cP (u, \theta)=\frac{(\eta \cos \theta)^{d/s}}{u^{(d/s)+1}\sin \theta}, \ \eta 
\equiv \frac{\bar{g}}{r_T^s}, \ r_T \equiv \frac{a_d}{(P_0\, k_B T)^{1/d}} .
\end{equation}
Here we have assumed that only the \fs with $E \lesssim k_B T$ are important for the decoherence since 
the \fs with $E \gtrsim k_B T$ are frozen in their ground states.  The typical distance between such \fs is 
$r_T \equiv a_d (P_0 k_B T)^{-1/d}$ where $a_d$ is a $d$-dependent dimensionless constant.  In the following, 
we will, for simplicity, assume that
\begin{equation} \label{as002}
r_{\min} \ll r_T \ll r_{\max}
\end{equation}
where $r_{\min}$ ($r_{\max}$)  are the distances between the qubit and the closest (most remote) fluctuator.
 In this case, $\eta \propto T^{s/d}$ is the typical value of the qubit-to-\fl coupling playing role of the coupling parameter $b$.
As soon as the  inequality (\ref{as002}) is violated the decoherence starts to depend on either $r_{\min}$ or $r_{\max}$,
  i.~e., becomes sensitive to particular mesoscopic details of the device. 
\begin{figure}[t]
\centerline{
\includegraphics[width=4.5cm]{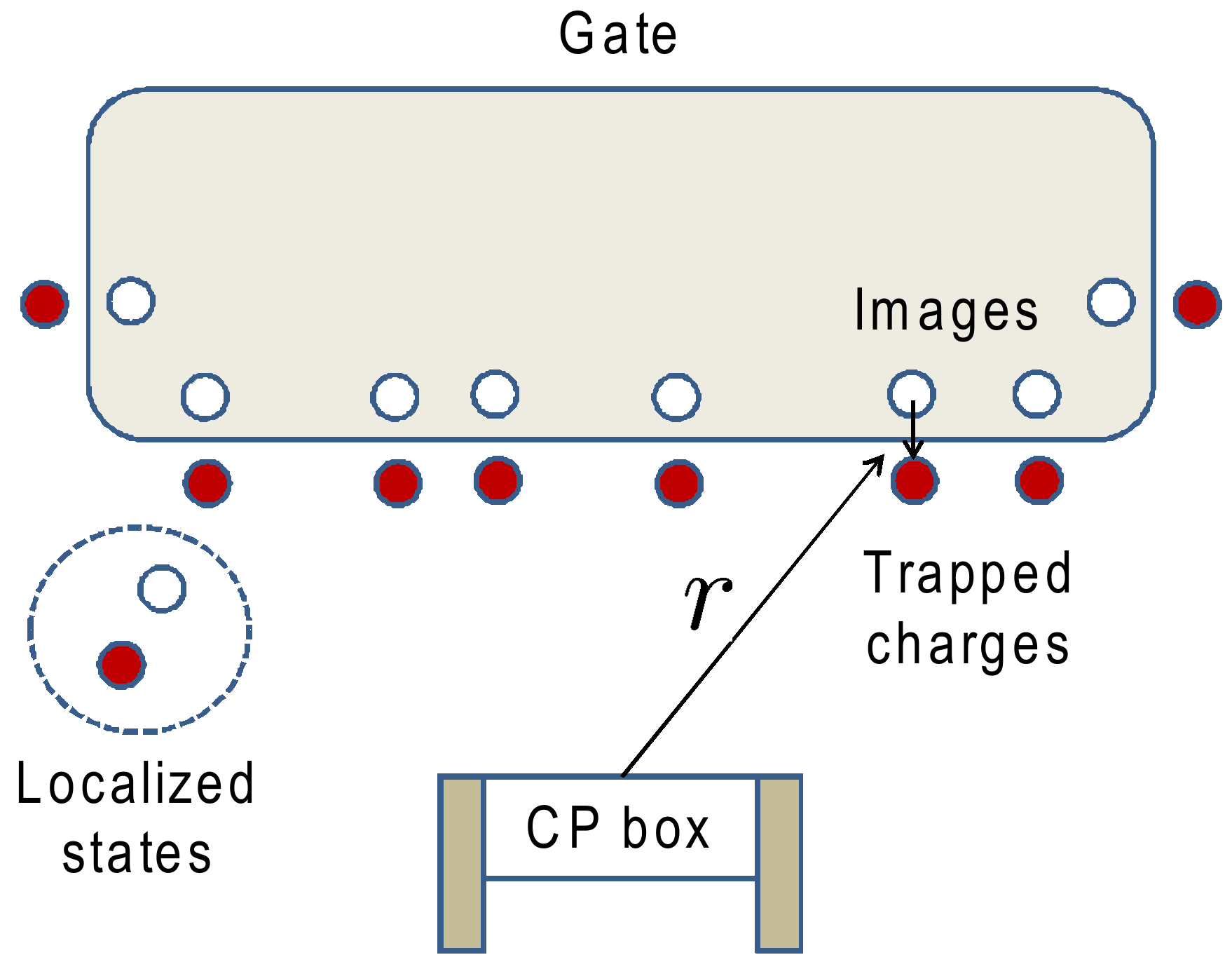}
}
\caption{(Color online) Sketch of localized charges near an electrode in a Cooper Pair box. 
Induced image charges create local dipoles that 
interact with the qubit.\label{figqb1}}
\end{figure}

Let us first consider the case when $d=s$, as it is for charged traps located near the gate 
electrode (where $s=d=2$ due to the dipole nature of the field produced by a charge and its 
induced image, as shown in Fig.~\ref{figqb1}).  In this case one can rewrite Eq.~(\ref{m+003}) as
\begin{equation} \label{m+004}
\cK_f(t)=\eta  \int \frac{du}{u^2} \int_0^{\pi/2} \! \! \! d\theta \,   \tan \theta\,  f(u\cos \theta, \gamma_0\sin^2 \theta|t).
\end{equation}
Here $f(v,\gamma|t) $ is equal either to $1-\av{m_+(v,\gamma|t)}$ or to $1-\av{m_+^{(e)}(v,\gamma|t)}$, 
depending on the manipulation protocol. 
Equation (\ref{m+003}) together with Eqs.~(\ref{FID02}) and (\ref{Ec001}) 
allows one to analyze various limiting cases.

As an example, let us consider the case of the two-pulse echo.  To estimate the integral in 
Eq.~(\ref{m+004}) let us look at asymptotic behaviors of the function $f$ following from Eq.~(\ref{Ec001}):
\begin{equation}\label{as003}
f^{(e)} \propto \left\{\begin{array}{ll}
t^3\gamma_0\sin \theta(u\cos \theta)^2, & t \ll (\gamma_0 \sin\theta)^{-1} , \\
t^2(u\cos \theta)^2, & (\gamma_0\sin\theta)^{-1} \! \! \ll t \ll u^{-1}, \\
tu\cos\theta, & u^{-1} \ll t\, .
\end{array} \right.
\end{equation}
Note, that these dependencies differ from those given by Eqs.~(\ref{echo_SF1}) and (\ref{echo_SF2}).
Splitting the regions of integration over $u$ and $\theta$ according to the domains (\ref{as003}) 
of different asymptotic behaviors, one obtains \cite{Galperin2003a}:
\begin{equation} \label{as004}
\cK^{(e)}(2\tau_{12}) \sim \eta \tau_{12} \min\{\gamma_0 \tau_{12},1\} \, .
\end{equation}
The dephasing time (defined for non-exponential decay as the time when $\cK \sim 1$) for the
 two-pulse echo  is thus equal to 
\begin{equation}\label{as005}
\tau_\varphi = \max \{ \eta^{-1}, (\eta \gamma_0)^{-1/2}\} \, .
\end{equation}

The result for $\gamma_0 \tau_{12} \ll 1$ has a clear physical meaning, cf. with \cite{Laikhtman1985}: 
the decoherence occurs only provided  \textit{at least one} of the \fs flips. Each flip provides a 
contribution $\sim \eta t$ to the phase while $\gamma_0 \tau_{12}$ is a probability for a flip during
 the observation  time $\sim \tau_{12}$. 

The result for long observation times, $\gamma_0 \tau_{12} \gg 1$, is less intuitive since in this
 domain the dephasing is non-Markovian, see \cite{Laikhtman1985} for more details. In this case the 
 decoherence is dominated by a set of \textit{optimal} (most harmful)  \fs
located at some distance, $r_{\text{opt}} (T)$, from the qubit.  This distance is determined by the condition
\begin{equation} \label{opt01}
g(r_{\text{opt}}) \approx \gamma_0 (T)\, .
\end{equation}
Though derivation of this estimate is rather tedious, see \cite{Galperin2003a} for details, it emerges naturally 
from the behavior of the decoherence in the limiting cases of strong ($b \gg \gamma$) and weak ($b \ll \gamma$) 
coupling. For strong coupling, the \fs are slow and the qubit's behavior is determined by quantum beatings 
between the states with $E \pm b$. Accordingly, the decoherence rate is $\sim \gamma$. In the opposite case, 
as we have already discussed, the decoherence rate is $\sim b^2/\gamma$. Matching these two limiting cases 
one arrives at the estimate (\ref{opt01}).

What happens is $d \ne s$? If the coupling decays as $r^{-s}$ and the \fs are distributed in a $d$-dimensional
 space, then $r^{d-1}dr$ is transformed to $du/u^{1+(d/s)}$. Therefore,  $\cP(u) \propto 1/u^{1+(d/s)}$. As a 
 result, at $d \le s$ the decoherence is controlled by the optimal \fs located at the distance $r_{\text{opt}}$ 
 \textit{provided they exist}. If $d \le s$, but the closest \fl has $b_{\max} \ll \gamma_0$, then it is the quantity 
 $b_{\max}$ that determines the decoherence. 
 At $d > s$ the decoherence at large time is dominated by most remote \fs with 
 $r=r_{\max}$. 
 In the last two cases, $\cK (t) \propto t^2$, and one can apply 
 the results of  \textcite{Paladino2002}, substituting for $b$ either $b_{\min}$ or $b_{\max}$.

Since $r_{\text{opt}}$ depends on the temperature, one can expect crossovers between different regimes as a 
function of temperature. A similar mesoscopic behavior of the decoherence rates was discussed for a 
microwave-irradiated Andreev interferometer~\cite{Lundin2001}.
It is worth emphasizing that the result (\ref{as004}) for the long-range interaction cannot be reproduced by 
the Gaussian approximation. 
Indeed, if we expand the Gaussian approximation, $e^{-\av{\psi^2}/2}$, with $\av{\psi^2}$ given by 
Eq.~(\ref{Gauss003}) in the same fashion as in Eq.~(\ref{as003}) and then substitute to Eq.~(\ref{m+004}) the 
resulting integral over $u$ will be divergent at its upper limit. Physically, this would mean the dominant 
role of nearest neighbors of the qubit. At the same time, the SF model implies that the most important \fs 
are those satisfying Eq.~(\ref{opt01}). 

The existence of selected groups of fluctuators, out of an ensemble of many,
which are responsible for decay of specific quantities, is closely related to the self-averaging 
property of the corresponding decay laws. In \cite{Schriefl2006} a class of distribution functions 
of the form (\ref{putheta}) with $0< d/s<2$ has been analyzed. 
In this case the average over the coupling constants
of the Lorentzian functions (\ref{ns2}) diverges at the upper limit and the noise is dominated by the most
strongly coupled fluctuators.
The free induction decay has been found non-self-averaging both at short and long 
times with respect to $\gamma_0^{-1}$.
Non-self-averaging of the
echo signal  for $2 \tau_{12} \gg \gamma_0^{-1}$ has also been demonstrated \cite{Schriefl2006}.

\paragraph{Dephasing according to other phenomenological models --}

\begin{figure}[t]
\centerline{
\includegraphics[width=0.5\columnwidth,angle=-90]{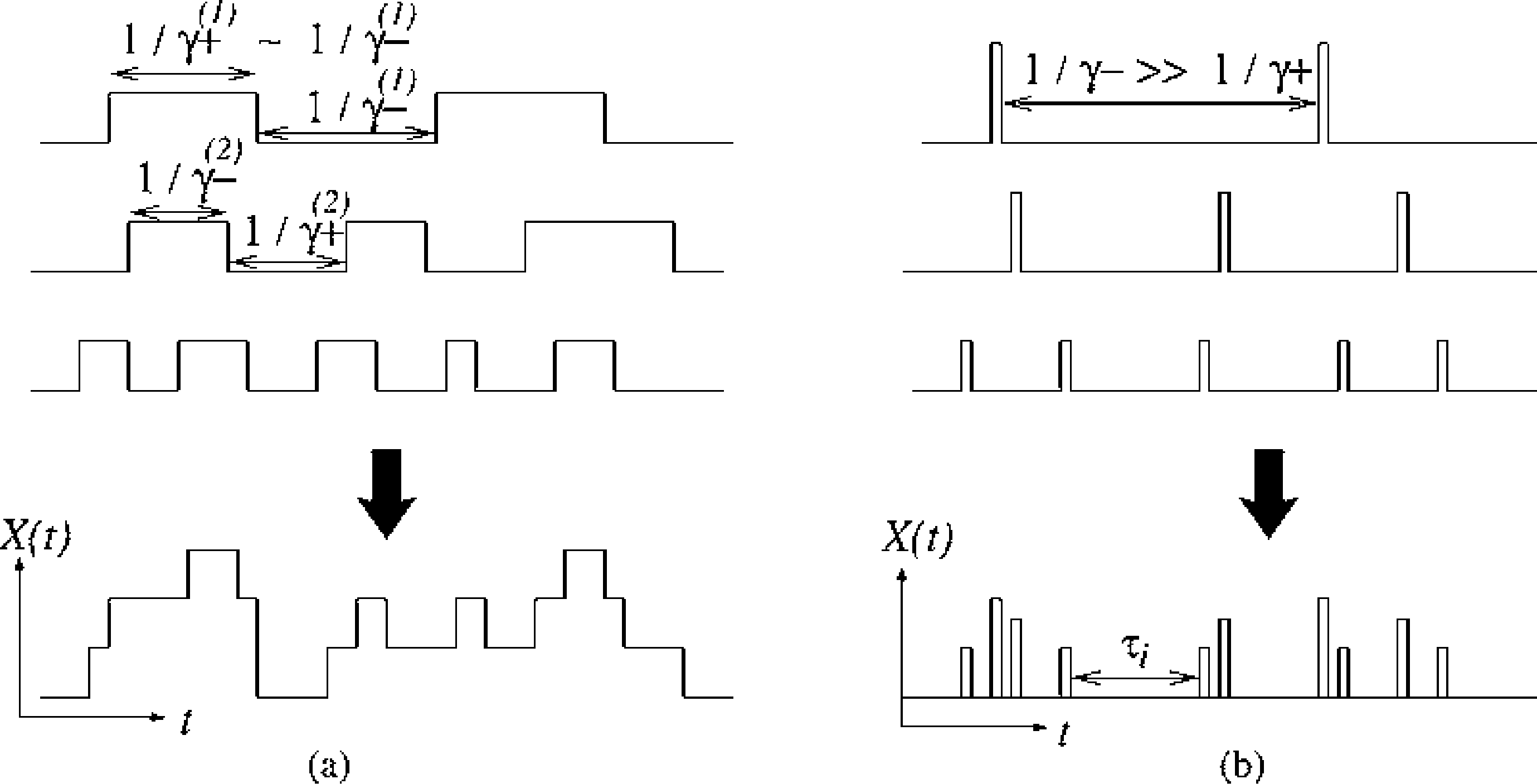}
}
\caption{(a) Low frequency noise as sum of RT signals. The switching rates for the ``up" and ``down" states are
comparable $\gamma_+ \approx \gamma_-$ and $1/f$ noise results from a superposition of RT signals 
distributed as $\propto 1/\gamma$. In (b) the intermittent limit corresponds to the limit where 
the noise stays in the down states most of the time ($\gamma_+ \gg \gamma_-$). The intermittent noise can 
be approximated by a spike field with independent random waiting times and spikes eighths. 
It gives a nonstationary $1/f^\mu$ spectrum depending on the plateaus distribution function.
 Adapted from \cite{Schriefl2005a}}
\label{fig1_SchrieflPRB05}
\end{figure}

We remark that, beside the spin-fluctuator model, other stochastic processes have been considered for the description of
fluctuations having $1/f$ spectral density. In this paragraph we mention an alternative
phenomenological model, which has been discussed in connection with the problem of qubit dephasing.
The motivation to consider these models comes from the observation of asymmetric telegraphic signals, with
longer stays in the ``down" state than in the ``up" state, reported in tunnel junctions~\cite{Zimmerlie1992}, in MOS 
tunnel diodes~\cite{Buehler2004}, 
and of a spike field detected in a SET electrometer \cite{Zorin1996}. In those cases the signal exhibits $1/f$ spectrum
at low frequencies. This suggests a description of the $1/f$ noise in terms of a single asymmetric RT signal.
In \cite{Schriefl2005a,Schriefl2005b} a phenomenological model for a $1/f^\mu$  classical intermittent noise
has been considered. The model can be viewed as the intermittent limit of the sum of RT signals where
the duration of each plateau of the RT signal, $\tau_{\text{av}}$, is assumed to be much shorter than waiting times
between the plateaus.
In this  limit, the noise is approximated by a spike field consisting of delta functions whose heights
$x$ follow a Gaussian distribution or, more generally, a distribution with 
finite first and second moments, $\bar x$ and $s \equiv (\overline{x^2})^{1/2}$, respectively, as illustrated in
Fig.~\ref{fig1_SchrieflPRB05} adapted from \cite{Schriefl2005a}. 
The variance $s$ plays the role of a coupling constant between the qubit and the stochastic process. 
 A $1/f$ spectrum for the intermittent noise is recovered for a distribution of waiting times $\tau$
 behaving as $\tau^{-2}$ at large times. Because the average waiting time is infinite, no time scale characterizes the
 evolution of the noise which is non-stationary. 
 In the pure dephasing regime the relative phase between the qubit states performs a continuous time random walk 
 (CTRW)~\cite{Haus1987} as time goes on.
 Using renewal theory \cite{Feller1962} authors find an exact expression for the Laplace transform of the
 dephasing factor  allowing analytical estimates in various limiting regimes and address the consequences of the 
 CTRWs nonstationarity.
The noise is initialized at time $t'=0$ and the coupling with the qubit is turned on at the preparation time, $t_p$. 
The two qubit states accumulate a random relative phase between $t_p$ and $t+t_p$ and nonstationarity
 manifests itself in the dependence of dephasing factor, 
 on the age of the noise, $t_p$.
This analysis shows that two dephasing regimes exist separated by a crossover coupling constant, 
$s_c(t_p)= [2 (\tau_{\text{av}}/t_p)\ln{(t_p/\tau_{\text{av}})}]^{1/2}$, strongly dependent on the preparation time $t_p$.
For $s < s_c(t_p)$ dephasing is exponential and independent on $t_p$, whereas when $s > s_c(t_p)$ the
 dephasing time depends algebraically on $t_p$. Since $s_c(t_p)$ decays to zero with the noise age, any qubit coupled
 to the non-stationary noise will eventually fall in the regime $s> s_c(t_p)$.
 
 We remark that in \textcite{Schriefl2005a,Schriefl2005b} the dephasing factor is defined as a configuration
 average over the noise, rather than a time average in a given configuration. The two averages do not
 coincide in general for nonstationary or aging phenomena.  Thus one should be careful in comparing these
 results with experiments.

\subsubsection{Decoherence due to the SF model at general working point}
\label{subsub:SFgeneral}

Here we discuss decoherence due to the spin-fluctuator model at general operating point where
\begin{equation}
\hat H_{tot} = \frac{\hbar}{2} \, \left[\Omega_x \sigma_x + \Omega_z(q) \sigma_z \right]
+ \frac{\hbar}{2} \,  b \chi(t) \sigma_z \, .
\label{H_SF_general}
\end{equation}
As a difference with the pure-dephasing regime, no exact analytic solution is available 
under these conditions. Different approaches have been introduced
to study the qubit dynamics leading to analytical approximations in specific limits and/or to
(exact) numerical results.

\begin{figure}[t!]
\centerline{
\includegraphics[width=0.9\columnwidth]{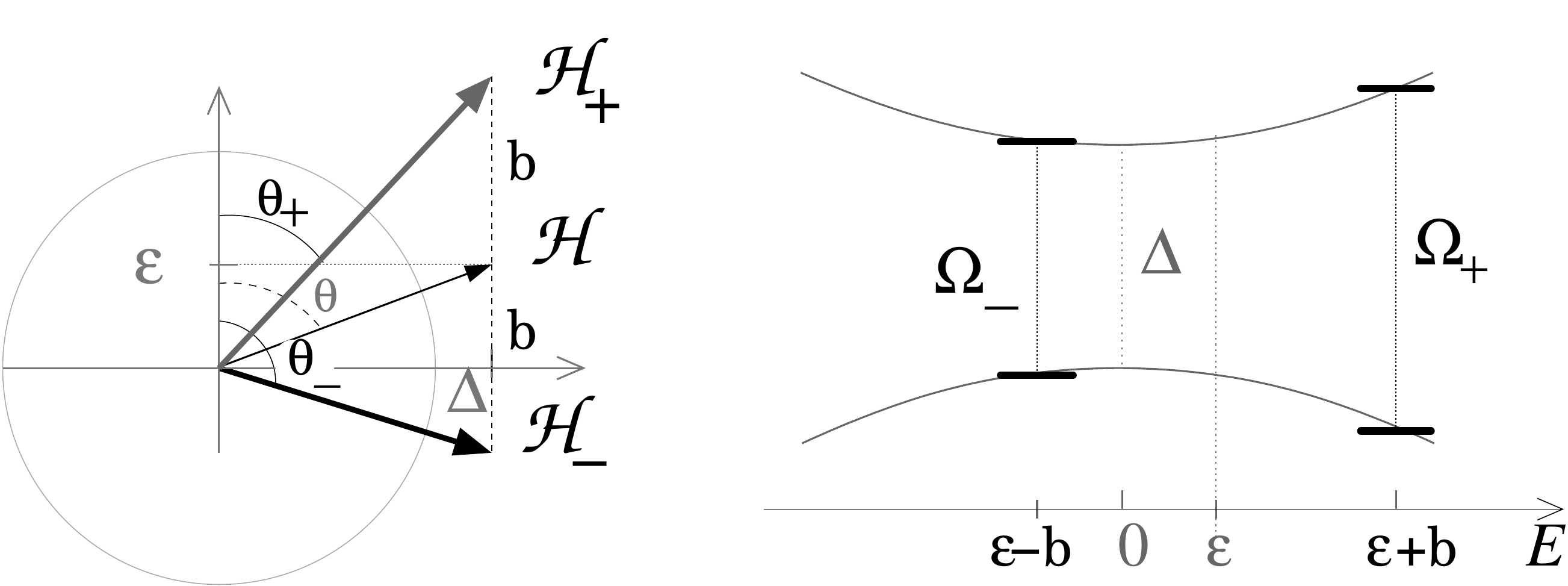}
}
\caption{
Left panel: Qubit Bloch sphere. An isolated qubit
defines the mixing angle $\theta = \arctan{\{\Delta/\epsilon\}}$,   
${\mathcal H}_\pm$ define 
$\theta_\pm = \arctan{ \{\Delta/(\epsilon \pm b)\}}$. 
Right panel: Qubit energy bands $\pm \sqrt{E^2 +\Delta^2}$: 
the energy splittings depend on the impurity state, 
$\Omega_{\pm}= \sqrt{(\epsilon \pm b)^2 + \Delta^2}$.}
\label{fig:splittings}
\end{figure}

A straightforward approach consists in solving the stochastic Schr{\"o}dinger equation. 
Using the theory of stochastic differential equations~\cite{Brissaud1974} a formal solution 
in Laplace space can be found which is hoverer difficult to invert analytically.
A numerical solution is instead feasible and the method can be extended to investigate the
effect of an ensemble of \fs with proper distribution of parameters to generate processes having 
$1/f$ power spectrum~\cite{Falci2005}. In \cite{Cheng2008} a generalized transfer-matrix method
has been proposed. It reduces to the algebraic problem of the diagonalization of a $6 \times 6$ matrix 
whose eigenvalues give the decoherence ``rates" entering the qubit dynamics. This approach also gives 
the system evolution on time scales shorter than the \fl correlation time.  The transfer
matrix method is suitable to address the dynamics under instantaneous dc-pulses like those required 
in the spin-echo protocol. The method can in principle be extended to the case of many fluctuators, 
though the size of the matrices soon becomes intractable.
Another approach consists in the evaluation of the quantum dynamics of a composite system composed 
of the qubit and one (or more) fluctuator(s), treated as quantum mechanical two-level systems~\cite{Paladino2002,Paladino2003}.
This requires either solving the Heisenberg equations of motion, or a master equation in the enlarged Hilbert
space. Impurities are traced out at the end of the calculation, when the high-temperature
approximation for the \fl is also performed~\cite{Paladino2002,Paladino2003}. 
Both methods lead to analytic forms in a limited parameter regime and a numerical analysis is 
required in the more general case.
Other approximate methods rely on the extension of the approach based on the evaluation of the
probability distribution $p(\phi,t)$ which is evaluated numerically \cite{Bergli2006}. 

The main effects of a RT fluctuator coupled to a qubit via an interaction term of the form (\ref{H_SF_general})
can be simply illustrated as follows. 
The noise term $- b \chi(t) \sigma_z/2$ induces two effective splittings, 
$\Omega_\pm= \sqrt{(\Omega_z \pm b)^2+ \Omega_x^2}$,  
and correspondingly two values of the polar angle, $\theta_\pm= \arctan(\Omega_x/(\Omega_z \pm b))$, 
as illustrated in Fig.~\ref{fig:splittings}.

\begin{figure}[t!]
\centerline{
\includegraphics[width=0.7\columnwidth]{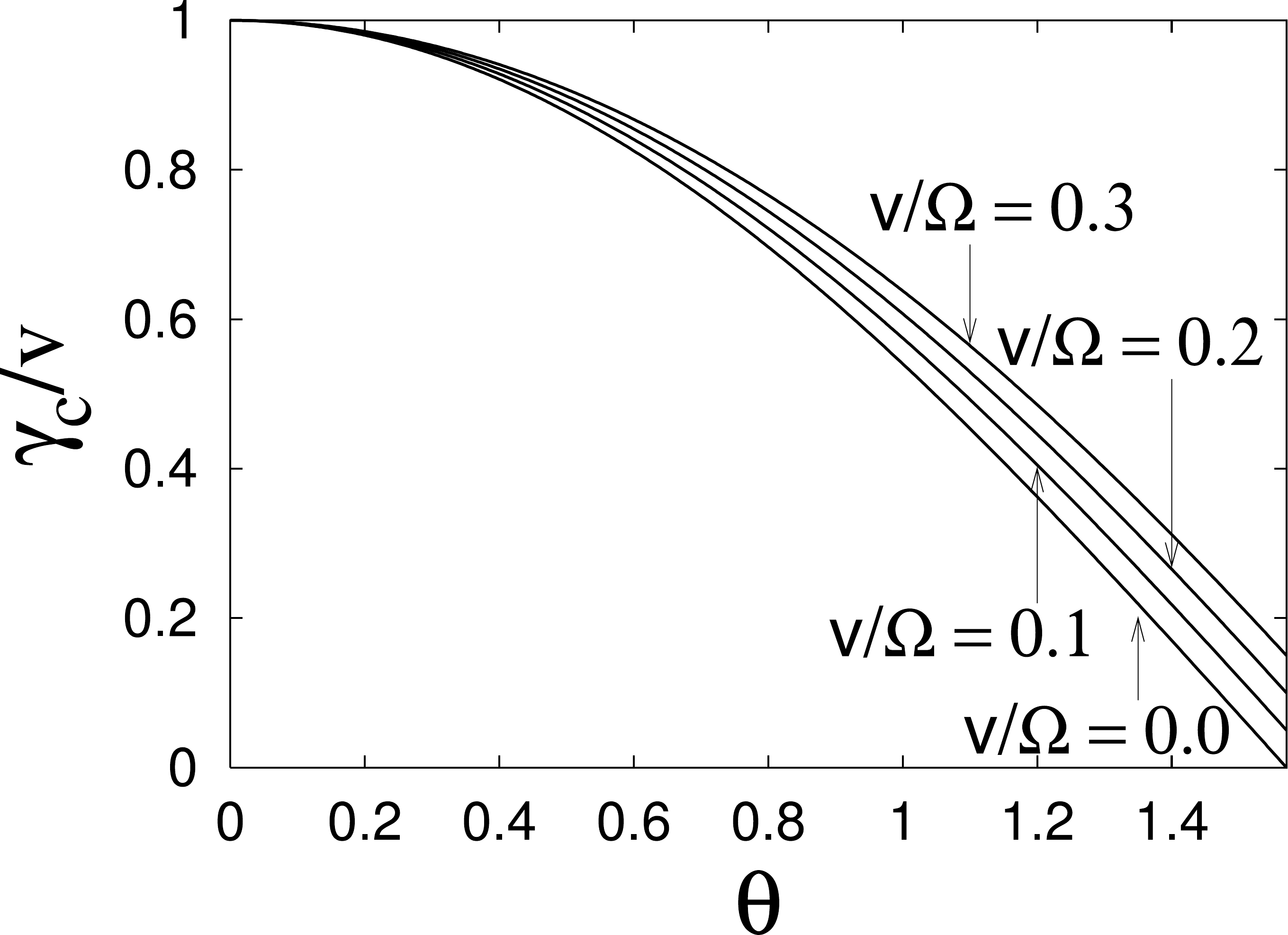}
}
\caption{The threshold value $\gamma_c/v\equiv(\Omega_+ - \Omega_-)/(2b \, c^2)$ 
for a \fl behaving as weakly coupled depends on the operating point ($v/\Omega \equiv 2b/\Omega$). 
Adapted from \cite{Paladino2003b}, with kind permission of Springer Science+Business Media.}
\label{threshold}
\end{figure}

\begin{figure}[t!]
\centerline{
\includegraphics[width=0.7\columnwidth]{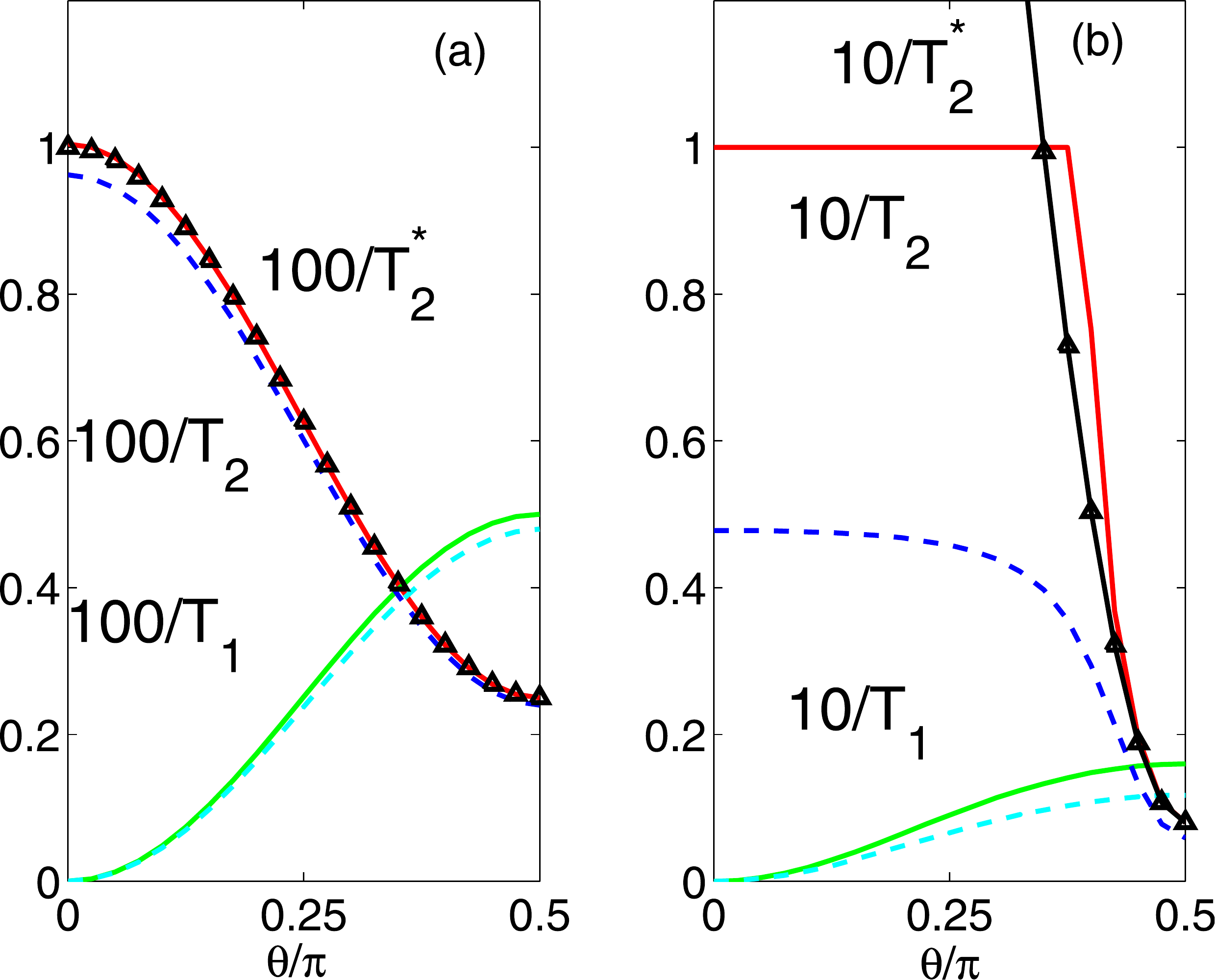}
}
\caption{(Color online) Relaxation $1/T_1$ and dephasing $1/T_2$ rates (rescaled for visibility)
 as functions of $\theta$. Triangles - $1/T_2^*= \cos^2 \theta S(0)/2$ for a weakly (a) and strongly (b) 
coupled fluctuator. $\gamma/\Omega$: (a) - 0.5, (b) - 0.1. $b/\Omega$: (a)=0.1, (b) - 0.3.
Solid (dashed) lines correspond to a symmetric (slightly asymmetric) fluctuator.
Adapted from \cite{Cheng2008}.}
\label{fig3Cheng2008}
\end{figure}

 In the adiabatic limit $\gamma \sim |\Omega_+ - \Omega_- | \ll \Omega_\pm $,
and neglecting any qubit back-action on the fluctuator, the qubit coherence takes a form similar to the pure
dephasing result Eq.~(\ref{FID02}). Here we report the average $\langle \sigma_y(t) \rangle$~\cite{Paladino2003}
\begin{eqnarray}
&& \langle \sigma_y(t) \rangle = - \mathrm {Im}\left[
\frac{e^{i(\Omega+ \gamma g/2)t}}{\alpha}
\sum_{\pm} A(\pm\alpha) e^{-\frac{1\mp \alpha}{2}\gamma t}\right] \label {sigmay} \\
&& \text{where} \ \alpha =  \sqrt{1-g^2-2ig \delta p_{\text{eq}} - (1-c^4)(1-\delta p_{\text{eq}}^2)}, \nonumber \\
&&\quad \quad c = \cos[(\theta_- - \theta_+)/2, \quad g= (\Omega_+ - \Omega_-)/\gamma,  \nonumber  \\
&&A(\alpha)= (\alpha + c^2 -i g^\prime) \rho_{+-}^{(-)}(0) p_0+ (\alpha + c^2 + i g^\prime) \rho_{+-}^{(+)}(0) p_1 \nonumber 
\end{eqnarray}
with $g^\prime= g + i \delta p_{eq}(1-c^2)$ and 
$\rho_{+-}^{(\pm)}(0)$ are values at $t=0$ of the qubit coherences in the eigenbasis $| \pm \rangle_{\theta_\pm}$ of the 
qubit conditional Hamiltonians corresponding to $\chi= \pm 1$.
In the  high-temperature regime, $\delta p_{\text{eq}} \approx 0$, and  $\alpha \approx \sqrt{c^4-g^2}$.
Thus the relevant parameter separating the weak- from the strong-coupling regime 
is $g/c^2=(\Omega_+ - \Omega_-)/(c^2\gamma)$.
A \fl is weakly-coupled if $g/c^2 \ll 1$ and strongly coupled otherwise. This condition depends on the qubit's 
operating point, and at $\theta=0$ we get $g/c^2=2b/\gamma$. 
Thus a \fl characterized by a set $(b, \gamma)$ affects in a qualitatively different way the qubit dynamics depending
on the qubit's operating point. In particular, a \fl turns from strongly to weakly coupled increasing 
$\theta$ from $0$ to $\pi/2$, as illustrated in Fig.~\ref{threshold}. From (\ref{sigmay}) it easy to see that a
 weakly coupled 
\fl induces an exponential decay with $T_2$ given by Eq.~(\ref{T2}), whereas  a strongly coupled \fl induces a saturation
effect which is less effective at $\theta=\pi/2$ than at $\theta=0$ \cite{Paladino2003,Bergli2006,Cheng2008,Zhou2010}.
The effect of strongly coupled \fs is more sensitive to deviations from the optimal point, $\theta= \pi/2$.
The dependence on the working point of the decoherence and relaxation times 
for a weakly and  a strongly coupled \fl is shown in Fig.~\ref{fig3Cheng2008}.
The decay rates at $\theta=\pi/2$ have also been obtained in \cite{Itakura2003} by solving the stochastic
differential equation.

\begin{figure}[t!]
\centerline{
\includegraphics[width=0.8\columnwidth]{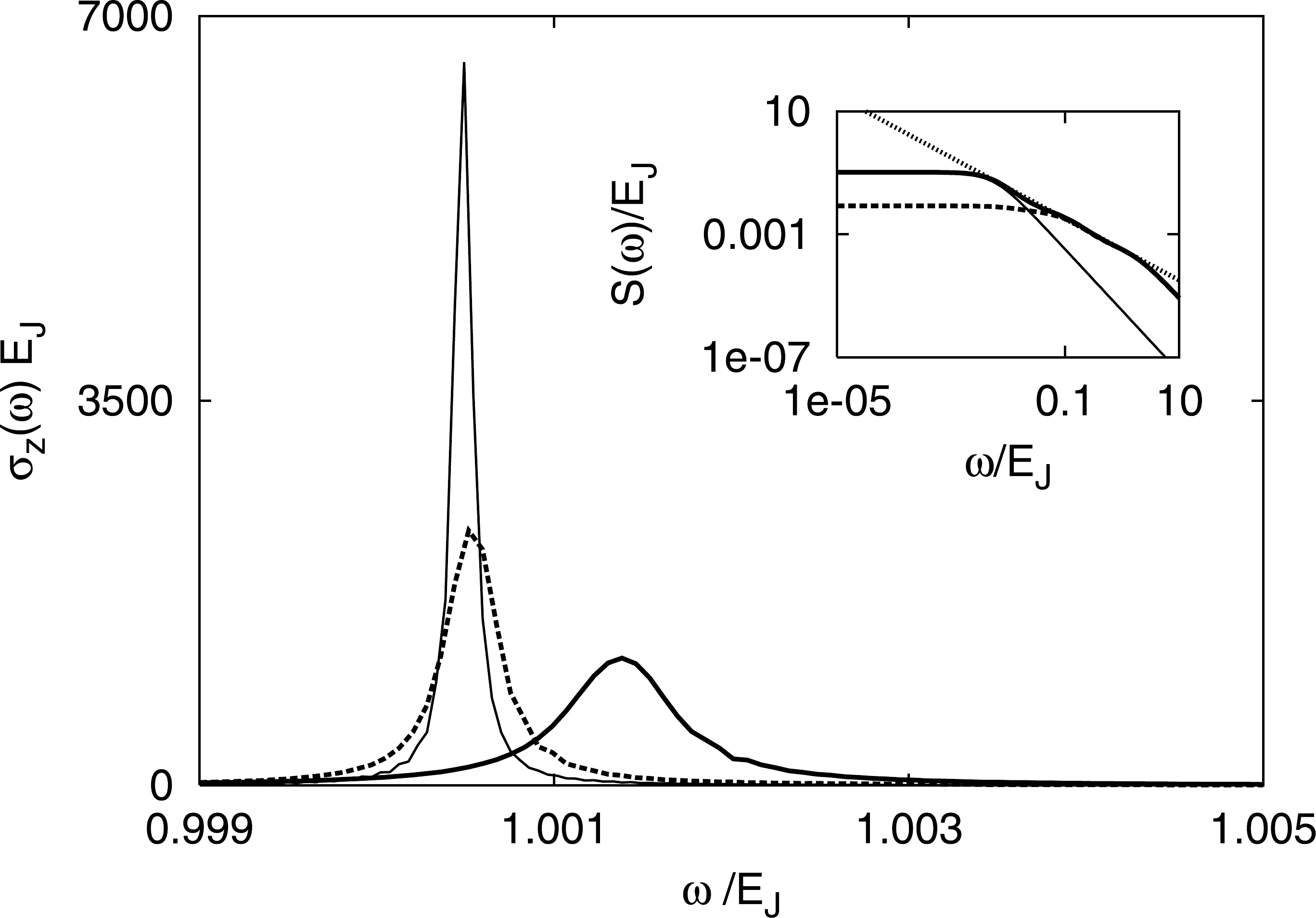}
}
\caption{The Fourier transform of $\langle \sigma_z(t) \rangle$ for a set of weakly coupled \fs plus a single
strongly coupled \fl (thick line). The separate effect of the coupled \fl ($g=8.3$, thin line) and
the set of weakly coupled \fs (dotted line) is shown for comparison. Inset: corresponding power
spectra. In all cases noise level at $\Delta= E_J$ is fixed to the value $S(E_J)/E_J= 3.18 \times 10^{-4}$
(from typical $1/f$ noise amplitude in charge qubits~\cite{Zorin1996,Nakamura2002,Covington2000} extrapolated 
at GHz frequencies). Adapted from \cite{Paladino2002}.}
\label{fig:1overf_optimal}
\end{figure}

\begin{figure}[t!]
\centerline{
\includegraphics[width=0.8\columnwidth]{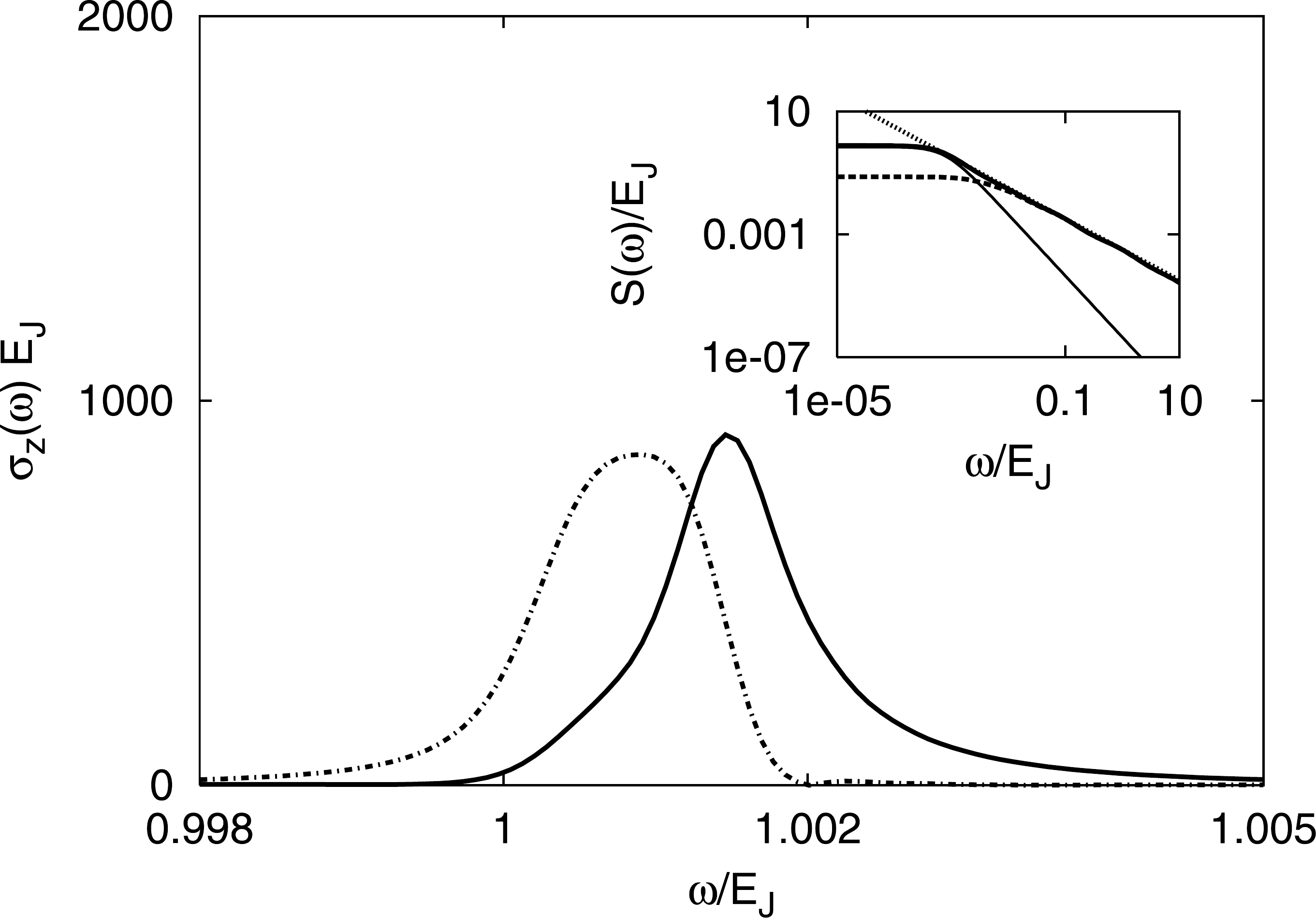}
}
\caption{The Fourier transform $\langle \sigma_z \rangle_\omega$ for a set of weakly coupled
\fs plus a strongly coupled \fl ($2b /\gamma= 61.25$) prepared
in the ground (dotted line) or in the excited state (thick line).
Inset: corresponding power spectra (the thin line corresponds to
the extra \fl alone). Adapted from \cite{Paladino2002}.}
\label{fig2_paladino2002}
\end{figure}

\paragraph{Ensemble of fluctuators: decoherence due to $1/f$ noise --}
\label{few-fs}
All the above mentioned approaches become intractable analytically as soon as the number of
\fs increases. Already in the presence of few fluctuators a numerical solution is required.
This difficulty and the physical scenario behind the mechanisms of dephasing due to $1/f$ noise 
at different operating points, can be traced back to the dependence of the qubit splitting on the induced
fluctuations, cf. with Eq.~(\ref{deU}), which we write here for $\Omega_\pm$:   
\begin{equation}
\delta \Omega_\pm \approx \pm \frac{\Omega_z}{\Omega}\, b +\frac{1}{2}\frac{\Omega_x^2}{\Omega^3}\, b^2 \, .
\label{expansion1}
\end{equation}
Far from the optimal point, the leading term 
is linear in the coupling $b$. This is the only effect left at pure dephasing, $\Omega_x=0$, and it
is also the reason way an exact formula is available under this condition both in the case of a single \fl
and in the presence of an ensemble of fluctuators. In this last case the phase memory functional for any number of \fs
is found by simply multiplying the phase-memory functionals for different fluctuators. 
At the optimal point instead, $\Omega_z =0$,  the first order term is quadratic. Thus the effect of the \fl is reduced with
respect to the pure dephasing regime. The effects of different \fs in this case do not simply add up independently and
no analytic solution is available. 
This simple observation suggests an important physical insight on the combined effect of various \fs
at different working points.
When the first order effect on the splitting is quadratic, even if the different \fs in themselves are independent, 
their effect on the qubit will be influenced by the position
of all the others.
At pure dephasing, since effects sum up independently, all slow \fs are ineffective at times $\gamma t \ll 1$. This
is no longer true when the leading term is quadratic, 
as they play a role in determining the effect of faster fluctuators.
In fact even if they do not have time to switch during the experiment, they contribute to the average effective operating
point the qubit is working as seen by the faster fluctuators. Thus very slow \fs may be of great importance at
the optimal point.
For example, in Fig.~\ref{fig:1overf_optimal} it is illustrated that even a single, strongly-coupled
fluctuator, out of the many \fs forming the $1/f$ spectrum, determines a dephasing more than twice the prediction of
the weak coupling theory, $T_2^*$, 
even when the \fl is not visible in the spectrum. 
Further slowing down the \fl produces a saturation of dephasing, as expected based on the
above considerations. In this regime effects related to the initial preparation of the fluctuator, or equivalently
effects of the measurement protocol, are visible as illustrated in Fig.~\ref{fig2_paladino2002}.
The presence of selected strongly coupled fluctuators in the ensemble leading to $1/f$ spectrum may  give rise to 
striking effects like beatings between the two frequencies $\Omega_\pm$ , see Fig.~\ref{fig_4_Falci2005} and 
the discussion in Sec.~\ref{paragraph:multistage}. 
Relevant effects of $1/f$ noise at general working point, especially for protocols requiring repeated measurements,
are captured by approximate approaches based on the adiabatic approximation which we discuss in the following Subsection.

\subsection{Approximate approaches for decoherence due to $1/f$ noise}
\label{subsection4B}

In the context of quantum computation, the effects of stochastic processes with long-time correlations,
like those characterized by $1/f$ spectral density, depend on the  
quantum operation performed and/or on the measurement protocol. 
Here we present approximate approaches proposed to predict dephasing due
to $1/f$ noise  and its interplay with quantum noise.  
An approach based on the adiabatic 
approximation \cite{Falci2005,Ithier2005} allows simple explanations of
peculiar non-exponential decay reported in different experiments with various 
setups. In some protocols a Gaussian approximation captures the main effects at least on a short time
scale~\cite{Falci2005,Makhlin2004,Rebenstein2004}.
Some other protocols, instead, clearly reveal the non-Gaussian nature of the noise.
The adiabatic approximation suggests  a route to identify  
operating conditions where leading order effects of $1/f$ noise are eliminated also for 
complex architectures.

\subsubsection{Approaches based on the adiabatic approximation}
\label{sub_adiabatic}

Our staring point is the phenomenological Hamiltonian Eq.~(\ref{H_qb_linearized})
where we separated the effect of 
quantum noise due to (high frequency) modes exchanging energy with the system and of
low frequency fluctuations, $\delta q(t)$, of the bias parameter $q$.
Environments with long-time memory, i.~e.,  correlated on a time scale much longer
than the inverse of the natural system frequencies, belong to the class of adiabatic noise.
Stochastic processes can be treated in the adiabatic approximation provided their contribution  
to spontaneous decay is negligible, a necessary condition being 
$t \ll T_1 \propto S(\Omega)^{-1}$.  This condition is satisfied for $1/f$ noise 
at short-enough times, considering that $S(\omega) \propto 1/\omega$ is substantially  different from zero 
only at frequencies  $\omega \ll \Omega$. 
For pure dephasing, $\theta=0$, relaxation processes are forbidden and the adiabatic approximation 
is exact for any $S(\omega)$.
In the adiabatic approximation the instantaneous Hamiltonian of a qubit, manipulated only with dc-pulses, 
reads
\begin{eqnarray}
 \hat H_{\text{tot}} & = &\frac{\hbar \Omega(q) }{2} (\cos \theta_q \, \sigma_z + \sin \theta_q \, \sigma_x)+ 
 \frac{\hbar E(t)}{2} \sigma_z \nonumber \\
&\equiv& 
\frac{\hbar}{2}\,\Omega[q,\delta q(t)] \, \sigma_{\tilde z} 
\label{H_adiabatic}
\end{eqnarray}
where we remind that $E(t) =   2 \delta q(t) (\partial  \Omega_z/\partial q) $, 
and the instantaneous splitting is 
\begin{equation}
\Omega(q,\delta q(t)) = \sqrt{\Omega^2(q) + E^2(t) +2E(t)\Omega (q) \cos \theta_q} \, .
\end{equation}
The $\sigma_{\tilde z}$ axis forms a time-dependent angle, 
$\tilde\theta_q(t)= \arctan [\Omega \sin \theta_q/(\Omega \cos \theta_q + E(t))]$, with $\sigma_z$.
In the qubit eigenbasis ($E\equiv 0$) the adiabatic Hamiltonian 
(\ref{H_adiabatic}) reads
\begin{equation}
\hat H= \frac{\hbar \Omega(q,\delta q)}{2} \left [ \cos{[\tilde\theta_q(t)- \theta_q]} \sigma_{z^\prime} 
+  \sin{[\tilde\theta_q(t)- \theta_q]} \sigma_{x^\prime} \right ] \, .
\label{H_adiabatic2}
\end{equation}
The effect of an adiabatic stochastic field $E(t)$ is to produce fluctuations both of the qubit splitting, 
$\Omega[q,\delta q(t)]$, and of the qubit eigenstates, or equivalently of the ``direction" of the qubit Hamiltonian,
expressed by the angle $\tilde\theta_q(t)$. Since we are interested to situations where 
$|E(t)| \ll \Omega$, both $\Omega[q,\delta q(t)]$
and $\tilde\theta_q(t)$ can be expanded in Taylor series about $\Omega(q,0)$ and $\theta_q$ respectively
\begin{eqnarray}
\Omega(q,\delta q) &\approx& \Omega(q,0) + \frac{\partial \Omega}{\partial q} (\delta q) 
+ \frac{1}{2} \frac{\partial^2\Omega}{\partial q^2} (\delta q)^2 + \cdots
\label{Omega_exp}\\
\tilde\theta_q &\approx& \theta_q + \frac{\partial \tilde\theta_q}{\partial q} (\delta q) 
+ \frac{1}{2} \frac{\partial^2\tilde\theta_q}{\partial q^2} (\delta q)^2 + \cdots \, ,
\end{eqnarray}
where $(\delta q) \equiv \delta q(t)$ and all derivatives are evaluated at $\delta q=0$.
The adiabatic Hamiltonian (\ref{H_adiabatic}) therefore can be cast in the form
\begin{equation}
\hat H= \frac{\hbar}{2} \left(\Omega(q,0) \sigma_{z^\prime} + \delta \Omega_\parallel(t) \sigma_{z^\prime} + 
\delta\Omega_\perp(t)\sigma_{\perp} \right) 
\label{H_adiabatic3}
\end{equation}
where, from Eq.(\ref{H_adiabatic2}), $\delta \Omega_\parallel$ includes the derivatives of 
$\Omega[q,\delta q(t)] \cos{[\tilde\theta_q(t)- \theta_q]}$ and $\delta\Omega_\perp$ the derivatives of 
$\Omega[q,\delta q(t)] \sin{[\tilde\theta_q(t)- \theta_q]}$. The Pauli matrix $\sigma_{\perp}$ in the case 
of Eq. (\ref{H_adiabatic2}) is $\sigma_{x^\prime}$, in general it can be a combination of 
$\sigma_{x^\prime}$ and $\sigma_{y^\prime}$.

\begin{figure}[h]
\centerline{
\includegraphics[width=0.85\columnwidth]{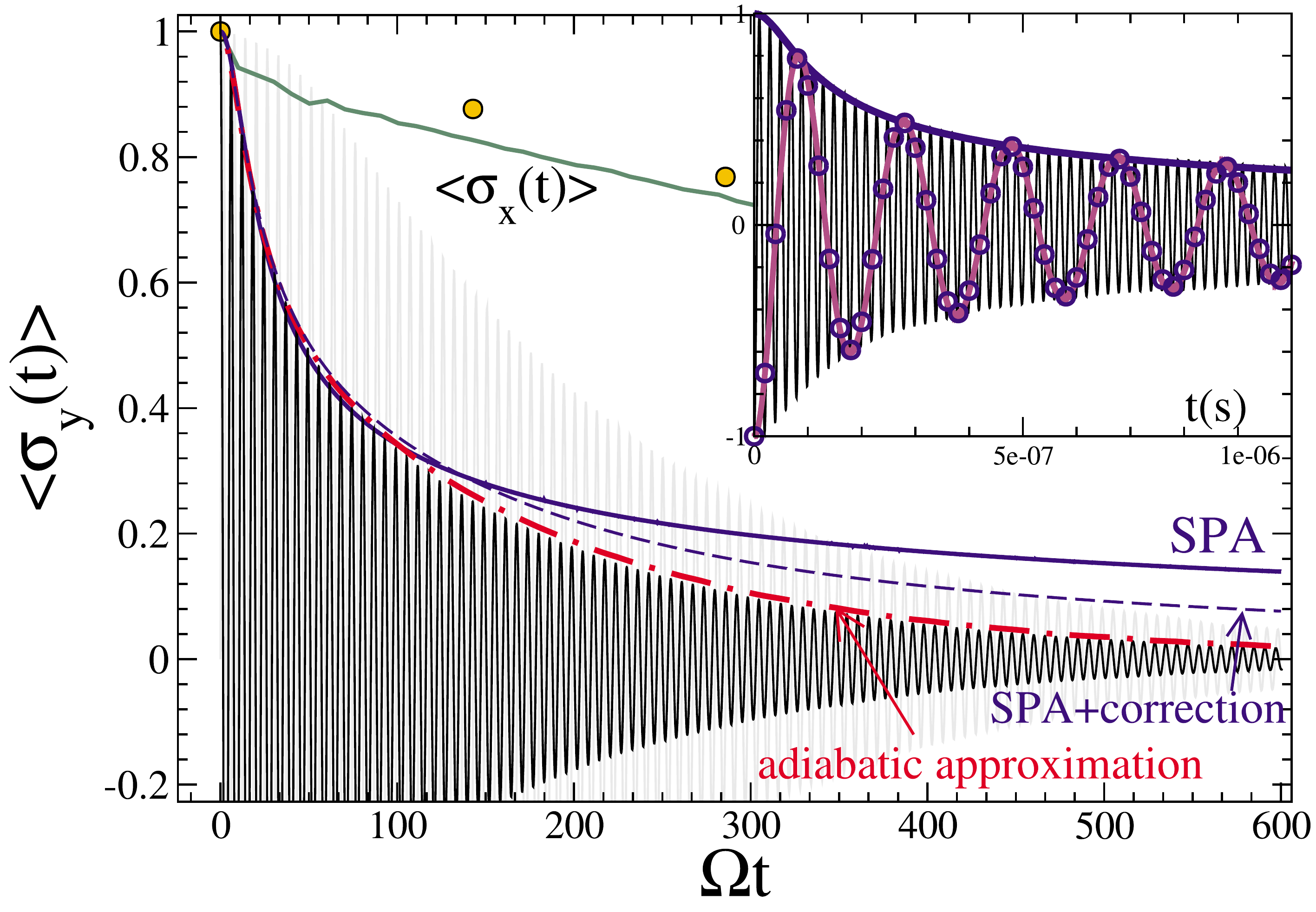}
}
\caption{(Color online) 
Simulations of an adiabatic \fs $1/f$
environment at $\theta=\pi/2$. Relaxation studied via $\langle \sigma_x \rangle$ (green
line) is well approximated by the weak coupling theory, $T_2$(dots).
Dephasing in repeated measurement damps the oscillations (thin
black line). Part of the signal is recovered if the environment is
re-calibrated (thin gray line). Noise is produced by $n_d= 250$ \fs
per decade, with $1/t_m =  10^5 {\rm rad/s} \leq \gamma_i \leq \gamma_M= 10^9 {\rm rad/s} <
\Omega= 10^{10} {\rm rad/s}$. The coupling $ \bar v = 0.02 \Omega$  is appropriate to
charge devices, and corresponds to  $S=16 \pi A E_C^2/\omega$
with $A = 10^{-6}$ \cite{Zorin1996}. The adiabatic approximation, 
Eq.~(\ref{eq:pathint-adiabatic-longitudinal}), fully accounts
for dephasing (red dotted-dash line). The static-path
approximation (SPA), Eq.~(\ref{ref:quadratic}), (blue solid line) and the first
correction (blue dashed line) account for the initial suppression,
and it is valid also for times $t \gg 1/ \gamma_M$. In the inset, Ramsey
fringes with parameters appropriate to the experiment \cite{Vion2002} (thin
black lines). The SPA (blue solid line), Eq.~(\ref{ref:quadratic}), is in excellent
agreement with observations, and also predicts the correct
phase shift of the Ramsey signal (blue dots, compared with
simulations for small detuning $\delta=  5$ MHz, violet line), which
tends to $\approx \pi/4$ for large times.
Adapted from \cite{Falci2005}}
\label{fig:adiabaticSPA}
\end{figure}

The effect on the qubit dynamics of adiabatic transverse fluctuations 
weakly depends on time, on the other side, longitudinal components are responsible for phase errors,
which accumulate in time. 
Thus adiabatic transverse noise has possibly some 
effect at very short times, but the phase damping channel eventually prevails. The relevance of these effects
quantitatively depends on the amplitude of the noise at low frequencies.
It has been demonstrated in \cite{Falci2005} (see Fig.~\ref{fig:adiabaticSPA}) that,
for the typical noise figures of superconducting devices and at least for short times scales relevant for quantum
computing, the leading effect of adiabatic noise is defocusing originated by the fluctuating splitting during the
repetitions of the measurement runs. 
This effect is captured by neglecting the transverse terms in
Eq.~(\ref{H_adiabatic3}), i.~e., making the {\em longitudinal approximation} of the Hamiltonian (\ref{H_adiabatic})
\begin{equation}
\hat H \approx \frac{\hbar \Omega(q,0)}{2} \sigma_{z^\prime} + \frac{\hbar \delta \Omega_\parallel(t)}{2} \sigma_{z^\prime}  \, .
\label{H_adiabatic_long}
\end{equation}
In addition, consistently with $\tilde\theta_q(t)\approx \theta_q$,  the splitting fluctuations are further 
approximated as~\cite{Falci2005,Ithier2005} 
\begin{equation}
\delta \Omega_\parallel(t) \approx  \frac{\partial \Omega}{\partial q} (\delta q) 
+ \frac{1}{2} \frac{\partial^2\Omega}{\partial q^2} ( \delta q)^2 +
\cdots  \, .
\label{splitting_adiab_exp}
\end{equation}
Considering the explicit dependence on $\delta q$ of $\Omega(q,\delta q)$, the first terms of the expansion read,
cf. with Eq. (\ref{deU}),
\begin{equation}
\Omega(q,\delta q(t)) \approx \Omega(q) + E(t) \cos \theta_q + \frac{1}{2} \frac{(E(t) \sin \theta_q)^2}{\Omega(q)} \, .
\label{Omega_expansion}
\end{equation}
For longitudinal noise, $\theta_q=0$, we recover the exact linear dependence on the noise of the instantaneous splitting.
For transverse noise instead, $\theta_q= \pi/2$, the first non-vanishing term of the expansion is quadratic.
It is common to refer to this regime as the ``quadratic coupling"~\cite{Ithier2005}, or
``quadratic longitudinal coupling" \cite{Makhlin2004} condition. 

We remark that Eq.~(\ref{H_adiabatic_long}) also applies when both components, 
$\Omega_x$ and $\Omega_z$,
fluctuate. For instance, this is the case of flux qubits where both flux and critical current fluctuate with 
$1/f$ spectrum and the physical fluctuating quantity $\delta q$ is  a function of  the magnetic flux or of
the critical current. The partial derivatives can be expressed in terms of noise sensitivities as follows
\begin{equation}
\frac{\partial \Omega}{\partial q} = \frac{\partial \Omega}{\partial \Omega_z} \frac{\partial \Omega_z}{\partial q}
+ \frac{\partial \Omega}{\partial \Omega_x} \frac{\partial \Omega_x}{\partial q}
=\frac{\Omega_z}{\Omega} \frac{\partial \Omega_z}{\partial q}
+ \frac{\Omega_x}{\Omega} \frac{\partial \Omega_x}{\partial q}
\label{sensitivities}
\end{equation}
where 
the noise sensitivities $\partial \Omega_z /\partial q$, $\partial \Omega_x /\partial q$ 
can be inferred from spectroscopy measurements as in the experiment \cite{Bylander2011}.

A formal expression for the qubit dynamics in the adiabatic approximation, introduced in \cite{Falci2005} has been
discussed in details and extended to more complex gates in \cite{Paladino2009}. Here we report the results in the
adiabatic and longitudinal approximations. In this regime populations of the qubit density matrix in the eigenbasis
do not evolve, whereas the off-diagonal elements are obtained by averaging over all the realizations of the stochastic
process $E(t)$  expressed by the path-integral
\begin{eqnarray}
\label{eq:pathint-adiabatic-longitudinal}
\frac{\rho_{mn}(t)}{\rho_{mn}(0)} 
&=&  
\int \hskip-2pt {\cal D}[ E(s)] \,
P[E(s)] \, {\mathrm e}^{- i \int_0^t \!\!d s \, 
\Omega_{m n}(q, \delta q(s))} .
\end{eqnarray}
 Here $P[E(s)]$ contains information both on the stochastic 
processes and on details of the specific protocol. It is convenient 
to split it as follows
$$
P[E(s)] \,=\, F[E(s)] \; p[E(s)] \, ,
$$
where $p[E(s)]$ is the probability of the realization $E(s)$. 
The filter function $F[E(s)]$ describes the specific operation. For 
most of present day experiments on solid-state qubits 
$F[E(s)]=1$. For an open-loop feedback protocol, which allows initial 
control of some collective variable of the environment,
say  $E_0 = 0$, we should put $F[E(s)] \propto \, \delta(E_0)$. 
The different decay of coherent oscillations in each protocol in the
presence of adiabatic noise originates from the specific filter function which
needs to be specified at this stage.

A critical issue is the identification of $p[E(s)]$ 
for the specific  noise sources, as those displaying $1/f$ power spectrum.
If we sample the stochastic process at times $t_k = k \Delta t$, with   
$\Delta t = t/n$, we can  identify
\begin{equation}
p[E(s)] = \lim_{n \to \infty} 
p_{n+1}(E_n, t;\dots; E_1, t_1; E_0, 0)
\label{sampling}
\end{equation}
where $p_{n+1}(\cdots)$ is a $n+1$ joint probability. 
In general, this is a formidable task. However, a systematic method can be found to 
select only the relevant statistical information on the stochastic 
process out of the full characterization included in $p[E(s)]$~\cite{Falci2005,Paladino2009}.

The signal decay in FID is obtained by performing in Eq.~(\ref{eq:pathint-adiabatic-longitudinal})
the Static Path Approximation (SPA), which
consists in neglecting the time dependence in the path, $E(s) = E_0$ and taking $F[E]=1$. 
In the SPA the problem reduces to ordinary integrations with
$p_1(E_0,0) \equiv p(E_0)$. The qubit coherences can be written as
$\rho_{01}(t) = \rho_{01}(0) \, 
\exp[- i  \,\Omega t - i \,\Phi(t)]$
with the average  phase shift 
\begin{equation} \label{eq:blur-static}
\Phi(t) \approx   i \ln  \left(\int  \! dE_0 \, p(E_0) \; e^{i t \sqrt{\Omega^2+E_0^2 +2\Omega E_0 \cos \theta }} \right). 
\end{equation}
Clearly, Eq.~(\ref{eq:blur-static}) describes the effect of a distribution of stray energy 
shifts $\hbar [\Omega(q, \delta q) - \Omega(q,0) ]$ and corresponds to the
rigid lattice breadth contribution to inhomogeneous 
broadening. In experiments with solid state devices this 
approximation describes the measurement procedure consisting in 
signal acquisition and averaging over a large number $N$ 
of repetitions of the protocol, for an overall time $t_m$ 
(minutes in actual experiments). 
Due to slow fluctuations of the 
environment calibration, the initial value, 
$\Omega \cos \theta+E_0$,  fluctuates during the repetitions
blurring the average signal, independently on 
the measurement being single-shot or not. 

The probability $p(E_0)$ describes the 
distribution of the random variable obtained by sampling 
the stochastic process $E(t)$ at the initial time of each repetition, i.~e.,
at times
$t_k = k \, t_m/N$, $k=0, \ldots , N-1$.
If $E_0$ results from many independent random variables 
of a multi-mode environment, the central limit theorem applies and
$p(E_0)$ is a Gaussian distribution with standard 
deviation $\sigma$,
$$
\sigma^2 \,=\, \langle E^2 \rangle \,=\, 2 \int_0^\infty d \omega \,
S(\omega) \, ,
$$
where  integration limits are intended as $1/t_{m}$, and the high-frequency cut-off of the
$1/f$ spectrum, $\gamma_M$.
In the SPA the distribution standard deviation,
$\sigma$, is the only adiabatic noise characteristic parameter.
If the equilibrium average of the stochastic process vanishes, 
Eq.~(\ref{eq:blur-static}) reduces to 
\begin{equation} 
\Phi(t) \approx i \ln \left[ \int \!  \! \frac{dE_0}{\sqrt{2 \pi \sigma^2}}\,  e^{-\frac{E_0^2}{ 2 \sigma^2 }}
e^{i t \sqrt{\Omega^2 +E_0^2  +2\Omega E_0\cos \theta}}\right] \! .
\label{SPA-oneqb}
\end{equation}
Expanding the square root in the above expression
we obtain~\cite{Falci2005}
\begin{equation}
\label{ref:quadratic}
\Phi(t) = - \frac{i}{2}\left[ \frac{\Omega (\cos \theta \, \sigma t)^2 }{ \Omega +  i 
\sin^2 \theta \sigma^2 t } + 
\ln \frac{\Omega +  i 
\sin^2 \theta \sigma^2 t }{\Omega} \right] .
\end{equation}
The short-times decay of coherent oscillations qualitatively depends on the
working point.
In fact, the suppression of the signal, $\exp[\Im \Phi(t)]$,  turns from an exponential behavior,
$\propto e^{- (\cos \theta \sigma t)^2/2}$,
at $\theta \approx 0$ to a power law,  
$[1 + (\sin \theta^2 \sigma^2 t/\Omega)^2 ]^{-1/4}$, 
at $\theta \approx \pi/2$. 
In these limits Eq.~(\ref{ref:quadratic}) reproduces the results for Gaussian $1/f$ environments
in the so called ``quasi-static case" reported in \cite{Ithier2005}. 
In particular, at $\theta = 0$ we obtain the short-times limit,
$t \ll 1/\gamma_M$, of the exact result of \textcite{Palma1996}, Eq.~(\ref{vf3}).
In fact, for very short times we can approximate $\sin^2(\omega t/2)/(\omega t/2)^2 \approx 1$ inside the
integral Eq.~(\ref{vf3}), obtaining the exponential quadratic decay law at pure dephasing predicted by
the Gaussian approximation.   
At $\theta = \pi/2$ the short and intermediate times result of \textcite{Makhlin2004}  for
a Gaussian noise and ``quadratic coupling"  is reproduced.

The fact that results of a diagrammatic approach with a quantum environment, 
as those of \textcite{Makhlin2004}, can be reproduced and 
generalized already at the simple SPA level makes 
the semi-classical approach quite promising. It shows that, at 
least for not too long times (but surely longer than times of interest
for quantum state processing), the quantum nature of the 
environment may not be relevant for the class of problems, 
which can be treated in the Born-Oppenheimer approximation. 
Notice also that the SPA itself has surely a wide validity 
since it does not require information about the {\em dynamics} 
of the noise sources, provided they are  
slow.
For Gaussian wide-band $1/f$ noise and for times $\gamma_m \ll 1/t < \gamma_M$, the contribution of frequencies
$\omega \ll 1/t$ can be approximated by Eq.~(\ref{ref:quadratic}) where the noise variance is evaluated 
integrating the power spectrum from $\gamma_m$ to $1/t$, that is 
$\sigma^2 = \mathcal{A} \ln{(1/\gamma_m t)}$~\cite{Cottet2001,Nakamura2002,Ithier2005}.
The diagrammatic approach of \textcite{Makhlin2004} for Gaussian noise and quadratic coupling 
also predicts a crossover from algebraic behavior to exponential decay at long times, $\mathcal A t \gg 1$,
with rate $\mathcal A/2$. This behavior is, however, hardly detectable in experiments, where the long-time
behavior is ruled by quantum noise (see discussion in the following section \ref{paragraph:multistage})

Equation (\ref{eq:pathint-adiabatic-longitudinal}) can be systematically approximated by proper sampling  
$E(t^\prime)$ in $[0,t]$.  
In this way also the dynamics of the noise sources during the runs can be systematically included.
For the first correction, $p[E(t^\prime)]$ can be approximated by the joint distribution
$p[E_t t; E_0,0]$, where $E_t \equiv  E(t)$.  At $\theta = \pi/2$ for generic Gaussian noise we find
$$
\Phi(t) =  -\frac{i }{2} 
\ln \left[\left(\frac{\Omega +  i 
\sigma^2 t [1 - \Pi(t)] } {\Omega} \right) 
\left(\frac{3\Omega +  i \sigma^2 t \Pi(t)} {3\Omega} \right)\right] $$
where $\Pi(t)  \equiv  \int_0^\infty (d \omega/ \sigma^2) S(\omega) \left(1- e^{-i \omega t} \right)$
is a transition probability, depending on the stochastic process. For
Ornstein-Uhlenbeck processes it reduces to the result of \textcite{Rebenstein2004}. 
The first correction suggests that the SPA, in principle valid for 
$t <  1/\gamma_M$, may have  a broader validity (see Fig.~\ref{fig:adiabaticSPA}). 
For $1/f$ noise due to a set of bistable impurities it is valid also for 
$t \gg 1/\gamma_M$,  if $\gamma_M \lesssim \Omega$. Of course, 
the adiabatic  approximation is tenable if $t < T_1 = 2/S(\Omega)$.

As already pointed out for the SF model, the signal decay during echo protocols is very sensitive 
to the dynamics of the noise sources within each pulse sequence.  The SPA cannot capture these effects
and would yield no decay for the echoes. On the other side, predictions critically depend on the specific
sampling of the stochastic process $E(t^\prime)$, making it difficult to obtain reliable estimates.
The decay of the echo signal is often approximated assuming Gaussian quasi-static  ($t< 1/\gamma_M$)  
noise  \cite{Ithier2005}.
 
\begin{figure}[t]
\centerline{
\includegraphics[width=1\columnwidth,angle=-90]{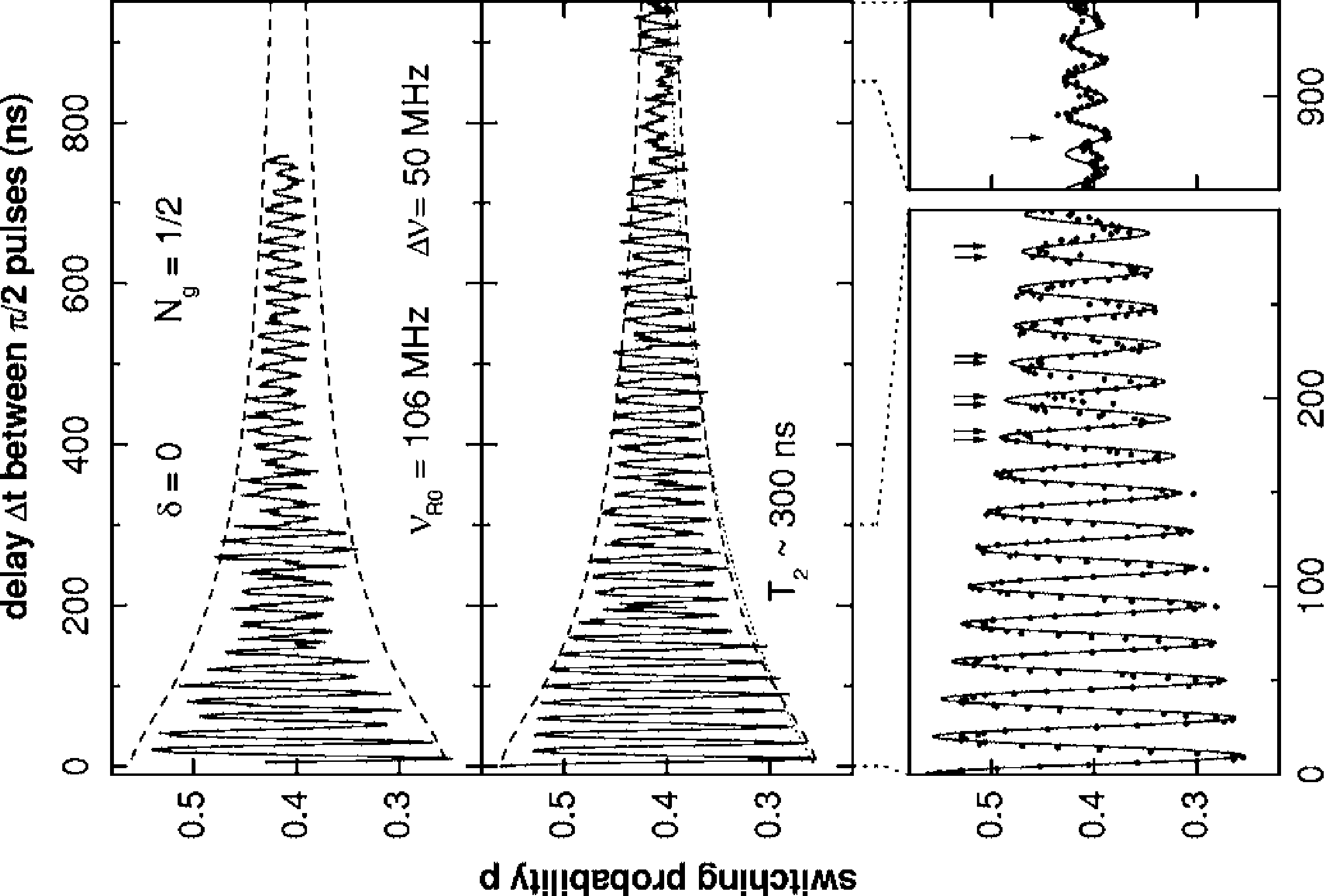}
}
\caption{Ramsey signals at the optimal point for 
$\omega_{R0}/2 \pi=106$ MHz and $ \Delta \nu =50$ MHz, as a function of the delay $\Delta t$ between
the two $\pi /2$ pulses. Top and middle panels: solid lines are
two successive records showing the partial reproducibility of the
experiment. Dashed lines are a fit of the envelope of the oscillations
in the middle panel  leading to $T_2=300$ ns. The dotted line
shows for comparison an exponential decay with the same $T_2$. Bottom
panels: zoom windows of the middle panel. The dots represent
now the experimental points whereas the solid line is a fit of the
whole oscillation with  $\Delta \omega/2 \pi =50.8$ MHz. Arrows point out a few
sudden jumps of the phase and amplitude of the oscillation, attributed
to strongly coupled charged TLFs. Adapted from \cite{Ithier2005}.}
\label{fig7_ithier2005}
\end{figure}

Several experiments confirmed that the effect of $1/f$ noise in repeated
measurements protocols can be described by the above simple
theory~\cite{Cottet2001,Ithier2005,VanHarlingen2004,Martinis2003,Chiarello2012,Sank2012,Yan2012}. 
Of course, the dominant $1/f$ noise sources are device and material dependent and the decay of the measured signal 
depends on the measurement procedure. An accurate estimate and comparison with Gaussian theory
of defocusing has been done in \cite{Ithier2005} for the quantronium.  A clear evidence of
an algebraic decay at the optimal point due to quadratic $1/f$ charge noise has been reported,
see Fig.~\ref{fig7_ithier2005}, in agreement with the prediction of \textcite{Falci2005}, 
Fig.~\ref{fig:adiabaticSPA}. 

In \cite{VanHarlingen2004} the effects of $1/f$ noise in the critical current in various superconducting
qubits incorporating Josephson junctions have been computed numerically showing that the envelope of
the coherent oscillations scales as $\exp{[-t^2/(2 \tau_\phi^2)]}$, in qualitative agreement with
the adiabatic and longitudinal approximation. Interestingly, the extrapolated decay time $\tau_\phi$
depends both on the elapsed time of the experiment and on the sequence in which the measurements are
taken. Two averaging methods were employed: a time-delay averaging (method A) which averages only high-frequency
fluctuations at each time-delay point, and  a time-sweep averaging (method B) which averages both 
high- and low-frequency components. Method A gives longer dephasing times than method B, since
the number of decades of $1/f$ noise that affect the qubit dynamics in method A is smaller than
the number of decades sampled in method B. For a large number of repetitions, $\tau_\phi$
for method B approaches the prediction of \textcite{Martinis2003} of the effect of critical current $1/f$ noise
on a phase qubit in the Gaussian approximation, i.~e., Eq.~(\ref{vf3}) with integration limits $0$ and $\Omega$.    
The dephasing time $\tau_\phi$ is found to scale as $I_0 \equiv \Omega \Lambda S_{I_0}^{1/2}(1 \, \text{Hz})$, 
where $S_{I_0}(1\, \text{Hz})$ is the spectral density of the critical-current noise 
at $1$ Hz, and $\Lambda = I_0 d\Omega / \Omega d I_0$ is computed for given parameters 
for each  type of qubit that specifies the sensitivity of the level splitting to critical-current 
fluctuations \cite{VanHarlingen2004}.

Of course, any setup also suffers from noise sources active at frequencies around GHz, which cannot be treated in the
adiabatic approximation. In 
Sec.~\ref{paragraph:multistage} we will discuss a convenient procedure
to deal with the interplay of noise sources responsible for spectral fluctuations in different frequency ranges.

\paragraph{$1/f$ noise during ac-driven evolution: decay of Rabi oscillations --}
\label{Inhom-Rabi}

The manipulation of superconducting qubits is often achieved by varying the control
parameters in a resonant way at a microwave angular frequency close to the qubit transition
frequency, $\Omega$. Rabi oscillations are routinely measured in different labs, e.~g., \cite{Nakamura2001,Vion2002,Yu2002,Martinis2002,Chiorescu2003}. 
The decohering effect of $1/f$ noise during driven evolution is actually weaker than in the
undriven case. The intuitive reason is that ac-field averages the effects
of noise~\cite{Ithier2005}. This is explained 
treating the ac-driven noisy system in the adiabatic
and longitudinal approximation~\cite{Falci2012}.
 
A qubit acted by an external ac-field can be modeled by the Hamiltonian (\ref{H_qb_linearized})
where the control is operated via $\hat H_c(t)= \hbar {\cal B} \, \cos(\omega t) \hat Q$. 
The device is nominally biased at $q$ and its low-frequency
fluctuations $\delta q(t)$ can be treated in the SPA. 
The problem is  solved in the Rotating Wave approximation
which keeps only control entries ``effective'' in triggering transitions between 
different states. This effective part of the control also depends on the random variable $\delta q$. 
The populations in the  rotating frame are readily found, 
e.~g., the population of the first excited state is 
$P_1(t|\delta q)= (\Omega_R/2 \Omega_{fl}) \, [1 - \cos (\Omega_{fl}t)]$
\noindent where the flopping frequency for Rabi oscillations is 
$\Omega_{fl}(q,\delta q)=\sqrt{\eta^2(q,\delta q)+\Omega_R^2(q,\delta q)}$. 
Here $\Omega_R(q,\delta q) = {\cal B} \, Q_{10}(q,\delta q)$ is the peak Rabi frequency,
$Q_{10}$ being the matrix element of $\hat Q$ in the instantaneous eigenbasis of $\hat H_0 (q + \delta q)$ and
the detuning is $\eta(q,\delta q) = \Omega(q,\delta q)-  \omega$.
Averages are evaluated by expanding $\Omega_{fl}$ to second order as in Eq.~(\ref{Omega_exp}),
\begin{equation}
\Omega_{fl}(q,\delta q) \approx \Omega_{fl}(q,0) +  
\Omega_{fl}^\prime(\delta q) 
+ \frac{1}{2} \Omega_{fl}^{\prime \prime} 
(\delta q)^2 + \cdots \, , 
\label{Omega_fl_exp}
\end{equation}
where $\Omega_{fl}^\prime =  \partial \Omega_{fl}/\partial q$,
$\Omega_{fl}^{\prime \prime}= \partial^2\Omega_{fl}/\partial q^2$.a
Assuming a Gaussian distribution of $\delta q$ with variance $\sigma_q$
one obtains
\begin{eqnarray}
&&\!\!\!\!\!\langle e^{-i \Omega_{fl}(\delta q) t} \rangle = 
e^{-i \Omega_{fl}(0) t} \,e^{-i \Phi(t)} , \nonumber \\ 
&&\!\!\!\!\!e^{-i \Phi(t)} =\frac{1}{\sqrt{1+i \, \Omega_{fl}^{\prime \prime} \sigma_q^2 t}}
\, \exp\left[- \frac{(\Omega_{fl}^\prime \sigma_q \,t)^2}{ 2 (1+i \, \Omega_{fl}^{\prime \prime} \sigma_q^2 t)}\right].
 \label{eq:static-path-approx}
\end{eqnarray}
This equation describes different regimes  for the decay of Rabi  oscillations, namely a Gaussian time decay 
$|\mathrm{e}^{-i \Phi(t)}|\sim \mathrm{e}^{-{1 \over 2} (\Omega_{fl}^\prime \sigma_q \, t)^2}$
when the linear term in the expansion dominates, and power-law behavior 
\mbox{$\sim 1/[\sigma_q (\Omega_{fl}^{\prime \prime}t)^{1/2}]$ }
 when $\Omega_{fl}^\prime \to 0$. 
In this regime Eq.~(\ref{eq:static-path-approx}) describes the initial suppression of the signal. 
The coefficients of the expansion  depend on several parameters such as
the amplitude of the control fiels and the nominal detuning $\eta_0 = \Omega(q,0)- \omega$. 
Further information, such as the dependence on 
$q$ of the energy spectrum and of matrix elements $Q_{ij}(q)$,  
comes from the characterization of the device. 
The variance can be related to the integrated  power spectrum,.  and can be extracted from FID
or Ramsey experiments~\cite{Ithier2005,Chiarello2012}.
A similar power law decay of Rabi oscillations has been observed for an electron spin in a 
QD, due to the interaction with a static nuclear-spin bath in \cite{Koppens2007}.

Notice that even if Eq.~(\ref{eq:static-path-approx}) 
describes the same  regimes of the SPA in the undriven case, Eq.~(\ref{ref:quadratic}), here the 
situation is different. In particular, Eq.~(\ref{eq:static-path-approx})
quantitatively accounts for the fact that ac-driving greatly reduces 
decoherence compared to undriven systems. 
In particular, at resonance,  $\eta_0 = 0$, non-vanishing linear fluctuations of the spectrum,  
$\partial \Omega / \partial q \neq 0$,  determine quadratic fluctuations of $\Omega_{fl}(q,\delta q)$
(neglecting fluctuations of $Q_{ij}$). 
Thus $\Omega_{fl}^\prime=0$ and Rabi oscillations undergo power-law decay, whereas in the absence of 
drive they determine the much stronger Gaussian decay 
$\sim \mathrm{e}^{-{1 \over 2}(\partial \Omega / \partial q)^2 \sigma^2 \,t^2}$ 
of coherent oscillations.
In this regime measurements of Rabi oscillations~\cite{Bylander2011}  have been used
to probe the environment of a flux qubit.
At symmetry points, $\partial \Omega / \partial q = 0$, coherent oscillations decay 
with a power law, whereas Rabi oscillations are almost unaffected by 
low-frequency noise. For non-vanishing detuning decay laws are equal being the system driven or not. 

This picture applies to many physical situations, 
since fluctuations of $Q_{ij}$ are usually small, corresponding to 
a fraction of $\Omega_R \neq 0$, whereas $\eta(q,\delta q)$ fluctuates 
on the scale of the Bohr splitting $\Omega \gg \Omega_R$ and may 
be particularly relevant for $\eta_0 =0$. 
The dependence $Q_{ij}(q)$ may have significant consequences in multilevel systems.

Recently, the influence of external driving on the noise
spectra of bistable \fs was investigated in \cite{Constantin2009}. 
The authors proposed an idea that external driving may saturate
the \fs thus decreasing their contribution to the dephasing. A calculation based
on the Bloch-Redfield formalism has shown that the
saturation of some \fs does not lead to significant decrease in decoherence.
\textcite{Brox2011} took into account  the effect of
driving on the dynamics of the fluctuators. The main result of this analysis is
the prediction that additional low-frequency driving may shift the noise spectrum 
towards high frequencies. Since the dephasing is influenced mostly by the 
low-frequency tail of the noise spectrum this shift decreases decoherence.  However, 
the predicted effect is not very strong.

\paragraph{Broadband noise: multi-stage approach --}
\label{paragraph:multistage}

In the last part of this subsection, we have to warn the reader that when comparing the above predictions
with experiments, one has to keep in mind
that in nanodevices noise is typically broadband and structured. 
In other words, the noise spectrum 
extends to several decades, it is non-monotonic, sometimes a few resonances are present.
We already mentioned that  typical $1/f$-noise measurements
extend from a few Hz to $\sim$$100$~Hz~\cite{Zorin1996}. 
The intrinsic high-frequency cut-off of  $1/f$ noise is hardly detectable.
Recently, charge noise up to $10$~MHz has been 
detected in a SET~\cite{Kafanov2008}. Flux and critical current 
noises with $1/f$ spectrum consistent with that in the $0.01$-$100$ Hz range
were measured  at considerably higher frequencies ($0.2$-$20$~MHz)~\cite{Bylander2011,Yan2012}. 
Incoherent energy exchanges between system and environment, leading to relaxation and
decoherence, occur at typical operating frequencies (about $10$~GHz).
Indirect measurements of noise spectrum in this frequency range often 
suggest a ``white" or Ohmic behavior~\cite{Astafiev2004,Ithier2005}.
In addition,  narrow resonances at selected frequencies (sometimes 
resonant with the nanodevice-relevant energy scales) have being
observed~\cite{Simmonds2004,Cooper2004,Eroms2006}.  
 
The various noise sources responsible for the above phenomenology
have a qualitative different influence on the system evolution. 
This naturally leads to a classification of the noise sources  
according to the effects produced rather than to their specific nature.
Environments with long-time memory  belong to the class of adiabatic noise for 
which the Born-Oppenheimer approximation is applicable.  
This part of the noise spectrum can be classified as ``adiabatic noise": $1/f$ noise falls in this noise class.
High-frequency noise is essentially responsible for spontaneous decay and can be classified as ``quantum noise".
Finally, resonances in the spectrum unveil the presence of discrete noise sources
which severely effect the system performances, in particular reliability of devices.
This is the case when classical impurities are slow enough to induce a visible
bistable instability in the system intrinsic frequency. This part of the
noise spectrum can be classified as ``strongly-coupled noise".
Each noise class requires a specific approximation scheme, which is not
appropriate for the other classes. The overall effect results from the interplay
of the three classes of noise. In order to deal with broadband and structured noise we
have to resort to a multi-scale theory which can be sketched as follows~\cite{Falci2005,Paladino2009,Taylor2006}.

Suppose we are interested to a reduced description of a $n$-qubit system, 
expressed by the reduced density matrix $\rho^{n}(t)$. It is formally obtained
by tracing out environmental degrees of freedom from the total density matrix 
$W^{Q,A,SC}(t)$, which depends on  quantum (Q), adiabatic (A) and strongly coupled 
(SC) variables.
The elimination procedure can be conveniently performed by separating
in the interaction Hamiltonian, which we write analogously to Eq.~(\ref{H_qb_linearized}), as 
$\sum_i \sigma_z^{(i)} \otimes \hat E_i$, various
noise classes, e.~g., by formally writing
\begin{equation}
\sigma_z^{(i)} \otimes \hat E_i = \sigma_z^{(i)} \otimes \hat E_i^Q + 
\sigma_z^{(i)} \otimes \hat E_i^A + \sigma_z^{(i)} \otimes \hat E_i^{SC} \, .
\label{int-plit}
\end{equation}
Adiabatic noise is typically correlated on a time scale much longer than 
the inverse of the natural frequencies $\Omega_i$, 
then application of the Born-Oppenheimer approximation
is equivalent to replace $\hat E_i^A$ 
with a classical stochastic field  $E_i^A(t)$.
This approach is valid when the contribution of adiabatic noise to 
spontaneous decay is negligible, a necessary condition being 
$t \ll T_1^A \propto S^A(\Omega)^{-1}$. This condition is usually satisfied at
short enough times  since $S^A(\omega) \propto 1/\omega$.
 
This fact  suggests a route to trace out different noise classes
in the appropriate order.  The total density matrix parametrically depends on 
the specific realization of the slow random drives $\vec E(t) \equiv \{E_i^A(t)\}$ and 
may be written as $W^{Q,A,SC}(t)= W^{Q,SC}[t | \vec E(t)]$. 
The first step is to trace out quantum noise. In the simplest cases this requires
solving a master equation. In a second stage, the average over all the realizations of
the stochastic processes, $\vec E(t)$, is performed. This leads to a reduced density
matrix for the $n$-qubit system plus the strongly coupled degrees of freedom.
These have to be traced out in a final stage  by solving the Heisenberg
equations of motion, or by approaches suitable to the specific microscopic 
Hamiltonian or interaction. In particular, this is the case discussed in the initial part of this Section, 
of the spin-fluctuator model at pure dephasing.
The ordered multi-stage elimination procedure can be formally written as
$$
\rho^{n}(t) = \mathrm{Tr}_{SC}  \! \left \{
\int \! \! {\mathcal D}[\vec E(t)]  P [\vec E(t)]  
\mathrm{Tr}_{Q} \! \left[ W^{Q,SC}\left(t | \vec E(t)\right) \right]  \right\}\! .
$$
\begin{figure}[t]
\centerline{
\includegraphics[width=1.00\columnwidth]{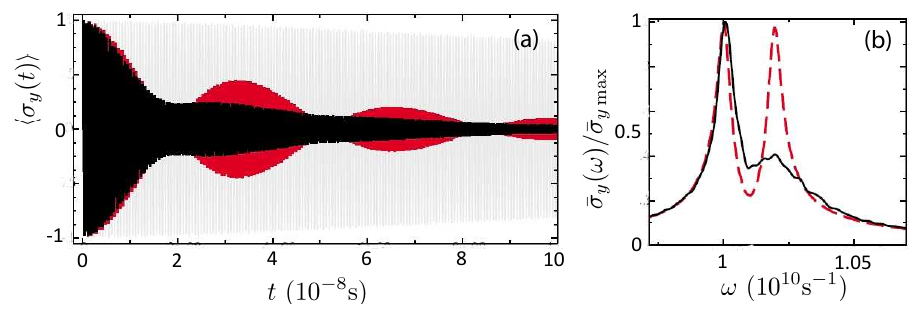}
}
\caption{(Color online) (a) $\langle \sigma_y \rangle$ at  $\theta = \pi/2$,  $\Omega/(2 \pi)= 10^{10}$~Hz. The
effect of weak adiabatic $1/f$ noise (light gray line)
($\gamma \in 2 \pi \times [10^5, 10^9]$~Hz, uniform $2b=  0.002 \Omega$, $n_d=  250$) is strongly
 enhanced by adding a single slow ($\gamma/\Omega = 0.005$) more strongly
coupled ($2b_0/\Omega=  0.2$) \fl (black line), which alone would give
rise to beats (red line). (b) When the bistable fluctuator is present the Fourier
transform of the signal may show a split-peak structure. Even if
peaks are symmetric for the single  bistable fluctuator alone (dashed line), $1/f$
noise broadens them in a different way (solid line). Adapted from~\cite{Falci2005}.
}
\label{fig_4_Falci2005}
\end{figure}
\noindent
The multi-stage approach allows making predictions in realistic situations
when the outcome of a measurement results from the effects of various noise classes.
For instance, we can address  the interplay of $1/f$ noise with RT noise produced by one \fl 
which is more strongly coupled with the qubit,
having $\gamma_0 \ll 1/t \ll \Omega$, but $b_0 \leq \Omega$. 
Even if the \fl is not resonant with the qubit it strongly affects the output signal. If $g_0 > 1$, it 
determines beats in the coherent oscillations and split peaks in
spectroscopy, which are signatures of a discrete environment. Because of the mechanism 
discussed in Sec.~\ref{few-fs}, the additional \fl makes bistable the working point
of the qubit and amplifies defocusing due to $1/f$ noise.
Even if the device is initially optimally polarized, during $t_m$
the \fl may switch it to a different working point. The line
shape of the signal will show two peaks, split by
and differently broadened by the $1/f$ noise in background.
The corresponding time evolution will show damped beats,
this phenomenology being entirely due to the non-Gaussian nature of the environment. 
Figure~\ref{fig_4_Falci2005}, adapted from \cite{Falci2005},
shows results of a simulation at the optimal point,
where $1/f$ noise is adiabatic and weaker than the typical
noise level in charge qubits, this picture
applies to smaller devices. The fact that even a single
impurity on a $1/f$ background causes a substantial suppression of the signal poses the problem of reliability of
charge based devices. We remark that the reported beatings
do not arise from a resonant coupling between the qubit and the
fluctuator. This situation will be address later in this Section.

Finally, we mention that a commonly used simplified version of the multi-stage approach consists in simply 
factorizing the effects  of different noise classes, in particular of adiabatic and quantum noise.
For some measurements protocols, when the responsible noise classes act on sufficiently different time scales
and the spectra are regular around the relevant frequency ranges, this approximation leads to reasonable
prediction of the signal decay~\cite{Martinis2003,Ithier2005}.

\subsubsection{1/f noise in complex architectures} 
\label{subsub_complex}

Realizing the promise of quantum computation requires implementing a universal set of
quantum gates, as they provide the building blocks for encoding complex algorithms and
operations. To this end, scalable qubit coupling and control schemes capable of realizing
gate errors small enough to achieve fault tolerance are required.  
Fluctuations with $1/f$ spectrum represent a limiting factor also for the controlled generation 
of entangled states (two-qubit gates) and for the reliable preservation two- (multi-) qubit quantum 
correlations (entanglement memory). 
In addition to fluctuations experienced by each single qubit, solid state coupled qubits, being usually 
built on-chip, may suffer from correlated noise sources acing simultaneously on both sub-units.
Already in 1996, measurements on SET circuits revealed $1/f$ behavior of voltage 
power spectra on two transistors and of the cross-spectrum power density \cite{Zorin1996}.\footnote{
The cross-spectrum of two stochastic processes $X_1(t)$ and $X_2(t)$ is defined as
$S_{X_1 X_2}(\omega)= (1/ \pi) \int_0^\infty dt \,  C_{X_1 X_2}(t) \, \cos \omega t $, where 
$ C_{X_1 X_2}(t)= \langle X_1(t) X_2(0) \rangle - \bar X_1 \bar X_2$ and $\bar X_\alpha = \langle X_\alpha(t)
\rangle$.}
In Josephson charge qubits,  \fs in the insulating substrate might influence several qubits
fabricated on the same chip, whereas fluctuating traps concentrated inside the oxide layer of the tunnel 
junctions are expected to act independently on the two qubits, due to screening by the junction electrodes. 
Background charge fluctuations could also lead to significant gate errors and/or decoherence
in semiconductor-based electron {\em spin-qubits} through inter-qubit exchange coupling~\cite{Hu2006}.

Recent investigations aimed at identifying operating conditions or control schemes allowing protection 
from $1/f$ fluctuations in complex architectures. 
On one side, passive protection strategies, like ``optimal tuning" of nanodevices
extending the single-qubit optimal point concept \cite{Paladino2010,Paladino2011,DArrigo2012} and the 
identification of symmetry protected subspaces~\cite{Storcz2005,Brox2012} have been proposed. 
Alternatively, or in combination with passive protection, resonant rf-pulses \cite{Rigetti2005} and 
pulse sequences eventually incorporating spin echoes \cite{Kerman2008} have been considered as well. 

In this  section we will focus on the first strategies; dynamical decoupling approaches will be presented 
in Sec.~\ref{DD}.
In addition, we will review the recent theoretical studies about the impact of $1/f$-noise correlations on 
the entanglement dynamics of coupled qubits.
We will refer to the strictly related experimental works, which have also provided important indications on the 
microscopic origin of the observed noise. 
Rather extensive literature on   decoherence of coupled qubits in the presence of correlated  \textit{quantum} noise
will be omitted, as well as 
the variety of relevant experimental works demonstrating the
feasibility of universal quantum gates and simple quantum algorithms based on  
superconductor and semiconductor  technologies.  It is worth mentioning that 
coupling schemes for superconductor-based qubits have been reviewed by \textcite{Clarke2008}; since
the first demonstration of quantum oscillations in superconducting charge qubits \cite{Pashkin2003},
several benchmarking results have been reached in different labs
\cite{Dicarlo2010, Neeley2010, Palacios2010, Mariantoni2011, Lucero2012, Reed2012,
Fedorov2012}.

\begin{figure}[t]
\centerline{
\includegraphics[width=0.5\columnwidth,angle=-90]{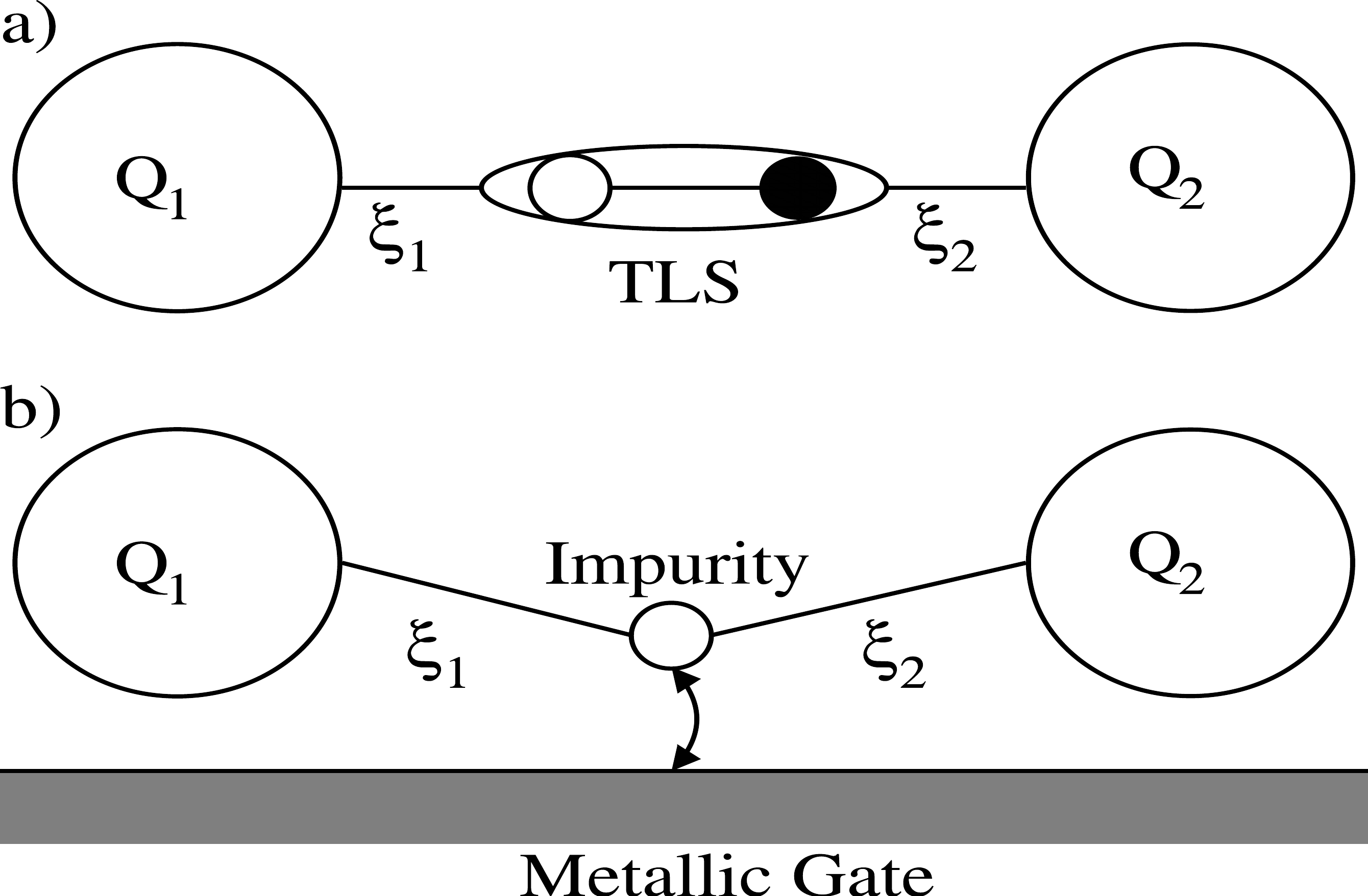}
}
\caption{ a) Two qubits, $\rm Q_1$ and $\rm Q_2$, coupled to a fluctuator in the substrate between
the two qubits, where the charge can tunnel between two sites. b) The qubits
are coupled to a charged impurity through its image charge on the metallic gate,
the charge can tunnel between the gate and the impurity. The coupling strengths
are given by $\xi_1$ and $\xi_2$. The configuration demonstrated in a) gives origin to
{\em anticorrelated} noise, while configuration b) gives origin to {\em correlated} noise.
Adapted from \cite{Brox2012}.
}
\label{fig_1_Brox2012}
\end{figure}

The core element of an entangling two-qubit gate can be modeled as
\begin{equation} 
\hat H= \sum_{\alpha=1,2} \hat H_{\rm tot}^{\alpha}[q_\alpha(t)] + \hat H_{\text{coupling}}[\{q_\alpha(t)\}] \, ,
\label{H_coupling_general}
\end{equation}
where for each qubit  $\hat H_{\rm tot}^{\alpha}$ is given by Eq.~(\ref{H_general}),
and we indicate that  the interaction term $\hat H_{\text{coupling}}$ may also depend on the control parameters 
$\{q_\alpha\}$. Following the same steps as
leading to Eq.~(\ref{H_tot_linearized}) 
and considering only classical bias fluctuations
one can cast  Eq.~(\ref{H_coupling_general}) as 
$\hat H + \delta \hat H$ where
\begin{eqnarray}
&& \!\!\!\!\! \hat H = \frac{\hbar}{2}\sum_\alpha  \vec \Omega_\alpha(q_\alpha) \cdot \vec \sigma^\alpha 
+ \hbar \sum_{ij}  \nu_{ij}(\{q_\alpha\}) \, \sigma_i^1 \, \sigma_j^2 \, , 
\label{H_coupled} \\
&& \!\!\!\!\! \delta  \hat H = \hbar \sum_{i} \sum_{\alpha=1,2}  E_{i}^\alpha(t) \, \sigma_i^\alpha
+ \hbar \sum_{i,j} E_{i,j}(t) \, \sigma_i^1  \, \sigma_j^2 \,  , 
\label{deltaH_coupled}
\end{eqnarray} 
$i,j \in \{x,y,z\}$.  
The stochastic processes $ E_{i}^\alpha(t)$ include fluctuations affecting each unit
and cross-talk effects due to the coupling element:
$ E_{i}^\alpha(t)= c_{\alpha i} \delta q_\alpha + d_{\alpha  i} \delta q_\beta$
 with  $c_{\alpha i}  \propto \partial \Omega_\alpha/\partial q_\alpha$
and $d_{\alpha i} \propto \nu_{ij}(\{q_\alpha\})$ ($\beta \neq \alpha$). 
Fluctuations of the interaction energy $\hbar \nu_{ij}(\{q_\alpha\})$ are included in 
$E_{i,j}(t) = (\partial \nu_{ij}/\partial q_1) \delta q_1 + (\partial \nu_{ij}/\partial q_2) \delta q_2 $.
Charge fluctuations affecting the exchange splitting of two electrons in a gate-defined double 
dot \cite{Hu2006} or background charge-induced fluctuations of the coupling capacitance of 
charge qubits \cite{Storcz2005} can be modeled by a term of this form.
The stochastic processes $\delta q_\alpha$ might originate from the same source, from different
sources or a combination. 
In the case of charge qubits for instance, random arrangement of background charges in the
substrate produce correlated gate-charge fluctuations to an extent depending on their
precise location (see Fig.~\ref{fig_1_Brox2012}), whereas impurities within tunnel junction $\alpha$ are expected to induce only
gate charge fluctuations $\delta q_\alpha$ \cite{Zorin1996}. These correlations are quantified by
the intrinsic correlation factor $\mu$, which for stationary and zero average processes with
same variance, follows from $\langle \delta q_\alpha(t)\delta q_\beta(t)\rangle=
[\delta_{\alpha \beta} + \mu (1-\delta_{\alpha \beta})]\bar \sigma^2$.
The overall degree of correlation between the processes $E_i^1(t)$ and $E_i^2(t)$
results both from intrinsic correlations and from cross-talk effects. It 
is expressed by the correlation coefficient $\mu_c$ defined, for two generic stochastic
processes $X_\alpha(t)$, as
\begin{equation}
\mu_c = \frac{\langle (X_1(t) - \bar X_1) (X_2(t) - \bar X_2) \rangle}{\sqrt{\langle (X_1(t) - \bar X_1)^2
\rangle \langle (X_2(t) - \bar X_2)^2\rangle}}
\label{correlation}
\end{equation}
where $\langle \dots \rangle$ denotes the ensemble average and 
$\bar X_\alpha \equiv \langle X_\alpha(t) \rangle$.

The adiabatic approximation scheme introduced in paragraph \ref{sub_adiabatic} 
can be easily extended to complex architectures to investigate
the short-times behavior relevant for quantum computing purposes. 
In this approximation the effect of stochastic processes with $1/f$ spectrum and/or cross-spectra on 
universal two-qubit gates has been studied in \cite{D'Arrigo2008,Paladino2009,Paladino2010,DArrigo2012,Brox2012} 
and on entanglement memory element in \cite{Bellomo2010}. In this case adiabatic 
noise does not induce the phenomenon of entanglement sudden-death, but it may reduce the 
amount of entanglement initially stored faster than quantum noise for noise figures typical for charge-phase
qubits.
 An extension  of the multi-stage  approach to complex architectures has been reported in \cite{Paladino2011}, where 
the characteristic time scales of entanglement decay in the presence of broad-band noise have been derived.
Analogously to single-qubit gates, low frequency noise induces fluctuations of the device eigenergies
resulting in a defocused averaged signal. 
One way to reduce inhomogeneous broadening effects is to ``optimally tune"
multi-qubit systems. In \cite{Paladino2010} a general route to reduce inhomogeneities due to
$1/f$ noise by exploiting tunabily of nanodevices has been proposed.
The basic idea is very simple: in the adiabatic and longitudinal approximation the system evolution  
is related to instantaneous eigen-frequencies, $\omega_l [\mathbf{E}(t)]$, which
depend on the noise realization $\mathbf{E}(t)$. For $\hat H$ given by Eq.~(\ref{H_coupled}), 
$ \mathbf{E}(t) \equiv \{E_i^\alpha(t), E_{ij}(t) \}$. 
The leading effect of low-frequency fluctuations in repeated measurements 
is given within the SPA. 
The frequencies $ \omega_{lm}(\mathbf{E})$ are  random variables, with standard deviation
$\Sigma_{lm}=\sqrt{\langle\delta \omega^2_{lm}\rangle-
\langle\delta \omega_{lm}\rangle^2}$, where $\delta \omega_{lm} = \omega_{lm}(\mathbf{E})-
\omega_{lm}$.   
``Optimal tuning" consists in fixing control parameters to values which minimize the 
variance $\Sigma_{lm}^2$
of the frequencies  $ \omega_{lm}(\mathbf{E})$.
This naturally results in a enhancement of the decay time of the corresponding  coherence 
due to inhomogeneous broadening. The short-times decay of the coherence in the SPA is in fact given by
\begin{equation}
|\langle  e^{- i \delta \omega_{lm}(\mathbf{E}) t}  \rangle | \approx \sqrt{1- (\Sigma_{lm} t)^2} \, ,
\end{equation}
resulting in reduced defocusing for minimal variance $\Sigma_{lm}$.
For a single-qubit gate, the optimal tuning recipe reduces to operating at the optimal point:
if $ \omega_{lm}(\mathbf{E})$  is monotonic, then  
$\Sigma_{lm}^2 \approx \sum_\alpha \left [\partial \omega_{lm}/\partial
q_\alpha \right ]^2 \sigma_{q_\alpha}^2$, and
the variance attains a minimum for vanishing differential dispersion.
When bands are non-monotonic in the control parameters, 
minimization of defocusing necessarily requires their 
tuning  to values depending on the noise variances.
For a multi-qubit gate, the optimal choice 
may be specific to the relevant coherence for the considered operation. 
By operating at an optimal coupling, considerable improvement of the efficiency of a 
$\sqrt{\rm{i-SWAP}}$ gate realized via a capacitive coupling of two quantronia
has been proved, even in the presence of moderate amplitude charge noise 
\cite{Paladino2010,Paladino2011}. Optimization against $1/f$ flux and critical current
noise of an entangling two-transmon gate has also been demonstrated \cite{DArrigo2012}.
Other passive protection strategies are based on the use of a ``coupler" element mediating
a controllable interaction between qubits. In \cite{Kerman2008}
a qubit mediates the controllable interaction between data qubits.
By Monte Carlo simulation, the feasibility of 
a set of universal gate operations with $\mathcal{O}(10^{-5})$ error 
probabilities in the presence of experimentally measured levels of $1/f$ flux noise
has been demonstrated. 
\begin{figure}[t]
\centerline{
\includegraphics[width=0.75\columnwidth]{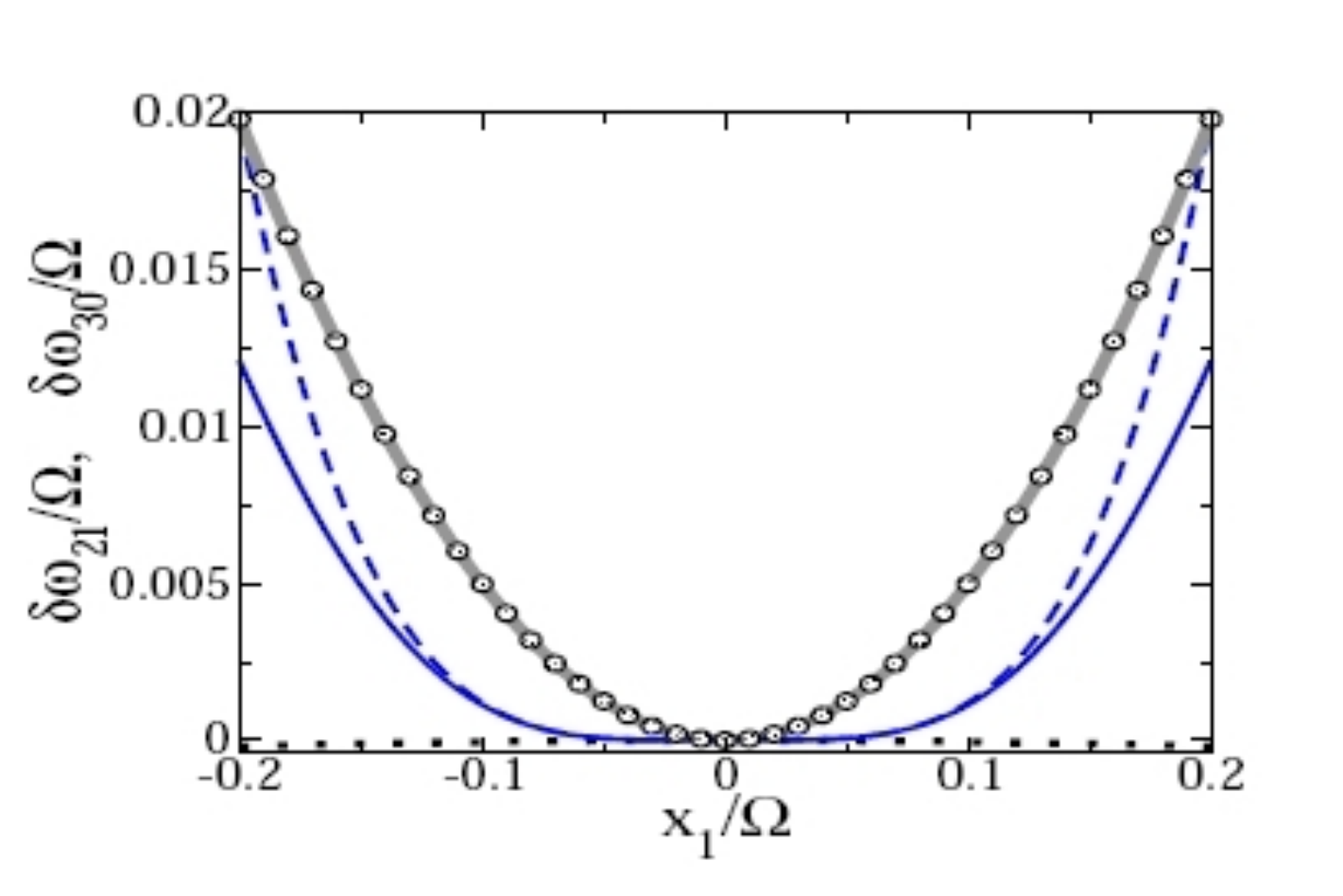}
}
\caption{(Color online)  Dispersion in the SWAP subspace $\delta \omega_{21}$ (blue) and in the orthogonal
subspace $\delta \omega_{30}$ (thick gray) as a function of transverse fluctuations on qubit 1 
$x_1 \equiv \delta q_1$  ($x_2\equiv \delta q_2=0$) for resonant qubits with 
$\nu_{zz}/\Omega= 0.01 $. The exact splitting (blue) is compared
with a second order expansion (dotted) and the single qubit dispersion (circles).
Adapted from \cite{Paladino2010}.
}
\label{fig_1b_Paladino2010}
\end{figure}

The effect of correlated or partially correlated low-frequency noise acting on two qubit gates has
been studied in \cite{Storcz2005,Hu2007,D'Arrigo2008,Faoro2010,Brox2012}. 
Due to the complexity of the Hilbert space of coupled qubits, the efforts have primarily 
resulted in numerical surveying of various situations. 
The natural question to ask is whether correlations between noise sources increase or suppress dephasing
of the coupled systems compared to uncorrelated noise.
The answer depends first of all on the symmetry of the system-environment
interaction, i.~e., on the existence of decoherence-free subspaces (DFS), possibly also one-dimensional, 
 {\em and} on the initial system state. 
In \cite{Brox2012}, introducing a generalized Bloch-sphere method
combined with the SPA, analytical expressions for the dephasing rates
of the two-qubit system as a function of the degree of correlation  $\mu_c$ have been derived.
For resonant qubits with a $\sigma_z^1 \sigma_z^2$-coupling in the presence of transverse noise,
two one-dimensional DFS are found, 
$|00 \rangle - |11 \rangle$ which does not decay in the presence
of correlated noise, but which is sensitive to anticorrelations (see Fig.~\ref{fig_1_Brox2012}), 
and $|00 \rangle + |11 \rangle$ showing the opposite behavior. 
In the absence of perfect symmetry (for instance if qubits are not resonant), the above symmetric
states are not eigentstates of the Hamiltonian and as a consequence are also less sensitive to noise
correlations. In other words, it is both the symmetry of the initial state and how much this state overlaps 
with an eigenstate of the Hamiltonian in the absence of noise that determines the decoherence rate.
This analysis suggests that for each setup, the most convenient subspace for two-qubit encoding  
should be based on preliminary investigation on the nature of noise correlations.
For instance, in the absence of correlations, the SWAP subspace generated by 
$(|01 \rangle \pm |10 \rangle)/\sqrt{2}$, is more resilient to transverse low-frequency fluctuations
with respect to the orthogonal subspace, this fact being ultimately due to the dependence of the corresponding 
eigenvalues on the deviations $\delta q_\alpha$ as illustrated in Fig.~\ref{fig_1b_Paladino2010}. 
Note that the SWAP subspace is expected to be also more stable for the single qubit. An analogous conclusion
was reached in \cite{De2011} for a pair of qubits coupled via the exchange interaction. 
In \cite{Hu2007,D'Arrigo2008} a phenomenological model for $1/f$ correlated noise affecting a 
two-qubit gate in a fixed coupling scheme has been considered.
The effect of noise correlations on entanglement generation in the SWAP subspace sensitively depends on the ratio 
$\sigma/\nu_{zz}$ between the amplitude of the low frequency noise and the qubits 
coupling strength \cite{D'Arrigo2008}.
For small amplitude noise, correlations increase dephasing at the relevant
short times scales (smaller than the dephasing time). On the other hand,
under strong amplitude noise, an increasing degree of correlations between 
noise sources acting on the two qubits always leads to reduced dephasing.
The reason for this behavior originates from the non-monotonic dependence of the SWAP splitting 
variance on the correlation coefficient $\mu_c$.
A numerical analysis has shown that the above features hold true for
adiabatic $1/f$ noise extending up to frequencies $10^9$~s$^{-1}$, 
which are about two orders of magnitude smaller that the qubit Bohr frequencies. At longer times, 
the entanglement decay time (defined as the time where the signal
is reduced by a factor $e^{-1}$) weakly increases with  $\mu_c$~\cite{Hu2007}.

Recent experiments on flux qubits quantified $1/f$ flux noise and flux-noise 
correlations providing relevant indications on its microscopic origin.
In \cite{Yoshihara2010} flux noise correlations have been studied in a system of
coupled qubits sharing parts of their loops, whereas in \cite{Gustavsson2011}
a single, two-loop qubit was used to investigate flux noise correlations
between different parts within a single qubit. In both experiments the qubit dephasing rate
was measured at different bias points.
A comparison of the data with the rate prediction in the Gaussian approximation 
based on the assumption of $1/f$-type behavior both of spectra and cross-spectrum,
provided indications of the noise amplitudes and on the sign of noise correlations. 
In the first experiment it was found that the flux fluctuations originating from the shared branch 
lead to correlations in the noise of the two qubits. In \cite{Gustavsson2011} flux fluctuations in the two loops
are found to be anticorrelated. Both experiments provided strong indication that the dominant
contribution to the noise comes from {\em local} fluctuations, in agreement with 
\cite{Bialczak2007, Martinis2002, Koch2007, Faoro2008b}. 
In particular, in the setup of 
\textcite{Gustavsson2011} a global fluctuating magnetic field would have given positive correlations,
which were not observed.  Results of both experiments are found to be consistent with a model where flux noise
is generated by local magnetic dipoles (randomly oriented unpaired spins) distributed on the metal surfaces. 
A similar conclusion on the local origin of flux noise was drawn in \cite{Lanting2010} from measurements of 
macroscopic resonant tunneling (MRT) between the lowest energy states of a pair of magnetically coupled rf-SQUID 
flux qubits. In this experiment, the MRT rate peak widths indicate that each qubit is coupled to a
local environment whose fluctuations are uncorrelated with that of the other qubit.
Indications of magnetic flux noise of local origin in two phase qubits separated by $500~\mu$m on the same 
chip has  been recently reported in \cite{Sank2012}.

In \cite{Sendelbach2009} the cross spectrum of inductance and flux fluctuations
in a dc-SQUID has been measured. In this experiment, the imaginary part of the
SQUID inductance  and the quasi-static flux threading the SQUID loop were monitored simultaneously as 
a function of time. From the two time series, the normalized cross spectral density has been computed. 
The inductance and flux fluctuations were found to be highly correlated at low temperature,
indicating a common underlying physical mechanism. The high degree of correlation provided 
evidence for a small number of dominant fluctuators. The data were
interpreted in terms of the reconfiguration of clusters of surface spins, with correlated fluctuations of
effective magnetic moments and relaxation times. 
The observed specific correlation between low-frequency flux noise and inductance fluctuation
suggests that the flux noise is related to the nonequilibrium dynamics of the
spin system, possibly described by spin glass models \cite{Chen2010} or
fractal spin clusters, which appear naturally in a random system of spins with wide distribution 
of spin-spin interactions \cite{Kechedzhi2011}.

\subsection{Quantum coherent impurities} \label{qcimp}

The model of two-level tunneling systems formulated by \textcite{Anderson1972} and \textcite{Phillips1972}, 
illustrated in Section \ref{sec:origin}, has been extensively tested experimentally by ensemble measurements performed
on samples having a large TLS density, such as structural glasses. Results reported in this Section demonstrated that ensembles of 
TLSs, sparsely present in the disordered oxide barrier of Josephson junctions or in the insulating substrates, 
induce fluctuations with $1/f$ spectrum which are a major source of decoherence in superconducting nanocircuits. 
However, the effects unambiguously proving quantum mechanical behavior of an individual fluctuator interacting with a qubit  were not observed.  
Only recently highly sensitive superconducting circuits could be used as ``microscopes" for probing spectral, spatial 
and coupling properties of selected TLSs. 
Understanding the origin of these spurious TLSs, their coherent quantum behavior, and their
connection to $1/f$ noise is important for any low-temperature application of Josephson junctions and
it is a challenge, which will be crucial to the future of superconducting quantum devices.

The first observations indicative of a considerable interaction of a superconducting circuit
with a strongly anharmonic quantum system were reported on a large-area Josephson junction ($\approx 10 \, \mu$m$^2$) 
phase qubit at Boulder~\cite{Simmonds2004,Cooper2004}.  
Microwave spectroscopy  revealed the presence of small unintended avoided crossings in the transition spectrum 
suggestive of the interaction between the device and individual coherent TLSs
resonantly coupled with the qubit.
In these experiments, a small number of spurious 
resonators with a distribution  of splitting size, the
largest being $\sim 25$~MHz and an approximate density of one major TLS per $\sim 60$~MHz, were observed.
Magnitude and frequency of the TLS considerably changed after thermal cycling to room
temperature, whereas cycling to $4$ K produced no apparent effect. Moreover,  different qubits in the same 
experimental setup displayed their own  unique ``fingerprint" of TLSs frequencies and splitting strengths. 
Qubit Rabi oscillations driven resonantly with a TLS showed considerably reduced visibility with respect to 
off-resonant driving. 
It has been observed that similar spectroscopic observations may also result from
macroscopic resonant tunneling in the extremely asymmetric double-well potential of the phase
qubits \cite{Johnson2005}.  
The TLS and MRT mechanisms could be distinguished measuring the low frequency voltage noise in a Josephson
junction in the dissipative (running phase) regime~\cite{Martin2005}.

Since these first experiments, similar avoided crossings in spectroscopy have been observed in different 
superconducting circuits. 
In phase qubits they have been reported in
\cite{Martinis2005,Shalibo2010,Palomaki2010,Lisenfeld2010,Lisenfeld2010b,Neeley2008,Bushev2010,Hoskinson2009}; 
in flux qubits -- in \cite{Lupascu2009,Plourde2005,Deppe2008};
in a Cooper-Pair-Box (ultrasmall Josephson junction with nominal area $120\times 120$~nm$^2$) -- in \cite{Kim2008};
in the quantronium -- in \cite{Ithier2005} and in the transmon -- in \cite{Schreier2008}. 

The close analogies among these observations, despite of the differences in the qubit setups, junctions size and 
materials, confirm that microscopic degrees of freedom located in tunnel barrier of Josephson junctions, 
usually  made of a 2- to 3-nm-thick layer of disordered oxide (usually AlO$_{x}$, $ x \approx 1$), are 
at least one common cause of these effects. These microscopic degrees of freedom are strongly anharmonic
systems and observations are fully consistent with coherent TLS behavior. 
A further confirmation comes from  multi-photon spectroscopy in phase~\cite{Bushev2010,Lisenfeld2010,Palomaki2010,Sun2010} and flux 
qubits \cite{Lupascu2009} where the hybridized states of the combined qubit-TLS systems have been probed
under strong microwave driving. For instance, an additional spectroscopic line in the middle of the qubit-TLS anticrossing
corresponding to a two-photon transition between the ground  state and the two excitations state of the qubit-TLS
system has been observed in \cite{Lupascu2009,Bushev2010,Sun2010}.  
Moreover, in some of these experiments, spontaneous changes of the resonator's frequency were observed.
The instability was observed  during many hours while the device was cold 
\cite{Simmonds2004},
whereas in  \cite{Lupascu2009} it was observed  in some samples during few
tens of minutes data acquisition time. Other samples were instead stable over the few months duration of the 
experiment. The instability supports the idea
that the coupled TLSs are of microscopic origin.
The qualitative trend is that small-area qubits show fewer splittings than do large area qubits, although larger
splittings are observed in the smaller junctions \cite{Martinis2005}.

Time-resolved experiments on phase qubits have demonstrated that an individual TLS can be manipulated
using the qubit as a tool to both fully control and read out its state.
The trajectory (i.~e., time record) of the switching current of a phase qubit revealed ``quantum jumps" between
macroscopic quantum states of the qubit coupled to a TLS in the Josephson tunnel junction, thus providing
a way to detect the TLS state \cite{Yu2008}.
Through the effective, qubit mediated, coupling between the TLS and an externally applied resonant electromagnetic
field ``direct" control the quantum state of individual TLSs 
been demonstrated in \cite{Lisenfeld2010b}. In this experiment the qubit always remained detuned during TLS 
operations, merely acting as a detector to measure its resulting state. 
A characterization of the TLS coherence properties was possible via detection of TLS Rabi oscillations, relaxation dynamics,
Ramsey fringes and spin echo. Measurements at different temperatures shown stable TLS resonance frequencies and 
qubit's coupling strenghts. Energy relaxation time is found to decrease quadratically with temperature,
whereas the TLSs dephasing times had a different behavior, only one of the measured TLSs being close to $2 T_1$. 
In \cite{Shalibo2010} relaxation and dephasing times of a large ensemble of TLSs in
a small area ($\sim 1\, \mu$m$^2$) phase qubit were measured (82 different TLSs obtained from 8 different 
cooling cycles of
the same sample). Decay times ranged almost 3 order of magnitudes, from 12 ns to more than 6000 ns, whereas coherence times
varied between 30 ns and 150 ns. The average $T_1$
followed a power-law dependence on the qubit-TLS coupling strength, whereas the average dephasing time was maximal
for intermediate coupling. Authors suggest that both time scales naturally result   from TLSs dipole phonon radiation 
and  anticorrelated dependence of the TLS tunneling amplitude and bias energy on low-frequency environmental fluctuations.
Non-monotonous dependencies of the qubit's decay time on the
qubit-TLS coupling and temperature were also predicted in \cite{Paladino2008} for a qubit longitudinally coupled to a coherent TLS.
In general, different experiments shown that some TLSs exhibit coherence times much longer than those of the 
superconducting qubits \cite{Lisenfeld2010b,Neeley2008,Palomaki2010}.\footnote{TLSs' decay times following 
from Ramsey fringes are of the order of few hundreds nanoseconds, and maximal relaxation times are about one microsecond.}
This remarkable fact, together with the ability to directly control selected TLSs, shed a new light
on these microscopic systems. Indeed, it has been proposed that TLSs in the barrier of a Josephson junction can 
themselves act 
as naturally formed qubits \cite{Zagoskin2006,Tian2009}. In \cite{Neeley2008} the first quantum memory
operation on a TLS in a phase qubit  has been demonstrated. An arbitrary quantum state was transferred  to a TLS, stored there for 
some time and then retrieved. In \cite{Sun2010} creation and coherent manipulation of quantum states of
a tripartite system  formed by a phase qubit coupled to two TLS has been demonstrated.
In this experiment, the avoided crossing due to the qubit-TLS interaction acted as a tunable quantum beam splitter of wave 
functions, which was used to precisely control the quantum states of the system and to demonstrate Landau-Zener-St\"uckelberg 
interference.
Although  TLSs were suitable for these initial proof-of-principle demonstrations, their use in a quantum computer 
still seems unlikely because of their intrinsically random nature and limited coherence times. A relevant step further 
along this direction has been done recently  by \textcite{Grabovskij2012} who report an experiment in which
the energy of coherent TLSs coupled resonantly to a phase qubit is tuned. When varying a static strain
field in situ and performing microwave spectroscopy of the junction, they observed continuously changing energies of 
individual coherent TLSs. Moreover, obtained results over 41 individual TLS between $11$ and $13.5$ GHz 
are explained readily by the tunneling model and, therefore, provide firm evidence of the hypothesis that atomic TLSs
are the cause of avoided level crossings in the spectra of Josephson junction qubits. Mechanical strain offers a
handle to control the properties of coherent TLSs, which is crucial for gaining knowledge about their
physical nature.

Alternative theoretical models of TLSs have been proposed to explain the avoided level crossings observed
in qubit spectroscopy data in phase and flux qubits. 
It has been suggested that the state of the TLS modulates the transparency of the junction and therefore its 
critical current, $I_c$ \cite{Simmonds2004,Faoro2006,Ku2005,Constantin2007}. 
In this case two-level defects could be formed by Andreev bound 
states \cite{Faoro2005,deSousa2009} or Kondo impurities \cite{Faoro2007,Faoro2008}. 
Alternatively the TLS may couple to the electric field inside the junction, which is consistent with
the TLS carrying a dipole moment located in the aluminum oxide tunnel barriers \cite{Martin2005,Martinis2005}.
Recently \textcite{Agarwal2013} analyzed the interaction with phonons of  individual electrons tunneling 
between two local minima of the potential well structure due to the electron  Coulomb interaction with the
 nearest atoms in the insulator. They concluded that the resulting strong polaronic effects  dramatically 
 change the TLS properties providing quantitative understanding of the TLS relaxation and dephasing observed 
in Josephson junctions. In particular, the strain effects observed by \textcite{Grabovskij2012} are quantitatively interpreted.

Finally, a TLS may modulate the  magnetic flux threading the superconducting loop \cite{Bluhm2009, Sendelbach2008}. In \cite{Cole2010} a direct comparison between these models and high precision spectroscopy data on a phase qubit   has been performed. Experimental data indicate a small or nonexistent longitudinal qubit-TLS coupling relative to the transverse term. In phase and flux qubits fluctuations of the critical current or magnetic flux  generate both transverse and longitudinal components, whereas the coupling to the electric field within the junction  is purely transverse. 
Although longitudinal coupling cannot be ruled out, no evident signatures of this coupling were 
observed in most of the experiments which have been consistently explained
considering purely transverse dipolar interaction  \cite{Bushev2010,Lisenfeld2010,Lupascu2009}. 
Other multilevel spectroscopy experiments did not uniquely pin down the coupling mechanism as well.
The similar features observed in phase qubits and in the flux qubit experiment \cite{Lupascu2009} suggest that 
strongly coupled TLS have the same origin  in flux and
phase qubits, even though the degrees of freedom manipulated are different.
A charge coupling model is also supported by spectroscopic observations in a Cooper-pair box \cite{Kim2008}.
A distribution of avoided splitting sizes consistent with the qubit coupling to charged ions tunneling
between random locations in the tunnel junction oxide and not directly interacting with each other has been
reported in  \cite{Palomaki2010}.
In \cite{Tian2007} a possible way to resolve the underlying coupling mechanism of TLSs to phase qubits through the use 
of a magnetic field applied along the plane of the tunnel barrier inside the junction was proposed. More generally,
one or two non-interacting qubits may be conveniently used as a probe of a coherent environment 
\cite{Paladino2008,Oxtoby2009,Jeske2012}.

The controllable interaction between a qubit and a microscopic coherent TLS
led to a number of interesting features also in the qubit  time evolution. 
One aspect is the reduced visibility of qubit Rabi oscillations driven resonantly with 
a TLS first observed by \textcite{Simmonds2004}.
This problem has been investigated using different approaches and under various driving conditions 
and TLS decoherence mechanisms in  \cite{Meier2005,Ku2005,Galperin2005,Ashhab2006,Sun2010}.
The main conclusion of \textcite{Meier2005} is that fluctuators are the dominant source
of visibility reduction at Rabi frequencies small compared to the qubit-TLS 
coupling strength, while leakage out of  the qubit computational subspace becomes increasingly important for the large Rabi
frequencies of experiments with phase qubits.
In \cite{Galperin2005} the quantum dynamics of the four-level system subject to an arbitrarily strong driving ac-field 
has been investigated including both phase and energy relaxation of the TLS in a phenomenological way.
It was demonstrated that if the fluctuator is close to resonance with the qubit, the Rabi oscillations of
the qubit are suppressed at short times and demonstrate beatings when damping is weak enough.
In addition, it was pointed out that if the read-out signal depends on the state of the fluctuator, 
the visibility of the Rabi oscillations can be substantially reduced, a possible scenario in \cite{Simmonds2004}.
Depending on the relative strength of the resonant ac-driving
and the qubit-TLS coupling, additional features in the qubit dynamics have been predicted 
as anomalous Rabi oscillations and
two-photon processes involving transitions between the four-level states of the coupled qubit-TLS \cite{Sun2010,Ashhab2006}.
Some of these effects have been observed in the experiments mentioned above.

The role of coherent TLSs on qubit relaxation processes has been investigated in \cite{Mueller2009}. 
In this article a qubit is considered as interacting with 
coherent TLSs each subject to relaxation and pure dephasing processes (in the 
underdamped regime) and in resonance or close to resonance with the qubit. Depending on  
the distribution of the TLSs energies (uniform or with strong local 
fluctuations), the qubit $T_1$ can either be a regular function of the qubit splitting or 
display an irregular behavior.
Authors suggest that this mechanism may explain the smooth $T_1$-versus-energy curve in large-area junction
phase qubits \cite{Cooper2004,Simmonds2004,Neeley2008} and the seemingly random dependence reported in smaller-area 
phase qubits and in flux or charge qubits \cite{Ithier2005,Astafiev2004}.  It is also speculated that
the large splittings observed in spectroscopy of the same phase qubits  may result from many weakly coupled spectrally 
dense TLSs.

A characterization of the effects of bistable coherent impurities in superconducting qubits
has been proposed by \textcite{Paladino2008}. Introducing an effective impurity description in terms of a 
tunable spin-boson environment, the qubit dynamics has been investigated for a longitudinal
qubit-TLS interaction. The asymptotic time limit is ruled by a dominant rate which depends
non-monotonically on the qubit-TLS coupling strength and reflects the TLS dissipative processes and
temperature. At the intermediate times relevant for quantum computing, different rates and frequencies
enter the qubit dynamics displaying clear signatures of non-Gaussian behavior of the quantum impurity.

Finally, the possibility to highlight the coherent interaction between a superconducting circuit and a 
microscopic quantum TLS, in principle, allows investigating the important question of the applicability 
domain of the classical RTN model. In \cite{Wold2012} the decoherence of a qubit coupled to either a TLS
again coupled to an environment, or a classical fluctuator modeled by RTN is investigated.
A model for the quantum TLS is introduced where the temperature
of its environment, and the decoherence rate  can be adjusted independently. The model has a well-defined classical
limit at any temperature and this corresponds to the appropriate asymmetric RT process.
The difference in the qubit decoherence rates predicted by the two models is found to depend on the ratio between
the qubit-TLS coupling and the decoherence rate in the pointer basis of the TLS. This is then the
relevant parameter which determines whether the TLS has to be treated quantum mechanically or
can be replaced by a classical RT process. This result validates the application
of the RT process model for the study of decoherence in qubits also when the coupling between the 
qubit and the fluctuator  is strong as long as the fluctuator couples even 
more strongly to its own environment.

%% file: Part_4_0721.tex
\subsection{Dynamical decoupling and $1/f$ noise spectroscopy}
\label{DD}

\subsubsection{Noise protection and dynamical decoupling}

In the last few years several strategies for coherence protection have been
developed, both for quantum information processing and in the broader
perspective of quantum control. Optimal bias point discussed earlier
is a passive stabilization (or error avoiding) code very   successful in solid-state nanodevices. 
Dynamical Decoupling (DD) relying on repeated application of pulsed or switched control
is an active stabilization (i.~e., error correcting) scheme  developed in the field of high-resolution 
NMR~\cite{kb:200-becker-highresNMR}. 
DD has been proposed as a method to extend  decoherence times in solid-state quantum hardware, 
and has been recently applied to decouple spin baths in semiconductor-based
qubits~\cite{Barthel2010,Bluhm2011,deLange2012,Ryan2010}  and $1/f$ noise in superconducting  
nanocircuits~\cite{Bylander2011,Gustavsson2012}.

Coherent averaging of unwanted couplings is at the heart of DD. The principle is illustrated
by the spin echo which is operated by a single $\pi$-pulse inducing a spin-flip transition.
Shining a pulse $X_{\pi}$ (evolution operator $U=\sigma_x$) 
at half of the evolution time $t$, say $\{t/2,X_{\pi},t/2\}$, dynamically suppresses terms
$\propto \sigma_y,\sigma_z$ in the qubit Hamiltonian. In NMR samples
unwanted terms $H_1 \propto \delta B\, \sigma_z$ are due to static randomly distributed local fields. 
The Bloch vector dynamics for the ensemble of spins is defocused resulting in inhomogeneous broadening. 
The Hahn echo $\{X_{\pi/2},t/2,X_{\pi},t/2,X_{\pi/2}\}$
is routinely used to achieve efficient refocusing~\cite{kb:200-becker-highresNMR}. 
Echoes also switch off ``dynamically'' two-qubit $J$-couplings of the nuclear spin Hamiltonian in 
liquid NMR quantum  computers~\cite{kr:205-vandersypen-rmp-NRM}.

\subsubsection{Pulsed control}

\begin{figure}[t]
\centerline{
\includegraphics[width=0.40\columnwidth,angle=-90]{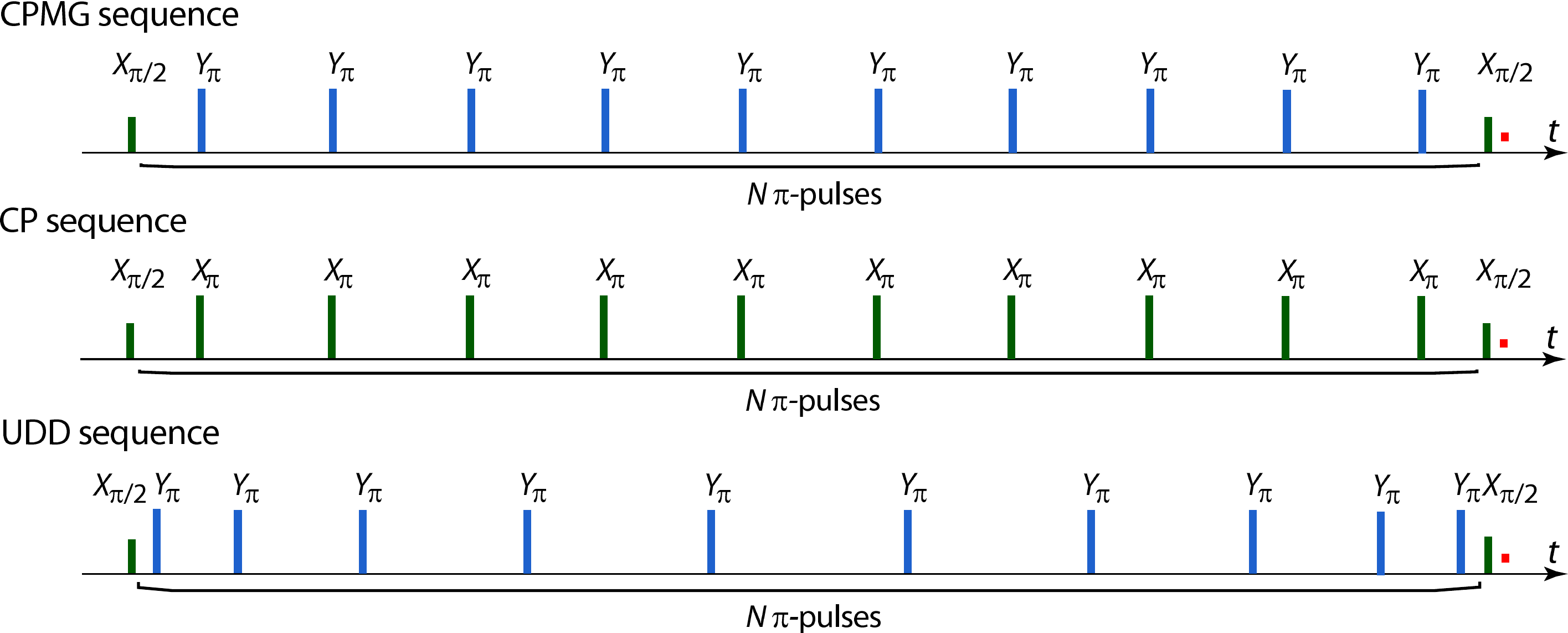}
}
\caption{Timing of the CPMG, CP, and UDD pulse sequences for N = 10. 
Adapted by permission from Macmillan Publishers Ltd.: \cite{Bylander2011}, copyright (2011).
\label{fig:pulses}}
\end{figure}

Coupling to a stochastic field $E(t)$ induces  diffusion in the free spin precession and decoherence, 
which mitigates Hahn  echoes.  \textcite{Carr1954} (CP)
recognized that sequences of $\pi$-pulses  may suppress spin diffusion since they coherently 
average out $E(t)$, Fig.~\ref{fig:pulses}. Composite pulses are also used in  NMR to stabilize 
a given quantum gate against errors in the  control~\cite{kb:200-becker-highresNMR,kr:205-vandersypen-rmp-NRM}. 
NMR pulse sequences beyond Hahn echo have been employed in  a superconducting qubit to demonstrate 
both gate  stabilization~\cite{204-collin-prl-nmrcontrolsqubit} and  dynamical reduction of decoherence
 due to $1/f$ noise~\cite{Ithier2005}.

\paragraph*{From Echo to DD --}
DD may selectively remove a noisy environment with a finite correlation time $\tau_c$.  
Control is operated via a time-dependent $\hat H_c(t)$ describing a sequence of $\pi$-pulses. We consider 
hard $X_{\pi}$ pulses, whose duration is very short $t_p \to 0$, at times 
$t_j$ ($j=1,\dots,N$). The time-evolution operator reads 
\begin{equation}
\label{eq:propagator-hard-pulses}
U_N(t,0) = \sigma_x U(t,t_{N-1}) \sigma_x \dots U(t_2,t_1) 
\sigma_x \, U(t_1,0) 
\end{equation}
where $U(t,t^\prime)$ describes the noisy free evolutions between pulses. Notice that pulses  
in $\sigma_x U(t_{n-1},t_{n})\sigma_x$ reverse the sign of operators $\sigma_{y,z}$ appearing in $U$. 
Therefore such ``orthogonal'' components flip in sequential steps of the protocol. It is convenient 
to use the language of  average Hamiltonian theory~\cite{kr:205-vandersypen-rmp-NRM}
and introduce the effective Hamiltonian $H_N(t)$, defined as 
$e^{i H_N(t) \,t/\hbar} := U_N(t,0)$. Then a periodic train of pulses
$t_{j+1}-t_j=\Delta t = t/N$, implementing a sequence named Periodic DD (PDD), with elementary block 
$\{\Delta t, X_\pi,\Delta t,X_\pi\}$, tends to average out orthogonal spin components (provided
 $N$ is even)~\cite{ka:199-violaknll-prl-ddopenquant}.
This emerges from the perturbative (Magnus) expansion~\cite{kr:205-vandersypen-rmp-NRM} of $H_N$, which
 washes these terms out in the limit of continuous flipping,
$\Delta t \to 0$.  In the simple case of a qubit coupled to pure dephasing 
classical noise,  $H = {\hbar \over 2} [\Omega + E(t)]  \sigma_z + \hat H_c(t)$, calculations can be carried out
 exactly~\cite{ka:211-biercuk-jpb-ddfilter}. Coherences decay as 
\begin{equation}
\label{eq:decay-pure-dephasing}
\rho_{01} (t) = \rho_{01} (0) \langle e^{- i \Phi_N(t)} \rangle
= \rho_{01} (0) e^{- \Gamma_N(t) - i \Sigma_N(t)}
\end{equation}
where the phase $\Phi_N(t) = \int_0^t ds \, y_N(s) E(s)$ is obtained by sampling a realization of 
noise with the piecewise constant $y_N(t)$ whose discontinuities reflect the effect of pulses  at $t_i$. 
The \textit{decay function} $\Gamma_N(t)$  obtained by noise averaging depends on the pulse sequence.  
For Gaussian noise with power spectrum $S(\omega)$ the averaging yields
\begin{equation}
\label{eq:decay-pure-dephasing-gaussian}
\Gamma_N(t) = \int {d \omega \over \omega^2} \, S(\omega)
\, F_N(\omega t)
\end{equation}
where the \textit{filter function} 
$F_{N}(\omega t)=|y_N(\omega t)|^2$
has been defined as~\cite{ka:207-uhrig-prl-UDD}:
$$
F_{N}(\omega t)=  |1+(-1)^{N+1} e^{i \omega t} + 2 \sum_{j=1}^N (-1)^j 
e^{i \omega t_j}|^2.
$$
In the absence of pulses the function $F_0(\omega)= 4 \sin^2 (\omega t/2)$ reproduces the decay in a 
FID protocol, whereas in the presence of pulses it  yields diffraction patterns induced by interference 
in the time domain~\cite{ka:211-ajoialvarezsuter-pra-opticalDD}
and, in particular, to coherent suppression of $F_N$ at low frequencies. As a consequence, $\Gamma_N(t)$ 
decreases and signal decay due to decoherence is effectively recovered.

\textcite{PhysRevA.58.2733} applied such techniques  to selectively decouple a pure dephasing quantum 
environment,  obtained by letting 
${1 \over 2} \sigma_z E(t) \to {1 \over 2} \sigma_z \hat{E} + H_R$,
where $H_R$ describes the environment alone. The structure of Eq.~(\ref{eq:decay-pure-dephasing}) is 
still valid,  $\Gamma_N(t)$ depending only on the dynamics ruled by $H_R$.
In particular,  \textcite{PhysRevA.58.2733} studied an environment of linearly coupled quantum oscillators, 
 $\hat{E} = \sum_\alpha g_\alpha (a^{\dagger}_\alpha +
a_\alpha)$. They found that Eq.~(\ref{eq:decay-pure-dephasing-gaussian}) holds
true, $S(\omega)$ being related to the symmetrized correlation  function of $\hat{E}$, uniquely expressed 
via the 
spectral density
$J(\omega) = \sum_\alpha g_\alpha^2 \,\delta(\omega-\omega_\alpha)$, namely
$$S(\omega) =  {1 \over 2} 
\langle \hat{E}(t)\hat{E}(0)+ \hat{E}(0)\hat{E}(t) \rangle_\omega = \coth \Big({\hbar \omega \over 2 k_B T}\Big) J(\omega).$$ 
The  ultraviolet (UV) cutoff of $J(\omega)$, $\omega_c$, sets the time scale of fastest response of the environment.
As for classical noise,  decoherence is washed out completely for
 $\Delta t \to 0$  and greatly suppressed for $\omega_c \Delta t \sim 1$.

In general, open loop schemes with a finite set of pulsed fields allow to perform 
fault-tolerant control~\cite{ka:199-violaknll-prl-ddopenquant}, i.~e., to design the dynamics of a 
quantum system to attain a given objective. The simplest goal is  the effective decoupling of the 
environment.  With respect to other active stabilization strategies, as
quantum error correction or closed loop (quantum feedback) schemes, DD has the
advantage that only unitary control of a small and well characterized  system is needed and it
 does not require additional  measurement resources.  Relying on coherent averaging, DD can suppress 
 errors  regardless of their amplitude. In the last few years  optimization of pulse 
 sequences~\cite{ka:211-biercuk-jpb-ddfilter} has been an active subject of investigation allowing  
 substantial improvement when dealing with real open quantum 
systems. 

\paragraph*{Optimized sequences and robust DD --}
Performances of sequences strongly depend on their details such as the parity of the number of pulses or
their symmetrization. For instance, in odd $N$ PDD the noise during the final $\Delta t$ remains uncompensated. 
Proper symmetrization of the sequences, such as CP sequence (Fig.~\ref{fig:pulses}) may lead to higher order 
cancellations in $H_N(t)$. \textcite{MeiboomGill}
proposed a refinement (CPMG sequence), which is usually  very efficient against spin
diffusion~\cite{kb:200-becker-highresNMR}, since it also averages errors due to
control field inhomogeneities. Indeed, if  pulses are implemented by resonant ac fields, the 
component producing spin-flip fluctuates in amplitude due to the same noise responsible for the 
spin diffusion. For a given initial state of the Bloch vector, CP  accumulates errors in $X_\pi$ at 
second order, while   in CPMG $Y_\pi$ errors appear  at  fourth order~\cite{ka:210-borneman-jmagres-cpmg}.

Recently \textcite{ka:207-uhrig-prl-UDD}  found that non-equidistant pulses  improve performances 
in a pure-dephasing spin-boson environment. In the Uhrig DD (UDD)
sequence $t_j/t = \sin^2[\pi j/(2N+2)]$ times are such that the first $N$ derivatives 
of the filter function vanish, $ \left[d^{j} F_N/dz^j\right]_{z=0}=0, \ j \in \{1,2,\ldots N\}.$
 This ensures that pure dephasing  is suppressed to $O(t^N)$ in the series expansion of $E(t)$.  
 \textcite{Lee2008} conjectured that  UDD is universal for generic pure dephasing model,  and 
 \textcite{ka:208-yang-prl-uniUDD} proved that  generalized UDD suppresses both pure dephasing and  
 relaxation to $O(t^N)$. In the spirit of UDD,  several  new pulse sequences were introduced in the 
 last few years,  achieving optimization for a 
given sequence duration~\cite{209-biercuk-nature-optDD}, or being nearly optimal for generic single-qubit  
decoherence~\cite{ka:210-westfonglidar-prl-optDD} or for specific
environments~\cite{ka:210-pasiniuhrig-pra-powernoise}.

Concatenated DD (CDD) proposed by~\textcite{ka:205-hkodjasteh-prl-CDD} is an alternative scheme based on 
the idea of recursively defined sequences, which guarantee to reduce decoherence below a pulse noise 
level. Within this framework high fidelity quantum gates have been demonstrated 
numerically~\cite{ka:210-west-prl-HiFigates}.

The quest for robust DD arises from the general problem of the tradeoff between the control
resources involved and efficient suppression of decoherence. Ideal DD requires available couplings 
allowing the synthesis of controlled  evolution~\cite{ka:199-violaknll-prl-ddopenquant}, large 
pulse repetition rate, and pulse hardness. Optimization can be used together with realistic bounded 
amplitude control or continuous always-on fields 
schemes~\cite{ka:203-violakn-prl-eulerian,ka:211-khodjastehviola-pra-limitsmultipulse,212-codyjladd-njp-alwayson} to allow 
a flexible use of resources needed to attain a
given decoupling error $\Gamma_N(t)$.  

\subsubsection{DD of $1/f$ noise}
DD has a large potential impact in solid state coherent nanosystems where noise has large low-frequency 
components. Indeed, DD of  Gaussian $1/f$ noise, does not always require ultrafast pulse 
rates~\cite{ka:204-lidar-pra-dd1overf}.   However, sources responsible for $1/f$ noise are often 
discrete producing non-Gaussian noise. For a proper treatment
Eq.~(\ref{eq:decay-pure-dephasing-gaussian}) must be  generalized accordingly. Initially addressed for 
 understanding charge noise in  superconducting qubits~\cite{Falci2004,Faoro2004},  
DD in non-Gaussian environments is important in other implementations since, independently on the 
microscopic origin, critical current noise and flux noise may also be described as due to a collection 
of discrete sources, and also resulting in a $1/f^\alpha$ spectrum. 
Recently this topic has attracted  large interest also for solid-state quantum hardware 
based on electron and nuclear spins~\cite{Lee2008,Witzel2007c,Witzel2007d}.
Besides determining the decay of coherences, non-Gaussian noises are responsible for additional
structure (splitting of spectroscopic peaks and beats)  observable in the qubit dynamics. This 
deteriorates the fidelity and must be washed out by stabilization.

\begin{figure}[t!]
\centerline{
\includegraphics[width=0.9\columnwidth]{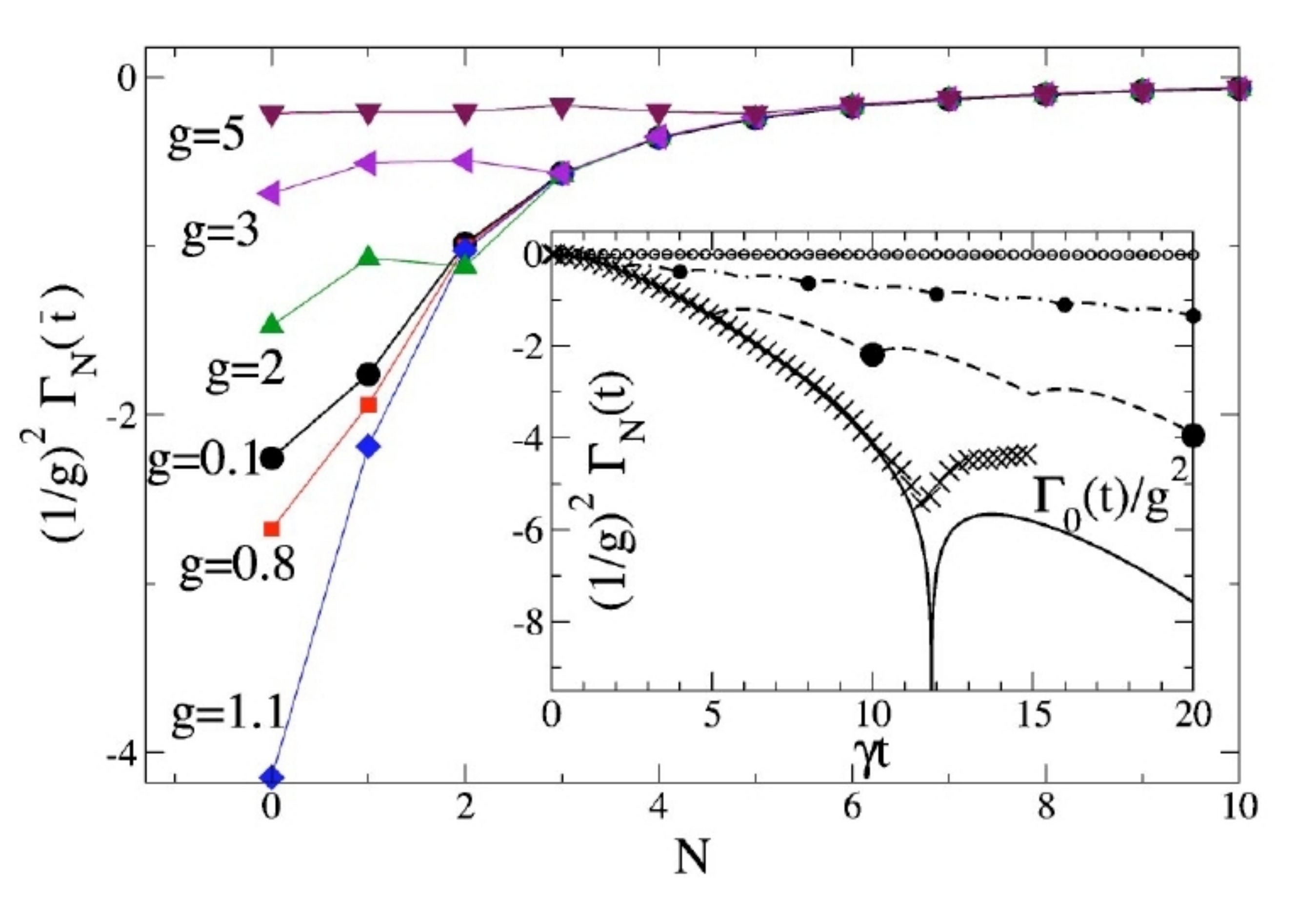}
}
\caption{(Color online) Scaled decay rate of the qubit coherence, $\Gamma_N(t)/g^2$, at fixed $t=10 \gamma^{-1}$ 
and DD (here $N$ enumerates echo pair of pulses). $N=0$ corresponds to FID. A Gaussian environment 
with the same power spectrum would give, for arbitrary $g$, the curve labeled with g=0.1, since 
$\Gamma_N(t) \propto g^2$. Inset:
$\Gamma_N(t)$ for $g =1.1$ for different intervals between pulses $\Delta t$  (lines with dots,
 $\Delta t= 5,2,0.2$) are compared with the FID $\Gamma_0(t)$ (thick line) and with results obtained 
 by numerical solution of the stochastic Schr\"odinger equation. Adapted from \cite{Falci2004}.
\label{fig:bbrtn-1}}
\end{figure}

\paragraph{RT noise  and quantum impurities --}
The simplest physical non-Gaussian environment is a single impurity coupled to the qubit. Models 
of quantum impurities were studied in~\cite{Falci2004,Lutchyn2008,Rebentrost2009},
 whereas the classical counterpart
was addressed in \cite{Faoro2004,Gutmann2005,Bergli2007,Cheng2008}.
\textcite{Falci2004} modeled the environment by an electron tunneling with switching rate $\gamma$  
from an impurity level to an electronic band~\cite{Paladino2002}. The parameter quantifying Gaussianity is 
$g= (\Omega_+ - \Omega_-)/\gamma$ (Sec. \ref{subsub:SFgeneral}).
The problem is tackled by studying the reduced dynamics of the  Qubit \textit{plus} Impurity system (QI) for arbitrary qubit bias. 
The band acts a \textit{Markovian}  environment for the QI reduced dynamics~\cite{Paladino2003} and can be 
treated exactly by a Master Equation. The key point is that  DD has no effect on a Markovian environment. 
Thus the quantum map in the presence of a number $N$ of pulses  can be written as 
$\rho_{QI}(t) = {\cal E}_{N}[\rho_{QI}(0)] = 
\{ {\cal P} \exp[{\cal L}\Delta t]\}^{N}\rho_{QI}(0)$.
Here the superoperators ${\cal L}$ and ${\cal P}$ describe, respectively, the QI reduced dynamics and 
$X_\pi$ pulses on the qubit, $\mathbbm{1}_I \otimes \sigma_x$.  
The reduced qubit  dynamics is obtained by tracing out the impurity 
$\rho_{Q}(t) =\mathrm{Tr}_I[\rho_{QI}(t)]$ 
\textit{at the end of the whole protocol}.
In this way non-Gaussianity and non-Markovianity of the impurity are accounted for exactly. 
At pure dephasing ($\Omega_x = 0$) a simple analytic form can be found.   
The different physics due to a weakly ($g<1$) or strongly coupled $g>1$ impurity is 
discussed in Sec. \ref{sec2E}. This difference is  washed out for large flipping 
rates ($N \gg \gamma t$)  where the environment becomes effectively Gaussian with  
universal behavior $\Gamma_N(t)\sim g^2$ (Fig.~\ref{fig:bbrtn-1}).
On the other hand, for  $N < \gamma t$ a crossover is clearly  observed between different domains of $g$. 
Notice that in the  intermediate regime, $N \lesssim \gamma t$, DD is still able to cancel fast noise, 
$g < 1$, and all the features of the qubit
dynamics appearing when $g \sim 1$.  A qualitatively similar behavior is found also for $\Omega_x \neq 0$ 
where the solution requires the diagonalization of ${\cal E}_{N}$. The new feature in this regime is that 
DD of slow fluctuators, $g > 1$,
 may be nonmonotonic with the  flipping rate, yielding  for  $N < \gamma t$ decoherence acceleration 
 which is reminiscent  of the anti-Zeno effect. 

Notice that this model reduces to a classical  RT fluctuator if mutual  QI back-action, described by 
frequency shifts, is dropped out. Numerical simulations in this limit
by \textcite{Gutmann2005} confirmed that decoherence is suppressed for large pulse rates $N \gg \gamma t$;
 \textcite{Gutmann2004}  addressed imperfect DD pulses and \textcite{Bergli2007} also found  analytic solutions 
 and showed that a train of $Y_\pi$ pulses
avoids decoherence acceleration.
A quantum impurity modeled by a ``rotating wave'' spin-boson model was recently studied by 
\textcite{Rebentrost2009} with the numerical  Gradient Ascent Pulse Engineering (GRAPE) algorithm.
Authors found decoherence acceleration at $t/N \approx \Omega/2\pi$ 
next to the optimal point,  and optimal pulses allowing for  relaxation-limited gates at larger pulse rates. 

\paragraph  {$1/f$ noise  --}
An environment composed by set of independent impurities can be treated along the same lines. At pure 
dephasing $\Gamma_N(t)$ is the sum of independent single impurity contributions,  and the analytic 
solution can be  found for arbitrary distribution of  parameters~\cite{Falci2004}.  An analytic expression
 valid  in the classical limit was also found by \textcite{Faoro2004} who pointed out that relatively slow DD
  control rates 
($\Delta t \sim 1/\gamma_M$ which is only a soft cutoff of the environment)
suffice for a drastic improvement.  For figures of noise typical for experiments, pulse rates
yielding recovery are insensitive to the average coupling strength of  the impurities~\cite{Falci2004}. 
The situation changes when  the distribution includes individual more strongly coupled 
impurities~\cite{Galperin2003,Paladino2003b}.  Time symmetric CP is found to perform better than  
PDD~\cite{Faoro2004}.   While no decoherence acceleration is found at pure dephasing, this may happen for 
low pulse rates ($\Omega \Delta t \sim 10$) when  $\Omega_x \neq 0$ and noise acquires a transverse 
part~\cite{Faoro2004}.

\paragraph {Robust DD  --}
Suppression of $1/f$ pure dephasing  longitudinal flux noise was demonstrated in a recent 
experiment~\cite{Bylander2011} using a 200 pulses CPMG sequence yielding a 50-fold improved $T_2$ over 
the baseline value, whereas the performance of UDD sequences was
slightly worse. Earlier work with few-pulse sequences has demonstrated partial suppression of low-frequency
 transverse charge  noise~\cite{Ithier2005}. 

Referring to superconducting qubits, \textcite{Cywinski2008}  studied CMPG, UDD and CDD for $1/f^\alpha$ 
($0.5\le \alpha \le 1.5$) Gaussian classical noise at pure dephasing  with UV-cutoff $\omega_c$.  For 
pulse rates larger than this cutoff, CPMG is the most effective sequences increasing $T_2$,  UDD keeping however higher 
fidelity. Instead, CDD does not give
relevant improvement even if it outperforms PDD for a wide range of  parameters~\cite{ka:207-hkodjasteh-pra-CDD}.
 For pulse rates smaller than $\omega_c$, 
 CPMG slightly outperforms all other sequences.  CPMG is also a better approach for strongly coupled 
RT noise, non-Gaussian features being suppressed in the large pulse rate limit. Similar conclusions were 
drawn by~\textcite{Lutchyn2008} for a quantum impurity
environment of Andreev fluctuators. \textcite{ka:210-pasiniuhrig-pra-powernoise} studied
sequences optimized for  specific power-law noise spectra  and found that they approach CPMG for soft 
UV-cutoff ($1/f$ noise) whereas for hard UV behavior (Ohmic) UDD is the limiting solution.

Bounded amplitude ``dynamical control by modulation"  was proposed by \textcite{Gordon2008} who studied 
optimization 
for Lorentzian and $1/f$ pure dephasing noise. 
A practical limitation of this optimal chirped  modulation is the sensitivity to the low-frequency cutoff.
 Design of GRAPE-optimized quantum gates in  the presence of $1/f$ noise and inhomogeneous dephasing was 
  recently investigated by \textcite{PhysRevA.86.012317}. 

Notice that in general solid state  nanodevices suffer from different noise sources with 
multi-axis structure of couplings. In these cases CDD~\cite{ka:205-hkodjasteh-prl-CDD} or concatenated UDD 
 sequences~\cite{ka:210-westfonglidar-prl-optDD,ka:209-uhrig-prl-CUDD} may give substantial advantages. 

\subsubsection{Spectroscopy}
The possibility that DD could be used as a spectroscopic tool was raised in a number of early 
works~\cite{Faoro2004,Falci2004}  and has been formalized using  the concept of filter 
function~\cite{ka:207-uhrig-prl-UDD,ka:211-biercuk-jpb-ddfilter}.
The key observation is that for a Gaussian process, the filter function in 
Eq.~(\ref{eq:decay-pure-dephasing-gaussian}) can be interpreted, at a fixed time $\bar{t}$,  as a 
linear filter~\cite{ka:211-biercuk-jpb-ddfilter},   transforming the input phase noise $E(t)$ to the 
output phase  $\Phi_N(\bar{t})$, yielding the decay function  $\Gamma_N(\bar{t})$ after noise averaging.
Each implementation of time-dependent control samples the noise
in a distinctive way determining  the form of the filter  $F_N(\omega \bar{t})$. 
For suitable sequences, we define  a filter frequency $\omega_{F1}$ such that 
$F(\omega_{F1}\bar{t}) \sim 1$, which roughly corresponds to the minimal inter-pulse $\Delta t$.
DD is described by a filter with negative gain for $\omega < \omega_{F1}$,  the steepness of the attenuation 
yielding a measure of the effectiveness of the given sequence.  This analysis allows  to develop a
filter-function-guided pulse design  suited to a particular noise spectrum.  

\begin{figure}[t]
\centerline{
\includegraphics[width=0.8\columnwidth]{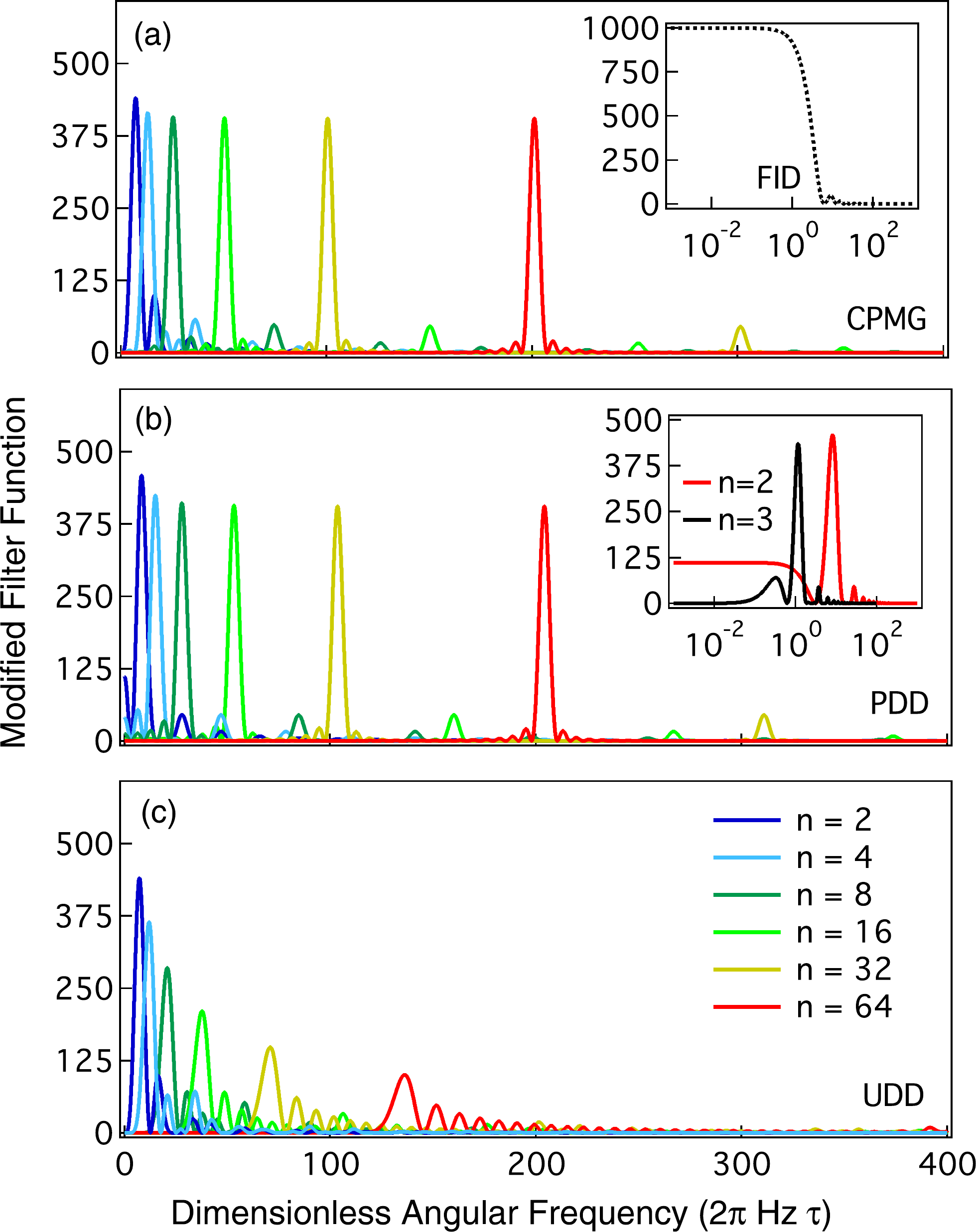}
}
\caption{(Color online) (a) - (c) Modified filter function $F_N(\omega
\bar{t})/\omega^2$, for the indicated pulse
sequences and $N$ values.
Dominant spectral contributions to the measured $\Gamma_N(t)$
appear as peaks
in the modified filter functions. (a) Modified filter function
for FID
showing large weight for low-frequency noise on a semilog scale,with arbitrary units. (b)
Demonstration of even-odd parity
through
the modified filter function of PDD on a semilog scale.
Adapted from \cite{ka:211-biercuk-jpb-ddfilter}.
\label{fig:filter-biercuk}
}
\end{figure}

Application to spectroscopy emerges  from the observation that, in addition to the decoupling regime, 
there exist spectral regions of positive gain about $\omega = \omega_{F}$,
where the effect of the corresponding spectral components of
noise is amplified. This is apparent from Fig.~\ref{fig:filter-biercuk} where the modified filter 
function $F(\omega \bar{t})/\omega^2$ is plotted, indicating the dominant spectral contributions to 
decoherence. For CPMG and PDD there is a single dominant peak. The shift of the peak towards larger 
$\omega$ for increasing $N$ indicates that effects of
sub-Ohmic noise are reduced by DD. 
\textcite{Cywinski2008} proposed that UDD spectroscopy   may also give informations on higher moments 
of noise via  the additional structure of the filter function
(Fig.~\ref{fig:filter-biercuk}).

In \cite{PhysRevLett.107.170504} a method for obtaining the noise spectrum from experimental data has 
been  proposed. The method, valid for the case of pure dephasing,  is based on the relationship 
between the spectrum and a generalized dephasing time, evaluated from the asymptotic
exponential decay in the presence of a sufficiently large number of $\pi$-pulses.

Recently \textcite{Bylander2011} have exploited the narrow-band filtering properties of CPMG to measure 
$1/f$ flux noise  in a persistent current qubit,   where flux noise is the main source of dephasing 
away from the optimal point, $\Omega_z\neq 0$.
Indeed, Eq.~(\ref{eq:decay-pure-dephasing-gaussian}) is approximated as
$\Gamma_N(\bar{t}) \propto \Delta \omega\, (\partial \Omega/\partial q)^2
S_q(\omega_{F})  F_N(\omega_{F} \bar{t})/ \omega_{F}^2$, where 
$\omega_{F} = \pi/2\Delta t$ is the peak frequency, $ \Delta \omega$ is the bandwidth of the filter, 
$S_q$  and $\partial \Omega/\partial q$  are related to the power spectrum of flux noise and to the  
qubit sensitivity, depending on the flux bias, Eq.~(\ref{sensitivities}).
This method has allowed to access the unexplored spectral region $0.2$-$20$~MHz where
$1/f^{0.9}$ noise was detected.  Decay of Rabi oscillations provides an alternative tool for
environment spectroscopy in the same frequency range~\cite{Bylander2011}. 
As explained in Sec.~\ref{Inhom-Rabi},  quasistatic noise is efficiently averaged out and the observed 
decay of Rabi oscillations is essentially exponential with a contribution from frequency components 
around $\Omega_R$ behaving as
$\Gamma_{R}  \propto [\Omega_z(q)/\Omega]^2\, S_q(\Omega_R)$  \cite{Ithier2005,ka:195-gevakoslov-jchph-genME}. 
 Extracting this contribution yields an independent information on the power spectrum at 
 $\omega = \Omega_R \sim \mathrm{MHz}$.  Remarkably, the very same power laws have been measured at much 
 lower frequencies  ($0.01-100\,\mathrm{Hz}$) by a direct method using   the noise sensitivity of a 
 free-induction Ramsey interference experiment \cite{Yan2012}. The peculiarity of this technique is that 
 it enables
measurements of noise spectra up to frequencies limited only by achievable  repetition rates of the
measurements.

\paragraph*{Two-qubits  --}
The extension of DD to entanglement protection from $1/f$ noise is a relevant issue, currently under 
investigation both theoretically  and experimentally. The first experimental demonstration of DD 
protection of pseudo two-qubit entangled
states of an electron-nucleus ensemble in a solid-state environment has been reported in \textcite{Wang2011}. 
DD control pulses operated on the electron spin suppressed inhomogeneous dephasing due to the static 
Overhauser field induced by the hyperfine interaction between electron spin and surrounding nuclear 
spins in a P:Si material.  Recently, in \textcite{Gustavsson2012}
single-qubit refocusing techniques have been extended to enhance the lifetime of an entangled state of 
a superconducting flux qubit coupled to a coherent TLS. Fluctuations of the qubit splitting due to 
$1/f$ flux noise   induce low frequency fluctuations of the qubit-TLS effective interaction. Authors demonstrated that rapidly
 changing the qubit's transition frequency relative to TLS, a refocusing pulse on the qubit 
improved the coherence time of the entangled state.
Further enhancement was demonstrated when applying multiple refocusing pulses. These results 
highlight the potential of DD techniques for improving two-qubit gate fidelities, an essential prerequisite 
for implementing fault-tolerant quantum computing. Quantum optimal control theory represents an alternative 
possibility to design high-fidelity quantum gates. \textcite{Montangero2007}, using the GRAPE numerical
algorithm,  demonstrated a stabilized two charge-qubits gate robust also to $1/f$ noise.
For realistic noise figures, errors of $10^{-3} - 10^{-4}$, crossing the fault tolerance threshold,
have been reached. A high-fidelity $\sqrt{\mathrm{SWAP}}$ has been studied 
in~\cite{PhysRevA.86.012317} using GRAPE optimization in the presence of $1/f$ noise.

%% file: conclusions.tex
\section{CONCLUSIONS AND PERSPECTIVES}
\label{sec:conclusions}

In this review we discussed the current state of theoretical work on $1/f$ noise in nanodevices
with emphasis on implications for solid state quantum information processing. Our focus was on 
superconducting systems and we referred to implementations  based on semiconductors only when
physical analogies and/or formal similarities were envisaged.
According to the existing literature, relevant mechanisms responsible for $1/f$ noise in 
superconducting nanocircuits have been largely identified. However, in many solid state nanodevices 
this problem  cannot be considered as totally settled and details of the interaction mechanisms
remain controversial (see Section \ref{sec:origin}). In some cases, available 
experiments do not allow drawing solid conclusions and further investigation is needed.

We have discussed the role of low-frequency noises in decoherence of quantum bits and gates. 
Various methods to address this problem have been presented.
A relevant issue in connection with quantum computation in the
solid state is decoherence control and the achievement of the high 
fidelities needed for the succesfull application of error correction codes. 
Various proposals have been put forward to limit the effects of 1/f noise. .  
Since such noises in solid-state devices are created by material-inherent sources, an
obvious way to improve performance of quantum devices is optimizing materials used 
for their fabrication.  In particular, it is very important to engineer ``dielectric" 
part of devices.  Enormous effort in this direction based on identification of the noise 
sources and properties has resulted in significant optimizing of existing devices and 
suggesting novel ones. The main focus of the review is relating the device performance  
along different protocols to the properties of the noise sources.  
We hope that understanding these relations may lead to improvement of the quantum devices. 

The integration of control tools, like dynamical decoupling sequences 
appropriate to 1/f noise, with other functionalities, such as quantum gates, 
in a scalable architecture is a non-trivial open problem.
Application of optimized pulse sequences to non-Markovian noise is subject 
of future investigation.
We reviewed the current status of the ongoing research along this 
direction, which represents an area for future development of the field.

Despite of the tremendous progress in this field, there is still a long way to go until a practically 
important quantum computer will be realized.  Many details have to be worked out, and at 
present time it is actually not clear which physical implementation of quantum devices 
and even which architecture will be the most advantageous.  However, it is fully clear that
ongoing research on design and studies of devices for quantum information processing will 
significantly improve our understanding of the quantum world and of interplay between 
classical and quantum physics.  It will certainly lead to significant development of 
modern mesoscopic physics and, in particular, of quantum electronics.

%% file: acknowledgments.tex
\acknowledgments
We thank Antonio D'Arrigo for careful reading of the manuscript
and suggestions to improve the paper.
We acknowledge very fruitful discussions with J. Bergli, R. Fazio, 
A. Mastellone, M. Sassetti, G. Sch\"on,  D. Vion, U. Weiss, A. Zorin. 
E. P. and G. F. acknowledge partial support from the
Centro Siciliano di Fisica Nucleare e Struttura della Materia,
Catania (I) and  by the European Community through grant 
no. ITN-2008-234970 NANOCTM 
and by PON02-00355-339123 - ENERGETIC.

\vspace{0.2cm}
\noindent E. P. and Y. M. G. contributed equally to the present work.

%% file: abbreviations.tex
\appendix
\section{ABBREVIATIONS}
\noindent
BCS -- Bardeen, Cooper, Schrieffer \\
CDD -- Concatenated Dynamical Decoupling\\
CPB -- Cooper Pair Box \\
CP -- Carr Purcell\\
CPMG -- Carr Purcell Meilboom Gill\\
CTRW -- Continuous Time Random Walk\\
CVD -- Chemical Vapor Deposition \\
DD -- Dynamical Decoupling\\
DQD -- Double  Quantum Dot\\
FID -- Free Induction Decay \\
GRAPE -- GRadient Ascent Pulse Engineering \\
LZ -- Landau-Zener \\
MRT - Macroscopic Resonant Tunneling \\
NMR -- Nuclear Magnetic Resonance \\
PDD -- Periodic Dynamical Decoupling\\
QED -- Quantum Electrodynamics \\
QI -- Qubit plus Impurity\\
QD -- Quantum Dot\\
QPC -- Quantum Point Contact \\
RKKY -- Ruderman-Kittel-Kasuya-Iosida\\
RT --Random Telegraph  \\
RTN -- Random Telegraph Noise\\
SEM -- Scanning Electron Microscopy\\
SET -- Single Electron Tunneling \\
SO -- Spin - Orbit \\
SQUID -- Superconducting Quantum Interference Device \\
TLS -- Two-Level System \\
UDD -- Uhrig Dynamical Decoupling\\
UV -- Ultra-Violet